\documentclass[a4paper,12pt,openany,twoside,notitlepage,english]{book}
\usepackage{geometry} 
\usepackage{amsmath,amssymb}
\usepackage{booktabs} 
\usepackage{bm}
\usepackage{ascmac,here}
\usepackage{braket} 
\usepackage{boites,boites_exemples}
\usepackage{fancybox}
\usepackage{amsfonts}
\usepackage{graphicx}
\usepackage{comment}
\usepackage{amsthm}
\usepackage{feynmp} 
\usepackage{setspace} 
\if{
\usepackage[uplatex,deluxe]{otf} 
\usepackage[prefernoncjk]{pxcjkcat}
\usepackage[T1]{fontenc} 
\usepackage[utf8]{inputenc}
}\fi
\usepackage{multirow}
\usepackage{cases}
\usepackage{url}
\makeatletter
\@addtoreset{footnote}{page}
\makeatother
\theoremstyle{definition}

\newtheorem*{theorem*}{Theorem}

\newtheorem*{definition*}{Definition}

\newcommand{\av}[1]{\overline{#1}}
\newcommand{\mc}[1]{\mathcal{#1}}
\newcommand{\mr}[1]{\mathrm{#1}}
\newcommand{\mbb}[1]{\mathbb{#1}}
\newcommand{\mbf}[1]{\mathbf{#1}}

\newcommand{\lrs}[1]{\left( #1 \right)}
\newcommand{\lrm}[1]{\left\{ #1 \right\}}
\newcommand{\lrl}[1]{\left[ #1 \right]}
\newcommand{\lrv}[1]{\left| #1 \right|}

\newcommand{\fracd}[2]{\frac{\mathrm{d} #1 }{\mathrm{d} #2 }}

\newcommand{\fracpd}[2]{\frac{\partial #1 }{\partial #2 }}

\newcommand{\aln}[1]{
\begin{align}
#1
\end{align}
}

\newcommand{\ra}{\rightarrow}

\newcommand{\Tr}{\mr{Tr}}
\newcommand{\hmo}{\hat{\mc{O}}}
\newcommand{\hmp}{\hat{\mc{P}}}
\newcommand{\hrho}{\hat{\rho}}
\newcommand{\hH}{\hat{H}}
\newcommand{\hT}{\hat{T}}
\newcommand{\hO}{\hat{O}}
\newcommand{\mooD}{\mc{O}_{\alpha\beta}}
\newcommand{\moD}{\mc{O}_{\alpha\alpha}}
\newcommand{\EQ}[1]{Eq. (\ref{#1})}
\newcommand{\CH}[1]{Chapter \ref{#1}}
\newcommand{\SEC}[1]{Sec. \ref{#1}}
\newcommand{\NON}{\nonumber \\}

\newcommand{\yabai}{@@@@@@@@@@@@
@@@@@@@@@@@@
@@@@@@@@@@@@
@@@@@@@@@@@@@@@@@
@@@@@@@@@@@@@@
@@@@@@@@@@@@@@@@@@@@@@
@@@@@@@@@@@@@@@@@@@@@@@@@
@@@@@@@@@@@@@@@@@
@@@@@@@@@@@@@@@@@@@
@@@@@@@@@@@@@@@@@@@@@@@@@@
@@@@@@@@@@@@@@@@@@@@@
@@@@@@@@@@@@@@@@@@
@@@@@@@@@@@@@@@@@@@@@@@@@@@@@
@@@@@@@@@@@@@@@@@@@@@@@@
@@@@@@@@@@@@@@@@@@@@@@@@
@@@@@@@@@@@@@@@@@@@
@@@@@@@@@@@@@@@@@@@@@
@@@@@@@@@@@@@@@@@@@@@@
@@@@@@@@@@@@@@@@@@
@@@@@@@@@@@@@@@@@@@@@@@@
@@@@@@@@@@@@@@@@@@@@@
@@@@@@@@@@@@@@@@@@@@@@
@@@@@@@@@@@@@@@@@@@@@@@@@@@
@@@@@@@@@@@@@@@
@@@@@@@@@@@@@@@@@@@@@@@
@@@@@@@@@@@@@@@@@@
@@@@@@@@@@@@@@@@@@@@@@
@@@@@@@@@@@@@@@@@@@@
@@@@@@@@@@@@@@@@
@@@@@@@@@@@@@@@@@@@@
@@@@@@@@@@@@@@@@@@@@@@@@@@@@
@@@@@@@@@@@@@@@@@
@@@@@@@@@@@@@@@@@@@@
@@@@@@@@@@@@@@@@@@@@@@@@@@@@@
@@@@@@@@@@@@@@@@@@@
@@@@@@@@@@@@@@@@@@@@
@@@@@@@@@@@@@@@@@@
@@@@@@@@@@@@@@@@@@@@@
@@@@@@@@@@@@@@@@@@@@@@
@@@@@@@@@@@@@@@@@@@@@@@@@
@@@@@@@@@@@@@@@@@@@
@@@@@@@@@@@@@@@@@@@@
@@@@@@@@@@@@@@@
@@@@@@@@@@@@@@@@@@@@@@@@@@@}

\renewcommand{\include}[1]{}
\begin{document}
\setlength\abovedisplayskip{6.5pt}
\setlength\belowdisplayskip{6.5pt}
\frontmatter
\begin{figure}[H]
\centering
\includegraphics[width=\linewidth]{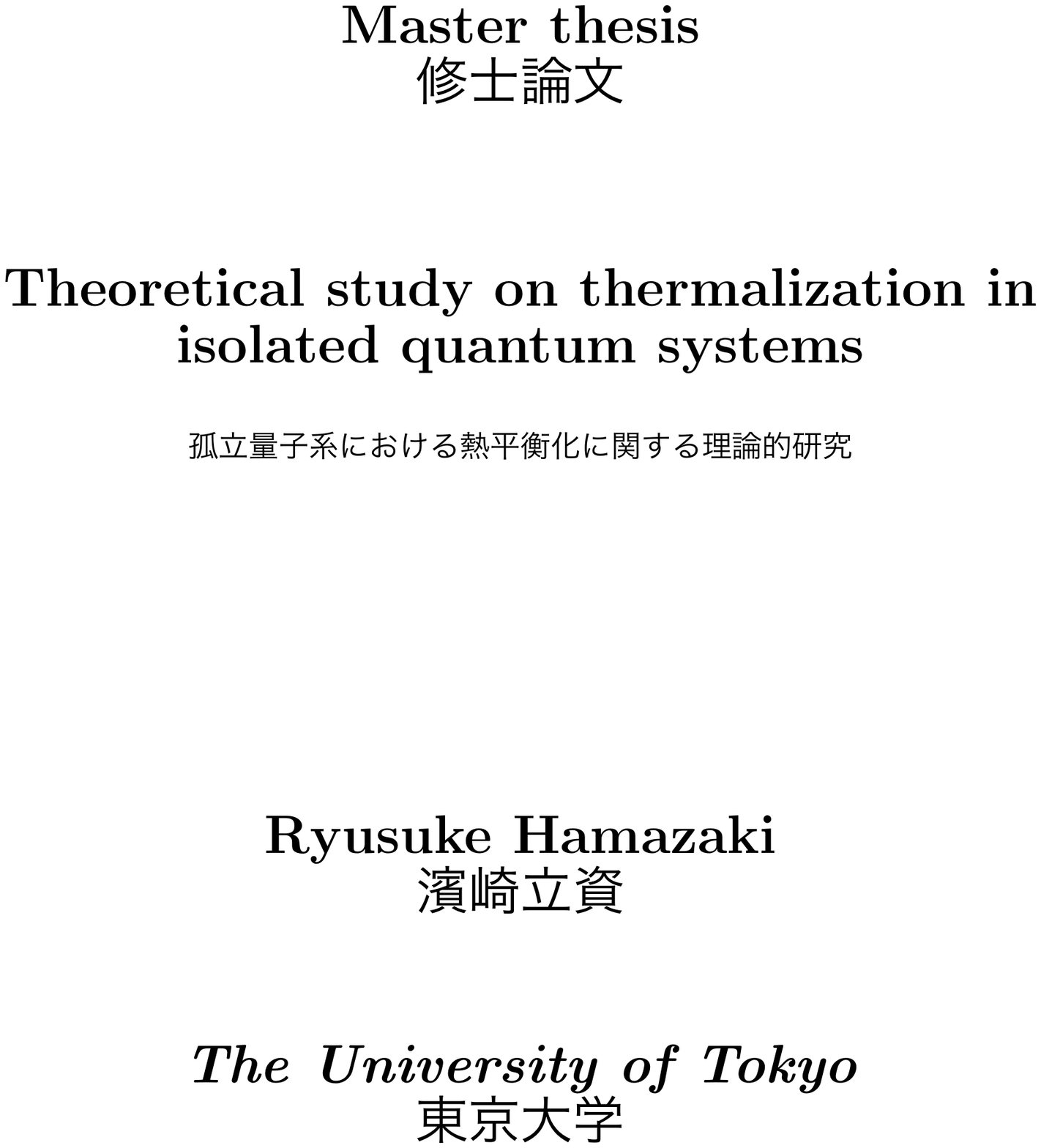}
\end{figure}
%
\addcontentsline{toc}{chapter}{Abstract}

\section*{Abstract}
Understanding how isolated quantum systems thermalize has recently gathered renewed interest almost 100 years after the first work by von Neumann, thanks to the experimental realizations of such systems.
Experimental and numerical pieces of evidence imply that nonintegrability of the system plays an important role in thermalization.
Nonintegrable systems that conserve energy alone are expected to be effectively described by the (micro)canonical ensemble due to the so-called eigenstate thermalization hypothesis (ETH) in the thermodynamic limit.
In contrast, it is expected that stationary states in integrable systems are described not by the canonical ensemble but by the generalized Gibbs ensemble (GGE) due to the existence of many nontrivial conserved quantities.
In this thesis, we study thermalization and its mechanism in nonintegrable systems from two perspectives.

First, we study how well the ETH and its finite-size corrections can be predicted by random matrix theory (RMT).
We first analytically calculate the finite-size corrections of the ETH using the RMT model and show that their statistics is universal, depending only on the symmetry (and what is called singularity) of the observables as well as the symmetry of the Hamiltonians. 
Then, we numerically show that, in nonintegrable systems (that conserve energy alone) and for a wide class of observables, the matrix elements are in good agreement with the prediction of RMT.
We also remark, however, that counterexamples of the RMT prediction always exist even among simple observables.

Next, we present our study on the emergence of the GGE in a nonintegrable system with an extensive number of local symmetries.
We have numerically investigated a nonintegrable model of hard-core bosons with an extensive number of local $\mbb{Z}_2$ symmetries. 
We find that the expectation values of macroscopic observables in the stationary state are described by the GGE and not by the canonical ensemble. 
We also show that, if the model has a less than extensive number of local symmetries, the stationary state is described by the canonical ensemble.

\clearpage

%
\tableofcontents
\mainmatter


\chapter{Statistical physics of isolated quantum systems: an overview}\label{ch:Stat}
\section{Foundations of quantum statistical mechanics and the notion of typicality}
Statistical mechanics, originally developed by Boltzmann, Gibbs and Einstein, has long been an indispensable tool for diverse areas of physics, ranging from cosmology to biology.
It tells us how to compute macroscopic variables and their fluctuations in thermodynamics from the knowledge of microscopic theories~\cite{Landau80,Callen06}.
Without explicit knowledge about the complex dynamics, we can calculate pressures or entropies of gases,  
once microscopic Hamiltonians of the atoms or molecules are given.

One of the most important assumptions in statistical mechanics is that macroscopic observables at thermal equilibrium can be computed using the microcanonical ensemble.
The microcanonical ensemble is a probabilistic model where each microstate is taken from a certain energy shell with a uniform probability distribution.
For a classical system, the microcanonical distribution function features a uniform density $\rho_\mr{mic}(\mbf{q},\mbf{p})$ over an energy shell
\aln{
\Gamma=\lrm{(\mbf{q},\mbf{p})\in\Gamma:E-\Delta E<H(\mbf{q},\mbf{p})\leq E+\Delta E}\:,
}
where $\Delta E$ is some small energy width.
For a quantum system, the microcanonical ensemble can be represented in the form of the following density matrix:
\aln{
\hrho_\mr{mic}(E):=\frac{1}{\dim[\mc{H}_{E,\Delta E}]}\hmp_{E,\Delta E}\:,
}
where 
$\mc{H}_{E,\Delta E}$ is a Hilbert space that is spanned by the set of eigenstates in an energy shell 
\aln{
\lrm{\ket{E_\alpha}:\hat{H}\ket{E_\alpha}=E_\alpha\ket{E_\alpha}, E-\Delta E<E_\alpha\leq E+\Delta E}
}
and 
\aln{
\hmp_{E,\Delta E}:=\sum_{\ket{E_\alpha}\in \mc{H}_{E,\Delta E}}\ket{E_\alpha}\bra{E_\alpha}
}
is a projection operator onto $\mc{H}_{E,\Delta E}$.
In this case, statistical mechanics tells us to calculate the expectation value of an observable $\hmo$ at thermal equilibrium as $\braket{\hmo}_\mr{mic}=\Tr[\hrho_\mr{mic}\hmo]$.
This equal \textit{a priori} probability postulate and the renowned Boltzmann's entropy formula, $S=k_B\log W$ (where $W=\dim[\mc{H}_{E,\Delta E}]$ for a quantum system), can be regarded as two fundamental assumptions of statistical mechanics~\cite{Shimizu}.

Although enormous previous studies have confirmed that statistical mechanics with the equal \textit{a priori} probability postulate successfully predicts many thermal equilibrium physical phenomena quite accurately, the complete justification of this principle has not yet been made.
The justification of the equal \textit{a priori} probability postulate boils down to the following two questions:
\begin{enumerate}
\item
What is the meaning of thermal equilibrium from a microscopic viewpoint? How is the microcanonical ensemble related to an actual microstate?
\item
Why does thermal equilibrium emerge as a macroscopically stationary state? Can we prove it only by assuming microscopic kinetics, even without considering any thermal bath?
\end{enumerate}
In fact, these questions were already actively investigated in the first half of the 20th century~\cite{Haar55} (note that Boltzmann himself proposed the H theorem in an attempt to solve the second question).
For isolated classical systems, the second question had been mainly studied in light of the ergodic theorem, which states that the long-time average of a physical quantity of a time-evolving microstate is equal to its phase-space average.
We note that this theorem itself is now considered as being irrelevant to the foundation of statistical mechanics for several reasons~\cite{Shimizu,田崎08}.
For isolated quantum systems, von Neumann tried to solve both of these questions in 1929~\cite{Neumann29}, which was just three years after the discovery of the Schr\"odinger equation.
Although von Neumann's discussion is worth notice even from the modern perspective, it had been forgotten until 2010 (see Chapter \ref{ch:Eq}).

As an answer to the first question, the notion of ``typicality" of thermal equilibrium has recently become popular~\cite{杉田06,Goldstein06,Popescu06,Reimann07,Sugiura12,Sugiura13,Goldstein15,Tasaki16,Goldstein16}.
The main idea of the typicality argument is that, if we can measure only a proper set $\mc{A}$ of restricted observables, almost all microstates in the energy shell are indistinguishable from the microcanonical ensemble.
In other words, under the assumption of the typicality, most of the pure states can describe thermal equilibrium if we are interested in only observables in $\mc{A}$.
As an example, consider a box containing $N\gg 1$ classical particles.
If we measure the number of particles in one half of the box, it is almost $N/2$ for most of the microstates, which is consistent with the prediction of the microcanonical distribution.

While in classical systems we may have to take a set of macroscopic observables for $\mc{A}$ in order to justify typicality, in quantum systems we are allowed to take a larger set of observables~\cite{Goldstein15,Goldstein16}.
In fact, we can rigorously show that most of the quantum pure states give the same expectation value of a  general few-body operator as the microcanonical ensemble (see Chapter \ref{ch:Eq} for details)~\cite{杉田06}. 
This difference implies that the applicability of quantum statistical mechanics is far wider than that of classical statistics mechanics.\footnote{Let us illustrate this with a simple example of $N$ (distinguishable) particles. For most of the microstates in the energy shell, the single-particle distribution of velocity is expected to obey the Maxwell-Boltzmann distribution if we make a histogram using $N$ particles. This is true both for classical and quantum cases, since we can write the single-particle distribution obtained from $N$ particles as macroscopic observables~\cite{Shimizu}. On the other hand, if we measure the velocity of a single particle (which is not a macroscopic observable), each microstate gives a definite value in the classical case and the Maxwell-Boltzmann distribution cannot be obtained from a single microstate. However, in the quantum case, the measurement results change because of quantum fluctuations, which allow us to obtain the Maxwell-Boltzmann distribution even from a single microstate~\cite{Srednicki94}.}
Due to this fact, we mainly focus on quantum systems throughout this thesis.

\begin{figure}
\begin{center}
\includegraphics[width=15cm]{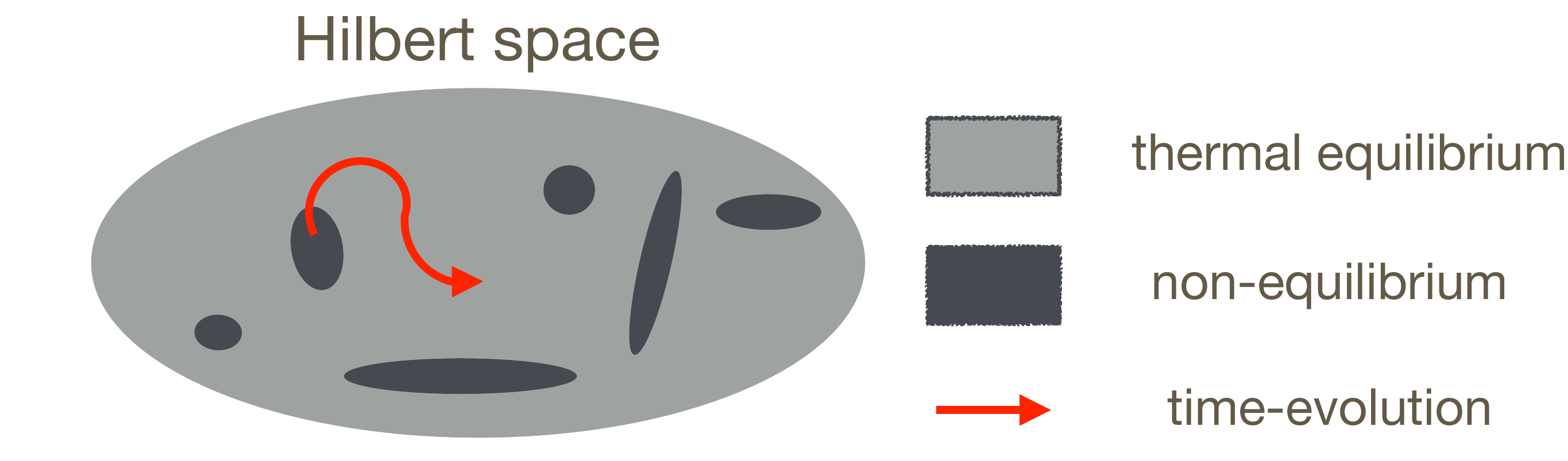}
\caption{Schematic illustration of typicality of thermal equilibrium and its relation to time evolution. The elliptic region shows the entire set of microstates in the Hilbert space. Under the typicality, most of the microstates are in thermal equilibrium (gray regions), and non-equilibrium states (black regions) occupy only a small fraction of it. Since thermal equilibrium covers most of the areas in the Hilbert space, a non-equilibrium initial state may rapidly develop into the thermal equilibrium by a unitary time evolution.}
\label{fig:typicality}
\end{center}
\end{figure}

Now, let us consider how the typicality is related to the second question, namely the approach to thermal equilibrium~\cite{Tasaki16}.
If the typicality holds true, most of the microstates in the Hilbert space are in thermal equilibrium, and non-equilibrium states are rare (see Fig. \ref{fig:typicality}).
From the figure, we expect that even if we prepare a non-equilibrium state as an initial state, it may rapidly develop into a thermal equilibrium by a unitary time evolution.

Though this argument of typicality seems natural, it is not quantitative enough.
Actually, as we will see later, it is known both theoretically and experimentally that some systems cannot reach thermal equilibrium by a unitary time evolution even after an infinite time.
Note that typicality of thermal equilibrium holds true in such systems, too.
In order to know whether some initial states approach thermal equilibrium, we should straightforwardly begin with unitary time evolutions.

\section{Approach to thermal equilibrium}
\subsection{Equilibration and thermalization}

In this section we briefly explain how to formulate the approach to thermal equilibrium starting from unitary time evolutions (the details are discussed in Chapter \ref{ch:Eq}).
Let us consider an initial state $\hrho_0$.
Under a unitary time evolution, the state develops into
\aln{
\hrho(t):=e^{-i\hH t}\hrho_0 e^{i\hH t}=\sum_{\alpha\beta}e^{i(E_\alpha-E_\beta)t}\braket{E_\beta|\hrho_0|E_\alpha}\ket{E_\beta}\bra{E_\alpha}
}
at time $t$ (note that we set $\hbar=1$ throughout this thesis).
We can immediately see that, as a microstate itself, $\hrho(t)$ will not be equivalent to a stationary (time-independent) microstate, much less a microcanonical ensemble $\hrho_\mr{mic}$.
In order to discuss if the system reaches thermal equilibrium, we have to restrict the set of observables $\mc{A}$, as discussed in the previous section.

Here, we consider the approach to thermal equilibrium by dividing the problem into the following two steps: 
\begin{enumerate}
\item
Why does the state look stationary after a certain time? In other words, we want to know if $\braket{\hmo}(t):=\Tr[\hrho(t)\hmo]$ is almost equal to $\braket{\hmo}_\mr{stat}:=\Tr[\hrho_\mr{stat}\hmo]$ for some stationary state $\hrho_\mr{stat}$ and $\hmo\in\mc{A}$. 
We will call this the problem of equilibration.

\item
Under the assumption of equilibration, can we justify the use of the microcanonical ensemble?
In other words, we want to know if $\braket{\hmo}_\mr{stat}:=\Tr[\hrho_\mr{stat}\hmo]$ is nearly equal to $\braket{\hmo}_\mr{mic}:=\Tr[\hrho_\mr{mic}\hmo]$.

\end{enumerate}
If we can show both of them, we will say that the system approaches thermal equilibrium. We will call this the problem of thermalization.

At first sight, it seems that the information about the initial state $\hrho_0$ is important to answer these questions.
In fact, by imposing certain conditions on $\hrho_0$, we can show that the system equilibrates~\cite{Tasaki98,Reimann08,Linden09,Short11,Short12} or thermalizes~\cite{Mueller15,Tasaki16,Mori16M}.
We will briefly discuss the recent development about these conditions in Section \ref{sec:role}.

However, for a sufficiently large quantum system, it is known that the approach to thermal equilibrium occurs for \textit{any} initial states, if the so-called eigenstate thermalization hypothesis (ETH) is satisfied~\cite{Neumann29,Jensen85,Deutsch91,Srednicki94,Tasaki98,Rigol08}.
The ETH focuses on matrix elements of an observable $\hmo$ with respect to the energy eigenstates $\ket{E_\alpha}$ in an energy shell $\mc{H}_{E,\Delta E}$.
Roughly speaking, the ETH states that, for any $\ket{E_\alpha},\ket{E_\beta}\in\mc{H}_{E,\Delta E}$, and in the thermodynamic limit, (a) off-diagonal matrix elements satisfy $\braket{E_\alpha|\hmo|E_\beta}\simeq 0\:(\alpha\neq\beta)$, and (b) diagonal matrix elements satisfy $\braket{E_\alpha|\hmo|E_\alpha}\simeq \braket{\hmo}_\mr{mic}(E_\alpha)$.
The first condition is related to the problem on the equilibration, and the second condition corresponds to the other problem.

The ETH is actively investigated recently, based mainly on numerics, because rigorous proofs are highly nontrivial for general systems and observables.
A number of numerical studies suggest that the ETH holds true for few-body observables in nonintegrable systems that conserve only energy~\cite{Rigol08,Rigol09Q,Santos10b,Ikeda11,Steinigeweg13,Beugeling14,Kim14,Beugeling15,Fratus15}, and that the ETH breaks down in integrable systems~\cite{Essler16Q,Vidmar16} or systems in many-body localized phases~\cite{Nandkishore14,Vasseur16,Parameswaran16} (see the next subsection).
Analytically, it is shown that the ETH holds true for most of the observables~\cite{Neumann29,Reimann15} (which may not be relevant to physical observables).
In addition, relations to random matrix theory (RMT) are also proposed for nonintegrable systems~\cite{DAlessio16}.
Details of the ETH and how it is related to thermalization are explained in Chapter \ref{ch:Eq} and Chapter \ref{ch:ETH}.

\subsection{Systems that approach non-thermal stationary states}\label{sec:dameda}
Although nonintegrable systems that conserve only energy are expected to approach thermal equilibrium, some systems are known to approach non-thermal stationary states.
Integrable systems and systems in many-body localized (MBL) phases are two such famous examples, as confirmed both theoretically~\cite{Rigol07,Rigol09,Pal10} and experimentally~\cite{Kinoshita06,Schreiber15}.
In those systems, the usual ETH does not hold true in general, unlike in ordinary nonintegrable systems.

For nonequilibrium dynamics of integrable systems, models that are mappable to free systems~\cite{Rigol07,Kollar08,Rigol09,Cassidy11,Calabrese11,Cazalilla12,Fagotti13,Langen15} or solvable by the Bethe ansatz~\cite{Mossel12,Caux12,Ikeda13,Kormos13,Mussardo13,Wouters14,Essler15,Alba15,Ilievski15} have intensively been investigated.
Every energy eigenstate of these systems can be determined by the set of $N$ quantum numbers, such as quasi-momentum occupation numbers or rapidities ($N$ is the size of the system).
Such peculiarity of eigenstates is one of the properties of integrable systems, though the notion of quantum integrability is rather ambiguous~\cite{Caux11}.
In such integrable systems, there exist an extensive number of local conserved quantities, which prevent the system from approaching thermal equilibrium.
Instead, stationary states are expected to be well described by the so-called generalized Gibbs ensemble (GGE)~\cite{Jaynes57,Jaynes57II,Rigol07,Rigol09}, which takes initial values of the conserved quantities into account.
We will review the integrable systems and the GGE in detail in Chapter \ref{ch:GGE}.

Many-body localized (MBL) systems are quantum interacting systems where energy eigenstates are localized in space, triggered by (effective) disorder~\cite{Basko06,Oganesyan07,Zunidaric08,Pal10,Gogolin11,Iyer13,Vosk13,Serbyn13,Huse14,Ponte15,Imbrie16D,Imbrie16O}.
Using perturbation theory, it was studied by Basko, Altschler, and Aleiner~\cite{Basko06}, who argued that the Anderson localization, which occurs in free systems with disordered potentials,  survives even if electrons have sufficiently weak interactions.
A few years later, Pal and Huse~\cite{Pal10} numerically showed that the MBL occurs in an interacting Heisenberg chain with strong disordered transverse fields:
\aln{\label{hei}
\hat{H}=\sum_{i=1}^Nh_i\hat{\sigma}_i^z+\sum_{i=1}^{N-1}J\hat{\vec{\sigma}}_i\cdot\hat{\vec{\sigma}}_{i+1}\:,
}
where $h_i$'s are random in $i$ and $\hat{\sigma}$ is a Pauli operator.
They also argued by the level-statistics analysis that delocalization-localization quantum phase transitions occur by changing the disorder strength, even at finite temperature.
Such a property of phase transitions, or its critical properties, are still one of the open questions concerning the MBL~\cite{Kjall14,Agarwal15,Potter15}.

In order to understand the property of the ``fully" many-body localized systems (i.e. we assume all of the eigenstates are localized due to strong disorder), some phenomenology is proposed~\cite{Vosk13,Serbyn13,Huse14}.
In this phenomenological argument, we note that a set of conserved quantities is almost localized in space, since there is no transport in localized systems.
For example, for a Heisenberg chain in Eq. (\ref{hei}) with a large disorder, we expect that the effective Hamiltonian can be written as
\aln{\label{FMBL}
\hat{H}_\mr{eff}=E_0+\sum_ih'_i\hat{\tau}_i^z+\sum_{ij}J'_{ij}\hat{\tau}_i^z\hat{\tau}_j^z+
\sum_{n=3}\sum_{i_1\cdots i_n}K_{i_1\cdots i_n}^{(n)}\hat{\tau}_{i_1}^z\cdots\hat{\tau}_{i_n}^z\:,
}
where $E_0, h'_i, J'_{ij}, K_{i_1\cdots i_n}^{(n)}$ are constants, and $J'_{ij} \:(K_{i_1\cdots i_n}^{(n)})$'s decay exponentially with $|i-j|\:\: (|i_1-i_n|)$.
The so-called localized bits (l-bits), $\hat{\vec{\tau}}_i$, are quasi-localized conserved quantities, which have a large overlap with the operator $\hat{\vec{\sigma}}_i$ and have an exponentially small overlap with $\hat{\vec{\sigma}}_j\:(|i-j|\gg 1)$.
In other words, if the interaction is sufficiently weak, l-bits are expected to be constructed by dressing $\hat{\vec{\sigma}}_i$ perturbedly.
This is in fact proven for some models~\cite{Imbrie16D,Imbrie16O}.

Since we have many (quasi-)local conserved quantities $\hat{\vec{\tau}}_i$, it is expected that, the ETH and the approach to thermal equilibrium do not hold true in MBL systems.
In fact, each energy eigenstate is determined by a set of $N$ conserved quantities, especially in fully many-body localized systems.
This feature makes MBL systems akin to integrable systems~\cite{Huse14}.
However, unlike usual integrable systems, the MBL is robust against integrability-breaking perturbations as long as the disorder is sufficiently strong.
For this reason, MBL systems are gathering attention as a useful phase that sustains the quantum order even at a finite temperature~\cite{Huse13,Bauer13,Chandran14}.

\section{Experiments of isolated quantum systems}
One of the reasons for the recent development of quantum isolated systems is the experimental realizations of such systems.
While it is extremely difficult to simulate quantum many-body systems with a classical computer, we may realize them using complex quantum systems themselves, as Feynman pointed out~\cite{Feynman82}.
In fact, current technologies allow us to control quantum models with various Hamiltonians using neutral atoms, ions, etc.~\cite{Buluta09,Georgescu14}.
The approach to stationary states that are (not) thermal equilibrium is also observed by these (analogue) quantum simulators.
In this section, we will briefly explain experiments that have addressed the issue of nonequilibrium dynamics in isolated systems.

\subsection{Ultracold atoms}
Ultracold atomic gases offer a suitable setting for analogue simulation of isolated quantum systems.
By magnetic fields or optical dipole interactions, an atomic gas whose temperature is less than a microkelvin is trapped and isolated in a high vacuum chamber.
The interactions between atoms can be tuned by a Feshbach resonance, which allows us to investigate novel phenomena of strongly correlated quantum matters~\cite{Bloch08,Bloch12}.
Moreover, by loading atoms onto optical lattices, various lattice models with controllable Hamiltonians are realized~\cite{Bloch05}.
Controlling optical lattices, we can vary dimensionality of the models, or the shape of the lattices.

\begin{figure}
\begin{center}
\includegraphics[width=10cm,angle=-90]{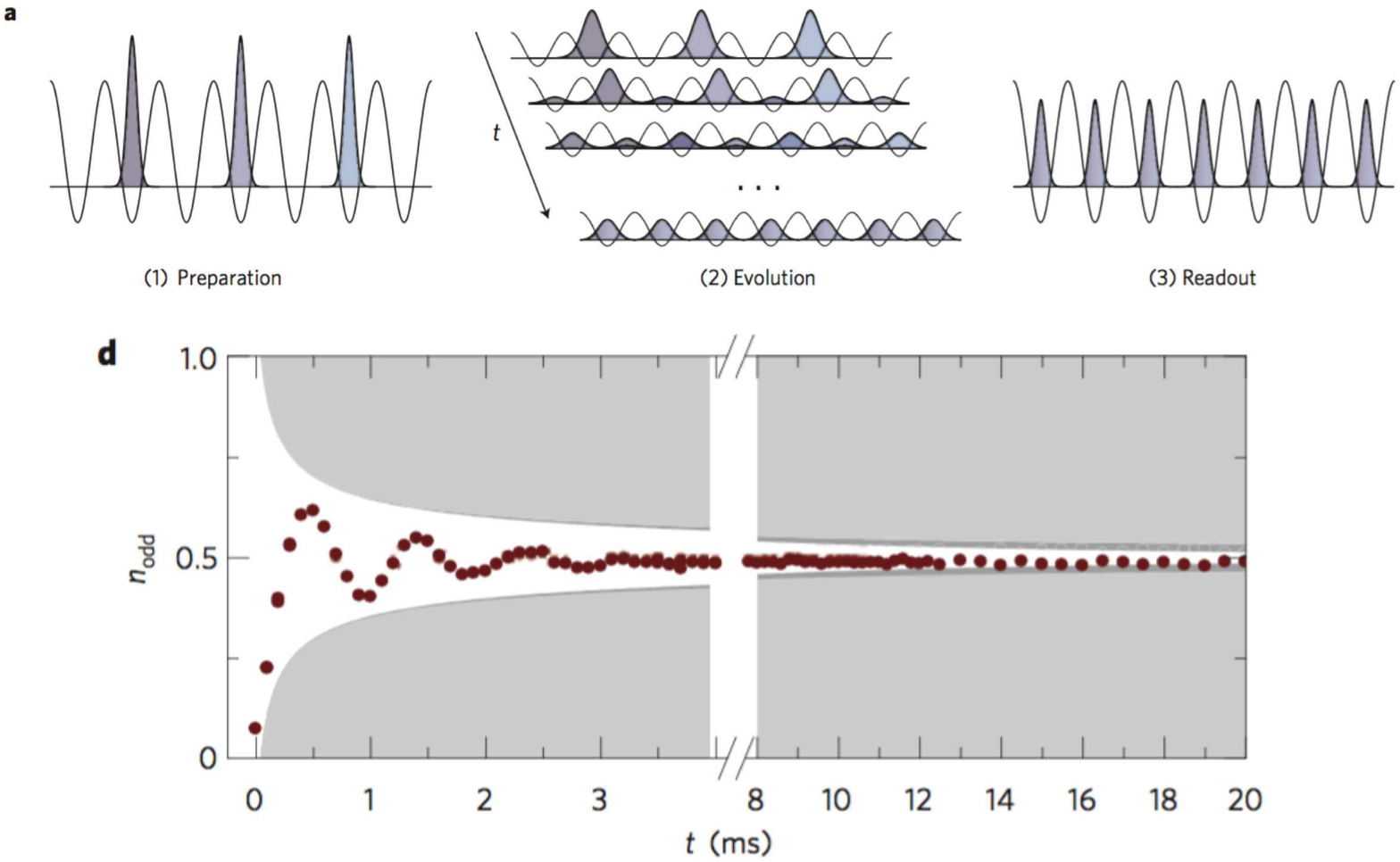}
\caption{Thermalization experiment by Trotzky \textit{et.}\textit{al.}~\cite{Trotzky12}. Initially, one $^{87}$Rb atom is placed at every second site, or ``even" sites (upper left). Then, the system undergoes a unitary time evolution (upper middle). After a certain time, the number density of odd sites is measured (upper right). After a few oscillations, the expectation value of the number density relaxes to the stationary state, which coincides with the prediction at thermal equilibrium (bottom). Reproduced from Fig. 1 of Ref.~\cite{Trotzky12}. \copyright 2012 Nature Publishing Group.}
\label{fig:tro}
\end{center}
\end{figure}

The approach to thermal equilibrium in an isolated nonintegrable system has been observed by Trotzky and coworkers~\cite{Trotzky12} using $^{87}$Rb atoms in a one-dimensional optical lattice (see Fig. \ref{fig:tro}).
By tuning a superlattice with bichromatic laser beams, they prepared a nonequilibrium initial state, where only one $^{87}$Rb atom resides at each ``even" site (upper left figure in Fig. \ref{fig:tro}).
Then, by quenching the height of the lattice potential, they let the system evolve according to the Bose-Hubbard Hamiltonian that is nonintegrable (upper middle figure).
After a certain time, they make the optical lattice higher again to suppress further evolution, and readout the number density of ``odd" sites (upper right figure).
The quantum expectation value of the density $n_\mr{odd}(t)$ is shown as a function of time in the bottom figure of Fig. \ref{fig:tro}.
We can see that $n_\mr{odd}(t)$ relaxes to $n_\mr{odd}(t)=0.5$, which is consistent with the prediction at thermal equilibrium.
They have also checked that the experimental results are in good agreement with the tDMRG numerical calculations, which confirms that the system actually undergoes a unitary time evolution.

Another notable example is the recent experiment that has demonstrated quantum thermalization in small systems~\cite{Kaufman16}.
In~\cite{Kaufman16}, Kaufman and coworkers have experimentally demonstrated that the approach to a thermal state takes place in a system with only six $^{87}$Rb atoms on six lattice sites.
By controlling potentials of individual lattice sites with a digital micromirror device (DMD), they experimentally created a one-dimensional Bose-Hubbard model with six particles on six sites.
They then succeeded in observing a unitary time evolution of the Renyi entanglement entropy or the local number density with the single-site microscopy.
These quantities relax to some stationary values after a certain time.
At that stage the entanglement entropy is nearly equal to thermal one, and local number densities coincide with the prediction at thermal equilibrium.
They have further confirmed that the reduced density matrix restricted to local sites is nearly equal to the thermal ensemble, even though the entire system is small.
These results are different from classical statistical mechanics that only considers macroscopic observables: they are expected to be genuine quantum thermalization that is related to the ETH.

Systems that do not approach thermal equilibrium are also realized.
Kinoshita, Wenger and Weiss~\cite{Kinoshita06} conducted a pioneering experiment that demonstrates the absence of thermalization in a near-integrable system.
They trapped 1D $^{87}$Rb gases in an anharmonic potential, and observed the time evolution of the momentum distribution of the gas.
They found that the momentum distribution relaxes to some stationary value that cannot be described by the thermal ensemble.
This result can be understood if we notice that the system is approximately described by the integrable Lieb-Liniger model.
The group in Vienna published several papers on prethermalization of a 1D Lieb-Liniger gas~\cite{Gring12,Langen13,Langen15}.
After suddenly splitting the gas into two halves, they studied a time evolution of the correlation function of bosonic fields by interfering these two halves.
In the experimentally observable timescale, the system seems to relax to a prethermalized state, which is a non-thermal quasi-stationary state emerging before complete thermalization~\cite{Berges04,Eckstein09,Kollar11}.
They argued in the most recent paper~\cite{Langen15} that the prethermalized state can be well fitted by the generalized Gibbs ensemble that considers occupation numbers of low-energy excited phonon modes.

\begin{figure}
\begin{center}
\includegraphics[width=10cm,angle=-90]{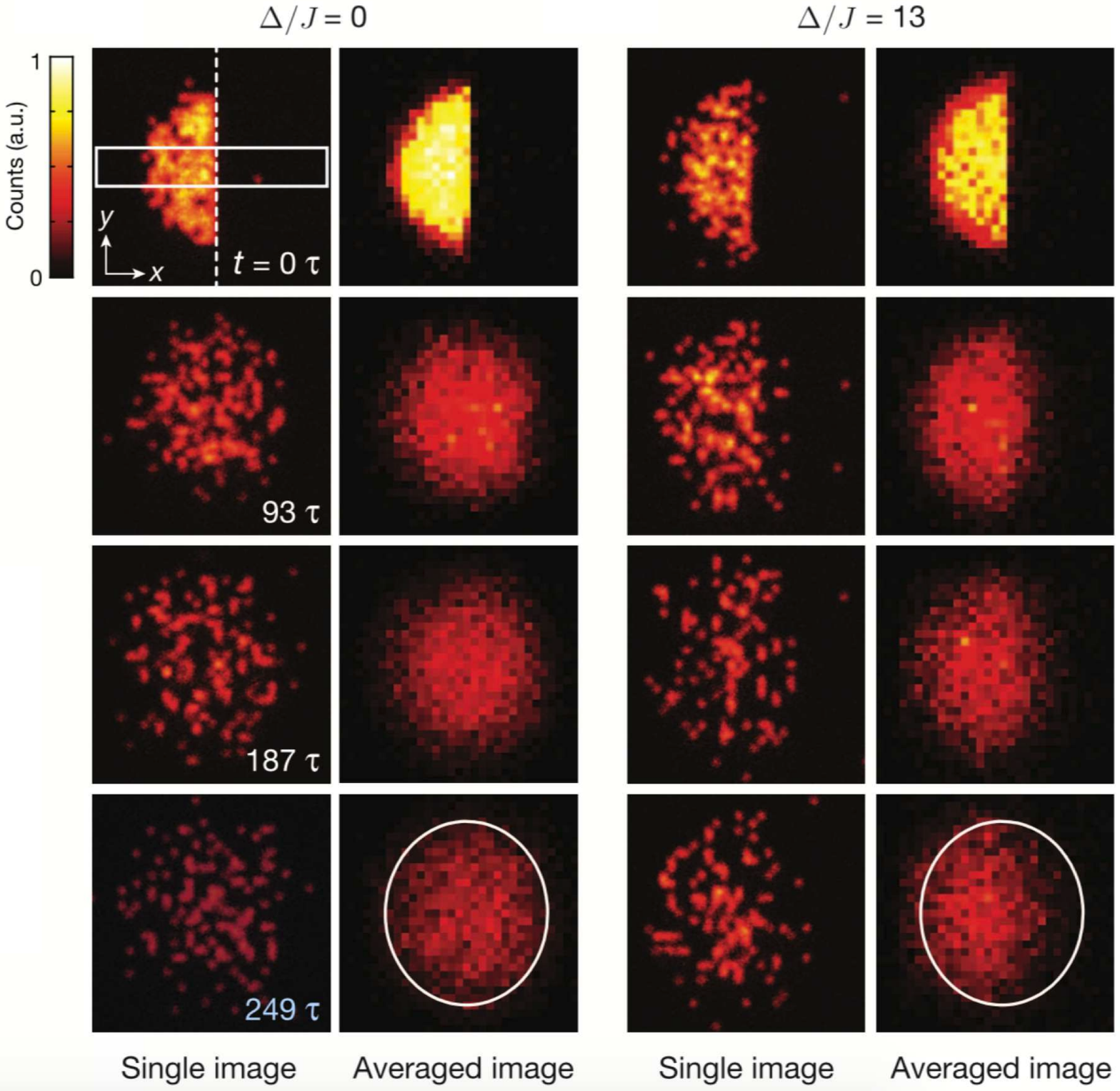}
\caption{Raw fluorescence images of the evolution of atoms~\cite{Choi16}. The left two columns show the case without disorder (single and averaged image, respectively). The right two columns show the case with disorder (single and averaged image, respectively). We can see that the population imbalance between left and right regions remains large in the case with the disorder. Reproduced from Fig. 1 of~\cite{Choi16}. \copyright 2016 by the American Association for the Advancement of Science.}
\label{fig:mbl}
\end{center}
\end{figure}

Many-body localization has also been observed by Immanuel Bloch's group~\cite{Schreiber15,Bordia16C,Choi16,Bordia16P,Luschen16}.
While the first experiment~\cite{Schreiber15} was done in a quasi-random optical lattice (i.e. the Aubry-Andr\'e model), genuine random potentials in two-dimensional lattices have recently been realized, too~\cite{Choi16}.
In~\cite{Choi16}, the authors have initially prepared $^{87}$Rb atoms in the Mott insulator phase in a left half of a 2D optical lattice (see Fig. \ref{fig:mbl}).
Then they allow the system to evolve by the Bose-Hubbard Hamiltonian with disorder $\Delta$, which can be generated by a DMD spatial light modulator.
As shown in the figure, they have found that the atom population imbalance between left and right regions remains significant after a long time if the disorder is present.
Using the population imbalance, they have observed the delocalization-localization phase transition as a function of the disorder strength.
They argue that the critical disorder strength of the phase transition gets smaller when the interactions between atoms are weaker.

We note that ultracold atom experiments also allow us to pursue universal nonequilibrium phenomena before reaching stationary states.
At a short timescale, light-cone-like spreading of two-point parity correlation functions ware observed~\cite{Cheneau12}, which indicates that the quasiparticle excitations propagate at a finite speed.
This is consistent with the famous Lieb-Robinson bound~\cite{Lieb72}, which imposes a bound on the speed of information spreading in systems with short-range interactions.
Transport phenomena have also been observed.
In Refs.~\cite{Schneider12,Ronzheimer13}, the authors have studied expansion dynamics of fermionic~\cite{Schneider12} or bosonic~\cite{Ronzheimer13} potassiums suddenly released from confining traps.
They have succeeded in changing the interaction or the dimensionality of the systems, especially in Ref.~\cite{Ronzheimer13}.
Their main finding is that, while atoms spread ballistically in the integrable limit (i.e. the non-interacting limit in 1D and 2D, or the hard-core limit in 1D), a diffusive core appears on top of a ballistic background if we break integrability.

\subsection{Trapped ions and other systems}
Although current experiments of many-body quantum dynamics have been mainly done using cold atoms, other systems may also be useful.
In fact, trapped ions offer a unique setting that is hard to obtain by ultracold atoms.
Due to the balance between the Coulomb repulsion and the confinement by electromagnetic potentials, laser-cooled trapped ions have vibrational degrees of freedom, in addition to internal states (which we model with two-level pseudospins)~\cite{Blatt12}.
Sideband transitions involving these internal and vibrational states lead to effective two-body interactions $J_{ij}^\gamma \:(\gamma=x,y,z)$ between distant spins at $i$ and $j$, after eliminating the vibrational degrees of freedom~\cite{Porras04}.
Moreover, by detuning the laser frequency of the sideband transitions, the range $\alpha$ of the interaction $J_{ij}^\gamma\propto \frac{1}{|i-j|^\alpha}$ can be practically controlled like $0\lesssim \alpha\lesssim 3$~\cite{Richerme14}.
We can also create an effective transverse field, and, in fact, transverse-field Ising chains or XY chains with various interaction ranges have been realized~\cite{Richerme14}.
By using such long-range spin models, the breakdown of the Lieb-Robinson bound were observed for the spreading of information, after a local quench with $^{40}$Ca$^+$ by Blatt's group~\cite{Jurcevic14} and a global quench with $^{171}$Yb$^+$ by Monroe's group~\cite{Richerme14}.
Monroe's group has also succeeded in observing the MBL using the Ising model with a disordered transverse field~\cite{Smith16}.
In Ref.~\cite{Neyenhuis16}, they also observed prethermalization in long-ranged Ising chain, where the quasi-stationary state cannot be described by the naive GGE.
We note that thermalization of spins due to the coupling with the vibrational mode is observed by Clos and coworkers~\cite{Clos16}.

Up to now, compared to neutral atoms or ions, there are not so many experiments of thermalization in isolated systems that use other potential (analogue or digital) quantum simulators, including Rydberg atoms, polar molecules, superconducting qubits, photons, or NV centers in diamonds.\footnote{We note that NV centers in diamonds have recently been used to demonstrate slow dynamics~\cite{Kucsko16} and time-crystalline order~\cite{Choi16O} in disordered quantum many-body systems.}
One notable exception is the work by Neill and coworkers~\cite{Neill16}, where they have observed  quantum thermalization of three superconducting qubits ($S=\frac{1}{2}$) that are periodically driven by pulse sequences.
Although the system does not conserve energy in this case, the dynamics can be written as unitary dynamics at each period of the cycles, if we neglect the decoherence from the environment.
They observed the entanglement entropy between one qubit and the others after a long time.
Then, what they found is follows:
initial states that give thermal/low stationary entanglement entropies is related to initial states that go into chaotic/regular trajectories on a phase space of the corresponding classical dynamics ($S\ra\infty$).
They also confirmed that the initial-state dependence of the stationary entanglement originates from the unitary dynamics of isolated quantum systems by checking that the decoherence due to the environment is independent of the initial states.

\section{Organization of this thesis}
In this thesis, we discuss the problem on thermalization by revisiting nonintegrable systems.
We are especially motivated by the following two questions:
\begin{enumerate}
\item
What is the underlying mechanism of the ETH, and that of the finite-size corrections from it in nonintegrable systems that conserve only energy?

\item
Do nonintegrable systems relax to non-thermal stationary states (possibly described by the GGE) if they have additional conserved quantities due to symmetries?

\end{enumerate}

\begin{figure}
\begin{center}
\includegraphics[width=14cm]{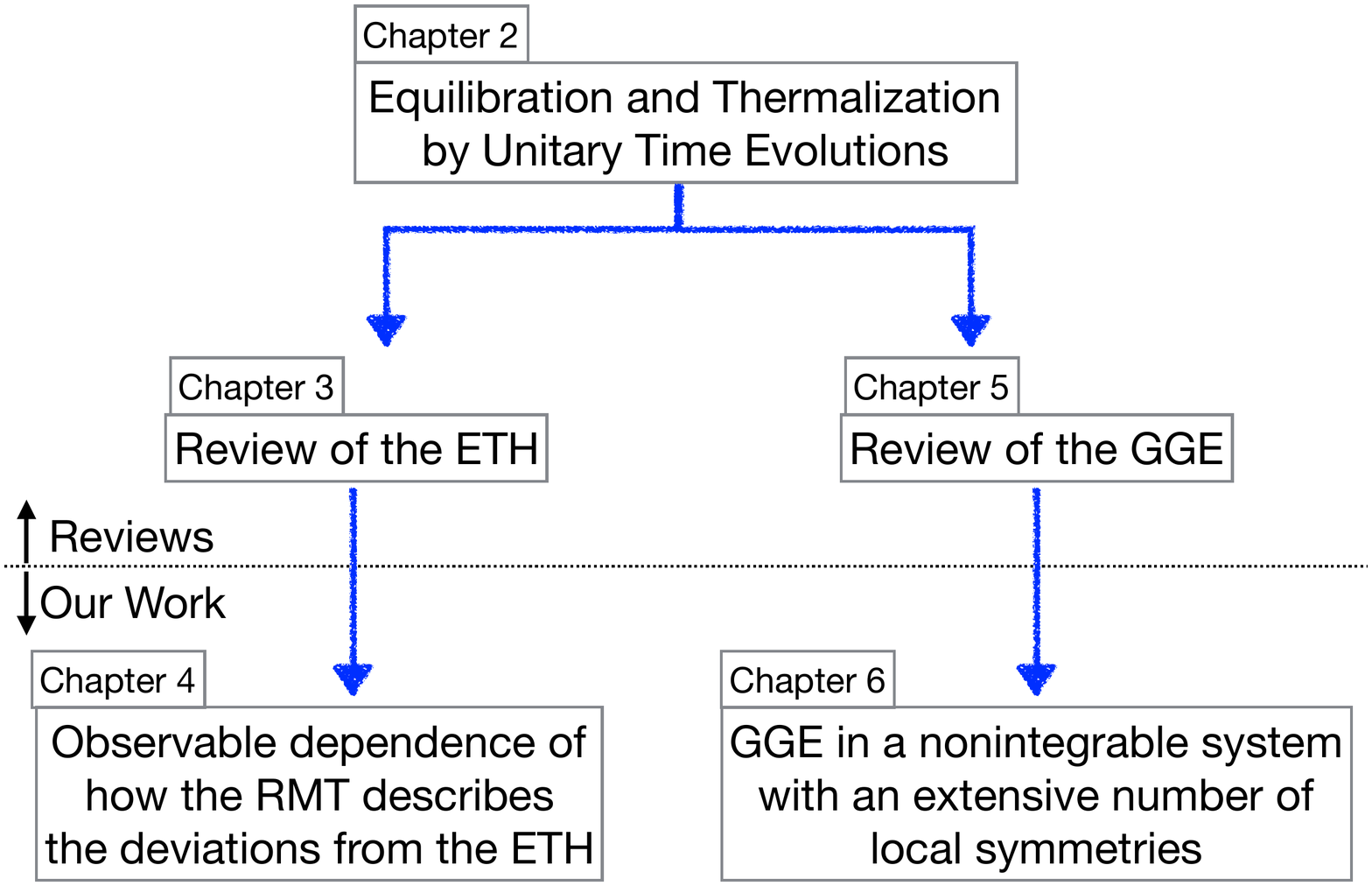}
\caption{Relations between the chapters in our thesis.}
\label{fig:outline}
\end{center}
\end{figure}

With these motivations in mind, we organize our thesis as follows.
In Chapter \ref{ch:Eq}, we review the current understanding of equilibration and thermalization in isolated quantum systems.
We especially explain how the ETH is relevant for equilibration and thermalization.
In Chapter \ref{ch:ETH}, we concentrate on the previous results that have addressed the first question raised above.
We review some early-days explanations of the ETH from the viewpoint of random matrix theory (RMT).
We also show some recent numerical simulations that have investigated the ETH and the finite-size corrections of it in quantum many-body systems.
In Chapter \ref{ch:Obs},  we show our first work related to the first question: are the ETH and its finite-size corrections predictable by the RMT in nonintegrable systems?
We will first analytically calculate the finite-size corrections of the ETH using the RMT model and show that their statistics is universal which depends on the anti-unitary symmetries of the Hamiltonians and the observables.\footnote{We will also see that the statistics will be further changed if the observable belongs to what we call the class of singular operators.} 
Then, we will numerically show that, in nonintegrable systems and for a wide class of observables (including many-body operators), the matrix elements are in good agreement with the prediction of the RMT model.
In Chapter \ref{ch:GGE}, we review the previous works on the generalized Gibbs ensemble in integrable systems.
We will stress the role of local conserved quantities.
In Chapter \ref{ch:Gn}, we show our second work (based on Ref.~\cite{Hamazaki16G}) related to the second question: emergence of a non-thermal stationary state in a nonintegrable system with an extensive number of local symmetries.
We have numerically investigated a nonintegrable model of hard-core bosons with an extensive number of local $\mbb{Z}_2$ symmetries.
We find that the expectation values of local observables in the stationary state are described by the GGE and not by the canonical ensemble.
We also show that, if the model has less local symmetries, the stationary state is described by the canonical ensemble.
In Chapter \ref{ch:Con}, we give the summary of the thesis with some remarks on the future prospect.
The relations between the chapters are shown in Fig. \ref{fig:outline}.

\chapter{Equilibration and thermalization by unitary time evolutions}\label{ch:Eq}
In this section, we review the current understanding of equilibration and thermalization in isolated quantum systems.
Before considering the dynamics, we first explain some of the possible definitions of thermal equilibrium, following Refs.~\cite{Goldstein15,Goldstein16}.
Then, we consider how to formulate the approach to thermal equilibrium.
We especially focus on the scenario of eigenstate thermalization hypothesis (ETH), which justifies thermalization for any initial state within the microcanonical energy shell.
Finally, we explain the role of initial states on equilibration and thermalization when we do not assume the ETH.

Let us briefly summarize the history of quantum thermalization. 
Von Neumann tackled the problem of quantum thermalization as early as in 1929~\cite{Neumann29}. 
He essentially showed that the ETH is a sufficient condition for the system to approach thermal equilibrium.
He also proved that the ETH holds true for most of the decompositions of what he called macrospaces (see Subsection \ref{sec:mate}).
Unfortunately, by 1950's, his theorems were severely criticized as being meaningless by several researchers~\cite{Fierz55,Farquhar57,Bocchieri58} because of the misunderstandings of the original statement.
Due to these misunderstandings, von Neumann's work had been forgotten for more than half a century until Goldstein and coworkers realized that it is of great value and published a commentary in 2010~\cite{Goldstein10L}. 
Note, however, that important progresses were made both analytically~\cite{Peres84E,Deutsch91,Srednicki94,Tasaki98,Reimann08,Linden09} and numerically~\cite{Jensen85,Rigol08} even before the rediscovery of von Neumann's work.

\section{Definitions of thermal equilibrium and typicality}\label{sec:Deft}
Here we review several definitions of thermal equilibrium, which slightly differ from one paper to another.
Historically, von Neumann originally considered macroscopic observables and defined phase spaces using them, which he called macrospaces.
Though he treated all macrospaces equally in order to discuss thermal equilibrium, Goldstein and coworkers realized that one special macrospace represents thermal equilibrium, and formulated  macroscopic thermal equilibrium (MATE)~\cite{Goldstein10,Goldstein10L}.
On the other hand, it turned out that the thermal ensemble can emerge by separating the entire system into a small subsystem and an environment through the quantum entanglement~\cite{Popescu06,Goldstein06}.
These works lead to the notion of microscopic thermal equilibrium (MITE), which is a stronger condition than MATE.
We note that these notions of thermal equilibrium are clearly reformulated only recently, with a more general framework~\cite{Goldstein15,Goldstein16}.
In this section, after discussing that general framework, we review MITE and MATE following Ref.~\cite{Goldstein16}, and then remark on a less strict view.\footnote{We note that some authors have proposed other formulations that are different from the formulation proposed in Refs.~\cite{Goldstein15,Goldstein16}. For example, in Ref.~\cite{Tasaki16}, the author uses some tricks to treat local or few-body observables as macroscopic observables with the help of translational invariance or fictitious copies of the original system.}

\subsection{General framework}
As we have mentioned in Chapter \ref{ch:Stat}, we need to specify a set of observables, $\mc{A}$, in order to discuss if a given (possibly pure) state $\hrho$ is close to thermal equilibrium.
In Ref.~\cite{Goldstein16}, the authors have defined ``thermal equilibrium \textit{relative to} $\mc{A}$" essentially as follows:
a state $\hrho$ is in thermal equilibrium \textit{relative to} $\mc{A}$ if and only if for any $\hmo\in\mc{A}$, the probability distribution over the spectrum of $\hmo$ with respect to $\hrho$ is approximately equal to that with respect to $\hrho_\mr{mic}(E)$, where $E=\Tr[\hat{H}\hrho]$.
In other words, if $\hmo=\sum_{f}f\hat{\mc{P}}_{f}$, where $\hat{\mc{P}}_{f}$ is a projection operator with eigenvalue $f$, then $\hrho$ satisfies
\aln{
\Tr[\hrho\hat{\mc{P}}_{f}]\simeq\Tr[\hrho_\mr{mic}(E)\hat{\mc{P}}_{f}]
}
for every $f$.
In the following, we see that we have to take operators on a small subsystem as $\mc{A}$ for MITE, and macroscopic observables as $\mc{A}$ for MATE.

\subsection{Microscopic thermal equilibrium (MITE)}
For MITE, we separate the entire system into a small subsystem and an environment, and take all observables that have supports on the subsystem.
To see this, we first decompose the whole Hilbert space into a tensor product,
\aln{
\mc{H}=\mc{H}_S\otimes\mc{H}_{S^c},
}
where $S$ is a small subsystem and $S^c$ is an environment ($c$ means the complement).
We consider all possible subsystems $S$ that are small.\footnote{For example, we can consider all spatially local regions $S$ that satisfy $\mr{diam}(S)\leq l_0\ll \mr{diam}(S\cup S^c)$ for some $l_0$, where $\mr{diam}(S)$ means the diameter of $S$.}
MITE is thermal equilibrium \textit{relative to} 
\aln{
\mc{A}_\mr{MITE}=\cup_S\mc{A}_S,
} 
where
\aln{
\mc{A}_S:=\lrm{\hmo_S\otimes \hat{\mbb{I}}_{S^c}\: :\: \hmo_S \:\text{is a Hermitian operator acting on }\mc{H}_S}.
}
MITE can also be simply written as follows: for all small $S$,
\aln{\label{onaji}
\hrho_S=\hrho_{\mr{mic},S}
}
is satisfied, where $\hrho_S:=\Tr_{S^c}[\hrho]$.

While we usually consider spatially local subsystems for the choice of $S$, we sometimes take a set of general few-body operators as $\mc{A}_S$, too.
As a simple example, let us consider a one-dimensional lattice quantum spin 1/2 chain of $N$ sites with a periodic boundary condition.
The whole Hilbert space $\mc{H}$ is a direct product of the Hilbert space at each site $\mc{H}_i$:
\aln{
\mc{H}=\bigotimes_{i=1}^N\mc{H}_i,
}
where $\dim[\mc{H}_i]=2$ and $\dim[\mc{H}]=2^N$.
If we are interested in spatially local observables with length smaller than $l_0\ll N$ for MITE,  
then 
\aln{
\mc{H}_{S_{l_0}}=\bigotimes_{i=i_0}^{i_0+l_0-1}\mc{H}_i
}
for some site $1\leq i_0\leq N$.
In this case, 
\aln{\label{locloc}
\prod_{i=i_0}^{i_0+l_0-1}\hat{\sigma}_i^{\mu_i}\:\:\:\:(\mu_i=0,x,y,z)
}
can be a basis set of the Hermitian operators $\hmo_S$ acting on $\mc{H}_S$, where $\hat{\sigma}_i^0:=\hat{\mbb{I}}_i$ is the $2\times 2$ identity operator. 
On the other hand, we can also consider a set of general $k$-body operators that satisfy $k\leq M$ for some $M$.
If $M\ll N$, then they are called few-body operators.
Here, we define $k$-body operators as those whose basis set can be written as
\aln{
\hat{\sigma}_{i_1}^{\alpha_{i_1}}\hat{\sigma}_{i_2}^{\alpha_{i_2}}\cdots \hat{\sigma}_{i_k}^{\alpha_{i_k}}\:\:\:\:(\alpha_i=x,y,z)
}
for $1\leq i_1<i_2<\cdots <i_k\leq N$.
In this case, the subsystem for MITE can be taken as 
\aln{
\mc{H}_{S_M}=\mc{H}_{i_1}\otimes\mc{H}_{i_2}\otimes\cdots\otimes\mc{H}_{i_M}\:\:(1\leq i_1<i_2<\cdots <i_M\leq N).
}
In both cases of local and few-body operators, we usually assume that $\dim[\mc{H}_S]\ll\dim[\mc{H}]$ in considering MITE.\footnote{We note, however, that the notion of MITE may be extended for subsystems that are as large as the half of the system size~\cite{Garrison15,Goldstein16,Geraedts16}.}

\subsubsection{Proof of typicality for MITE}
We can show the typicality of pure states in the microcanonical energy shell, $\mc{H}_\mr{mic}=\mc{H}_{E,\Delta E}$ for MITE that considers general few-body operators~\cite{Popescu06,杉田06,Goldstein06}.
Note that $\mc{H}_\mr{mic}$ is explicitly written as
\aln{
\mc{H}_\mr{mic}=\lrm{\ket{\psi}=\sum_{\alpha\in\mc{S}} z_\alpha\ket{E_\alpha}: z_\alpha\in\mbb{C},\sum_{\alpha\in\mc{S}}|z_\alpha|^2=1},
}
where $\alpha$ labels energy eigenvalues that fall within the energy shell, and 
\aln{
\mc{S}=\lrm{\alpha:|E-E_\alpha|<\Delta E}.
}
We consider picking up a pure state $\ket{\psi}\in\mc{H}_\mr{mic}$ randomly, and assume that  $z_\alpha$'s are taken from the following probability distribution:
\aln{\label{eq213}
P(\lrm{z_\alpha})\prod_{\alpha\in\mc{S}} d\mr{Re}z_\alpha d\mr{Im}z_\alpha= c\times\delta\lrs{\sum_{\alpha\in\mc{S}}|z_\alpha|^2-1}\prod_{\alpha\in\mc{S}} d\mr{Re}z_\alpha d\mr{Im}z_\alpha,
}
where the constant $c$ is determined from $\int P(\lrm{z_\alpha})\prod_{\alpha\in\mc{S}} d\mr{Re}z_\alpha d\mr{Im}z_\alpha=1$.

We first show that, for any operator $\hmo$, 
\aln{\label{ev}
\mbb{E}[\braket{\psi|\hmo|\psi}]&=\braket{\hmo}_\mr{mic},\\
\mbb{V}[\braket{\psi|\hmo|\psi}]&=\frac{\mc{V}_\mr{mic}(\hmo)}{d+1}\leq \frac{||\hmo||_\mr{op}^2}{d+1},
}
where $\mbb{E}$ and $\mbb{V}$ are the expectation value and the variance over $P(\lrm{z_\alpha})$, respectively, $||\hmo||_\mr{op}$ is an operator norm of $\hmo$,\footnote [1]{The operator norm $||\hmo||_\mr{op}$ is defined as the square root of the maximum eigenvalue of $\hmo^\dag\hmo$. If $\hmo$ is Hermitian, it is the largest absolute eigenvalue of $\hmo$.}
 $d=\dim[\mc{H}_\mr{mic}]$, and
\aln{\label{uhyahya}
\mc{V}_\mr{mic}(\hmo)&:=\frac{1}{d}\sum_{\alpha,\beta\in\mc{S}}|\mc{O}_{\alpha\beta}|^2-\lrl{\frac{1}{d}\sum_{\alpha\in\mc{S}}\mc{O}_{\alpha\alpha}}^2\nonumber\\
&=\braket{\hmo\mc{\hat{P}}_{E,\Delta E}\hmo}_\mr{mic}-\braket{\hmo}_\mr{mic}^2,
}
where $\mc{O}_{\alpha\beta}=\braket{E_\alpha|\hmo|E_\beta}$.
To show Eq. (\ref{ev}), we use expectation values of the moments of $\lrm{z_\alpha}$.
For second moments, we have
\aln{
\mbb{E}[z_\alpha^*z_\beta]=\frac{\delta_{\alpha\beta}}{d}
}
and the only non-vanishing fourth moments are
\aln{
\mbb{E}[|z_\alpha|^2|z_\beta|^2]=\frac{1+\delta_{\alpha\beta}}{d(d+1)}.
}
The first and third moments are all zero.
Then we obtain
\aln{
\mbb{E}[\braket{\psi|\hmo|\psi}]&=\sum_{\alpha,\beta\in \mc{S}}\mbb{E}[z_\alpha^*z_\beta]\mc{O}_{\alpha\beta}\nonumber\\
&=\frac{1}{d}\sum_{\alpha\in\mc{S}}\mc{O}_{\alpha\alpha}\nonumber\\
&=\braket{\hmo}_\mr{mic}
}
and
\aln{
\mbb{V}[\braket{\psi|\hmo|\psi}]&=\mbb{E}[|\braket{\psi|\hmo|\psi}|^2]-|\mbb{E}[\braket{\psi|\hmo|\psi}]|^2\nonumber\\
&=\lrs{\sum_{\alpha,\beta,\gamma,\delta\in\mc{S}}\mbb{E}[z_\alpha^*z_\beta z_\gamma^*z_\delta]\mc{O}_{\alpha\beta}\mc{O}_{\gamma\delta}}-\braket{\hmo}_\mr{mic}^2\nonumber\\
&=\sum_{\alpha\in\mc{S}}\frac{2\mc{O}_{\alpha\alpha}^2}{d(d+1)}+\sum_{\alpha,\gamma\in\mc{S},\alpha\neq\gamma}\frac{\mc{O}_{\alpha\alpha}\mc{O}_{\gamma\gamma}}{d(d+1)}+
\sum_{\alpha,\beta\in\mc{S},\alpha\neq\beta}\frac{|\mc{O}_{\alpha\beta}|^2}{d(d+1)}-\braket{\hmo}_\mr{mic}^2\nonumber\\
&=\sum_{\alpha,\beta\in\mc{S}}\frac{|\mc{O}_{\alpha\beta}|^2}{d(d+1)}-\frac{1}{d^2(d+1)}\lrl{\sum_{\alpha\in\mc{S}}\mc{O}_{\alpha\alpha}}^2\nonumber\\
&=\frac{\mc{V}_\mr{mic}(\hmo)}{d+1}.
}
Moreover, we obtain
\aln{
\frac{\mc{V}_\mr{mic}(\hmo)}{d+1}
&\leq\sum_{\alpha,\beta\in\mc{S}}\frac{|\mc{O}_{\alpha\beta}|^2}{d(d+1)}\nonumber\\
&\leq\sum_{\alpha\in\mc{S}}\sum_\beta\frac{|\mc{O}_{\alpha\beta}|^2}{d(d+1)}\nonumber\\ 
&=\frac{1}{d(d+1)}\sum_{\alpha\in\mc{S}}(\hmo^2)_{\alpha\alpha}\nonumber\\
&\leq \frac{||\hmo^2||_\mr{op}}{d+1}\nonumber\\
&\leq \frac{||\hmo||_\mr{op}^2}{d+1}.
}

From Chebyshev's inequality, Eq. (\ref{ev}) implies
\aln{
\mbb{P}\lrl{\lrv{\braket{\psi|\hmo|\psi}-\braket{\hmo}_\mr{mic}}\geq \epsilon}\leq \frac{||\hmo||_\mr{op}^2}{\epsilon^2(d+1)}
}
for any $\epsilon>0$, where $\mbb{P}$ denotes the probability with respect to $P(\lrm{z_\alpha})$.
By taking $\epsilon=d^{-1/3}$, we obtain 
\aln{\label{cheb}
\mbb{P}\lrl{\lrv{\braket{\psi|\hmo|\psi}-\braket{\hmo}_\mr{mic}}\geq d^{-1/3}}\leq \frac{||\hmo||_\mr{op}^2}{d^{-2/3}(d+1)}.
}
Since we are discussing the typicality of MITE, we can assume that $\hmo$ is an $M$-body operator for some number $M$, which we assume is independent of $N$.
In fact, however, we can also show the typicality for an operator that is written as the sum of the $M$-body operators.
Thus, we will consider such a general operator, which can be written for the case of a spin 1/2 model as
\aln{
\hmo=\sum_{1\leq i_1,\cdots ,i_M\leq N}\sum_{\alpha_{i_1},\cdots,\alpha_{i_M}=x,y,z}f_{i_1\cdots i_M;\alpha_{i_1}\cdots\alpha_{i_M}}\hat{\sigma}_{i_1}^{\alpha_{i_1}}\hat{\sigma}_{i_2}^{\alpha_{i_2}}\cdots \hat{\sigma}_{i_M}^{\alpha_{i_M}}.
}
The generalization to other models is straightforward.
Here $f_{i_1\cdots i_M;\alpha_{i_1}\cdots\alpha_{i_M}}$'s are constants whose absolute values are bounded by $f$, which is independent of $N$.
Then, the operator norm of $\hmo$ is bounded as 
\aln{
||\hmo||_\mr{op}&\leq\sum_{1\leq i_1,\cdots ,i_M\leq N}\sum_{\alpha_{i_1},\cdots,\alpha_{i_M}=x,y,z}|f_{i_1\cdots i_M;\alpha_{i_1}\cdots\alpha_{i_M}}|\times||\hat{\sigma}_{i_1}^{\alpha_{i_1}}\hat{\sigma}_{i_2}^{\alpha_{i_2}}\cdots \hat{\sigma}_{i_M}^{\alpha_{i_M}}||_\mr{op}\nonumber\\
&\leq  \sum_{1\leq i_1,\cdots ,i_M\leq N}\sum_{\alpha_{i_1},\cdots,\alpha_{i_M}=x,y,z}f\nonumber\\
&= \frac{3^MfN!}{M!(N-M)!}\sim N^M.
}
Thus, $||\hmo||_\mr{op}$ does not increase faster than the polynomial of $N$.
Since $d$ increases exponentially with $N$, $d^{-1/3}$ and $\frac{||\hmo||_\mr{op}^2}{d^{-2/3}(d+1)}$ appearing in Eq. (\ref{cheb}) decrease rapidly with $N$.
Then, Eq. (\ref{cheb}) means that, when $N$ is sufficiently large, most of the pure states $\ket{\psi}$ with respect to \EQ{eq213} give $\braket{\psi|\hmo|\psi}\simeq\braket{\hmo}_\mr{mic}$ for any  operator $\hmo$ that can be written as the sum of few-body operators.

We can prove the typicality of MITE in the form of Eq. (\ref{onaji}), too.
Namely, we can prove
\aln{\label{typMITE}
\mbb{P}\lrl{||\hrho_S-\hrho_{\mr{mic},S}||_\mr{op}\geq \epsilon}\leq \frac{(\dim[\mc{H}_S])^2}{\epsilon^2(d+1)}
}
for any $\epsilon>0$.
The proof is given in Appendix \ref{sec:typMITE}.
Here, we remember the equivalence of ensembles between the microcanonical ensemble and the canonical ensemble in a thermodynamically normal system, which has recently been proven rigorously under certain conditions~\cite{Mueller15,Brandao15,Mori16M}.
Then, Eq. (\ref{typMITE}) further indicates that $\hrho_S\simeq \hrho_{\mr{can},S}$ holds true for most  pure states, where $\hrho_\mr{can}=\frac{e^{-\beta\hat{H}}}{Z}$ is the canonical ensemble with $\beta$ being determined from the condition $\Tr[{\hat{H}\hrho}]=\Tr[{\hat{H}\hrho_\mr{can}}]$.
This type of the typicality is called the canonical typicality~\cite{Goldstein06}.

\subsection{Macroscopic thermal equilibrium (MATE)}\label{sec:mate}
For MATE,  we consider only a set of observables $\{\hat{M}'_1, \hat{M}'_2,\cdots, \hat{M}'_K\}$ that are measured macroscopically, such as the magnetization density or the number of particles (invoking the usual thermodynamics).
In general, $\{\hat{M}'_1, \hat{M}'_2,\cdots, \hat{M}'_K\}$ are not commutable with one another.
However, we expect that we can construct a set of commuting operators $\{\hat{M}_1, \hat{M}_2,\cdots, \hat{M}_K\}$ from $\{\hat{M}'_1, \hat{M}'_2,\cdots, \hat{M}'_K\}$, with $||\hat{M}_l-\hat{M}'_l|| \:(1\leq l \leq K)$ being small, if the commutators among $\{\hat{M}'_1, \hat{M}'_2,\cdots, \hat{M}'_K\}$ are sufficiently small.
Since this conjecture has been proven for several situations~\cite{Ogata13}, we will use $\{\hat{M}_1, \hat{M}_2,\cdots, \hat{M}_K\}$ as macroscopic observables (we remark that the formalism that uses $\{\hat{M}'_1, \hat{M}'_2,\cdots, \hat{M}'_K\}$ is proposed by Tasaki~\cite{Tasaki16}. See Sec. \ref{sec:tmate}). 

Since $\{\hat{M}_1, \hat{M}_2,\cdots, \hat{M}_K\}$ are commutable with one another, we can decompose the Hilbert space using the simultaneous eigenstates of these macroscopic observables.
We denote such eigenstates by $\ket{\lrm{\mu},\lambda}=\ket{\mu_1\cdots\mu_K,\lambda}\:(\hat{M}_l\ket{\lrm{\mu},\lambda}=\mu_l\ket{\lrm{\mu},\lambda})$, where $\mu_l$'s are determined only macroscopically (i.e., only macroscopically different $\mu_l$'s can be distinguished), and $\lambda$ labels the degeneracies. 
Then the orthogonal decomposition of the Hilbert space is
\aln{
\mc{H}=\bigoplus_{\lrm{\mu}}\mc{H}_{\lrm{\mu}},
}
where the projection onto $\mc{H}_{\lrm{\mu}}$ can be written as
\aln{
\hat{\mc{P}}_{\lrm{\mu}}=\sum_{\lambda=1}^{\dim[\mc{H}_{\lrm{\mu}}]}\ket{\lrm{\mu},\lambda}\bra{\lrm{\mu},\lambda}.
}
We call $\mc{H}_{\lrm{\mu}}$ as a macrospace.

We now consider that $\hat{M}_1$ is a coarse-grained Hamiltonian ($\hat{M}_1'=\hat{H}$).
Then, $\mu_1$ denotes the coarse-grained energy, $\mu_1\sim E\pm\Delta E$,  where $\Delta E$ represents the inaccuracy due to the coarse-graining.
We consider the Hilbert space with $\mu_1\simeq E$ as the microcanonical energy shell at energy $E$.
Then this energy shell can be decomposed by the other macroscopic observables, $\{\hat{M}_2,\cdots, \hat{M}_K\}$, as
\aln{
\mc{H}_\mr{mic}(E)=\bigoplus_{\mu_2,\cdots,\mu_K}\mc{H}_{\mu_1\simeq E, \mu_2\cdots\mu_K}.
} 
We note that
\aln{
\dim[\mc{H}_\mr{mic}(E)]=\sum_{\mu_2,\cdots,\mu_K}\dim[\mc{H}_{\mu_1\simeq E, \mu_2\cdots\mu_K}].
}

An important observation, which von Neumann did not realize, is that for one special set of $(\mu_2,\cdots,\mu_K)$, we should usually expect 
\aln{
\dim[\mc{H}_\mr{mic}(E)]\simeq\dim[\mc{H}_{\mu_1\simeq E, \mu_2\cdots\mu_K}].
}
In other words, the dimension of only one macrospace dominates, and we will call that macrospace as the thermal equilibrium subspace, $\mc{H}_\mr{eq}$.
As a result, for a small $\epsilon>0$, we can decompose $\mc{H}_\mr{mic}$ as
\aln{\label{wake}
\mc{H}_\mr{mic}=\mc{H}_\mr{eq}\oplus\mc{H}_\mr{neq},\nonumber\\
\frac{\dim[\mc{H}_\mr{eq}]}{\dim[\mc{H}_\mr{mic}]}> 1-\epsilon,
}
where we define a nonequilibrium subspace, $\mc{H}_\mr{neq}$, as the direct sum of non-thermal equilibrium macrospaces.

We define that a state $\hrho$ is in MATE if and only if
\aln{
\Tr[\hrho\hat{P}_\mr{eq}]>1-\delta
}
for small $\delta$.
MATE is also regarded as thermal equilibrium \textit{relative to} $\mc{A}_\mr{MATE}=\{\hat{M_1},\hat{M}_2,\cdots, \hat{M}_K\}$.
As for MITE, we can show the typicality of pure states for MATE.
In fact, MITE generally implies MATE, since a macroscopic observable can be written as a sum of local operators.
In that sense, MITE is a stronger assumption for thermal equilibrium than MATE.\footnote{As an example of a state that seems to be in MATE but not in MITE, consider a noninteracting spin 1/2 chain of $N$ sites, where $\hH=0$.
Let us take a product state $\ket{\psi}=\bigotimes_{i=1}^N\ket{\psi_i}$, where $\ket{\psi_i}\in\mc{H}_i$ (the local Hilbert space at each site).
If each $\ket{\psi_i}$ is randomly chosen from $\mc{H}_i$, we expect that $\hrho=\ket{\psi}\bra{\psi}$ and the microcanonical ensemble $\hrho_\mr{mic}=\frac{1}{D}\hat{\mbb{I}}_{D\times D}\:(D=2^N)$ are indistinguishable for macroscopic observables such as $\hat{M}_z=\sum_{i=1}^N \hat{\sigma}_i^z$: $\Tr[\hrho\hat{M}_z]\simeq\Tr[\hrho_\mr{mic}\hat{M}_z]=0$.
On the other hand, if we only consider the first spin, we have $\hrho_1=\ket{\psi_1}\bra{\psi_1}\neq \frac{1}{2}\hat{\mbb{I}}_{2\times 2}=(\hrho_\mr{mic})_1$.
This shows that $\ket{\psi}$ is not in MITE.
}
Although MATE seems a natural situation for considering thermodynamics (because it only considers macroscopic observables), we know that MITE is also meaningful in quantum statistical mechanics, so we will more often pay attention to MITE than MATE.

\subsection{A looser way to consider thermal equilibrium}
We have seen how to formulate thermal equilibrium by restricting observables, but it is, in general, not easy to test whether a state satisfies these criteria.
Instead, many previous studies have investigated the expectation values of certain observables~\cite{Srednicki94,Reimann08,Rigol08,Reimann15}, and simply checked if they are approximately equal to the prediction of the thermal ensemble:
\aln{\label{hitoshi}
\Tr[\hrho\hmo]\simeq \Tr[\hrho_\mr{mic}\hmo].
}
Although the expectation value of a single observable $\hmo\in\mc{A}_\mr{MITE/MATE}$ would not tell us if the state is in MITE/MATE, at least it implies.
Moreover, the definition of MITE/MATE may be too restrictive, and $\hmo\not\in\mc{A}_\mr{MITE/MATE}$ may satisfy the condition (\ref{hitoshi}), which implies that statistical mechanics is applicable even beyond MITE.
This possibility has recently been pointed out in Refs.~\cite{Garrison15,Hosur16,Geraedts16}.
For these reasons, we will mainly consider (\ref{hitoshi}) for certain observables in the following discussions, including our works in Chapters \ref{ch:Obs} and \ref{ch:Gn}.

\section{Conditions for equilibration and thermalization}
Now, we consider the meanings and conditions for a nonequilibrium initial state $\ket{\psi_0}$ to approach thermal equilibrium under unitary time evolutions (for simplicity, we will consider pure states in this section).
As we have mentioned in the overview, it is convenient to separate the problem into two.
We first consider equilibration, namely, a phenomenon where a state seemingly relaxes to a stationary state.
Then, we will discuss when the stationary state is indistinguishable to a thermal state when we measure $\hmo$.

\subsection{Equilibration}
\begin{figure}
\centering
\includegraphics[angle=-90,width=13cm]{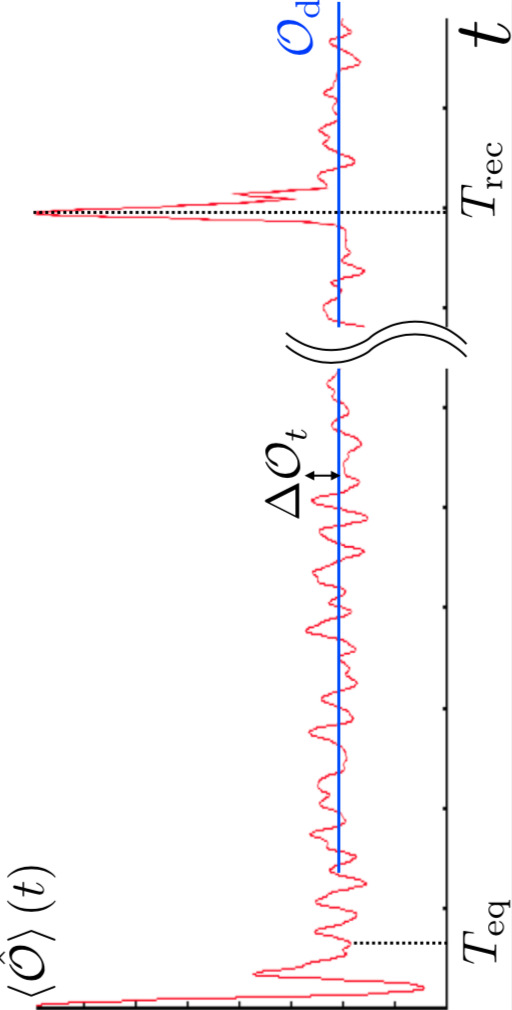}
\caption{Schematic illustration of equilibration of an observable $\hmo$ (note that the figure is not produced by actual simulations). Though the expectation value $\braket{\hmo}(t)$ approaches the stationary value $\mc{O}_\mr{d}$ after a timescale $T_\mr{eq}$, there appears the quantum recurrence at time $T_\mr{rec}$. 
However, for most of the times, $\braket{\hmo}(t)$ is close to $\mc{O}_\mr{d}$, and the temporal fluctuation $\Delta\mc{O}_t$ is small.}
\label{fig:time}
\end{figure}
We first consider the meaning of equilibration.
Naively, we might expect that after some equilibration time, $T_\mr{eq}$, the expectation value of an observable $\hmo$ is almost equal to some stationary value $\mc{O}_\mr{d}$ (the meaning of the subscript ``d" is explained later):
\aln{\label{rec}
\braket{\hmo}(t)\simeq \mc{O}_\mr{d} \text{ for any }t>T_\mr{eq} \:\:\:\text{(wrong)},
}
where $\braket{\hmo}(t)=\braket{\psi(t)|\hmo|\psi(t)}$ with $\ket{\psi(t)}=e^{-i\hat{H}t}\ket{\psi_0}$.
This equation is wrong in general because of the quantum recurrence theorem~\cite{Bocchieri57}; we can rigorously show that, for an arbitrary small $\epsilon>0$, there exists an infinite sequence $0<T_i\:(i=1,2,\cdots)$ such that
\aln{
||\ket{\psi(T_i)}-\ket{\psi_0}||<\epsilon.
}
This theorem also implies that there always exists $t=T_\mr{rec}$ that makes $\braket{\hmo}(t)$ arbitrary close to $\braket{\hmo}(0)$.
Then, if $|\braket{\hmo}(0)-\mc{O}_\mr{d}|$ is large, Equation (\ref{rec}) does not hold true.
The important thing to notice here is that the recurrence times become super-exponential of the system size, and that we rarely expect such recurrences.
In other words, it is reasonable to regard equilibration as a phenomenon where the expectation value of an observable stays close to some stationary value for \textit{almost all times} (see Fig. \ref{fig:time}):
\aln{\label{ealt}
\braket{\hmo}(t)\simeq \mc{O}_\mr{d} \text{ for almost all }t.
}

In order to justify Eq. (\ref{ealt}), we consider the average and the variance of $\braket{\hmo}(t)$ over time $t$.
In doing this, we assume two assumptions about the energy eigenvalues $E_\alpha$ of the Hamiltonian $\hat{H}$, namely,
\begin{enumerate}
\item
(non-degeneracies)\\
\aln{\label{nd}
 E_{\alpha}= E_{\beta}\Rightarrow \alpha=\beta
}
and
\item
(non-resonances)
\aln{\label{ndg}
 E_{\alpha}- E_{\beta}= E_{\gamma}-E_{\delta} \neq 0\Rightarrow \alpha=\gamma,\beta=\delta.
}

\end{enumerate}
We expand the initial state by the energy eigenstates as
\aln{
\ket{\psi_0}=\sum_\alpha c_\alpha\ket{E_\alpha},
}
where $c_\alpha=\braket{E_\alpha|\psi_0}$.
Then the long-time average of $\braket{\hmo}(t)$ becomes
\aln{\label{avet}
\av{\braket{\hmo}(t)}&=\sum_{\alpha\beta}c_\alpha^*c_\beta \av{e^{i(E_\alpha-E_\beta)t}}\mc{O}_{\alpha\beta}\nonumber\\
&=\sum_\alpha |c_\alpha|^2\mc{O}_{\alpha\alpha}\nonumber\\
&=\Tr[\hrho_\mr{d}\hmo]\nonumber\\
&=\mc{O}_\mr{d},
}
where $\av{\cdots}=\lim_{T\ra\infty}\frac{1}{T}\int_0^T\cdots dt$ denotes the average over time.
We have also introduced 
\aln{
\hrho_\mr{d}=\sum_\alpha |c_\alpha|^2\ket{E_\alpha}\bra{E_\alpha},
}
which is called the diagonal ensemble because it is diagonal in the basis of the energy eigenstates (the subscript ``d" stands for ``diagonal").
Similarly, the variance over time can be calculated as
\aln{\label{vart}
\Delta \mc{O}_t^2&:=\av{[{\braket{\hmo}(t)}-\av{\braket{\hmo}(t)}]^2}\nonumber\\
&=\av{\lrl{\sum_{\alpha\neq\beta}c_\alpha^*c_\beta {e^{i(E_\alpha-E_\beta)t}}\mc{O}_{\alpha\beta}}^2}\nonumber\\
&=\sum_{\alpha\neq\beta,\gamma\neq\delta}c_\alpha^*c_\beta c_\gamma^*c_\delta \av{e^{i(E_\alpha-E_\beta+E_\gamma-E_\delta)t}}\mc{O}_{\alpha\beta}\mc{O}_{\gamma\delta}\nonumber\\
&=\sum_{\alpha\neq\beta}|c_\alpha|^2|c_\beta|^2|\mc{O}_{\alpha\beta}|^2.
}

Using Chebyshev's inequality, Eq. (\ref{avet}) and Eq. (\ref{vart}) lead to
\aln{\label{cht}
\mbb{P}\lrl{\lrv{\braket{\hmo}(t)-\mc{O}_\mr{d}}>\epsilon}\leq \frac{\Delta \mc{O}_t^2}{\epsilon^2}
}
for any $\epsilon>0$, where $\mbb{P}$ denotes the probability with respect to the uniform measure $t\in[0,\infty)$.\footnote{This probability distribution is not normalizable in a rigorous sense, since $\int_0^\infty dt =\infty$. We regard it as the uniform measure in $[0,T]$ for a sufficiently large $T$. Under certain conditions, we can show that \aln{
\frac{1}{T}\int_0^Tdt[{\braket{\hmo}(t)}-\mc{O}_\mr{d}]^2
}
is small when $T$ is sufficiently large. Using Markov's inequality, we can justify equilibration~\cite{Short12}.
}
Then, when $\Delta \mc{O}_t^2$ is sufficiently smal, we can justify Eq. (\ref{ealt}) with $\mc{O}_\mr{d}$ being the long-time average of $\braket{\hmo}(t)$.
In Sec. \ref{sec:appe} and Sec. \ref{sec:role}, we will consider in what situations $\Delta \mc{O}_t^2$ is small in order to justify equilibration.

\subsubsection{Timescales}
Before considering thermalization, we shall make a remark about timescales of equilibration.
Although we have taken the long-time average in Eq. (\ref{avet}) and Eq. (\ref{vart}), we usually do not have to wait for such a long time to observe equilibration.
Indeed, we know that systems approach stationary state in an accessible time $T_\mr{eq}$ by experiments~\cite{Trotzky12} and numerics~\cite{Rigol08}.
Many authors have tried to estimate such timescales of equilibration.
In Ref.~\cite{Goldstein15E}, Goldstein, Hara and Tasaki estimated the timescale with which the state gets out of a typically chosen nonequilibrium subspace of MATE (see Eq. (\ref{wake})).
The obtained timescale is $\sim\frac{\hbar}{k_BT}$, which is unphysically short in general.
The lesson from this observation is that we should not rely on the typicality argument in considering the dynamics (note, however, that the typicality argument might be useful to describe quick prethermalization~\cite{Reimann16}).
Another notable proposition is to consider the operator norm of the commutator between the Hamiltonian and an observable $\hmo$ in order to estimate the slowest timescale of equilibration~\cite{Kim15}.
If $||[\hmo,\hH]||_\mr{op}\leq \chi$, then using the Heisenberg representation for $\hmo$, we obtain
\aln{
\lrv{\lrv{\fracd{\hmo(t)}{t}}}_\mr{op}&= ||e^{i\hH t}[\hmo,\hH]e^{-i\hH t}||_\mr{op}\nonumber\\
&=||[\hmo,\hH]||_\mr{op}\nonumber\\
&\leq\chi,\\
||\hmo(t)-\hmo(0)||_\mr{op}&=\lrv{\lrv{\int_0^td\tau\fracd{\hmo(\tau)}{\tau}}}_\mr{op}\nonumber\\
&\leq \int_0^td\tau\lrv{\lrv{\fracd{\hmo(\tau)}{\tau}}}_\mr{op}\nonumber\\
&\leq \chi t,
}
which implies that $|\braket{\hmo}(t)-\braket{\hmo}(0)|$ remains small for $t\ll\frac{1}{\chi}$.
Thus we expect that equilibration will not occur until $T_\mr{eq}\sim\frac{1}{\chi}$ if $|\mc{O}_\mr{d}-\braket{\hmo}(0)|$ is large.
In Ref.~\cite{Kim15}, the authors investigate $\chi$ to find that there are operators that decay slower than the diffusive modes.
Finally, we notice that the Lieb-Robinson bound~\cite{Lieb72} can be used to estimate the timescale in a system with short-range interactions.
Indeed, we can rigorously show that the effect of local operation in region A is negligible in region B till the time $t\sim L/v$, where $L$ is the distance of regions A and B, and $v$ is the finite velocity that depends on the interactions~\cite{Bravyi06}.

Although many works are present, estimating timescales of equilibration is still an open question.
One of the difficulties is that timescales of equilibration are highly observable- and system-dependent.\footnote{We note that some systems may not show equilibration within accessible timescales~\cite{Banuls11,Kim15}}
Moreover, complex systems show prethermalization, which means that the equilibration cannot be characterized by a single timescale.
The issue of timescales is interesting, but still much remains to be understood, and we will set the issue aside in the following discussions.

\subsection{Thermalization}
Next, we consider if the stationary state is close to the thermal state.
Since we have seen that the expectation values at the stationary state is regarded as the expectation value with respect to the diagonal ensemble, what we have to prove is that $\Tr[\hrho_\mr{d}\hmo]\simeq\Tr[\hrho_\mr{mic}\hmo]$, or more explicitly
\aln{\label{mic}
\sum_\alpha |c_\alpha|^2\mc{O}_{\alpha\alpha}\simeq \frac{1}{d}\sum_{\alpha\in\mc{S}} \mc{O}_{\alpha\alpha},
}
where $d=\dim[\mc{H}_\mr{mic}]$.
We will consider when Eq. (\ref{mic}) is justified in the following sections.
We note, however, that the approach to MITE can also be formulated at the ensemble level: 
\aln{
\hrho_{\mr{d},S}=\hrho_{\mr{mic},S},
}
which requires that all of the operators acting on $S$ should thermalize.

\section{Approach to thermal equilibrium from any initial state: the eigenstate thermalization hypothesis}\label{sec:appe}
In this section, we introduce the eigenstate thermalization hypothesis (ETH), which is one of the main subjects in this thesis.
We will see that the ETH for off-diagonal terms is related to equilibration, and the ETH for diagonal terms is related to thermalization.

The ETH is a statement for matrix elements $\mooD$ in the thermodynamic limit. 
We say that the ETH holds true if for \textit{any} $\alpha,\beta$ in a certain energy range,\footnote{In many nonintegrable systems, \EQ{ETH} seems to hold true for most of the eigenstates except for the edge of the spectrum.}
\aln{\label{ETH}
\mooD\ra \mc{O}_\mr{m}(E_\alpha/V)\:\delta_{\alpha\beta},
}
where the arrow denotes the thermodynamic limit, and $\mc{O}_\mr{m}(x)$ is a smooth function of $x$.
We note that there are some small deviations from the ETH in finite-size systems, and that they decrease with  system sizes (see Chapter \ref{ch:ETH} for details).

\subsection{Off-diagonal matrix elements and equilibration}
First, Eq. (\ref{ETH}) implies that every off-diagonal term vanishes in the thermodynamic limit, namely $\mooD\ra 0\:(\alpha\neq\beta)$.
Then,  we can see that equilibration is justified under the assumption of the ETH, since
\aln{
\Delta\mc{O}_t^2&=\sum_{\alpha\neq\beta}|c_\alpha|^2|c_\beta|^2|\mc{O}_{\alpha\beta}|^2\nonumber\\
&\leq\sum_{\alpha\neq\beta}|c_\alpha|^2|c_\beta|^2[\max_{\alpha\neq\beta}|\mc{O}_{\alpha\beta}|]^2\nonumber\\
&\leq[\max_{\alpha\neq\beta}|\mc{O}_{\alpha\beta}|]^2\sum_{\alpha\beta}|c_\alpha|^2|c_\beta|^2\nonumber\\
&=[\max_{\alpha\neq\beta}|\mc{O}_{\alpha\beta}|]^2\ra 0
}
in the thermodynamic limit.
This argument is applicable for \textit{any} initial state $\ket{\psi_0}$.

\subsection{Diagonal matrix elements and thermalization}
Next, we consider the validity of Eq. (\ref{mic}) from diagonal matrix elements in Eq. (\ref{ETH}).
In order to do this, we make one more assumption: the spread of the energy of $\ket{\psi_0}$ is macroscopically negligible, namely, 
\aln{\label{init}
\delta E:=\sqrt{\braket{\psi_0|\hH^2|\psi_0}-\braket{\psi_0|\hH|\psi_0}^2}=\sum_\alpha |c_\alpha|^2(E-E_\alpha)^2=\mr{o}(V),
}
where $E=\braket{\psi_0|\hH|\psi_0}$, and $V$ is a system size.
This condition is satisfied for usual quench setups~\cite{Rigol08}.
To see one example, we consider a quench where we change the Hamiltonian from $\hat{H}_0$ to $\hH=\hat{H}_0+\hat{V}$, with $\hH_0$ and $\hat{V}$ expressed as the sums of local operators.
We also assume that $\ket{\psi_0}$ is an eigenstate (e.g., a ground state) of $\hH_0$ for simplicity.
Then
\aln{
\delta E^2&={\braket{\psi_0|\hat{V}^2|\psi_0}-\braket{\psi_0|\hat{V}|\psi_0}^2}.
}
If we write $\hat{V}$ as $\hat{V}=\sum_i\hat{v}_i$, where $i$ denotes a lattice site and $\hat{v}_i$ is a localized operator near $i$, then
\aln{
\delta E^2=\sum_{ij}\lrl{\braket{\psi_0|\hat{v}_i\hat{v}_j|\psi_0}-\braket{\psi_0|\hat{v}_i|\psi_0}\braket{\psi_0|\hat{v}_j|\psi_0}}.
}
If $\ket{\psi_0}$ has a cluster property, $\braket{\psi_0|\hat{v}_i\hat{v}_j|\psi_0}-\braket{\psi_0|\hat{v}_i|\psi_0}\braket{\psi_0|\hat{v}_j|\psi_0}$ is sufficiently small when $|i-j|$ becomes large.
Therefore, $\delta E$ scales only subextensively with the system size.

Under the assumption of Eq. (\ref{init}), we can show that, if the system is large enough,
\aln{\label{henkeida}
\sum_\alpha|c_\alpha|^2\moD&\simeq \sum_\alpha|c_\alpha|^2\mc{O}_\mr{m}(E_\alpha/V)\nonumber\\
&=\sum_\alpha|c_\alpha|^2\lrl{\mc{O}_\mr{m}(E/V)+\frac{E_\alpha-E}{V}\mc{O}_\mr{m}'+\frac{1}{2}\lrs{\frac{E_\alpha-E}{V}}^2\mc{O}_\mr{m}''+\cdots}\nonumber\\
&\simeq\mc{O}_\mr{m}(E/V)+\frac{\delta E^2}{2V^2}\mc{O}_\mr{m}''\nonumber\\
&\simeq\mc{O}_\mr{m}(E/V),
}
where $\mc{O}_\mr{m}'(x)=\fracd{\mc{O}_\mr{m}(x)}{x}$.
Similarly, we can show
\aln{
\frac{1}{d}\sum_{\alpha\in\mc{S}} \mc{O}_{\alpha\alpha}&\simeq\mc{O}_\mr{m}(E/V)+\frac{\Delta E^2}{2V^2}\mc{O}_\mr{m}''\nonumber\\
&\simeq\mc{O}_\mr{m}(E/V),
}
which leads to Eq. (\ref{mic}) in the thermodynamic limit ($\Delta E$ is a microcanonical energy shell).
This argument is applicable for \textit{any} initial state $\ket{\psi_0}$ that satisfies Eq. (\ref{init}).

We remark that the ETH for diagonal matrix elements is in a sense a necessary condition for the system to approach thermal equilibrium from any initial state as well.
Consider a composite system with a small system and a bath, $\mc{H}_S\otimes\mc{H}_{S^c}$, which is the setup for MITE.
Roughly speaking, in Ref.~\cite{Palma15}, the authors showed the following:
If all of the product states in the energy shell, $\rho_S\otimes\rho_{S^c}\in\mc{H}_\mr{mic}$, relax to MITE as
\aln{
||\hrho_{\mr{d},S}-\hrho_{\mr{mic},S}||\simeq 0.
}
Then we obtain the ETH for diagonal matrix elements (in the MITE form) as
\aln{
||\hat{\tau}_{\alpha,S}-\hat{\tau}_{\beta,S}||\simeq 0\:\:(\text{for any } \alpha,\beta\in\mc{S}),
}
where $\hat{\tau}_\alpha=\ket{\alpha}\bra{\alpha}$.\footnote{They used the trace norm ($||\hmo||:=\Tr[\sqrt{\hmo^\dag\hmo}]$) for the proof.}

\section{Roles of initial states for equilibration and thermalization}\label{sec:role}
In this section, we review several conditions of initial states, with which equilibration or thermalization is justified.
We have seen that equilibration and thermalization occur for any nonequilibrium initial state in the energy shell, if we admit the ETH.
However, equilibration and thermalization is also shown to occur if the initial state satisfies certain conditions.
For example, integrable systems often equilibrate, though the ETH breaks down in such systems~\cite{Rigol07,Langen15}.
We will see that the so-called effective dimension,
\aln{
d_\mr{eff}:=\frac{1}{\sum_\alpha|c_\alpha|^4}=e^{S_2(\hrho_\mr{d})},
}
where $S_2(\hrho_\mr{d})=-\ln\lrl{\Tr[\hrho_\mr{d}^2]}$ is a R\'enyi-2 entropy, plays important roles for equilibration and thermalization.

First, we consider the condition for equilibration of a Hermitian operator $\hmo$, following Refs.~\cite{Reimann08,Short12}.
We can bound Eq. (\ref{vart}) from above as
\aln{
\Delta \mc{O}_t^2&=\sum_{\alpha\neq\beta}|c_\alpha|^2|c_\beta|^2|\mc{O}_{\alpha\beta}|^2\nonumber\\
&\leq\sum_{\alpha\beta}|c_\alpha|^2|c_\beta|^2|\mc{O}_{\alpha\beta}|^2\nonumber\\
&=\Tr[\hrho_\mr{d}\hmo\hrho_\mr{d}\hmo]\nonumber\\
&=\Tr[\hrho_\mr{d}\hmo(\hmo\hrho_\mr{d})^\dag]\nonumber\\
&\leq\sqrt{\Tr[\hrho_\mr{d}\hmo(\hrho_\mr{d}\hmo)^\dag]\Tr[(\hmo\hrho_\mr{d})^\dag\hmo\hrho_\mr{d}]}\nonumber\\
&=\Tr[\hrho_\mr{d}^2\hmo^2]\nonumber\\
&\leq ||\hmo^2||_\mr{op}\Tr[\hrho_\mr{d}^2]\nonumber\\
&\leq ||\hmo||_\mr{op}^2\Tr[\hrho_\mr{d}^2]\nonumber\\
&=\frac{||\hmo||_\mr{op}^2}{d_\mr{eff}},
}
where we have used the Cauchy-Schwartz inequality $\lrm{\Tr[\hat{A}\hat{B}^\dag]}^2\leq\Tr[\hat{A}\hat{A}^\dag]\Tr[\hat{B}\hat{B}^\dag]$, and the fact that $\Tr[\hat{A}\hat{B}]\leq ||\hat{A}||_\mr{op}\Tr[\hat{B}]$ for positive operator $\hat{A}$ and $\hat{B}$.
We can see that, if the effective dimension $d_\mr{eff}$ is much larger than $||\hmo||_\mr{op}^2$, equilibration occurs thanks to Eq. (\ref{cht}).
In fact, $d_\mr{eff}$ increases exponentially with the system size in many setups regardless of integrability of the systems.
Since we have seen that $||\hmo||_\mr{op}^2$ increases at most as a polynomial in the system size if $\hmo$ is a sum of few-body operators (see Sec. \ref{sec:Deft}), we expect equilibration of such an operator.

Though the proof above relies on several assumptions, namely Eq. (\ref{nd}) and Eq. (\ref{ndg}), we remark that these assumptions can be abandoned or weakened~\cite{Short11,Short12}.
In Ref.~\cite{Short12}, Short and Farrelly discussed equilibration without assuming the non-degenerate and the strict non-resonance conditions.
They considered a Hamiltonian that can be written as $\hH=\sum_\alpha E_\alpha\hat{\mc{P}}_\alpha$, where $\hat{\mc{P}}_\alpha$ is a projection operator onto the Hilbert space with an associated energy $E_\alpha$ (note that we allow degeneracies).
They also introduced the so-called density of energy gaps, which is defined as
\aln{
N(\epsilon):=\max_E\#\lrm{(E_\alpha,E_\beta):E\leq E_\alpha-E_\beta<E+\epsilon,E_\alpha\neq E_\beta},
}
where $\#\lrm{}$ means the number of the elements in the set $\lrm{ }$.
We note that Eq. (\ref{ndg}) corresponds to assuming $\lim_{\epsilon\ra 0+}N(\epsilon)=1$.
Then they showed that 
\aln{
\Delta \mc{O}_t^2\leq \frac{N(\epsilon)||\hmo||_\mr{op}^2}{\tilde{d}_\mr{eff}}
}
for any $\epsilon>0$, where $\tilde{d}_\mr{eff}^{-1}:=\sum_\alpha\lrm{\braket{\psi_0|\hat{\mc{P}}_\alpha|\psi_0}}^2$.\footnote{They derived a more general inequality concerning the finite time average, but we will not discuss it here.}

Next, we comment on the condition about initial states with which the system thermalizes.
Here let us consider a macroscopic observable, which can be written as the average of local operators like $\hmo=\frac{1}{V}\sum_i\hat{o}_i$.
In Ref.~\cite{Mori16W}, Mori showed that, in certain systems (e.g., 1D short-range interacting lattice systems), 
an exponentially small fraction of the energy eigenstates in $\mc{H}_\mr{mic}$ give the different expectation value from the microcanonical ensemble.
In other words, for an arbitrary $\delta>0$, there exists $\gamma>0$ such that
\aln{
\mbb{P}\lrl{|\moD-\braket{\hmo}_\mr{mic}|>\delta}\leq e^{-N\gamma},
}
where $\mbb{P}$ denotes the probability in the uniform distribution of $\alpha$ with $\ket{E_\alpha}\in\mc{H}_\mr{mic}$, and $N$ is the system size.
This is a stronger form of what is called a weak ETH (see Sec. \ref{sec:wETH}).
Using this, the difference between $\braket{\hmo}_\mr{d}$ and $\braket{\hmo}_\mr{mic}$ can be estimated.
Let us denote ${\tilde{\mc{S}}}(\subset\mc{S})$ as the set of $\alpha$'s that satisfy $|\moD-\braket{\hmo}_\mr{mic}|>\delta$.
In a large system, we can approximate $\ket{\psi_0}\in\mc{H}_\mr{mic}$ and thus
\aln{
|\braket{\hmo}_\mr{d}-\braket{\hmo}_\mr{mic}|&\simeq \lrv{\sum_{\alpha\in\mc{S}}|c_\alpha|^2(\moD-\braket{\hmo}_\mr{mic})}\nonumber\\
&\leq \sum_{\alpha\in\mc{S}}|c_\alpha|^2\lrv{\moD-\braket{\hmo}_\mr{mic}}\nonumber\\
&=\sum_{\alpha\in{\tilde{\mc{S}}}}|c_\alpha|^2\lrv{\moD-\braket{\hmo}_\mr{mic}}+\sum_{\alpha\in\mc{S}-{\tilde{\mc{S}}}}|c_\alpha|^2\lrv{\moD-\braket{\hmo}_\mr{mic}}\nonumber\\
&\leq\sqrt{\sum_{\alpha\in{\tilde{\mc{S}}}}|c_\alpha|^4\sum_{\alpha\in{\tilde{\mc{S}}}}\lrv{\moD-\braket{\hmo}_\mr{mic}}^2}+\delta\nonumber\\
&\leq\sqrt{\sum_{\alpha\in\mc{S}}|c_\alpha|^4\sum_{\alpha\in{\tilde{\mc{S}}}}\lrv{\moD-\braket{\hmo}_\mr{mic}}^2}+\delta\nonumber\\
&\leq 2||\hmo||_\mr{op}\sqrt{\frac{1}{d_\mr{eff}}\sum_{\alpha\in{\tilde{\mc{S}}}}1}+\delta.
}
Since $\sum_{\alpha\in{\tilde{\mc{S}}}}1\ra de^{-\gamma N}$ for a large $N$, the condition
\aln{\label{mori}
d_\mr{eff}>e^{-\eta N}d\:\:(0<\eta<\gamma)
}
justifies that $|\braket{\hmo}_\mr{d}-\braket{\hmo}_\mr{mic}|$ is small for a large $N$.
The condition of thermalization in Eq. (\ref{mori}), or similar conditions to it~\cite{Tasaki16,Mori16M}, are tighter than the condition of equilibration (which requires exponentially large $d_\mr{eff}$).
In fact, it is not easy to show in what conditions Eq. (\ref{mori}) is achieved.

\chapter{Review of the eigenstate thermalization hypothesis (ETH)}\label{ch:ETH}
As we have seen in Chapter \ref{ch:Eq}, the ETH is expected to play a crucial role in quantum thermalization of nonintegrable isolated systems.
Though many numerical simulations suggest that the ETH holds true in nonintegrable systems for few-body observables, there are hardly any mathematical proofs of the ETH for a given set of a Hamiltonian and an observable.
We do not have definite criteria of when the ETH holds true, either.
Despite the lack of the complete understanding, possible analytical explanations of the ETH have been proposed since the notable work by von Neumann~\cite{Neumann29}.
These explanations and numerical verifications will provide important clues for mathematical proofs and definite criteria.

\begin{figure}
\begin{center}
\includegraphics[width=14cm]{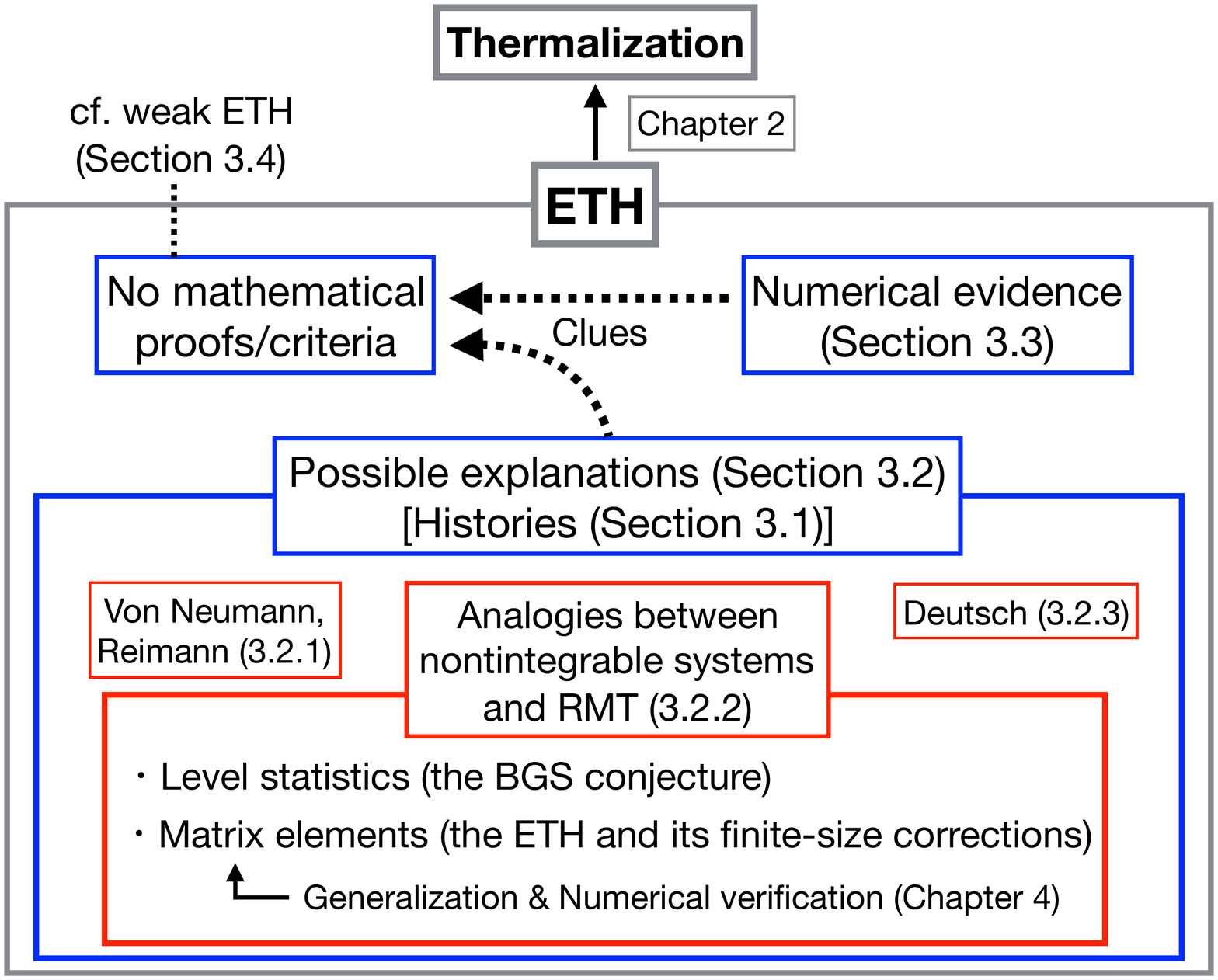}
\caption{The structure of Chapter \ref{ch:ETH}.}
\label{chap3s}
\end{center}
\end{figure}

In this chapter, we review the previous detailed studies on the ETH.
After looking back into the history of the ETH, we review three possible analytical explanations of why the ETH seems true for a wide variety of systems.
In particular, some important relations between the ETH in nonintegrable systems and random matrix theory (RMT) will be discussed in Subsection \ref{sec:sre}.
We also introduce an ansatz that describes the finite-size deviations from the ETH.
Then, we will show some previous numerical results that tested the ETH.
Finally, we remark on what is called the weak ETH.
The structure of this chapter is summarized in Fig. \ref{chap3s}.

\section{Histories}
Historically, von Neumann first tried to justify what is now essentially regarded as the ETH~\cite{Neumann29}.
He originally showed that for almost all decompositions of macrospaces, the ETH holds true.
Thus, his argument is similar (but not equivalent) to the typicality argument that we have reviewed in Chapter \ref{ch:Eq}.
Since the ETH holds true, thermalization also occurs for almost all decompositions of macrospaces.
Von Neumann called this fact the quantum ergodic theorem, but it is actually irrelevant to the classical ergodicity, which incurred the misunderstandings of his work in 1950's.
We note that, though von Neumann originally considered only macroscopic observables, his argument has recently been extended to arbitrary observables by Reimann~\cite{Reimann15}.

The study of the ETH greatly developed from the late 1970's to the 1990's, motivated by the relation between the quantum chaos theory and random matrix theory (RMT).
This relation was especially investigated for quantum systems that have semiclassical limits ($\hbar\ra 0$).
One important achievement is the establishment of two conjectures about the statistics of the energy eigenvalues of the Hamiltonian.
If the corresponding classical system is chaotic, the level-spacing distribution of the eigenvalues shows the Wigner-Dyson distribution, which is predicted by RMT. 
This conjecture is called the Bohigas-Giannoni-Schmit (BGS) conjecture~\cite{Bohigas84}.
On the other hand, if the corresponding classical system is integrable, the level-spacing distribution of the eigenvalues shows the Poisson distribution.
This conjecture is called the Berry-Tabor conjecture~\cite{Berry77}.
Similarly, statistics of the energy eigenstates was investigated especially for semiclassical models.
It is conjectured that, the energy eigenstates are delocalized in phase space if the corresponding classical system is chaotic (Berry's conjecture~\cite{Berry77R}),
which is also consistent with RMT.\footnote{In fact, physical systems may have a negligible fraction of non-delocalized energy eigenstates even in a classically chaotic system, which is called a scar. However, the presence of scars will not be important in the following discussions~\cite{Srednicki94}.}
These observations lead to the work by Peres~\cite{Peres84E}, who conjectured that matrix elements of observables in the basis of the Hamiltonian look random and suggested the notion of the ETH and its finite-size corrections.
Several authors~\cite{Alhassid89,Kus91} also numerically verified that, matrix elements of observables are distributed as Gaussian in systems whose corresponding classical systems are chaotic.
Srednicki also applied Berry's conjecture to show that thermalization occurs thanks to the ETH~\cite{Srednicki94}.

\begin{figure}
\begin{center}
\includegraphics[width=15cm]{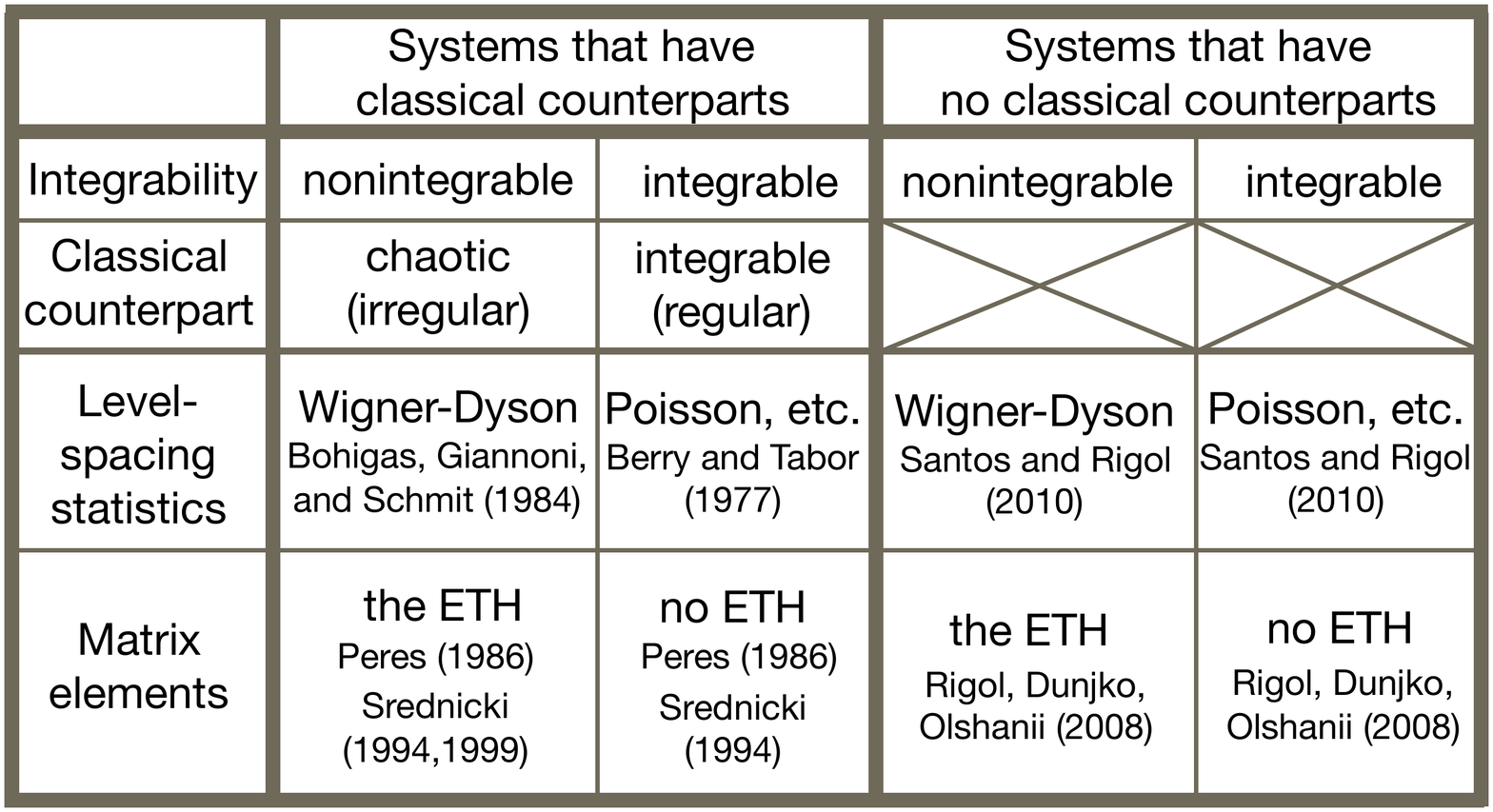}
\caption{Relations between RMT and (non)integrability of systems. We have referred some relevant works on each issue.}
\label{semimatome}
\end{center}
\end{figure}

The connection to RMT also encouraged researchers to investigate the ETH from the viewpoint of (non)integrability, even if many-body quantum systems have no classical counterparts (we have summarized the relations between RMT and (non)integrability in Fig. \ref{semimatome}).
In fact, analogously to the semiclassical situations, level distributions of the eigenvalues are different depending on the integrability of the Hamiltonian.
If it is a nonintegrable system that conserves only energy, the level spacings show the Wigner-Dyson distribution; if it is integrable (e.g., mappable to free systems or solvable by the Bethe ansatz), the level spacings show the Poisson distribution.
A similar classification in light of nonintegrability was sought for eigenstates and matrix elements in general many-body quantum systems.
In 1985, Jensen and Shanker numerically investigated the ETH for nonintegrable and integrable transverse spin chains~\cite{Jensen85}.
In 1991, Deutsch proposed the origin of the ETH by considering the integrable model perturbed by random interactions~\cite{Deutsch91} (see Sec. \ref{sec:deu}).
In 1999, Srednicki developed his semiclassical arguments, and conjectured a general form of matrix elements in nonintegrable systems~\cite{Srednicki99}, which describes the ETH and its finite-size corrections (see Sec. \ref{sec:sre}).
We note that his conjecture is testable for general many-body quantum systems.

These works were rediscovered after Rigol, Dunjko, and Olshanii numerically demonstrated the validity of the ETH in a nonintegrable many-body quantum system and its breakdown in an integrable system~\cite{Rigol08}.
After that, many numerical simulations appeared that tested the ETH for both diagonal and off-diagonal matrix elements of (a sum of) few-body operators~\cite{Rigol09Q,Biroli10,Genway12,Khatami13,Steinigeweg13,Beugeling14,Beugeling15}.
Some of them investigated how the finite-size corrections of the ETH behave~\cite{Khatami13,Steinigeweg13,Beugeling14,Beugeling15,Luitz16}, sometimes referring to the Srednicki's argument (see \EQ{corre}).

Though many works have been done, the complete understanding of the ETH is yet to be made.
Rigorous proofs of the ETH for a given set of a Hamiltonian and an observable are hardly obtained.
We do not have definite criteria of when the ETH holds true, either.\footnote{For example, even though most of the numerical simulations have tested the ETH only for (a sum of) few-body observables in nonintegrable systems that conserve only energy, several studies indicate that the ETH may hold true in wider situations~\cite{Garrison15,Hosur16,Geraedts16}.}
In seeking for mathematical proofs and definite criteria, it is important to understand underlying mechanisms of why the ETH is valid (note that even qualitative understanding has not sufficiently been obtained).
In that sense, the possible analytical explanations and numerical verifications have the meanings of providing clues for understanding such mechanisms of the ETH.

\section{Possible explanations of the ETH}\label{sec:pete}
In this section, we review three possible explanations of the ETH.
First, we review the argument by von Neumann~\cite{Neumann29} and Reimann~\cite{Reimann15}; they investigated the ETH using the notion similar to the typicality.
Second, in Subsection \ref{sec:sre}, we review the analogy between nonintegrable systems and random matrices; this analogy leads to the explanations of the ETH.
We introduce a general form of matrix elements which describes the ETH and its finite-size corrections, which are predicted by Srednicki~\cite{Srednicki99}.
Topics treated in this subsection is especially relevant to Chapter \ref{ch:Obs}.
Third, we briefly explain the argument by Deutsch, who modeled a nonintegrable Hamiltonian as the sum of an integrable Hamiltonian and a random perturbation.

\subsection{Arguments by von Neumann and Reimann}\label{sec:nr}
In this section, we review von Neumann's original work on the justification of the ETH, using the typicality-like argument.
He only considered macroscopic observables, but Reimann has recently extended Von Neumann's argument to an arbitrary observable with a more modern method.
Thus, we follow the paper by Reimann~\cite{Reimann15}.

\subsubsection{The setup and the statement}
As a setup, we consider only one microcanonical energy shell and assume that the initial state is in this energy shell ($\ket{\psi_0}\in\mc{H}_\mr{mic}$).
Then, we need to show the ETH for matrix elements $\mooD$ with $\ket{E_\alpha},\ket{E_\beta}\in\mc{H}_\mr{mic}$ to justify thermalization.
Thus, we can consider an observable $\hat{O}:=\mc{\hat{P}}_\mr{mic}\hmo\mc{\hat{P}}_\mr{mic}$, instead of $\hmo$ itself (note that $\mooD={O}_{\alpha\beta}$, where $\hat{\mathcal{P}}_\mathrm{mic}$ is the projection into $\mathcal{H}_\mathrm{mic}$).
If we diagonalize $\hat{O}$ as $\hat{O}=\sum_{i=1}^da_i\ket{a_i}\bra{a_i}$, where $d=\dim[\mc{H}_\mr{mic}]$, we can define the transformation $U:\mc{H}_\mr{mic}\ra\mc{H}_\mr{mic}$ between the bases $\lrm{\ket{E_\alpha}}$ and $\lrm{\ket{a_i}}$, whose matrix elements are
$U_{\alpha i}:=\braket{E_\alpha|a_i}$.
We note that $U$ can be an arbitrary $d\times d$ unitary matrix if the Hamiltonian has neither unitary nor anti-unitary symmetry.

Von Neumann and Reimann showed that, for almost all $U$ (with respective to the unitary Haar measure), the ETH holds true.
Namely, they showed that for any $\epsilon>0$,
\aln{
&\mbb{P}\lrl{\max_{\alpha\in\mc{S}} |\moD-\braket{\hmo}_\mr{mic}|\geq\epsilon}\leq 2d\exp\lrl{-\frac{2\epsilon^2 d}{9\pi^3\Delta_{\hat{O}}^2}} \label{ddayo},\\
&\mbb{P}\lrl{\max_{\alpha,\beta\in\mc{S},\alpha\neq\beta} |\mooD|\geq\epsilon}\leq 4d(d-1)\exp\lrl{-\frac{\epsilon^2 d}{18\pi^3\Delta_{\hat{O}}^2}}\label{oddayo},
}
where $\mbb{P}$ denotes the probability over $U$ with respect to the unitary Haar measure, and
\aln{
\Delta_{\hat{O}}=\max_ia_i-\min_ia_i.
}
Before proving the inequalities, let us explain the meanings of these inequalities.
When $d$ is large enough, the right-hand sides of Eq. (\ref{ddayo}) and Eq. (\ref{oddayo}) become negligibly small, and the ETH occurs for almost all $U$.
Thus, for a physically relevant set of a Hamiltonian and an observable as well, it is not unnatural to consider $U$ as being ``typical," which leads to the ETH.
We note that the uniform Haar distribution of $U$ can be formally regarded as taking a randomly sampled matrix for $\mc{\hat{P}}_\mr{mic}\hH\mc{\hat{P}}_\mr{mic}$ with a fixed observable.\footnote{In fact, a random matrix whose probability distribution is invariant under arbitrary unitary transformations has eigenstates that are distributed uniformly with respect to the unitary Haar measure. See Appendix \ref{ch:RMT}.}
Instead, it can also be regarded as taking a randomly sampled $\hat{O}$ with a fixed Hamiltonian, which is close to the original argument of macrospaces by von Neumann.

\subsubsection{Proof}
Here we prove Eq. (\ref{ddayo}) and Eq. (\ref{oddayo}).
In order to deal with the probability with respect to the unitary Haar measure, 
we use Levy's lemma~\cite{Popescu05,Popescu06}:
\aln{
\mr{Prob}\lrl{|g(\phi)-\braket{g(\phi)}_\phi|\geq\epsilon}\leq 2\exp\lrl{-\frac{\epsilon^2(d'+1)}{9\pi^3\eta^2}}
}
for any $\epsilon>0$.
Here ``$\mr{Prob}$" means the probability with respect to uniformly distributed random points $\phi$ on a  $d'$-dimensional unit sphere $\mbb{S}^{d'}\subset\mbb{R}^{d'+1}$, and $\braket{\cdots}_\phi$ denotes the average over $\phi$.
A function $g(\phi): \mbb{S}^{d'}\ra\mbb{R}$ is a Lipshitz continuous function with a Lipshitz constant $\eta$.
In our case, $\ket{\phi}\in\mc{H}_\mr{mic}$ represents a point $\phi$ on a $(2d-1)$-dimensional unit sphere ($d'=2d-1$).
Moreover, we can show that $g(\phi)=\braket{\phi|\hat{O}|\phi}$ is a Lipschitz continuous with $\eta=\Delta_{\hat{O}}$ because~\cite{Popescu05}
\aln{
|\braket{\phi|\hat{O}|\phi}-\braket{\psi|\hat{O}|\psi}|&=|\braket{\phi|\hat{O}-X_{\hat{O}}/2|\phi}-\braket{\psi|\hat{O}-X_{\hat{O}}/2|\psi}|\nonumber\\
&=\frac{1}{2}|(\bra{\phi}+\bra{\psi})\hat{O'}(\ket{\phi}-\ket{\psi})+(\bra{\phi}-\bra{\psi})\hat{O'}(\ket{\phi}+\ket{\psi})|\nonumber\\
&\leq ||\hat{O'}||_\mr{op}|\ket{\phi}-\ket{\psi}||\ket{\phi}+\ket{\psi}|\nonumber\\
&\leq 2||\hat{O'}||_\mr{op}|\ket{\phi}-\ket{\psi}|\nonumber\\
&=\Delta_{\hat{O}}|\ket{\phi}-\ket{\psi}|,
}
where $\hat{O'}:=\hat{O}-X_{\hat{O}}/2$ and $X_{\hat{O}}:=\max_ia_i+\min_ia_i$.
We also note that $\braket{g}_\phi=\braket{\hat{O}}_\mr{mic}$ which can be obtained by the same calculation done in obtaining Eq. (\ref{ev}).
Then, observing that randomizing $\ket{\phi}$ and randomizing $U$ are equivalent, we obtain
\aln{
\mbb{P}\lrl{\lrv{\braket{\phi|\hat{O}|\phi}-\braket{\hat{O}}_\mr{mic}}\geq \epsilon}\leq 2\exp\lrl{-\frac{2\epsilon^2 d}{9\pi^3 \Delta_{\hat{O}}^2}}.
}

Now, we will prove Eq. (\ref{ddayo}).
We have
\aln{
\mbb{P}\lrl{\lrv{{O}_{\alpha\alpha}-\braket{\hat{O}}_\mr{mic}}\geq \epsilon}\leq 2\exp\lrl{-\frac{2\epsilon^2 d}{9\pi^3 \Delta_{\hat{O}}^2}}
}
for any $\ket{E_\alpha}\in \mc{H}_\mr{mic}$.
To deal with ``max" in Eq. (\ref{ddayo}), we note that for an arbitrary set of functions $\{f_\rho\}_\rho$, we have
\aln{\label{tasuyo}
\mbb{P}\lrl{\max_\rho\lrs{f_\rho}\geq a} &=\mbb{E}\lrl{\theta\lrs{\max_\rho\lrs{f_\rho}- a}} \nonumber\\
&\leq \mbb{E}\lrl{\sum_\rho\theta\lrs{{f_\rho}- a}} \nonumber\\
&= \sum_\rho\mbb{E}\lrl{\theta\lrs{{f_\rho}- a}}\nonumber\\
&= \sum_\rho \mbb{P}\lrl{f_\rho \geq a}.
}
Here $\theta(\cdot)$ denotes a step function.
Applying this, we obtain
\aln{
\mbb{P}\lrl{\max_{\alpha\in\mc{S}}\lrv{{O}_{\alpha\alpha}-\braket{\hat{O}}_\mr{mic}}\geq \epsilon}&\leq \sum_{\alpha\in\mc{S}}2\exp\lrl{-\frac{2\epsilon^2 d}{9\pi^3 \Delta_{\hat{O}}^2}}\nonumber\\
&=2d\exp\lrl{-\frac{2\epsilon^2 d}{9\pi^3 \Delta_{\hat{O}}^2}}.
}
This is equivalent to Eq. (\ref{ddayo}).

To prove Eq. (\ref{oddayo}), we use the following inequality which is proven in Appendix \ref{sec:dtood}. Namely, for any $\epsilon>0$,
\aln{\label{ukky}
\mbb{P}\lrl{|{O}_{\alpha\beta}|\geq \epsilon}&\leq 4\mbb{P}\lrl{|\braket{\phi|\hat{O}|\phi}-\braket{\hat{O}}_\mr{mic}|\geq \frac{\epsilon}{2}}\nonumber\\
&\leq 8\exp\lrl{-\frac{\epsilon^2 d}{18\pi^3 \Delta_{\hat{O}}^2}},
}
where $\ket{\phi}$ is some state in the energy shell, and $\alpha\neq\beta$.
Applying this and Eq. (\ref{tasuyo}), we obtain
\aln{
\mbb{P}\lrl{\max_{\alpha,\beta\in\mc{S},\alpha\neq\beta} |O_{\alpha\beta}|\geq\epsilon}\leq 4d(d-1)\exp\lrl{-\frac{\epsilon^2 d}{18\pi^3\Delta_{\hat{O}}^2}},
}
which is equivalent to Eq. (\ref{oddayo}).

\subsection{Some predictions from random matrix theory in nonintegrable systems}\label{sec:sre}
In the previous subsection, we saw that we can rigorously show the ETH for almost all $U$'s.
However, there is no \textit{a priori} reason to believe that the typicality of $U$ with respect to the unitary Haar measure is physically meaningful.
In fact, we can easily find physical systems that have an atypical $U$ by taking them integrable or many-body localized systems, as we have seen in the overview.
From this observation, it is reasonable to attribute the validity of the ETH to nonintegrability of the system.
Actually, many previous studies suggested that nonintegrable systems have in  common with random matrix theory (RMT), which also indicates how matrix elements of an observable behave.

In this section, we explain how nonintegrability of the system is related to RMT and the ETH.
We first review the BGS conjecture, which connects the level-spacing statistics of nonintegrable systems with those of RMT.
Next, we consider eigenvectors and matrix elements.
We explain how the ETH is derived from a model of RMT, with a brief review of the works by Srednicki~\cite{Srednicki94} and others.
Finally, we introduce a general form of matrix elements in nonintegrable systems conjectured by Srednicki~\cite{Srednicki99}, which describes the ETH and its finite-size corrections.
We summarize the results of this subsection in Fig. \ref{taiomatome}.
For the sake of self-containedness, we summarize the basics of RMT in Appendix \ref{ch:RMT}.

\begin{figure}
\begin{center}
\includegraphics[width=13cm]{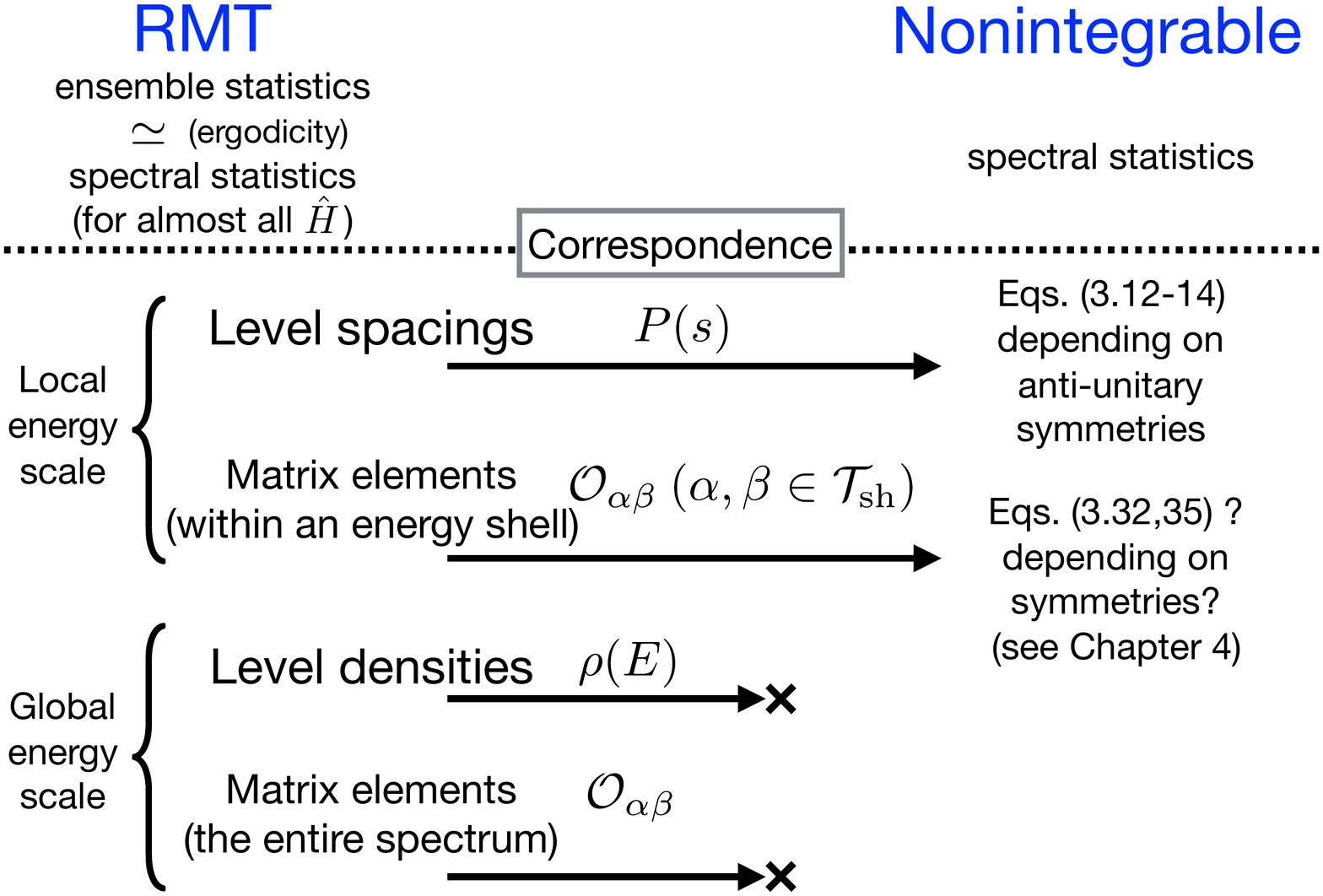}
\caption{A summary of Subsection \ref{sec:sre}. Statistics of certain quantities for RMT and for the nonintegrable systems is related. Note that the statistics for RMT means either the ensemble statistics or the spectral statistics if we assume the ergodicity of random matrices (see the main text). We have a correspondence of level-spacing distributions depending on anti-unitary symmetries of the nonintegrable systems.
It is also expected that there exists a correspondence of matrix-element statistics within a small energy shell, though it is less confirmed than that of the level-spacing statistics.
We note that there is no correspondence if we consider quantities that are related to the global energy scale.}
\label{taiomatome}
\end{center}
\end{figure}

\subsubsection{Level-spacing statistics of random matrices and nonintegrable systems}
We first consider level-spacing distributions of the Gaussian random matrices introduced by Dyson, namely the Gaussian unitary ensemble (GUE), the Gaussian orthogonal ensemble (GOE), and the Gaussian symplectic ensemble (GSE) (see Appendix \ref{ch:RMT}). 
Roughly speaking, matrices without any anti-unitary symmetry $\hat{T}$ belong to the GOE,
matrices with only one anti-unitary symmetry $\hat{T}\:(\hat{T}^2=1)$ belong to the GOE, and
matrices with only one anti-unitary symmetry $\hat{T}\:(\hat{T}^2=-1)$ belong to the GSE.
A level-spacing distribution $P(s)$ is defined as the probability density for two neighboring energy levels $S_{\alpha+1}$ and $S_\alpha$ to have a spacing equal to $s$.
Here we assume that the spectrum $\lrm{S_\alpha}$ is obtained by renormalizing the original spectrum $\lrm{E_\alpha}$ by the so-called unfolding procedure~\cite{Guhr98}, and that the mean level density of $\lrm{S_\alpha}$ is set to unity.
It is known that the level-spacing distributions for $D\times D$ Gaussian random matrices are well approximated by those of $2\times 2$ Gaussian random matrices.
For each of the three ensembles, the level-spacing distribution is given by the following Wigner-Dyson distribution:\footnote{As we will see, a matrix in the GSE has a doubly degenerate spectrum (the Kramers degeneracy). In this case, the level-spacing distribution $P_\mr{GSE}(s)$ is defined from the non-degenerate neighboring levels.}
\aln{\label{stat}
P_\mr{GUE}(s)&=\frac{32s^2}{\pi^2}e^{-\frac{4s^2}{\pi}},\\
P_\mr{GOE}(s)&=\frac{\pi s}{2}e^{-\frac{\pi s^2}{4}},\\
P_\mr{GSE}(s)&=\frac{2^{18}s^4}{3^6\pi^3}e^{-\frac{64s^2}{9\pi}},
}
where we have assumed normalization conditions for a probability density as $\int_0^\infty ds P(s)=1$, and for a first moment of $s$ as $\int_0^\infty ds sP(s)=1$.
The important feature of these distributions is that they all show $p(s\ra 0+)=0$, which means the level repulsions.
This feature cannot be seen in the uncorrelated level statistics that leads to the following Poissonian form:
\aln{\label{pois}
P_\mr{P}(s)=e^{-s}.
}

To clarify the meaning of the statistics, we introduce the notion of what is called the ``ergodicity of random matrices," which relates the spectral statistics and the ensemble statistics.\footnote[2]{Note that this notion is rather different from the usual ergodicity, which relates the long-time average and the phase-space average.}
Although the discussion of the previous paragraph considered the ensemble statistics of random Hamiltonians, we can also consider the spectral statistics for a randomly sampled single Hamiltonian, where we make a histogram of $S_{\alpha+1}-S_\alpha$ for $d_s$ different $\alpha$'s.
The statement of the ergodicity of random matrices is the following:
when the dimension of the random matrices $D$ and the number of the samplings $d_s$ is sufficiently large, the ensemble statistics and the spectral statistics coincide to a certain accuracy, for almost all fixed Hamiltonians randomly sampled from the ensemble.
The ergodicity is proven for certain quantities, which include level-spacing statistics~\cite{Pandley79,Brody81}.
Thus, we assume that we can also regard \EQ{stat} and the other statistics as the spectral statistics of a fixed Hamiltonian.

The BGS conjecture states that the level spacings of quantum systems whose classical counterparts exhibit chaos show the Wigner-Dyson statistics.
According to the symmetry of the system, the level statistics change to those of the random matrix with the same symmetry class.
Despite the absence of the complete proof, the BGS conjecture is verified in many concrete situations.
We note that for quantum systems whose classical counterparts are completely integrable, the Poisson statistics in \EQ{pois} are expected to be applicable as implied by the Berry-Tabor conjecture~\cite{Berry77}.

Though the BGS conjecture was originally proposed for quantum systems that have classical counterparts,
it is now known that the level-spacing statistics seems to be related to nonintegrability of general quantum systems that may not have classical counterparts. 
Many integrable systems, which include noninteracting systems, systems mappable to free systems and systems solvable by the Bethe ansatz, show the level-spacing statistics that is Poissonian or more degenerate.\footnote{For example, $P(s)$ has a delta-function peak at $s=1$ for a single-mode harmonic oscillator.}
On the other hand, nonintegrable systems that conserve only energy are expected to show the Wigner-Dyson statistics.
We note that if the nonintegrable system has unitary symmetries, the Hamiltonian is block-diagonalized and the level repulsions become unclear because eigenstates from different symmetry sectors are uncorrelated.
In that case, by restricting the symmetry sectors, the Wigner-Dyson distributions are obtained within the sectors~\cite{Santos10a}.

We comment on the analogy of the level statistics of Gaussian random matrices and nonintegrable systems.
We have seen that the level-spacing distributions are similar between the two, but they only measure level  correlations in local energy scale.
On the other hand, if we consider quantities related to the global energy scale such as the level density $\rho(E)=\frac{1}{D}\sum_{\alpha=1}^D\delta(E-E_\alpha)$, these two are different.
In fact, the Gaussian random matrices predict the following the ``semicircle law"~\cite{Haake,Brody81,Guhr98}:
\aln{
\av{\rho(E)}=
 \begin{cases}
     \frac{1}{\lambda\pi}\sqrt{1-\lrs{\frac{E}{2\lambda}}^2} & \text{for  }|E|<2\lambda;\\
     0 & \text{for }|E|>2\lambda,
  \end{cases}
}
where $\av{\cdots}$ denotes the ensemble average, and $\lambda=\sqrt{D\av{|H_{ij}|^2}}$ for the GUE.
This expression is not valid in realistic nonintegrable systems.
One of the reasons for the discrepancy is that the Hamiltonian in physical systems consists of few-body and local interactions, unlike the Hamiltonian of Gaussian random matrices.
Some other random matrices are proposed to deal with such physical structures~\cite{Brody81}, but we will not discuss them because it is difficult to analyze them in general (see, however, Subsection \ref{sec:deu}).

\subsubsection{Matrix elements from the viewpoint of RMT}
Next, let us examine how the eigenstates of random matrices predict matrix elements of observables, following the discussion similar to Ref.~\cite{DAlessio16}.
We calculate the ensemble average of diagonal and off-diagonal matrix elements of an observable $\hat{O}=\sum_ia_i\ket{a_i}\bra{a_i}$.
The matrix elements can be written as
\aln{
{O}_{\alpha\beta}=\sum_ia_iU_{\alpha i}U_{i\beta},
}
where $U_{\alpha i}:=\braket{E_\alpha|a_i}$ denotes a basis transformation.
Let us assume that the Hamiltonian belongs to the GUE and that the observable is fixed.
Then, it is known that $U$ is distributed uniformly with respect to the unitary Haar measure (see Appendix \ref{ch:RMT}).
In this case, we have the following moments of $U$~\cite{Brody81}:
\aln{
\av{U_{\alpha i}U_{i\beta}}&=\frac{1}{d}\delta_{\alpha\beta},\\
\av{|U_{\alpha i}|^2|U_{i\beta}|^2}&=\frac{1+\delta_{\alpha\beta}}{d(d+1)},\\
\av{|U_{\alpha i}|^2|U_{j\alpha}|^2}&=\frac{1+\delta_{ij}}{d(d+1)},\\
\av{U_{\alpha i}U_{i\beta}U_{\beta j}U_{j\alpha}}&=-\frac{1}{d(d-1)(d+1)}\:\:\:(\alpha\neq\beta,i\neq j)\label{highyo},
}
where $\av{\cdots}$ denotes the ensemble average (the average with respect to the unitary Haar measure), and $d$ denotes the dimension of the matrices.
These lead to the following average and the variance of the matrix elements:\footnote{We note that in Ref.~\cite{DAlessio16} the variance is different from our calculations because the authors of Ref.~\cite{DAlessio16} ignore some correlations such as in \EQ{highyo}, which contribute to the lowest-order term after the summation of $i$ and $j$.}
\aln{\label{sim}
\av{O_{\alpha\beta}}&=\frac{\delta_{\alpha\beta}}{d}\sum_ia_i,\\
\av{O_{\alpha\alpha}^2}-\av{O_{\alpha\alpha}}^2&=\sum_{ij}a_ia_j\av{|U_{\alpha i}|^2|U_{j\alpha}|^2}-\lrs{\frac{1}{d}\sum_ia_i}^2\NON
&=\frac{1}{d(d+1)}\sum_ia_i^2+\frac{1}{d(d+1)}\lrs{\sum_{i}a_i}^2-\lrs{\frac{1}{d}\sum_ia_i}^2\NON
&=\frac{1}{d+1}\lrl{\frac{1}{d}\sum_ia_i^2-\lrs{\frac{1}{d}\sum_ia_i}^2},
}
and
\aln{
\av{|O_{\alpha\beta}|^2}&=\sum_{ij}a_ia_j\av{U_{\alpha i}U_{i\beta}U_{\beta j}U_{j\alpha}}\NON
&=\frac{1}{d(d+1)}\sum_ia_i^2-\frac{1}{d(d-1)(d+1)}\sum_{i\neq j}a_ia_j\NON
&=\frac{d}{(d+1)(d-1)}\lrl{\frac{1}{d}\sum_ia_i^2-\lrs{\frac{1}{d}\sum_ia_i}^2}\:\:\:(\alpha\neq\beta).
}
If $d$ is sufficiently large, matrix elements are written as
\aln{\label{rmtcal}
{O}_{\alpha\beta}\simeq\frac{\delta_{\alpha\beta}}{d}\sum_ia_i+\sqrt{\frac{1}{d}\lrl{\frac{1}{d}\sum_ia_i^2-\lrs{\frac{1}{d}\sum_ia_i}^2}}R_{\alpha\beta},
}
where $R_{\alpha\beta}$ is a random variable that satisfies $\av{R_{\alpha\beta}}=0$ and $\av{|R_{\alpha\beta}|^2}=1$.
Note that the second term in \EQ{rmtcal} is much smaller than the first term due to the factor $\frac{1}{\sqrt{d}}$.

Though the discussion above is based on the ensemble average, we can reinterpret this as the spectral average, if we assume the ergodicity of random matrices for a function $g$ of the matrix elements.
Here we assume that we make samplings from the eigenstates in some Hilbert space $\mc{H}_s$ with $\dim[\mc{H}_s]=d_s\:(1\ll d_s\leq d)$. We define $\mc{T}$ as a set of labels of the eigenstates in $\mc{H}_s$. 
In this case, the ergodicity states that, for most of the fixed Hamiltonians randomly sampled from the ensemble, we have
\aln{
\braket{g({O}_{\alpha\alpha})}_\mc{T}\simeq\av{g(O_{\alpha\alpha})},\\
\braket{g({O}_{\alpha\beta})}_\mc{TT}\simeq\av{g(O_{\alpha\beta})},
}
where 
\aln{\label{save}
\braket{g(O_{\alpha\alpha})}_\mc{T}:=\frac{1}{d_s}\sum_{\alpha\in \mc{T}}g(O_{\alpha\alpha})\\
}
and
\aln{
\braket{g(O_{\alpha\beta})}_\mc{TT}:=\frac{1}{d_s(d_s-1)}\sum_{\alpha,\beta\in \mc{T};E_\alpha\neq E_\beta}g(O_{\alpha\beta})
}
denotes the spectral average for the diagonal and the off-diagonal matrix elements, respectively.
The ergodicity is proven for a wide class of $g$~\cite{Brody81}, so we will assume it in the following discussions.
Then, \EQ{rmtcal} can be regarded as the matrix elements for a fixed Hamiltonian which fluctuate from eigenstates to eigenstates, satisfying $\braket{R_{\alpha\alpha}}_\mc{T}=\braket{R_{\alpha\beta}}_\mc{TT}=0$ and 
$\braket{|R_{\alpha\alpha}|^2}_\mc{T}=\braket{|R_{\alpha\beta}|^2}_\mc{TT}=1$.
Note, however, that the Hermiticity of the observable requires that $R_{\alpha\beta}=R_{\beta\alpha}^*$.

\subsubsection{Relations to nointegrable systems}
The statistics of energy eigenstates and that of matrix elements of observables in physical systems have been investigated especially in systems that have classical counterparts.
In Ref.~\cite{Berry77R}, Berry conjectured that, in quantum systems whose classical counterparts are chaotic, an energy eigenfunction $\psi_\alpha(\mbf{x})$ is approximated as a Gaussian random function of $\mbf{x}$.
He also suggested that such a Gaussian structure does not arise in systems whose classical counterparts are integrable.
As Peres~\cite{Peres84E} and Srednicki~\cite{Srednicki94} pointed out, the randomness of the eigenstates in quantum chaotic systems leads to the randomness of matrix elements of observables.
In particular, Srednicki replaced the statistics of $\mbf{x}$ used by Berry with the spectral statistics within the energy shell\footnote{In Srednicki's 1994 paper, he used the term ``eigenstate ensemble" for the statistics. However, his subsequent papers~\cite{Srednicki96,Srednicki99} suggest that he also considered the spectral statistics.}
and derived the ETH for momentum distributions of a single particle in a dilute gas.

We expect that the RMT predictions for matrix elements of an observable $\hmo$ also apply to general nonintegrable systems in analogy with the BGS conjecture.
In this case, we may have to consider a sufficiently narrow energy shell in applying \EQ{rmtcal}.
We thus consider projecting an observable onto an energy shell $\mc{H}_\mr{sh}$ as $\hat{O}:=\mc{\hat{P}}_\mr{sh}\hmo\mc{\hat{P}}_\mr{sh}=\sum_{i=1}^{d_\mr{sh}}a_i\ket{a_i}\bra{a_i}$, where $d_\mr{sh}=\dim[\mc{H}_\mr{sh}]$.
Here $\mc{H}_\mr{sh}$ is a Hilbert space spanned by the energy eigenstates $\lrm{\ket{E_\alpha}}_{\alpha\in\mc{T}_\mr{sh}}$, where
\aln{
\mc{T}_\mr{sh}=\lrm{\alpha:|E-E_\alpha|<\omega_\mr{sh}},
}
and $\mc{\hat{P}}_\mr{sh}$ is a projection onto $\mc{H}_\mr{sh}$.
In nonintegrable systems, the transformation $U$ between $\lrm{\ket{E_\alpha}}_{\alpha\in\mc{T}_\mr{sh}}$ and 
$\lrm{\ket{a_i}}$ is expected to be so complex that we can conjecture that the RMT model \EQ{rmtcal} applies within this energy shell.
Rewriting \EQ{rmtcal}, we obtain
\aln{\label{semai}
\mooD\simeq\braket{\hmo}_\mr{sh}(E_\alpha)\delta_{\alpha\beta}+\sqrt{\frac{\mc{V}_\mr{sh}(\hmo)}{d_\mr{sh}}}R_{\alpha\beta}\:\:\:(\alpha,\beta\in\mc{T}_\mr{sh}),
}
where 
\aln{
\braket{\hmo}_\mr{sh}(E)&:=\frac{1}{d_\mr{sh}}\sum_{\alpha\in\mc{T}_\mr{sh}}\mc{O}_{\alpha\alpha},\\
\mc{V}_\mr{sh}(\hmo)&:=\frac{1}{d_\mr{sh}}\sum_{\alpha,\beta\in\mc{T}_\mr{sh}}|\mc{O}_{\alpha\beta}|^2-\lrl{\frac{1}{d_\mr{sh}}\sum_{\alpha\in\mc{T}}\mc{O}_{\alpha\alpha}}^2\nonumber\\
&=\braket{\hmo\mc{\hat{P}}_\mr{sh}\hmo}_\mr{sh}-\braket{\hmo}_\mr{sh}^2.
}
Note that $\omega_\mr{sh}$ is expected to be determined from the Hamiltonian and the observables.
In numerics, we take samplings from $d_s\:(1\ll d_s\leq d_\mr{sh})$ energy eigenstates that satisfy $\mc{T}=\lrm{\alpha:|E-E_\alpha|<\omega_s(<\omega_\mr{sh})}$.

The important point in \EQ{semai} is that RMT predicts that nonintegrable systems have matrix elements with the following properties:
\begin{enumerate}
\item
The diagonal matrix elements fluctuate around $\braket{\hmo}_\mr{sh}(E_\alpha)$, and the off-diagonal matrix elements fluctuate around zero. The fluctuations decrease exponentially with the system size (because of the factor $\frac{1}{\sqrt{d_\mr{sh}}}$).
\item  
The statistics of the fluctuations is the same for any choice of $\alpha$ and $\beta$ within the energy shell.
\item
For the GUE, the ratio $r$ of variances between diagonal and off-diagonal matrix elements is universally one. (We will see in Chapter \ref{ch:Obs} that the change of the symmetry class leads to different values of $r$.)

\end{enumerate}

\subsubsection{The ETH and its finite-size corrections}
Though \EQ{semai} only concerns the matrix elements within the energy shell $\mc{H}_\mr{sh}$,
Srednicki~\cite{Srednicki99} predicted that the matrix elements for the entire spectrum can be written down as
\aln{\label{corre}
\mooD\simeq O_\mr{m}(E)\delta_{\alpha\beta}+e^{-S(E)/2}f(E,\omega)R_{\alpha\beta},
}
where $E:=\frac{E_\alpha+E_\beta}{2},\omega:=E_\alpha-E_\beta,O_\mr{m}(E):=\braket{\hmo}_\mr{sh}(E)=\mc{O}_\mr{m}(E/V)$ (see \CH{ch:Eq}), $S(E)$ is the microcanonical entropy at energy $E$, and $f(E,\omega)$ is a mildly varying function of $E$ and $\omega$.
In addition, Hermiticity requires that $R_{\alpha\beta}=R_{\beta\alpha}^*$ and $f(E,\omega)={f(E,-\omega)}^*$.
The factor $e^{-S(E)/2}$ ensures that the second term vanishes exponentially with the system size, which is also the case in \EQ{semai}.\footnote{We note that the number of states $e^{S(E)}$ has some ambiguity because it depends on the width of the energy shell. In this thesis, we do not care about the exact value of $e^{S(E)}$ and just notice that $e^{S(E)}$ is expected to increase exponentially with the system size.}
Thus, if $\mr{Prob}[|R_{\alpha\beta}|\gg 1]$ is sufficiently small, the ETH is satisfied.
In other words, the second term describes the finite-size corrections from the ETH.
We also note that RMT requires that $f(E,\omega)$ is almost constant for $\omega<\omega_\mr{sh}$.

Although Eqs. (\ref{semai}) and (\ref{corre}) can be tested in general quantum many-body systems, there are less numerical or analytical studies on them than the ETH (for numerical studies, see \SEC{sec:num}).
In particular, it is not yet clear how universally Eqs. (\ref{semai}) and (\ref{corre}) describe the matrix elements of observables in nonintegrable systems.
We will investigate these problems in Chapter \ref{ch:Obs}.

\subsection{Argument by Deutsch}\label{sec:deu}
Finally, we briefly explain the essence of Deutsch's argument, following his original paper~\cite{Deutsch91} and recent developments~\cite{Reimann15E,Borgonovi16}.
In his formulation, we model a nonintegrable system as a Hamiltonian $\hH$, which can be written as an integrable Hamiltonian $\hH_0$ plus an integrability-breaking perturbation $\hat{V}$:
\aln{
\hH=\hH_0+\hat{V}.
}
For example, we can consider a situation in which $\hH$ is the Hamiltonian of a non-interacting gas and $\hat{V}$ is a weak interaction between the particles.
Let $\ket{n}{}_0$ be an eigenstate of $\hH_0$ with an eigenvalue $E_m^0$.
Then
\aln{
V_{nm}^0:={}_0\braket{n|\hat{V}|m}{}_0
}
is expected to be a sparse, banded matrix whose matrix elements rapidly decay with increasing $|n-m|$.
We will treat $V_{nm}^0$ as being sampled from an ensemble of random matrices that imitate certain physical properties such as the banded structure or sparsity.
If we can show the ETH for almost all $\hat{V}$, then we expect that it is true for physical perturbations, which is the spirit of the typicality (this is similar to what we saw in Subsection \ref{sec:nr}).

The randomness of $\hat{V}$ leads to the randomness of $\ket{E_\alpha}$ which is the eigenstate of $\hH$.
Thus the transformation of the basis between perturbed and unperturbed eigenstates 
\aln{
\mc{U}_{\alpha m}=\braket{E_\alpha|m}{}_0
}
is also randomized.
We define matrix elements of $\hmo$ in the basis of the unperturbed energy eigenstates
\aln{
\mc{O}_{nm}^0:={}_0\braket{n|\hmo|m}{}_0,
}
in addition to those in the basis of the perturbed eigenstates,
$\mooD:=\braket{E_\alpha|\hmo|E_\beta}$.
Note that $\mc{O}_{nm}^0$ is a non-random quantity.
Using $\mc{U}$, these matrix elements can be related to each other as
\aln{
\mooD=\sum_{nm}\mc{U}_{\alpha n}\mc{O}_{nm}^0\mc{U}_{\beta m}^*.
}

In order to justify the ETH (for diagonal matrix elements) for most $\hat{V}$, we have to show that $\moD-\mc{O}_{\beta\beta}$ is sufficiently small for most of the $\hat{V}$'s, if $E_\alpha$ and $E_\beta$ are close to each other (note that we do not consider an explicit energy shell in contrast to the previous subsections).
In Ref.~\cite{Reimann15E},  Reimann justified this by the following two steps.
First, he showed that the difference between the expectation values with respect to neighboring eigenstates
\aln{\label{tonari}
|\braket{\mc{O}_{\alpha+1,\alpha+1}}_V-\braket{\moD}_V|
}
is sufficiently small, where $\braket{\cdots}_V$ denotes the average over $\hat{V}$.
Second, he showed that the variance over $\hat{V}$, 
\aln{\label{bunbun}
\sigma_\alpha^2:=\braket{(\moD-\braket{\moD}_V)^2}_V=\braket{\moD^2}_V-\braket{\moD}_V^2
}
is sufficiently small.
From the smallness of Eq. (\ref{tonari}), we can say that $|\braket{\moD}_V-\braket{\mc{O}_{\beta\beta}}_V|$ is small if $|\alpha-\beta|$ is sufficiently small.
Moreover, $\moD\simeq \braket{\moD}_V$ from the smallness of Eq. (\ref{bunbun}) for almost all $\hat{V}$'s, so
$\moD\simeq\mc{O}_{\beta\beta}$ is concluded.

In order to prove that Eq. (\ref{tonari}) and Eq. (\ref{bunbun}) are small, we need to calculate the moments of $\mc{U}$,
\aln{
\braket{\mc{U}_{\alpha_1 n_1}\mc{U}_{\alpha_2 n_2}\cdots}_V.
}
Calculating the averages over a banded random matrix $\hat{V}$ is much more difficult than calculating the full random matrix average that we used in the previous subsections.
Therefore, here we only mention some known results for the second moments. These results enable us to believe that 
Eq. (\ref{tonari}) is small.
For simplicity, we model a banded random matrix $\hat{H}=\hH_0+\hat{V}$ as follows.
Diagonal elements can be written as $H_{nn}^0=E_{n}^0=n \delta$, where $\delta$ is a mean level spacing.
Off-diagonal matrix elements are random and banded: we have $\braket{H^0_{nm}}_V=\braket{V^0_{nm}}_V=0$, and
\aln{
\braket{(H^0_{nm})^2}_V=\braket{(V^0_{nm})^2}_V\sim
     v^2e^{-|E_n^0-E_m^0|/T},
}
where $v$ is the strength of the perturbation and we assume that a cutoff of the band is determined by the temperature $T^{-1}=\fracpd{S}{E}$.
In this model, if the dimension of the matrix is infinitely large, we can expect that the second moment of $\mc{U}$ takes the following Breit-Wigner form~\cite{Deutsch91,Deutsch10,Reimann15}\footnote{The exact forms of $\braket{|\mc{U}_{\alpha n}|^2}_V$ slightly differ depending on models one assumes. However, it is expected that the following discussions are qualitatively correct for such models as well.}
\aln{\label{BW}
\braket{|\mc{U}_{\alpha n}|^2}_V&\simeq\frac{v^2}{(\alpha \delta-n\delta)^2+\lrs{\frac{\pi v^2}{\delta}}^2}\nonumber\\
&=:u_2(\alpha-n)
}
for a relatively small $v$ (i.e., $\delta \ll\frac{2\pi v^2}{\delta}\ll T$).
Here we note that $u_2(x)$ takes the maximum value at $x=0$:
\aln{
\max_xu_2(x)=u_2(0)=\frac{\delta^2}{\pi^2v^2}\ll 1.
}
Moreover, $u_2(x)$ monotonically increases (decreases) with $x$ when $x<0\:(x>0)$.

Now, assuming Eq. (\ref{BW}), we consider the smallness of Eq. (\ref{tonari}).
We also assume that other second moments and a first moment vanish.
First, we note
\aln{
\braket{\moD}_V&=\sum_n\braket{|\mc{U}_{\alpha n}|^2}_V\mc{O}_{nn}^0\nonumber\\
&=\sum_nu_2(\alpha-n)\mc{O}_{nn}^0.
}
Then
\aln{
|\braket{\mc{O}_{\alpha+1,\alpha+1}}_V-\braket{\moD}_V|&\leq \sum_n|u_2(\alpha+1-n)-u_2(\alpha-n)||\mc{O}_{nn}^0|\nonumber\\
&\leq 2||\hmo||_\mr{op}u_2(0)\:.
}
Since $u_2(0)$ is expected to be small, Eq. (\ref{tonari}) is small.
We remark that the smallness of Eq. (\ref{bunbun}) or the ETH for the off-diagonal matrix elements can also be justified using higher moments of $\mc{U}$~\cite{Reimann15E}.

\section{Numerical simulations of the ETH}\label{sec:num}
Here, we review some recent numerical simulations that investigate the ETH and its finite-size corrections for few-body operators in quantum many-body systems.
First, we show some results found in Ref.~\cite{Santos10a} which discusses the relation between random matrices and nonintegrable systems of hardcore bosons or spinless fermions.
Then, we review the previous works on the ETH and its finite-size corrections both for diagonal matrix elements and off-diagonal matrix elements.

\subsection{Level-spacing statistics of hardcore-particle systems}
In Ref.~\cite{Santos10a}, Santos and Rigol demonstrate that the level-spacing distributions change in the course of integrable-nonintegrable transitions in systems of hardcore bosons or spinless fermions.
Let us consider hardcore bosons here (the results are almost the same for spinless fermions).
They consider $M$ hardcore bosons on a one-dimensional lattice with $N$ sites.
The Hamiltonian of the system is
\aln{\label{hamaham}
\hH&=\hH_0+\hat{V},\NON
\hH_0&=\sum_{i=1}^N\lrl{-t\lrs{\hat{b}_i^\dag\hat{b}_{i+1}+\mr{h.c.}}+V\lrs{\hat{b}_i^\dag\hat{b}_{i}-\frac{1}{2}}\lrs{\hat{b}_{i+1}^\dag\hat{b}_{i+1}-\frac{1}{2}}},\NON
\hat{V}&=\sum_{i=1}^N\lrl{-t'\lrs{\hat{b}_i^\dag\hat{b}_{i+2}+\mr{h.c.}}+V'\lrs{\hat{b}_i^\dag\hat{b}_{i}-\frac{1}{2}}\lrs{\hat{b}_{i+2}^\dag\hat{b}_{i+2}-\frac{1}{2}}},
}
where $\hat{b}_i$ is an annihilation operator of a hardcore boson at the site $i$, and the periodic boundary condition is imposed.
If $t'=J'=0$, $\hH$ is integrable since $\hH_0$ can be mapped to the spin 1/2 XXZ model.
If $\hat{V}$ becomes comparable to $\hH_0$, $\hH$ is expected to be nonintegrable.

\begin{figure}
\centering
\includegraphics[angle=-90,width=12cm]{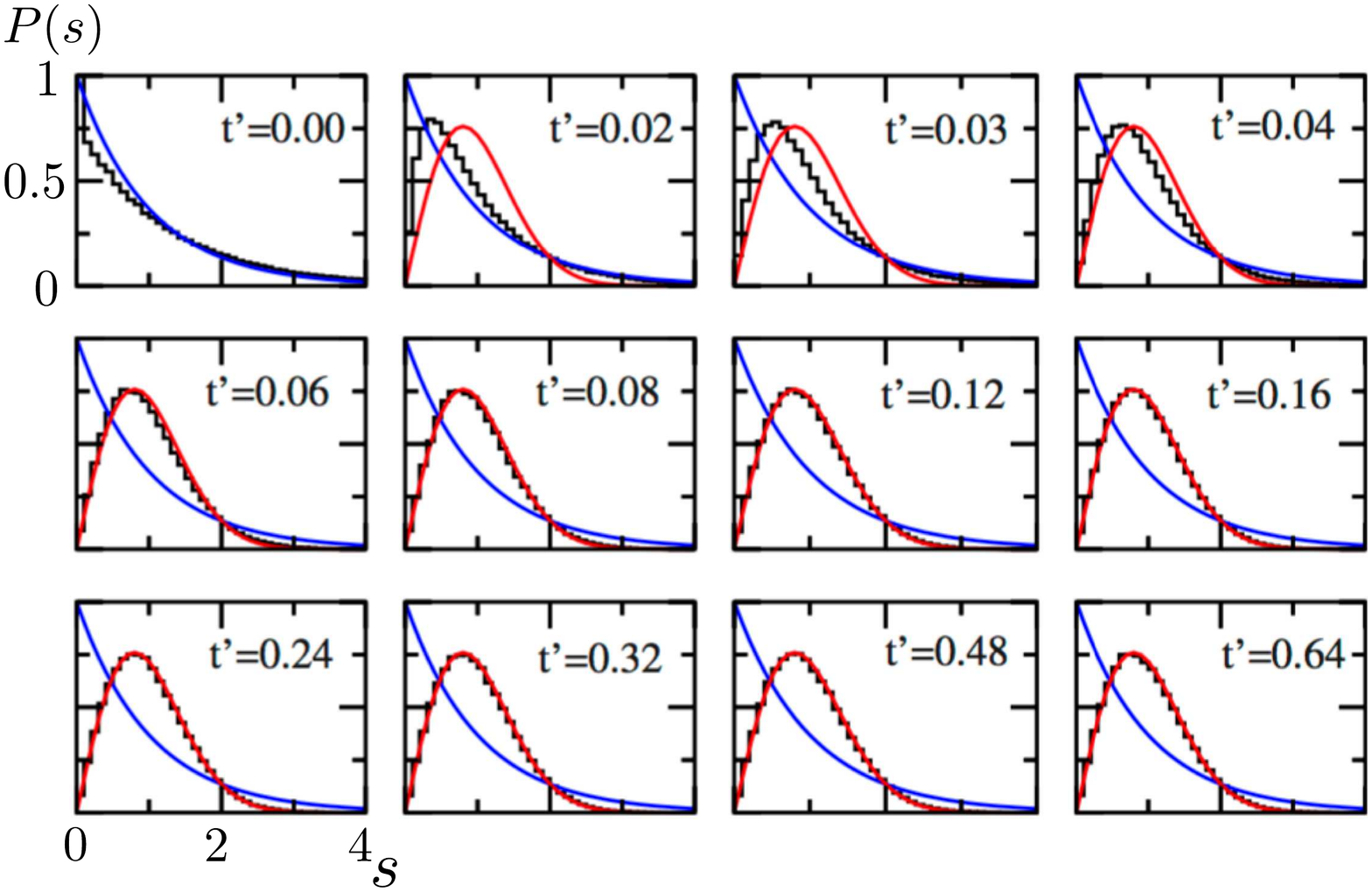}
\caption{Level-spacing distributions of the Hamiltonian in \EQ{hamaham} after the separation of the quasi-momentum sectors~\cite{Santos10a}. The result is obtained for $N=24,M=8, t=V=1, t'=V'$, and by averaging the results of all quasi-momentum sectors with no further symmetries. As $t'$ becomes large, $P(s)$ changes from the Poisson distribution $P_\mr{P}(s)$ to the Wigner-Dyson distribution $P_\mr{GOE}(s)$. Reproduced from Fig. 1 of Ref.~\cite{Santos10a}. \copyright 2010 American Physical Society.}
\label{fig:rigollevel}
\end{figure}

Santos and Rigol investigate the level-spacing distibutions $P(s)$ for various values of $t'=V'$ (by setting $t=V=1$).
Since the system is translationally invariant, the Hamiltonian is decomposed into several quasi-momentum sectors.
This means that we have to calculate the level spacings for each sector, not for the entire spectrum.\footnote{In fact, there are some sectors that have further symmetries. For example, the sector with zero quasi-momentum has a parity symmetry. We avoid using such sectors to obtain the level distributions.}
The obtained level-spacing distributions are shown in Fig. \ref{fig:rigollevel}.
We can see that, as $t'$ becomes larger, $P(s)$ changes from the Poisson distribution $P_\mr{P}(s)$ to the Wigner-Dyson distribution $P_\mr{GOE}(s)$.
This result clearly shows that the nonintegrability of the system is well captured by the level-spacing distributions predicted by RMT.
We note that the obtained distribution $P_\mr{GOE}(s)$ reflects the time-reversal symmetry of the Hamiltonian in \EQ{hamaham}.

\subsection{Diagonal matrix elements}
The ETH for diagonal matrix elements has extensively been investigated recently especially for few-body observables.
After the notable work by Rigol~\cite{Rigol08} that uses hardcore bosons on a 2D lattice, the ETH has been numerically verified in various nonintegrable systems, including systems with spinless~\cite{Rigol09Q} or spinful~\cite{Genway12} fermions, interacting spin chains~\cite{Steinigeweg13}, and Bose-Hubbard models~\cite{Biroli10}.
It has also been known that the ETH breaks down in integrable~\cite{Rigol08} and MBL systems~\cite{Nandkishore14}.
Recently, the coexistence of the energy ranges that do or do not satisfy the ETH is also gathering attention.
Such phenomena are expected to be observed in the mobility edge of MBL systems~\cite{Kjall14}, excited-state quantum phase transitions in Dicke and other models~\cite{Lobez16}, and spontaneous symmetry breaking in 2D transverse Ising models~\cite{Fratus15,Mondaini16}.

\begin{figure}
\centering
\includegraphics[angle=-90,width=12.5cm]{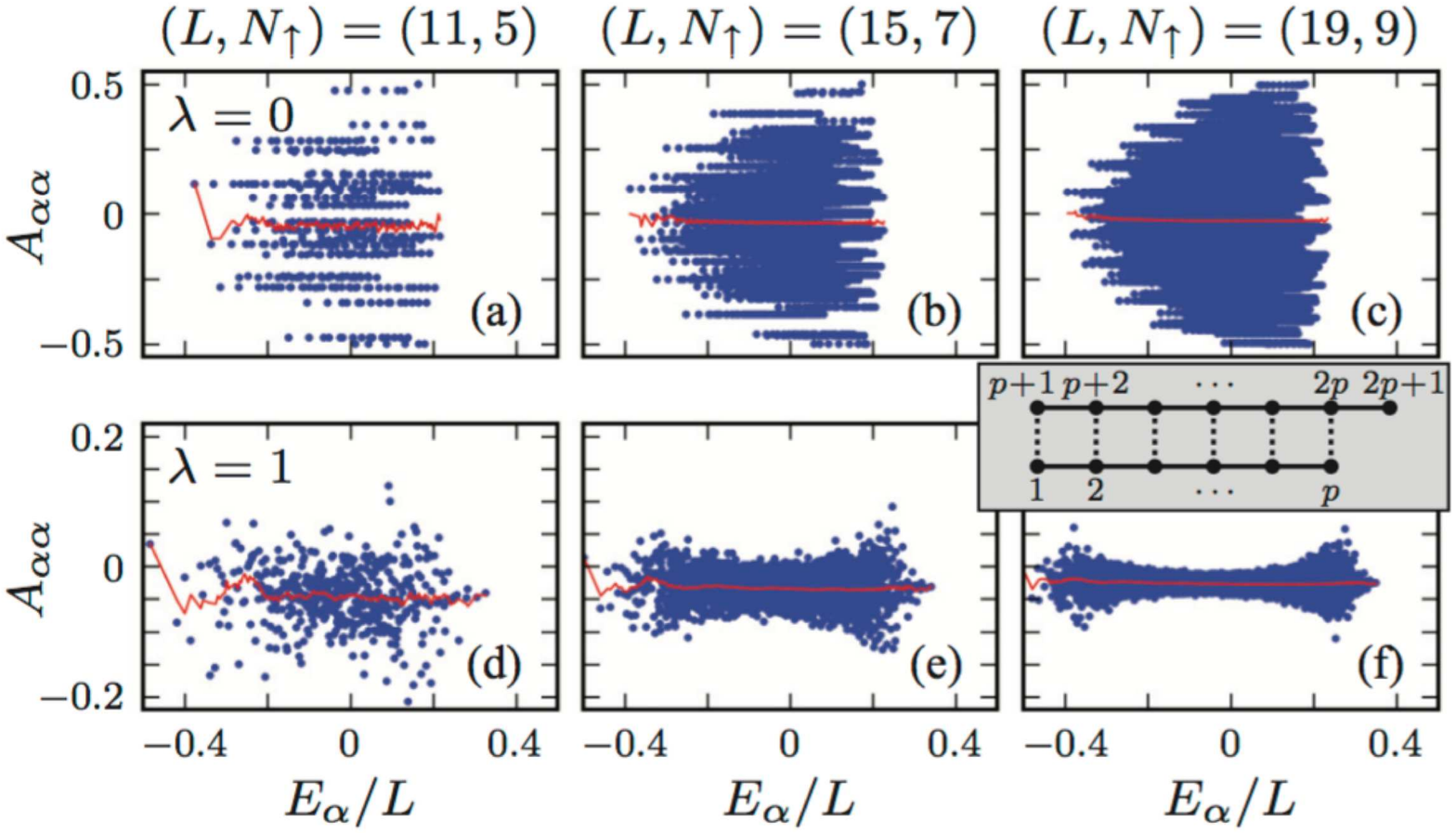}
\caption{Diagonal matrix elements $A_{\alpha\alpha}=\braket{E_\alpha|\hat{S}_2^z|E_\alpha}$ plotted for all of the eigenstates shows the results a function of $E_\alpha/L$~\cite{Beugeling14}.
The upper row for $\lambda=0$ (integrable) and the lower row is for $\lambda=1$ (nonintegrable).
This figure shows that if we increase the system, the fluctuations of $A_{\alpha\alpha}$ rapidly decay only when the system is nonintegrable.
(Inset) A ladder composed of $(L=2p+1)$ spins with neighboring XXZ interactions.
The interactions of the dotted bonds are $\lambda$ times stronger than those of the solid bonds.
Reproduced from Fig. 1 of Ref.~\cite{Beugeling14}. \copyright 2014 American Physical Society.}
\label{fig:dia}
\end{figure}

In Ref.~\cite{Beugeling14}, the authors investigate the ETH and its finite-size corrections for diagonal matrix elements of few-body operators.
They consider a ladder composed of $(L=2p+1)$ spins with neighboring XXZ interactions (see the inset of Fig. \ref{fig:dia}).
In the figure, the interactions of the dotted bonds are $\lambda$ times stronger than those of the solid bonds.
If $\lambda=0$, the ladder is decoupled to two spin chains and thus integrable.
We also note that $\lambda\ra\infty$ again makes the system integrable.
If $\lambda\sim 1$, the ladder becomes nonintegrable.
Since the total magnetization is conserved, they use the fixed sector with $N_{\uparrow}=p$ up spins.

In Fig. \ref{fig:dia}, the diagonal matrix elements $A_{\alpha\alpha}=\braket{E_\alpha|\hat{S}_2^z|E_\alpha}$ are plotted for all of the eigenstates as a function of $E_\alpha/L$.
The upper row is for $\lambda=0$ (integrable) and the lower row is for $\lambda=1$ (nonintegrable).
This figure shows that if we increase the system, the fluctuations of $A_{\alpha\alpha}$ rapidly decay for the entire spectrum (except for the edge) only when the system is nonintegrable.
The authors in Ref.~\cite{Beugeling14} investigated the (spectral) variance of the matrix elements and showed that it decays proportionally to $\frac{1}{\sqrt{D}}$, where $D$ is the dimension of the Hilbert space.
In another work~\cite{Steinigeweg13}, the Gaussian distribution of the fluctuations for the diagonal matrix elements of current operators has been reported using nonintegrable spin chains.
This means that $R_{\alpha\alpha}$ in \EQ{corre} obeys a Gaussian distribution for these operators.

\subsection{Off-diagonal matrix elements}
The ETH for off-diagonal matrix elements in many-body quantum systems has been less investigated than the ETH for diagonal matrix elements.
We note, though, that it was already pointed out in Rigol's paper~\cite{Rigol08} that the off-diagonal terms are very small.

In Ref.~\cite{Beugeling15}, the authors investigate the ETH and its finite-size corrections for off-diagonal matrix elements of few-body operators, similarly to Ref.~\cite{Beugeling14}.
They use the same model as shown in Fig. \ref{fig:dia} and investigate off-diagonal matrix elements.

\begin{figure}
\centering
\includegraphics[angle=-90,width=12.5cm]{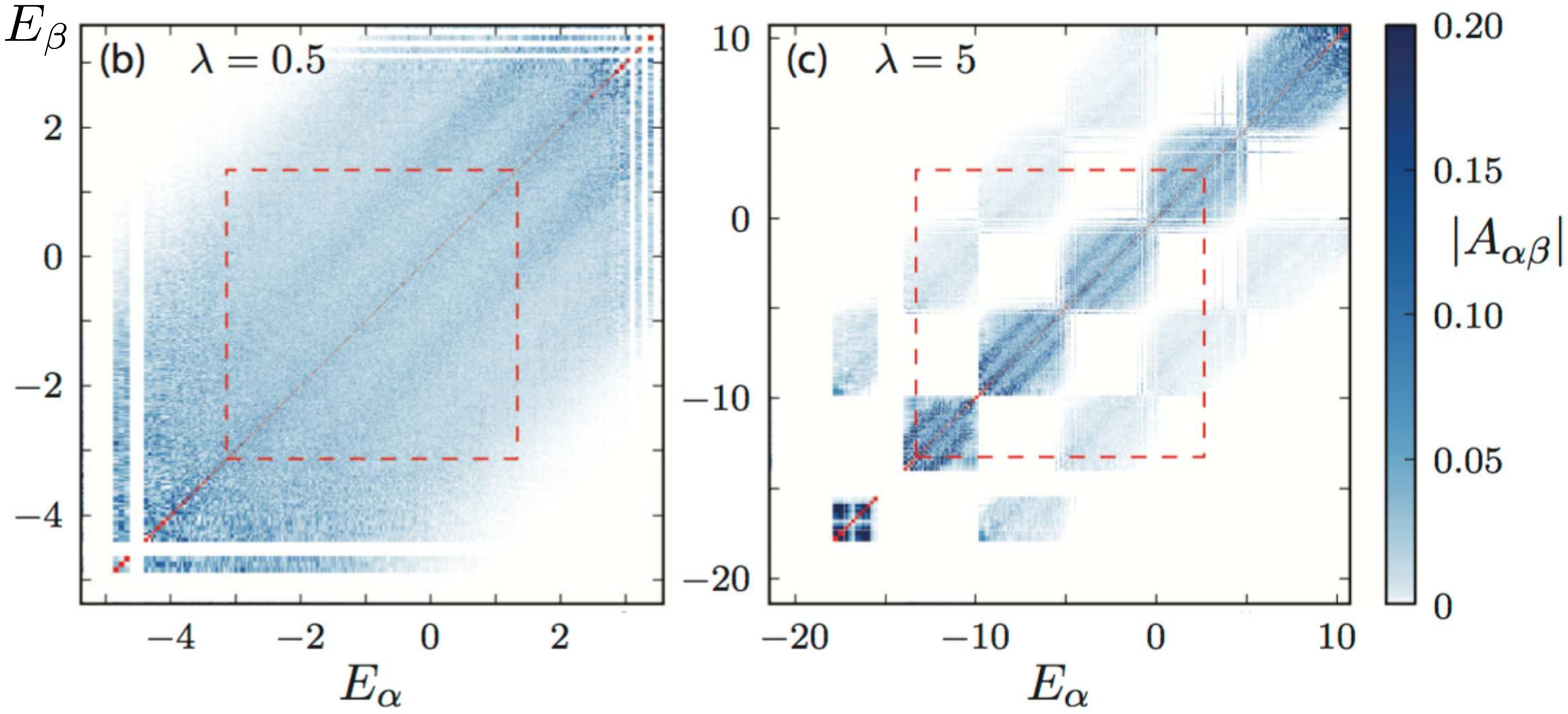}
\caption{Off-diagonal matrix elements $|A_{\alpha\beta}|=|\braket{E_\alpha|\hat{S}_2^z\hat{S}_{p+2}^z|E_\beta}|$ plotted for all of the eigenstates as a function of $E_\alpha$ and $E_\beta$~\cite{Beugeling15} (diagonal matrix elements are not shown).
The case with $L=13$ and $N_\uparrow=6$ are shown.
The left figure is for $\lambda=0.5$ (nonintegrable) and the right figure is for $\lambda=5$ (near-integrable).
If the system is nonintegrable, the behavior of the matrix elements seems to change mildly with the change of the energy.
On the other hand, if the system is integrable, we can see the block-like structure.
Reproduced from Fig. 1 of Ref.~\cite{Beugeling15}. \copyright 2015 American Physical Society.}
\label{fig:odia}
\end{figure}

In Fig. \ref{fig:odia}, the off-diagonal matrix elements $|A_{\alpha\beta}|=|\braket{E_\alpha|\hat{S}_2^z\hat{S}_{p+2}^z|E_\beta}|$ are plotted for all of the eigenstates as a function of $E_\alpha$ and $E_\beta$.
The left figure is for $\lambda=0.5$ (nonintegrable) and the right figure is for $\lambda=5$ (near-integrable).
If the system is nonintegrable, the behavior of the matrix elements seems to change mildly with the change of the energy.
In other words, within a small energy shell $\lrm{(E_\alpha,E_\beta):|E_\alpha-E_1|<\omega_{s,1},|E_\beta-E_2|<\omega_{s,2}}$, typical magnitude of $A_{\alpha\beta}$'s seems constant.
We note that the entire structure depends on the global energy, such as the bandlike structure as a function of $|E_\alpha-E_\beta|$ as shown in Fig. \ref{fig:odia}.
On the other hand, if the system is integrable, we can see the block-like structure.
The authors in Ref.~\cite{Beugeling14} showed that the variance of the matrix elements decays proportionally to $\frac{1}{\sqrt{D}}$ only in the nonintegrable case, as is the case with diagonal matrix elements.
They also found the Gaussian distributions of the matrix elements within the small energy shells in that case (similar results are found in Ref.~\cite{Steinigeweg13}).
This means that $R_{\alpha\beta}$ in \EQ{corre} obeys a Gaussian distribution for these operators.
We note that the Gaussian distributions have been investigated using systems that have classical counterparts in Refs.~\cite{Alhassid89,Kus91}.

Several studies have investigated off-diagonal matrix elements motivated by \EQ{corre}~\cite{Khatami13,DAlessio16,Luitz16,Serbyn16}.
In Ref.~\cite{Khatami13}, the authors have investigated the $\omega$-dependence of off-diagonal matrix elements (namely, the behavior of $f(E,\omega)$ in \EQ{corre}).
The important findings are that $f(E,\omega)$ has a plateau for $\omega\lesssim\omega_\mr{sh}$, which indicates the validity of \EQ{corre}, and that $f(E,\omega)$ rapidly decays when $\omega$ is large.
We note, however, that this result is not enough to justify the mechanism of RMT in the energy shell.
For example, it has been reported~\cite{Steinigeweg13} that the variances of diagonal and off-diagonal matrix elements of current operators do not seem to be related to each other, contrary to the prediction of RMT.
The mechanism and the justification of \EQ{corre} (or \EQ{semai}) have been still under investigation.

\section{Weak ETH}\label{sec:wETH}
We have mainly discussed the ETH that requires \EQ{ETH} for every eigenstate within certain energy ranges, which is sometimes called a strong ETH.
However, some authors investigate a bit weaker statement, which is sometimes called a weak ETH~\cite{Biroli10,Iyoda16}.
The weak ETH states that, the variance of diagonal matrix elments\footnote{The weak ETH is mostly discussed only for diagonal matrix elements.} within some energy shell vanishes in the thermodynamic limit:
\aln{\label{weakdayo}
\Delta \mc{O}_\mr{d}^2:=\braket{\moD^2}_\mc{T}-\braket{\moD}_\mc{T}^2\ra 0.
}
The condition in \EQ{weakdayo} implies that most of the eigenstates satisfy
\aln{\label{aaaaaa}
\moD\simeq \braket{\moD}_\mc{T},
}
because of Chebyshev's inequality.

Although the strong ETH holds true only in nonintegrable systems, the weak ETH holds true for a wider class of systems.
Several numerical studies show that the weak ETH holds true even in interacting integrable systems~\cite{Ikeda13,Alba15}. In this case, $\Delta \mc{O}_\mr{d}$ decreases as a polynomial with system size $N$, in contrast to nonintegrable systems, where $\Delta \mc{O}_\mr{d}$ is expected to decrease exponentially with $N$.
Moreover, it is rigorously shown that, in certain systems (e.g., 1D short-range interacting lattice systems) and for macroscopic observables that can be written as the average of local operators, 
an exponentially small fraction of the energy eigenstates violates \EQ{aaaaaa}~\cite{Mori16W} (see \SEC{sec:role}).
This is a refined statement of the weak ETH.

We note that the weak ETH does not justify thermalization from all initial states, unlike the strong ETH~\cite{Biroli10,Mori16W}. 
When the weak ETH holds true and the strong ETH does not, we can find  an eigenstate $\ket{E_\alpha}$ that violates \EQ{aaaaaa} even in the thermodynamic limit.
Then, if we take an initial state that has a peak at $\rho_{\alpha\alpha}$, we expect to obtain a non-thermal stationary state.
A crucial point here is that, for integrable systems, we can actually prepare such initial states that do not relax to thermal equilibrium with physically accessible protocols.
Therefore, the relation between the weak ETH and thermalization is not simple.\footnote[1]{As reviewed in \SEC{sec:role}, we impose the condition on initial states, namely \EQ{mori}, for systems to approach thermal equilibrium. This condition does not seem to hold true for integrable systems.}

\section{Summary and remarks}
Let us summarize this chapter and make some critical comments.
Though the rigorous proofs and definite criteria for the validity of the ETH have hardly been found, possible analytical explanations and  numerical simulations provide clues for understanding the mechanisms of the ETH.
We have reviewed such explanations in \SEC{sec:pete} and numerical simulations in \SEC{sec:num}.
In particular, we explained the analogies between nonintegrable systems and random matrices in Subsection \ref{sec:sre}.

If we assume the analogy for matrix elements of observables, it predicts the ETH and its finite-size corrections within a small energy shell (see Eq. (\ref{semai})), but the validity of this analogy is nontrivial.
In fact, there are not many verifications of this analogy in actual nonintegrable systems: even though some numerical studies have investigated the finite-size fluctuations of the matrix elements in nonintegrable systems, their relations to the RMT predictions are not clear~\cite{Steinigeweg13,Beugeling14,Beugeling15}.
It is also unclear how the RMT prediction is relevant to other previous works concerning the validity of the ETH.
For example, few-body properties of observables are often stressed as the validity of the ETH\footnote{Such arguments are made as follows: small subsystems can be regarded as being thermal through the quantum entanglement of the energy eigenstate; it is related to the ETH in the form of microscopic thermal equilibrium (MITE).}; however, the relation to the RMT predictions remains to be clarified, as we did not use such properties in Subsection \ref{sec:sre}.

\chapter{Observable-dependence of how random matrix theory can predict deviations from the ETH}\label{ch:Obs}

\section{Motivations}\label{sec:mot}
The ETH is the possible scenario for thermalization as we have seen in \CH{ch:Eq} and it is expected to be related to RMT as reviewed in \CH{ch:ETH}.
If we assume the analogy between nonintegrable systems and RMT for matrix elements of observables, it predicts not only the ETH but also its finite-size corrections within a small energy shell (see Eqs. (\ref{semai}) and (\ref{corre})).

However, the conjecture of \EQ{corre} is not well verified in actual nonintegrable systems;
matrix elements in such systems have been investigated, but the evidences of the RMT conjecture are few.
For example, the Gaussian distributions found in Refs.~\cite{Steinigeweg13,Beugeling14,Beugeling15} have not been attributed to the conjecture of RMT, yet.
Moreover, the result of Ref.~\cite{Steinigeweg13} does not seem consistent with the RMT conjecture.
One of the reasons of the lack of the evidences is that the statistics $R_{\alpha\beta}$ in \EQ{corre} has not been completely obtained yet.

To clarify to what extent RMT can predict actual situations, we should first refine and generalize \EQ{corre} (or \EQ{semai}) to make it applicable for a wide variety of situations, and then verify it by thorough numerics.
We especially consider how the conjecture of RMT is influenced by observables we take,
which has not been well investigated.
Since the distributions of the matrix elements are determined not by $\ket{E_\alpha}$ alone but by the transformation of the basis $U_{\alpha i}=\braket{E_\alpha|a_i}$ and the behavior of $\{a_i\}$, the change of observables may alter the statistics of finite-size deviations of the ETH.
We thus need to generalize the RMT prediction for arbitrary observables as well as for Hamiltonians (we note that previous studies focused only on nonintegrability of the Hamiltonians~\cite{Beugeling14,Beugeling15,Luitz16}).

The importance of observables also suggests that the nonintegrability of the Hamiltonian is not enough to justify that the matrix elements in the actual systems are predicted by those of RMT.
Actually, we can easily find an observable for which the RMT conjecture and the ETH break down even in nonintegrable systems.
To see this, take an energy eigenstate $\ket{E_\delta}$ of the Hamiltonian of the system and define $\hmo=\ket{E_\delta}\bra{E_\delta}$.
Then, we trivially obtain 
\aln{
\mc{O}_{\alpha\alpha}&=\begin{cases}
 1 &  \alpha=\delta,\\
 0 &  \alpha\neq\delta.\\
  \end{cases}
}
Since $\braket{\hmo}_\mr{sh}(E_\delta)\ra 0$ is expected in the thermodynamic limit, the ETH does not hold true in this case.
Moreover, we can find an observable for which the ETH is not valid in a sufficiently large subsystem.\footnote{
This is understood by the following example. We take a one-dimensional spin 1/2 system with $N\gg 1$ sites with a local Hamiltonian.
Consider a subsystem $\mc{M}$ with $M(>N/2)$ sites and a reduced density matrix of an energy eigenstate $\ket{E_\alpha}$ as $\hrho_\mc{M}=\Tr_{\mc{M}^c}[\ket{E_\alpha}\bra{E_\alpha}]$.
If all observables on $\mc{M}$ satisfied the ETH, $\hrho_\mc{M}$ would be written as 
\aln{
\hrho_\mc{M}=\Tr_{\mc{M}^c}\lrl{\frac{e^{-\beta\hH}}{Z}}\simeq \frac{e^{-\beta\hH_\mc{M}}}{Z_\mc{M}}\:\:\:(\mr{wrong}),
}
where $\hH_\mc{M}$ denotes a Hamiltonian restricted onto $\mc{M}$, $Z=\Tr[e^{-\beta\hH}]$ and $Z_\mc{M}=\Tr_\mc{M}[e^{-\beta\hH_\mc{M}}]$.
The first equality is the ETH in the MITE form and the second approximation comes from the locality of the Hamiltonian.
Thus, this equation would lead to an extensive von Neumann entropy $S_\mr{vN}(\hrho_\mc{M})\simeq S_\mr{vN}(\frac{e^{-\beta\hH_\mc{M}}}{Z_\mc{M}})\propto M$. However, the property of pure states leads to $S_\mr{vN}(\hrho_\mc{M})=S_\mr{vN}(\hrho_{\mc{M}^{c}})\leq (N-M)\ln 2$.
We can see an apparent contradiction of these two representations by taking $M=\frac{3}{4}N$.
This contradiction arises from the false assumption of the ETH.
}
The unsolved question is to what type of observables the RMT conjecture does not apply.
Such observables might be many-body observables, or something else.

\begin{figure}
\centering
\includegraphics[width=15.5cm]{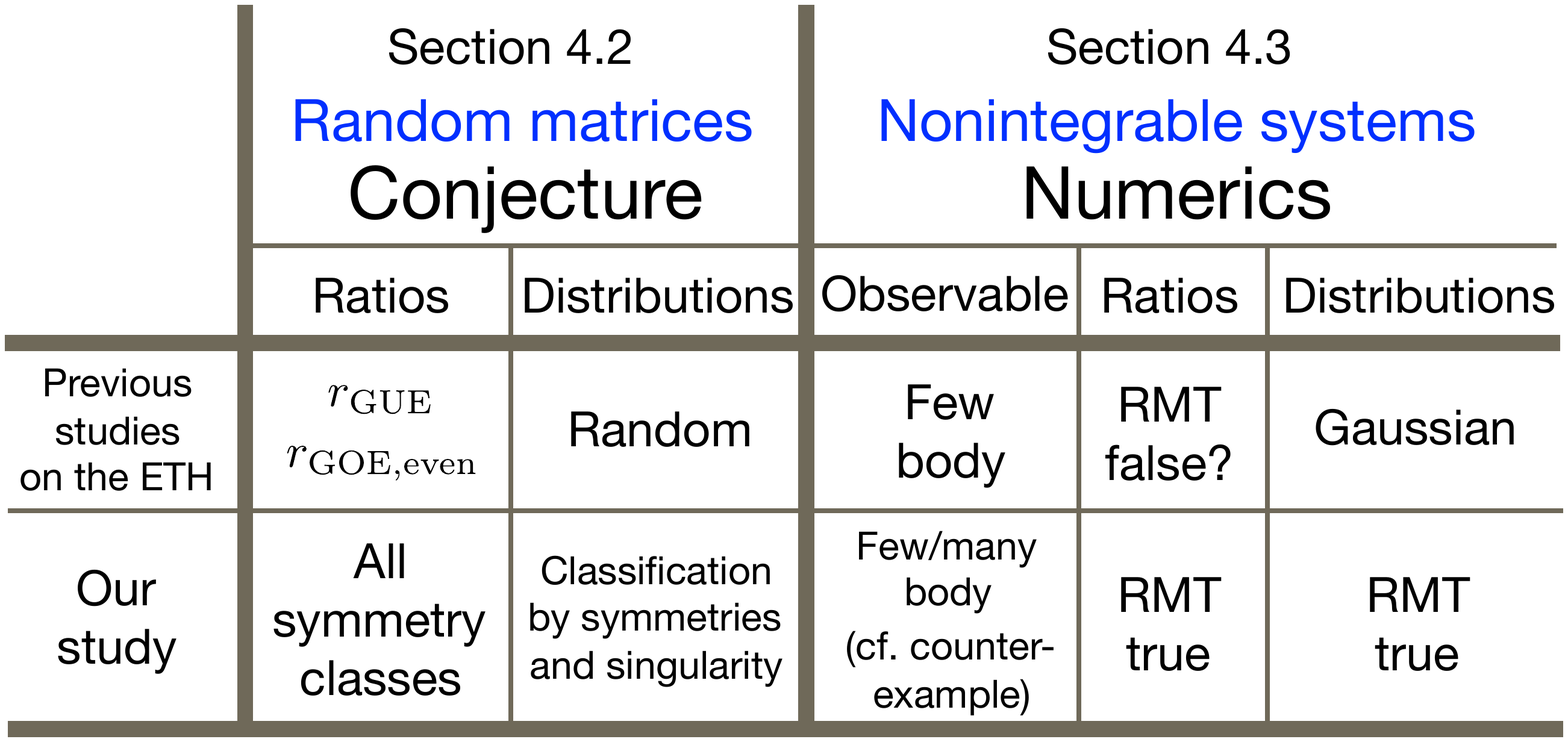}
\caption{Comparison between previous studies on the ETH and our study.
We show that the finite-size corrections of the ETH are in excellent agreement with the predictions of RMT (i.e., the ratios and probability densities)  for a wide class of observables with various symmetries, including many-body correlations and singular operators.
We also note that  counterexamples always exist even for simple observables.}
\label{fig:hyou}
\end{figure}

In this chapter, we show that RMT can predict the finite-size corrections of the ETH in nonintegrable systems and for a wide class of observables, including many-body operators.
In Section \ref{sec:staff}, we first refine and generalize the finite-size corrections of the ETH from the random matrix model.
We will especially see that that the ratios between standard deviations of diagonal and off-diagonal matrix elements become universal ones that depend only on anti-unitary symmetries of the Hamiltonian and those of the observable.
We also show that the probability densities of off-diagonal matrix elements obey the statistics that is determined by what we call singularity of observables as well as anti-unitary symmetries.
In Section \ref{sec:nvrmt}, we numerically investigate matrix elements of various observables in nonintegrable systems that only conserve energy.
We will demonstrate that the finite-size corrections of the ETH are in excellent agreement with the predictions of RMT for a wide class of observables with various symmetries, including many-body correlations and singular operators.
We also remark, though, that counterexamples always exist even for simple observables.
We compare previous studies and our results in Fig. \ref{fig:hyou}.

\section{Statistics of the finite-size corrections of the ETH for the random matrix model}\label{sec:staff}
\begin{figure}
\begin{center}
\includegraphics[width=14cm]{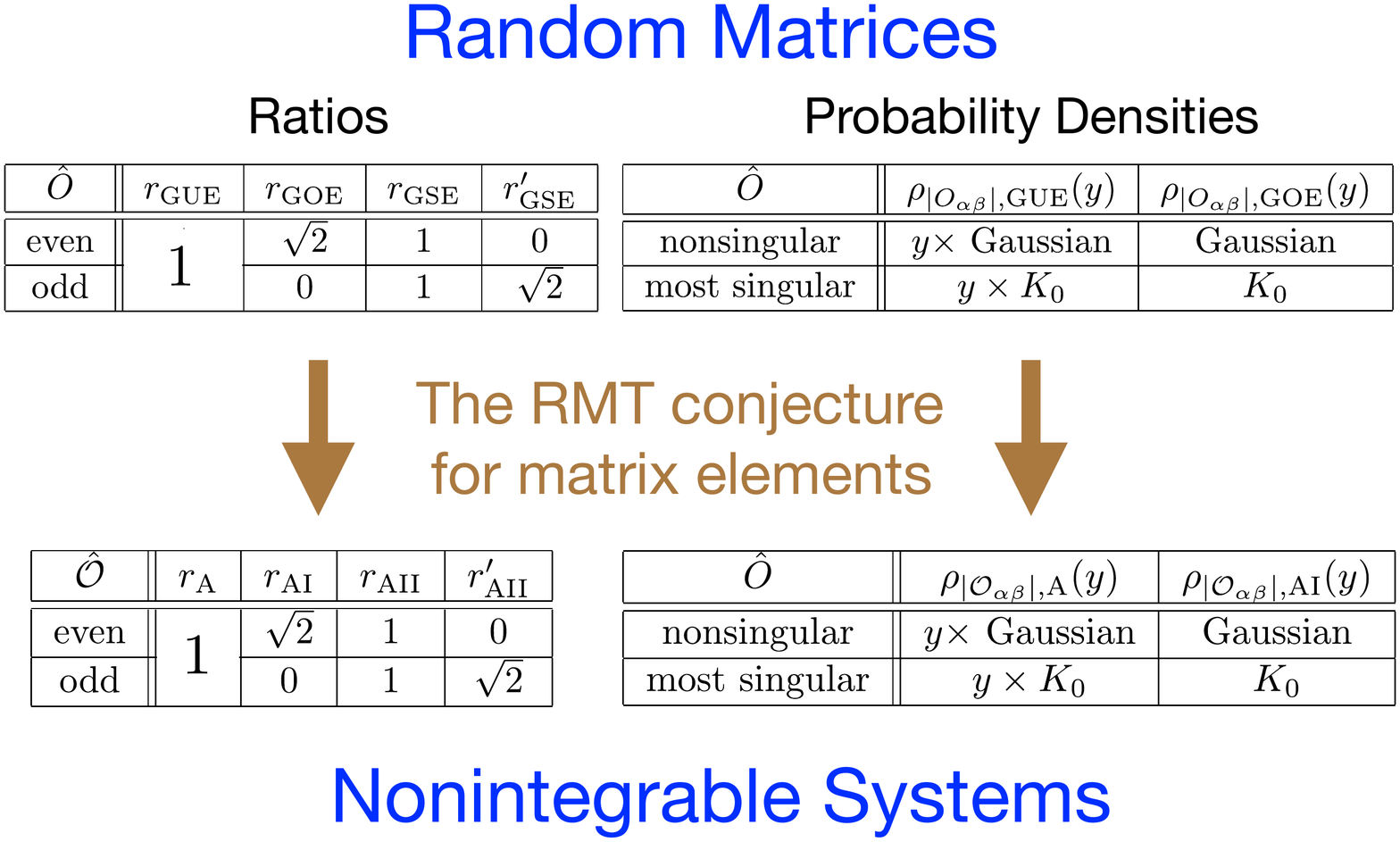}
\caption{Two conjectures from the random matrix model: (Left column) the ratio of standard deviations between the diagonal and the off-diagonal matrix elements within $\mc{H}_\mr{sh}$ become universal in nonintegrable systems. The universal values are determined from those of the random matrix models with the same symmetry class. (Right column) The probability densities $\rho_{|\mc{O_{\alpha\beta}}|}(y)$ within $\mc{H}_\mr{sh}$ are those predicted by random matrix models. We note that $K_0$ means a modified Bessel function of the second kind.}
\label{conjecture}
\end{center}
\end{figure}

In this section, we make a refined RMT conjecture about the matrix elements within an energy shell, focusing on a change of the statistics $R_{\alpha\beta}$.
By calculating the ratios of standard deviations between diagonal and off-diagonal matrix elements,
we show that RMT predicts the universal ratios that depend only on anti-unitary symmetries of the Hamiltonian and those of the observable.
Next, we examine the probability densities of the off-diagonal matrix elements,
and find that RMT predicts Gaussian statistics for a wide class of observables in consistent with previous numerical studies~\cite{Steinigeweg13,Beugeling15} (in the case of the GOE), but that it predicts other statistics if observables are ``singular."
We summarize our results and conjectures in Fig. \ref{conjecture} (see Subsection \ref{sec:conj}).

\subsection{Universal ratios between diagonal and off-diagonal matrix elements}
First, let us consider the ratio of standard deviations between diagonal and off-diagonal matrix elements that is defined as
\aln{\label{eqratio}
r:=\frac{\Delta \mc{O}_\mr{d}}{\Delta \mc{O}_\mr{od}},
}
where 
\aln{\label{vnokei}
\Delta \mc{O}_\mr{d}^2&:=\braket{\moD^2}_\mc{T}-\braket{\moD}_\mc{T}^2,\\
\Delta \mc{O}_\mr{od}^2&:=\braket{|\mooD|^2}_\mc{TT}\:\:\:(E_\alpha\neq E_\beta).
}
If the Hamiltonian has Kramers degeneracies, the degenerate eigenstates can be written as $\ket{E_\alpha}$ and $\ket{\tilde{E_\alpha}}:=\hat{T}\ket{E_\alpha}$.
Because we can consider $\mc{O}_{\alpha\tilde{\alpha}}:=\braket{E_\alpha|\hmo|\tilde{E_\alpha}}$ in this case, we introduce the corresponding ratio as\footnote{We note that in the case of the GSE, we have a freedom to choose two orthogonal energy eigenstates in the Kramers degenerate space. In the numerical calculation in \SEC{sec:nvrmt}, we use two eigenstates that are directly obtained by the exact-diagonalization programming.}
\aln{
r':=\frac{\Delta \mc{O}_\mr{K}}{\Delta \mc{O}_\mr{od}},
}
where 
\aln{
\Delta \mc{O}_\mr{K}^2&:=\braket{|\mc{O}_{\alpha\tilde{\alpha}}|^2}_\mc{T}.
}

We assume that the symmetry of the system is at most one anti-unitary symmetry, the corresponding operator of which commutes with the Hamiltonian.\footnote{In other words, we assume that the system has no unitary symmetry, whose corresponding operator (anti)commutes with the Hamiltonian. 
We also assume that the system has no anti-unitary symmetry, the corresponding operator of which anticommutes with the Hamiltonian.}
Then, the Hamiltonian in RMT belongs to the GUE, the GOE, or the GSE.
The Hamiltonian that belongs to the GOE has an anti-unitary symmetry $\hat{T}$ that satisfies $\hat{T}^2=1$.
In contrast, the Hamiltonian that belongs to the GSE has an anti-unitary symmetry $\hat{T}$ that satisfies $\hat{T}^2=-1$.
In these two cases, we can consider two types of observables that satisfy either $\hat{T}\hmo\hat{T}^{-1}=\hmo$ or $\hat{T}\hmo\hat{T}^{-1}=-\hmo$.
We will call the former and latter observables as the even and odd operators, respectively.\footnote{Though operators that are neither even nor odd exist, we will not consider such operators for simplicity.}

RMT predicts that $r$ and $r'$ become universal values that depend only on the symmetries of the Hamiltonian and those of the observable.
If the Hamiltonian belongs to the GUE, $r_\mr{GUE}\ra  1 \:(d\ra \infty)$ is expected as we have seen in \SEC{sec:sre}.
However, we will show that the change in the symmetry affects the values of $r$ and $r'$ as illustrated in the upper left table in Fig. \ref{conjecture}.

\subsubsection{The GOE}

First consider the case in which the Hamiltonian belongs to the GOE and the observables are even under $\hT$.
We assume that neither the Hamiltonian nor the observable has a degeneracy.
As in \SEC{sec:sre}, we define 
$\hat{O}=\sum_{i=1}^{d}a_i\ket{a_i}\bra{a_i}$.
We note that we can assume that $\hat{T}\ket{a_i}=\ket{a_i}$ and $\hat{T}\ket{E_\alpha}=\ket{E_\alpha}$ without loss of generality.\footnote[1]{Let us consider $\ket{E_\alpha}$. Since $\hH\hT\ket{E_\alpha}=\hT\hH\ket{E_\alpha}=E_\alpha\hT\ket{E_\alpha}$ and no degeneracy exists, we obtain $\hT\ket{E_\alpha}=e^{i\theta}\ket{E_\alpha}$ for some $\theta\in [0,2\pi)$. If we redefine the eigenstate as $\ket{E_\alpha'}:=e^{i\theta/2}\ket{E_\alpha}$,  we obtain $\hT\ket{E_\alpha'}=\ket{E_\alpha'}$.}
Then the matrix elements can be taken as being real because
\aln{
O_{\alpha\beta}&=(\ket{E_\alpha},\hat{O}\ket{E_\beta})\NON
&=(\hT\ket{E_\alpha},\hT\hat{O}\hT^{-1}\hT\ket{E_\beta})^*\NON
&=(\ket{E_\alpha},(+\hat{O})\ket{E_\beta})^*=O_{\alpha\beta}^*,
}
where $(\vec{a},\vec{b})$ denotes an inner product of $\vec{a}$ and $\vec{b}$ and we have used $(\hT\vec{a},\hT\vec{b})^*=(\vec{a},\vec{b})$.

In this case, we can assume that the basis transformation $U_{\alpha i}:=\braket{E_\alpha|a_i}$ is distributed uniformly with respect to the \textit{orthogonal} Haar measure from RMT.\footnote[2]{This reflects the fact that $U_{\alpha i}$ can be taken as being real: $\braket{E_\alpha|a_i}=(\hT\ket{E_\alpha},\hT\ket{a_i})^*=\braket{E_\alpha|a_i}^*$.}
Some of the moments of $U$ can be written as
\aln{
\av{U_{\alpha i}U_{i\beta}}&=\frac{1}{d}\delta_{\alpha\beta},\label{highO0}\\
\av{U_{\alpha i}^2U_{i\beta}^2}&=\frac{1+2\delta_{\alpha\beta}}{d(d+2)},\\
\av{U_{\alpha i}^2U_{j\alpha}^2}&=\frac{1+2\delta_{ij}}{d(d+2)},\\
\av{U_{\alpha i}U_{i\beta}U_{\beta j}U_{j\alpha}}&=-\frac{1}{d(d-1)(d+2)}\:\:\:(\alpha\neq\beta,i\neq j)\label{highO}.
}

Using Eqs. (\ref{highO0})-(\ref{highO}), $r_\mr{GOE,even}=\sqrt{2}$ is obtained as follows.
The averages and the variances of diagonal and off-diagonal matrix elements can be calculated as
\aln{
\av{O_{\alpha\beta}}&=\frac{\delta_{\alpha\beta}}{d}\sum_ia_i,\\
\av{(O_{\alpha\alpha}-\av{O_{\alpha\alpha}})^2}&=\frac{2}{d+2}\lrl{\frac{1}{d}\sum_ia_i^2-\lrs{\frac{1}{d}\sum_ia_i}^2},\\
\av{O_{\alpha\beta}^2}
&=\frac{d}{(d+2)(d-1)}\lrl{\frac{1}{d}\sum_ia_i^2-\lrs{\frac{1}{d}\sum_ia_i}^2}\:\:\:(\alpha\neq\beta),
}
which can be obtained with calculations similar to those made in \SEC{sec:sre}.
If we assume the ergodicity of random matrices, we can regard the ensemble average above as the spectral average for most of the randomly sampled Hamiltonians.
Therefore, if $d$ is sufficiently large, we obtain
\aln{
r_\mr{GOE, even}=\sqrt{2},
}
which is different from $r_\mr{GUE}=1$.
We note that this ratio has been predicted in several studies~\cite{Eckhardt95,Hotikar98,DAlessio16}.

Next, if the Hamiltonian belongs to the GOE and the observable is odd under $\hT$, $r_\mr{GOE,odd}=0$ is obtained.
This results comes from the vanishing diagonal matrix elements:
\aln{{O}_{\alpha\alpha}&=(\hT\ket{E_\alpha},\hT\hO\hT^{-1}\hT\ket{E_\alpha})^*\NON
&=(\ket{E_\alpha},(-\hO)\ket{E_\alpha})^*\NON
&=-O_{\alpha\alpha}\NON
&= 0.
}

\subsubsection{The GSE}
If the Hamiltonian belongs to the GSE, it can be written as $\hH=\sum_{\alpha=1}^{d/2}E_\alpha(\ket{E_\alpha}\bra{E_\alpha}+\ket{\tilde{E_\alpha}}\bra{\tilde{E_\alpha}})$.
Here, we note that $d$ is always even number in the GSE.
Let us first consider that the observable is even.
In this case, the observable can be written as $\hO=\sum_{i=1}^da_i\ket{a_i}\bra{a_i}=\sum_{i'=1}^{d/2}a_{i'}(\ket{a_{i'}}\bra{a_{i'}}+\ket{\tilde{a_{i'}}}\bra{\tilde{a_{i'}}})$, where $\ket{\tilde{a_{i'}}}:=\hT\ket{a_{i'}}$ and $\hO\ket{\tilde{a_{i'}}}=a_{i'}\ket{\tilde{a_{i'}}}$.
We assume that the Hamiltonian and the observable have no degeneracy except for Kramers degeneracies.
By calculating (higher) moments of $\braket{E_\alpha|a_{i'}}$ and $\braket{E_\alpha|\tilde{a_{i'}}}$ using RMT, we obtain the averages and the variances of diagonal and off-diagonal matrix elements as follows (see Appendix \ref{sec:gseE}):
\aln{\label{gseE}
\av{O_{\alpha\beta}}&=\frac{\delta_{\alpha\beta}}{d}\sum_ia_i\:\:\:(E_\alpha\neq E_\beta),\\
\av{(O_{\alpha\alpha}-\av{O_{\alpha\alpha}})^2}&=\frac{1}{d+1}\lrl{\frac{1}{d}\sum_ia_i^2-\lrs{\frac{1}{d}\sum_ia_i}^2},\\
\av{|O_{\alpha\beta}|^2}
&=\frac{d}{(d-2)(d+1)}\lrl{\frac{1}{d}\sum_ia_i^2-\lrs{\frac{1}{d}\sum_ia_i}^2}\:\:\:(E_\alpha\neq E_\beta),
}
Assuming the ergodicity of random matrices, we can regard the ensemble averages above as the spectral averages. Thus we obtain 
\aln{
r_\mr{GSE,even}\ra 1
}
if $d$ is sufficiently large.
For the ratio concerning the Kramers pair, we obtain $r'_\mr{GSE,even}=0$.
This ratio comes from the vanishing ${O}_{\alpha\tilde{\alpha}}$:
\aln{{O}_{\alpha\tilde{\alpha}}&=(\hT\ket{E_\alpha},\hT\hO\hT^{-1}\hT\hT\ket{E_\alpha})^*\NON
&=(\ket{\tilde{E_\alpha}},\hO(-1)\ket{E_\alpha})^*\NON
&=-O_{\alpha\tilde{\alpha}}\NON
&= 0.
}
Here we have used $\hat{T}^2=-1$.

Finally, we consider the case in which the Hamiltonian belongs to the GSE and the observable is odd under $\hT$.
In this case, the observable can be written as $\hO=\sum_{i=1}^da_i\ket{a_i}\bra{a_i}=\sum_{a_{i'}>0}a_{i'}(\ket{a_{i'}}\bra{a_{i'}}-\ket{\tilde{a_{i'}}}\bra{\tilde{a_{i'}}})$, where $\ket{\tilde{a_{i'}}}:=\hT\ket{a_{i'}}$ and $\hO\ket{\tilde{a_{i'}}}=-a_{i'}\ket{\tilde{a_{i'}}}$.
By calculating the (higher) moments of inner products such as $\braket{E_\alpha|a_{i'}}$ and $\braket{E_\alpha|\tilde{a_{i'}}}$ using RMT, we obtain the averages and the variances of matrix elements (see Appendix \ref{sec:gseE}).
The averages of the matrix elements are all zero since $\sum_ia_i=0$:
\aln{\label{gseE2}
\av{O_{\alpha\beta}}=0.
}
For the variances, we obtain
\aln{
\av{(O_{\alpha\alpha})^2}&=\frac{1}{d+1}\lrl{\frac{1}{d}\sum_ia_i^2},\\
\av{|O_{\alpha\beta}|^2}
&=\frac{1}{d+1}\lrl{\frac{1}{d}\sum_ia_i^2}\:\:\:(E_\alpha\neq E_\beta),\\
\av{|O_{\alpha\tilde{\alpha}}|^2}&=\frac{2}{d+1}\lrl{\frac{1}{d}\sum_ia_i^2}.\\
}
Assuming the ergodicity, we expect that
\aln{
r_\mr{GSE,odd}&\ra 1,\\
r'_\mr{GSE,odd}&\ra\sqrt{2},\label{gseE3}
}
if $d$ is sufficiently large.

\subsection{Observable-dependent probability densities of the off-diagonal matrix elements}\label{sec:distd}
Next, we consider the probability densities of the off-diagonal matrix elements $|O_{\alpha\beta}|$ using RMT.
To do this, we apply the method by Brody et al.~\cite{Brody81} to the GUE and the GOE (in Ref.~\cite{Brody81}, the calculation is done only for the GOE).\footnote{We will not consider the probability densities of the matrix elements in the case for the GSE for simplicity.}
We only consider off-diagonal matrix elements, since they are suitable for obtaining many samplings in later numerical calculations.

To calculate the probability densities of $|\braket{E_\alpha|\hO|E_\beta}|$, we first move $\ket{E_\alpha}$ uniformly in the $(d-1)$-dimensional Hilbert space that is orthogonal to $\ket{E_\beta}$ and then move $\ket{E_\beta}$ uniformly in the $d$-dimensional Hilbert space.
We note that
\aln{
\braket{E_\alpha|\hO|E_\beta}&=\braket{E_\alpha|\mc{\hat{P}}_{(d-1)}\hO|E_\beta}\\
&=\bra{E_\alpha}\frac{\mc{\hat{P}}_{(d-1)}\hO\ket{E_\beta}}{||\mc{\hat{P}}_{(d-1)}\hO\ket{E_\beta}||}\times ||\mc{\hat{P}}_{(d-1)}\hO\ket{E_\beta}||,
}
where $\mc{\hat{P}}_{(d-1)}$ is a projection operator onto a Hilbert space that is orthogonal to $\ket{E_\beta}$.
Since $\ket{v}:=\frac{\mc{\hat{P}}_{(d-1)}\hO\ket{E_\beta}}{||\mc{\hat{P}}_{(d-1)}\hO\ket{E_\beta}||}$ is a fixed normalized vector in the $(d-1)$-dimensional Hilbert space, $\braket{E_\alpha|v}$ is uniformly distributed on a high-dimensional unit sphere when we move $\ket{E_\alpha}$.
Consequently, the probability densities of $x:=|\braket{E_\alpha|v}|^2$ obeys the following Porter-Thomas distribution in the large $d$ limit~\cite{Brody81}:\footnote[1]{The following results of the GOE hold true whether observables are even or odd. However, if observables are neither even nor odd, the results may not be valid.}
\aln{\label{PTd}
 \rho_{|\braket{E_\alpha|v}|^2}(x)=\begin{cases}
    \sqrt{\frac{d}{2\pi x}}e^{-xd/2} &  \text{for the GOE},\\
    de^{-xd} &  \text{for the GUE}.\\
  \end{cases}
}
Since $|O_{\alpha\beta}|^2=||\mc{\hat{P}}_{(d-1)}\hO\ket{E_\beta}||^2x:=s_\beta^2 x$, the probability densities of $y=|O_{\alpha\beta}|^2$ with a fixed $s_\beta$ is
\aln{
 \rho'_{|O_{\alpha\beta}|^2}(y;s_\beta)=\begin{cases}
    \sqrt{\frac{d}{2s_\beta^2\pi y}}e^{-yd/2s_\beta^2} &  \text{for the GOE},\\
    \frac{d}{s_\beta^2}e^{-yd/s_\beta^2} &  \text{for the GUE}.\\
  \end{cases}
}
If we denote the probability densities of $s_\beta$ by $\rho_{s_\beta}(z)$, the probability densities of $|O_{\alpha\beta}|^2$ can be written as 
\aln{
 \rho_{|O_{\alpha\beta}|^2}(y)=\int dz  \rho'_{|O_{\alpha\beta}|^2}(y;z)\rho_{s_\beta}(z).
}
We note that 
\aln{
s_\beta^2&=\braket{E_\beta|\hO\mc{\hat{P}}_{(d-1)}\hO|E_\beta}\NON
&=\braket{E_\beta|\hO^2|E_\beta}-\braket{E_\beta|\hO|E_\beta}^2\NON
&=\braket{E_\beta|\hO'^2|E_\beta}-\braket{E_\beta|\hO'|E_\beta}^2,
}
where $\hO':=\hO-\frac{1}{d}\sum_\alpha O_{\alpha\alpha}$.

\subsubsection{Nonsingular operators}
Next, we consider the probability densities of $s_\beta$ by moving $\ket{E_\beta}$.
We first diagonalize $\hO'$ as
\aln{
\hO'=\sum_{i=1}^d a_i'\ket{a_i}\bra{a_i},
}
where $a_i'=a_i-\frac{1}{d}\sum_\alpha O_{\alpha\alpha}$.
Then
\aln{\label{aaaaa}
s_\beta^2=\sum_i{a'}_i^2|\braket{E_\beta|a_i}|^2-\lrs{\sum_i{a'}_i|\braket{E_\beta|a_i}|^2}^2.
}
The second moment of $s_\beta$ is thus calculated as
\aln{\label{ohayo}
\av{s_\beta^2}=\frac{1}{d}\sum_{i=1}^d{a'}_i^2\times (1+\mr{O}(d^{-1})),
}
where $\mr{O}(\cdot)$ denotes Landau's symbol and we have used $\frac{1}{d}\sum_ia_i'=0$.
Similarly, we obtain the fourth moment as
\aln{
\av{s_\beta^4}=  
\begin{cases}
  \lrl{\lrs{\frac{1}{d}\sum_{i=1}^d{a'}_i^2}^2+\frac{1}{d}\lrs{\frac{1}{d}\sum_{i=1}^d{a'}_i^4}}(1+\mr{O}(d^{-1})) &  \text{for the GOE},\\
  \lrl{\lrs{\frac{1}{d}\sum_{i=1}^d{a'}_i^2}^2+\frac{2}{d}\lrs{\frac{1}{d}\sum_{i=1}^d{a'}_i^4}}(1+\mr{O}(d^{-1})) &  \text{for the GUE}.\\
  \end{cases}
}
From this expression, we can expect that if 
\aln{\label{norio}
\frac{\frac{1}{d}\lrs{\frac{1}{d}\sum_{i=1}^d{a'}_i^4}}{\lrs{\frac{1}{d}\sum_{i=1}^d{a'}_i^2}^2}\ra 0\:\:\:(d\ra\infty),
}
$s_\beta^2$ fluctuates little around the average value $\av{s_\beta^2}=\frac{1}{d}\sum_{i=1}^d{a'}_i^2$ because $\frac{\av{s_\beta^4}-(\av{s_\beta^2})^2}{(\av{s_\beta^2})^2}\ra 0$.
We will call $\hO$ that satisfies \EQ{norio} as nonsingular operators, following Ref.~\cite{Brody81}.

For nonsingular operators, $\rho_{|O_{\alpha\beta}|^2}(y)$ can be calculated by noticing $\rho_{s_\beta}(z)\ra\delta(z-\sqrt{\mc{V}})$, where $\mc{V}:=\av{s_\beta^2}$.
The result is
\aln{
 \rho_{|O_{\alpha\beta}|^2}(y)=\begin{cases}
    \sqrt{\frac{d}{2\mc{V}\pi y}}e^{-yd/2\mc{V}} &  \text{for the GOE},\\
    \frac{d}{\mc{V}}e^{-yd/\mc{V}} &  \text{for the GUE}.\\
  \end{cases}
}
We can also write down the probability densities for $|O_{\alpha\beta}|$ as follows:
\aln{\label{gaussynise}
 \rho_{|O_{\alpha\beta}|}(y)=\begin{cases}
    \sqrt{\frac{2d}{\mc{V}\pi}}e^{-y^2d/2\mc{V}} &  \text{for the GOE},\\
    \frac{2dy}{\mc{V}}e^{-y^2d/\mc{V}} &  \text{for the GUE},\\
  \end{cases}
}
where $0<y$.
Moreover, we assume that we can replace $\mc{V}/d$ with the spectral variance of the off-diagonal matrix elements:\aln{
\sigma^2:=\frac{1}{d_s(d_s-1)}\sum_{\alpha,\beta\in\mc{T};\alpha\neq\beta}|{O}_{\alpha\beta}|^2=\Delta O_\mr{od}^2
}
for sufficiently large  $d$ and $d_s$ (see Appendix \ref{sec:bunsan}).
Therefore, we obtain the following expression:
\aln{\label{gaussy}
 \rho_{|O_{\alpha\beta}|}(y)=\begin{cases}
    \sqrt{\frac{2}{\sigma^2\pi}}e^{-y^2/2\sigma^2} &  \text{for the GOE},\\
    \frac{2y}{\sigma^2}e^{-y^2/\sigma^2} &  \text{for the GUE}.\\
  \end{cases}
}
We note that the expression for the GOE is Gaussian, as numerically indicated in Refs.~\cite{Steinigeweg13,Beugeling15}.
We remark, though, that the probability densities of $|O_{\alpha\beta}|$ are not Gaussian for the GUE.\footnote{It is expected that $\mr{Re}[O_{\alpha\beta}]$ and $\mr{Im}[O_{\alpha\beta}]$ independently obey Gaussian distributions.}
Assuming the ergodicity of random matrices, we can reconsider the probability densities in \EQ{gaussy} as spectral statistics.

\subsubsection{Singular operators}
Here we consider an example that does not satisfy \EQ{norio}.
The simplest example is an observable whose (modified) spectrum $\lrm{a'_i}$ can be written as
\aln{
 a'_i=\begin{cases}
    1 &  \text{(for a single $i=i_0$)},\\
    0 &  \text{(other $i$'s)}.\\
  \end{cases}
}
In this case, 
\aln{
\frac{\frac{1}{d}\lrs{\frac{1}{d}\sum_{i=1}^d{a'}_i^4}}{\lrs{\frac{1}{d}\sum_{i=1}^d{a'}_i^2}^2}=1
}
and the fluctuations of $s_\beta$ are not negligible.
We will call such operators that do not satisfy \EQ{norio} as singular operators, following Ref.~\cite{Brody81}.

Let us calculate the probability densities of a singular operator that can be written as follows:
\aln{\label{katati}
\hO=a\ket{\psi}\bra{\psi},
}
where $a$ is some real number.
We will call this type of operators as the ``most singular."
We first calculate the probability densities of $|O_{\alpha\beta}|^2=a^2|\braket{E_\alpha|\psi}|^2|\braket{E_\beta|\psi}|^2$ for $E_\alpha\neq E_\beta$.
If $d$ is large, $|\braket{E_\alpha|\psi}|^2$ and $|\braket{E_\beta|\psi}|^2$ become independent of each other~\cite{Brody81}.
Indeed, each of them follows the Porter-Thomas distributions in \EQ{PTd}.
The probability densities of $|O_{\alpha\beta}|^2$ can be calculated as
\aln{
\rho_{|O_{\alpha\beta}|^2}(y)&=\iint dxdz\rho_{|\braket{E_\alpha|\psi}|^2}(x)\rho_{|\braket{E_\beta|\psi}|^2}(z)\delta(y-a^2xz)\NON
&=\int_0^\infty \frac{dx}{a^2x}\rho_{|\braket{E_\alpha|\psi}|^2}(x)\rho_{|\braket{E_\beta|\psi}|^2}(y/a^2x)\NON
&=\begin{cases}
    \frac{d}{a\pi\sqrt{y}}K_0(0,d\sqrt{y}/a) &  \text{for the GOE},\\
    2(d/a)^2K_0(0,2d\sqrt{y}/a) &  \text{for the GUE},\\
  \end{cases}
}
where $K_0(0,y)=\int_0^\infty dze^{-y\mr{cosh} z}$ is a modified Bessel function of the second kind.
Changing the variables, we obtain
\aln{\label{aiueokaki}
\rho_{|O_{\alpha\beta}|}(y)&=\begin{cases}
    \frac{2d}{a\pi}K_0(0,d{y}/a) &  \text{for the GOE},\\
    4(d/a)^2yK_0(0,2d{y}/a) &  \text{for the GUE}.\\
  \end{cases}
}
We note that 
\aln{
\mc{V}&:=\av{s_\beta^2}\NON
&= a^2\av{|\braket{E_\beta|\psi}|^2}-a^4\av{|\braket{E_\beta|\psi}|^4}\NON
&\simeq\frac{a^2}{d},
}
which allows us to rewrite \EQ{aiueokaki} as
\aln{\label{k0ynise}
\rho_{|O_{\alpha\beta}|}(y)&\simeq\begin{cases}
    \frac{2\sqrt{d/\mc{V}}}{\pi}K_0(0,{y}\sqrt{d/\mc{V}}) &  \text{for the GOE},\\
    \frac{4yd}{\mc{V}}K_0(0,2{y}\sqrt{d/\mc{V}}) &  \text{for the GUE}.\\
  \end{cases}
}
Finally, in terms of $\sigma^2$, we can express \EQ{k0ynise} as
\aln{\label{k0y}
\rho_{|O_{\alpha\beta}|}(y)&\simeq\begin{cases}
    \frac{2}{\pi\sigma}K_0(0,{y}/\sigma) &  \text{for the GOE},\\
    \frac{4y}{\sigma^2}K_0(0,2{y}/\sigma) &  \text{for the GUE}.\\
  \end{cases}
}
Thus, we obtain the non-Gaussian distributions for singular operators.
Assuming the ergodicity of random matrices, we can reconsider the probability densities in \EQ{k0y} as spectral statistics.

\subsection{Conjectures from the random matrix model}\label{sec:conj}
We summarize the results of the matrix-element statistics for the random matrix models and clarify the conjecture about the statistics for actual nonintegrable models that conserve only energy.
As we have seen in Sec. \ref{sec:sre}, the analogy between RMT and the nonintegrable systems seems to hold true in a small energy shell $\mc{H}_\mr{sh}$.
Thus, we consider $\hat{O}:=\hat{\mc{P}}_\mr{sh}\hmo\hat{\mc{P}}_\mr{sh}$ for an observable $\hmo$ in nonintegrable systems.

\begin{figure}
\begin{center}
\includegraphics[width=14cm]{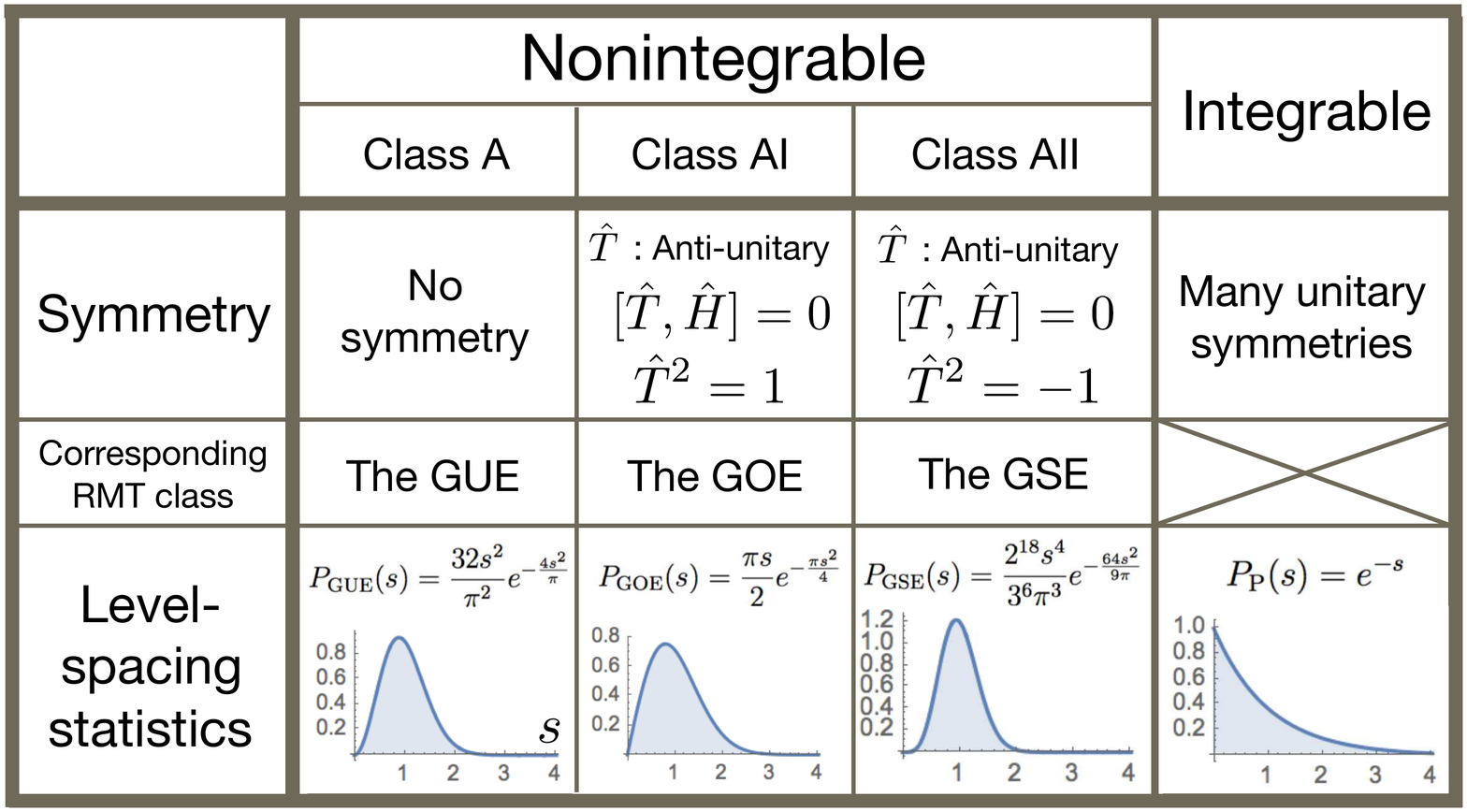}
\caption{Symmetry classifications of systems and corresponding level-spacing statistics.
}\label{symhyou}
\end{center}
\end{figure}

As is the case with the level-spacing statistics, the system is expected to be related to the random matrices with the same anti-unitary symmetry class.
Hamiltonians without any anti-unitary symmetry are said to belong to  ``Class A," the term adapted from the mathematical terminology.
Similarly, Hamiltonians with only one anti-unitary symmetry $\hat{T}$ that satisfies $\hat{T}^2=1$ belong to 
``Class AI," and 
Hamiltonians with only one anti-unitary symmetry $\hat{T}$ that satisfies $\hat{T}^2=-1$ belong to 
``Class AII."
We note that matrices in the GUE, the GOE and the GSE belong to Class A, Class AI, and Class AII, respectively (see Fig. \ref{symhyou}).

As illustrated in the left column of Fig. \ref{conjecture}, we conjecture that the ratios of standard deviations between diagonal and off-diagonal matrix elements within $\mc{H}_\mr{sh}$ become universal in nonintegrable systems, and that the universal values are determined from those of the random matrix models with the same symmetry class.
We note that even and odd properties of $\hmo$ are the same as those of $\hO$.
In the language of Srednicki's conjecture in \EQ{corre}, $\braket{|R_{\alpha\alpha}|^2}_\mc{T}=r^2\braket{|R_{\alpha\beta}|^2}_\mc{TT}\:(E_\alpha\neq E_\beta)$ and $\braket{|R_{\alpha\tilde{\alpha}}|}_\mc{T}^2=r'^2\braket{|R_{\alpha\beta}|^2}_\mc{TT}$ for $\omega<\omega_\mr{sh}$, where $r$ and $r'$ are determined from the symmetries of the Hamiltonian and those of the observable.

Moreover, we conjecture that the probability densities $\rho_{|\mc{O_{\alpha\beta}}|}(y)$ within $\mc{H}_\mr{sh}$ are those predicted by random matrix models, as shown in the right column of Fig. \ref{conjecture}.
In the language of Srednicki's conjecture in \EQ{corre}, $R_{\alpha\beta}$ is Gaussian (or $y\times$ Gaussian) only when the observable is nonsingular for $\omega<\omega_\mr{sh}$; if the observable is the most singular, $K_0$ functions appear.

\section{Numerical verifications of the random matrix predictions}\label{sec:nvrmt}
In this section, we show some numerical results that investigate the statistics of the matrix elements of various observables in nonintegrable systems.
In particular, focusing on the ratio $r,r'$ and the statistics of off-diagonal matrix elements, we ask if the RMT predictions in the previous section hold true in actual situations.

\subsection{Models}
We first introduce one-dimensional spin chain models that contain Ising interactions, transverse fields and Dzyaloshinskii-Moriya interactions as follows:
\aln{\label{spinchain}
\hat{H}&=\hat{H}_I+\hat{H}_{TF}+\hat{H}_{DM},\\
\hat{H}_\mr{I}&= -\sum_{i=1}^{N-1}J_{i}\hat{\sigma}_i^z\hat{\sigma}_{i+1}^z,\\
\hat{H}_\mr{TF}&= -\sum_{i=1}^{N}(h'\hat{\sigma}_i^x+h\hat{\sigma}_i^z),\\
\hat{H}_\mr{DM}&=\sum_{i=1}^{N-1}\vec{D}\cdot (\vec{\hat{\sigma}}_i\times\vec{\hat{\sigma}}_{i+1})\NON
&=\frac{D}{\sqrt{2}}\sum_{i=1}^{N-1}\lrl{\lrs{\hat{\sigma}_i^y\hat{\sigma}_{i+1}^z-\hat{\sigma}_i^z\hat{\sigma}_{i+1}^y}+
\lrs{\hat{\sigma}_i^x\hat{\sigma}_{i+1}^y-\hat{\sigma}_i^y\hat{\sigma}_{i+1}^x}
},
}
where $N$ denotes the number of spins, $\vec{D}=D\frac{1}{\sqrt{2}}(\vec{e}_x+\vec{e}_z)$, and we impose the open boundary condition.
In addition, we assume that $J_i=J(1+\epsilon_i)$ is a random variable that breaks the reflection symmetry of sites ($i\ra N-i$), where $\epsilon_i$ is uniformly chosen from $[-0.1,0.1]$ at each site.\footnote{As we will see in the following discussions, the randomness is sufficiently weak and no localization arises.}

The model in \EQ{spinchain} can be a nonintegrable model that conserves only energy by changing the strength of $\hH_\mr{TF}$ and $\hH_\mr{DM}$.
Further, by changing the parameter of the interactions and $N$,  nonintegrable systems that belong to Class A, AI, and AII are obtained.
We note that our model is unique in a sense that all these three classes are achievable by changing only a few parameters.
In the followings, we assume $J=1, h'=-2.1h$ and consider three nonintegrable models (a), (b), and (c), which are determined by the parameters $h$ and $D$.

\begin{figure}
\begin{center}
\includegraphics[width=14cm]{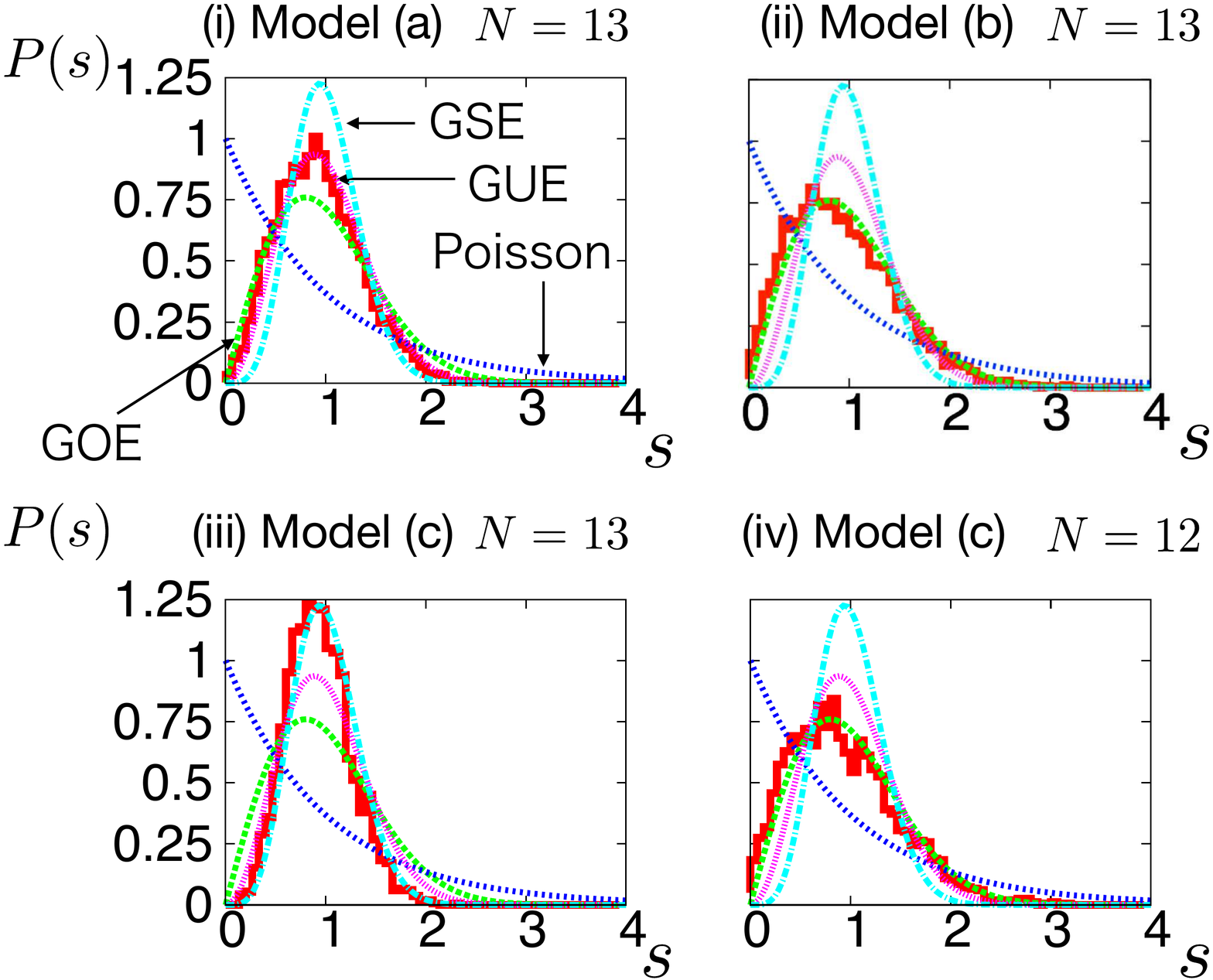}
\caption{Level-spacing statistics $P(s)$ for models (a), (b), (c) with the Hamiltonians in \EQ{spinchain}.
Superimposed are the level-spacing statistics for the Poisson, the GUE, the GOE, and the GSE is also shown.
(i) Model (a) ($N=13,h=0.5,D=0.9$). The statistics resembles that of the GUE.
(ii) Model (b) ($N=13,h=0.5,D=0$). The statistics resembles that of the GOE.
(iii) Model (c) ($N=13,h=0,D=0.9$). The statistics resembles that of the GSE.
(iv) Model (c) ($N=12,h=0,D=0.9$). The statistics resembles that of the GOE.
}
\label{level_spacings}
\end{center}
\end{figure}

First, model (a) is a model without any anti-unitary symmetry, which is obtained by taking $h=0.5$ and $D=0.9$ (i.e., model (a) belongs to Class A).
If we calculate the level-spacing statistics of model (a), it obeys  statistics similar to the level statistics of the GUE, $P_\mr{GUE}(s)=\frac{32s^2}{\pi^2}e^{-\frac{4s^2}{\pi}}$, as shown in Fig. \ref{level_spacings}(i).
Here, we also show the Poisson statistics $P_\mr{P}(s)=e^{-s}$, the GOE statistics $P_\mr{GOE}(s)=\frac{\pi s}{2}e^{-\frac{\pi s^2}{4}}$, and the GSE statistics $P_\mr{GSE}(s)=\frac{2^{18}s^4}{3^6\pi^3}e^{-\frac{64s^2}{9\pi}}$ for comparison (see Fig. \ref{symhyou}).

Next, model (b) is a model with one anti-unitary symmetry $\hat{T}=\hat{K}\:(\hat{T}^2=1)$, which is obtained by taking $h=0.5$ and $D=0$ (i.e., model (b) belongs to Class AI).
Here, $\hat{K}$ denotes the complex conjugate operator.
Note that $\hat{K}\hat{\sigma}_i^x\hat{K}^{-1}=\hat{\sigma}_i^x, \hat{K}\hat{\sigma}_i^y\hat{K}^{-1}=-\hat{\sigma}_i^y$, and $\hat{K}\hat{\sigma}_i^z\hat{K}^{-1}=\hat{\sigma}_i^z$ are satisfied.
If we calculate the level-spacing statistics of model (b), it obeys  statistics similar to the level statistics of the GOE, as shown in Fig. \ref{level_spacings}(ii).

Finally, model (c) is a model with one anti-unitary symmetry $\hat{T}=\hat{T}_0:=\lrs{\prod_{i=1}^{N}[i\hat{\sigma}_i^y]}\hat{K}$, which is obtained by taking $h=0$ and $D=0.9$.
Since $\hat{T}_0^2=(-1)^N$, model (c) belongs to Class AI if $N$ is even and Class AII if $N$ is odd.
Indeed, when $N=13$, the model obeys statistics similar to the level statistics of the GSE, as shown in Fig. \ref{level_spacings}(iii).
On the other hand, when $N=12$, the level statistics resembles that of the GOE, as shown in Fig. \ref{level_spacings}(iv).

\subsection{Few-body observables}\label{tatatatata}
Using the models defined above, we first investigate matrix elements of few-body observables.
We consider the $z$-component of a spin at a certain site $\hmo_1:=\hat{\sigma}^z_{\left[N/2\right]+1}$
and the correlation of two spins $\hat{\mc{O}}_2:=\hat{\sigma}^z_{\left[N\right/2]+1}\hat{\sigma}^z_{\left[N\right/2]+2}$, where $[x]$ denotes the maximum integer that does not exceed $x$.
Since $\hmo_1$ satisfies $\hat{K}\hat{\mc{O}}_1\hat{K}^{-1}=\hat{\mc{O}}_1$, it is an even operator for models (a) and (b).
On the other hand, since $\hat{T}_0\hat{\mc{O}}_1\hat{T}_0^{-1}=-\hat{\mc{O}}_1$, it is odd for model (c).
As for $\hat{\mc{O}}_2$, it is an even operator for all of the models because $\hat{K}\hat{\mc{O}}_2\hat{K}^{-1}=\hat{\mc{O}}_2,\hat{T}_0\hat{\mc{O}}_2\hat{T}_0^{-1}=\hat{\mc{O}}_2$.

\begin{figure}
\begin{center}
\includegraphics[width=12cm]{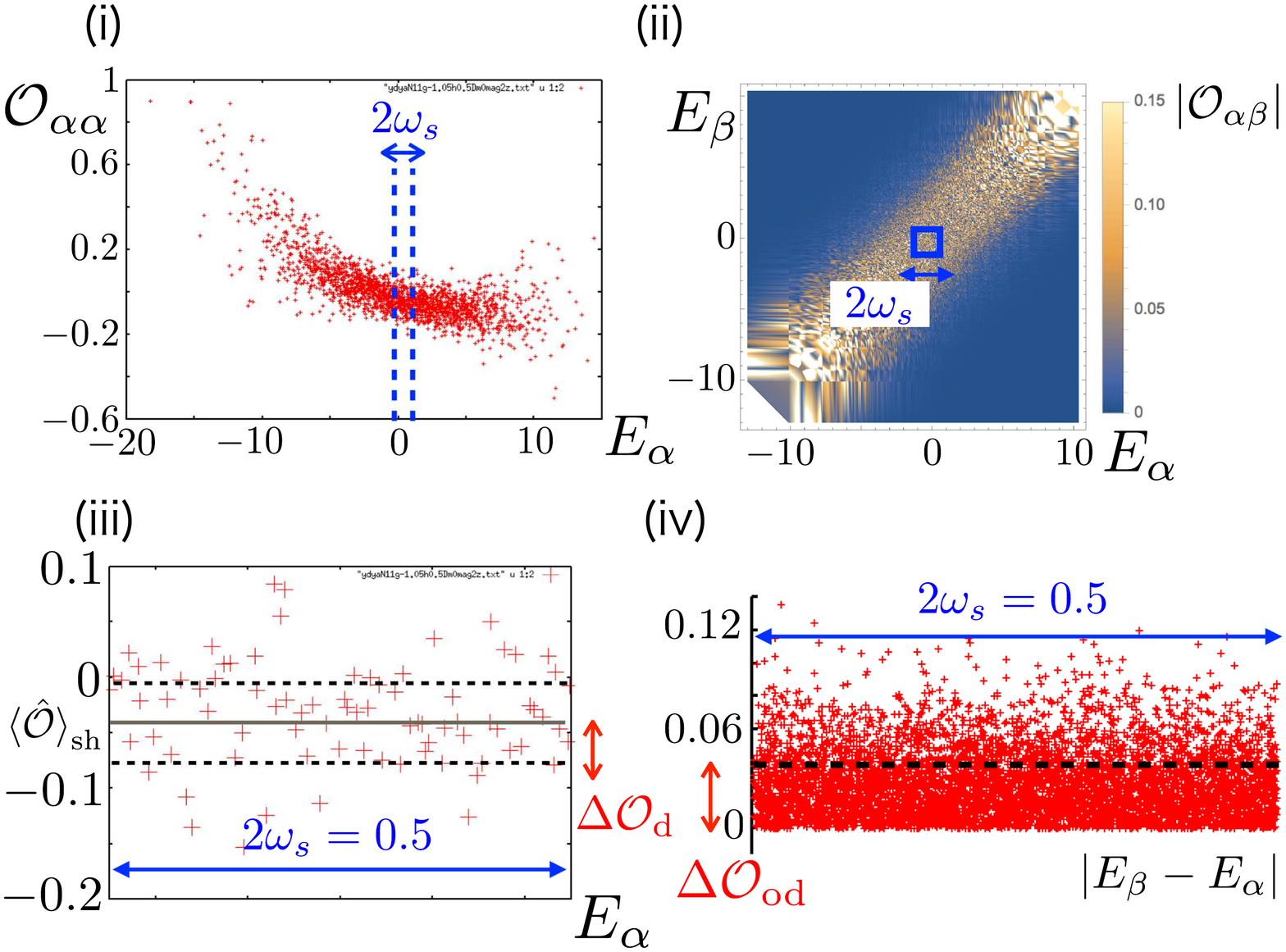}
\caption{Diagonal and off-diagonal matrix elements of $\hmo=\hmo_1$ for model (b). We use $N=11$ for (i), (iii), (iv), and $N=8$ for (ii).
(i) Diagonal matrix elements $\mc{O}_{\alpha\alpha}$ for all of the eigenstates as a function of $E_\alpha$.
(ii) Density plot of the absolute value of the off-diagonal matrix elements as a function of $E_\alpha$ and $E_\beta$.
(iii) Plot of diagonal matrix elements in an energy shell with width $2\omega_{s}=0.5$.
(iv) Plot of off-diagonal matrix elements in an energy shell with width $2\omega_{s}=0.5$.
}\label{o1_b}
\end{center}
\end{figure}

We first show the example of diagonal and off-diagonal matrix elements of $\hmo=\hmo_1$ for model (b) in Figs. \ref{o1_b}(i)-(iv).
Figure (i) shows the diagonal matrix elements $\mc{O}_{\alpha\alpha}$ for all of the eigenstates as a function of $E_\alpha$.
Similarly, Figure (ii) shows the density plot of the absolute value of the off-diagonal matrix elements as a function of $E_\alpha$ and $E_\beta$.
Both of these figures show that the behavior of the matrix elements depends on the global energy: for example, Figure (ii) shows that the typical magnitude of $|\mc{O}_{\alpha\beta}|$ vanishes as $|E_\alpha-E_\beta|$ becomes large.
However, if we stick to some small energy shell $\mc{H}_{s}$, the matrix elements $\mc{O}_{\alpha\beta}=O_{\alpha\beta}\:(\alpha,\beta\in\mc{T})$ seems to fluctuate randomly from one eigenstate to another eigenstate with a constant amplitude.
Indeed, if we take an energy shell with width $2\omega_s=0.5$ and plot matrix elements in that energy shell, we obtain Figure (iii) for the diagonal matrix elements and Figure (iv) for the off-diagonal matrix elements as a function of $|E_\beta-E_\alpha|$.
Using the eigenstates within the energy shell, we can consider $r=\frac{\Delta\mc{O}_\mr{d}}{\Delta\mc{O}_\mr{od}},r'=\frac{\Delta\mc{O}_\mr{K}}{\Delta\mc{O}_\mr{od}}$ and the probability densities of $\mc{O}_{\alpha\beta}$.\footnote{We note that $\Delta\mc{O}_\mr{d}$ is numerically calculated from modified fluctuations where the effect of the energy shell is reduced. We make a linear fitting $O_\mr{m}(\tilde{E})=a\tilde{E}+b\:(|\tilde{E}-E|<\omega_s)$ within a small energy shell instead of a constant $O_\mr{m}(E)$. Then we consider the variance of $\mc{O}_{\alpha\alpha}-O_\mr{m}(\tilde{E}=E_\alpha)$.}

\subsubsection{The universal ratios}
\begin{figure}
\begin{center}
\includegraphics[width=11cm,angle=-90]{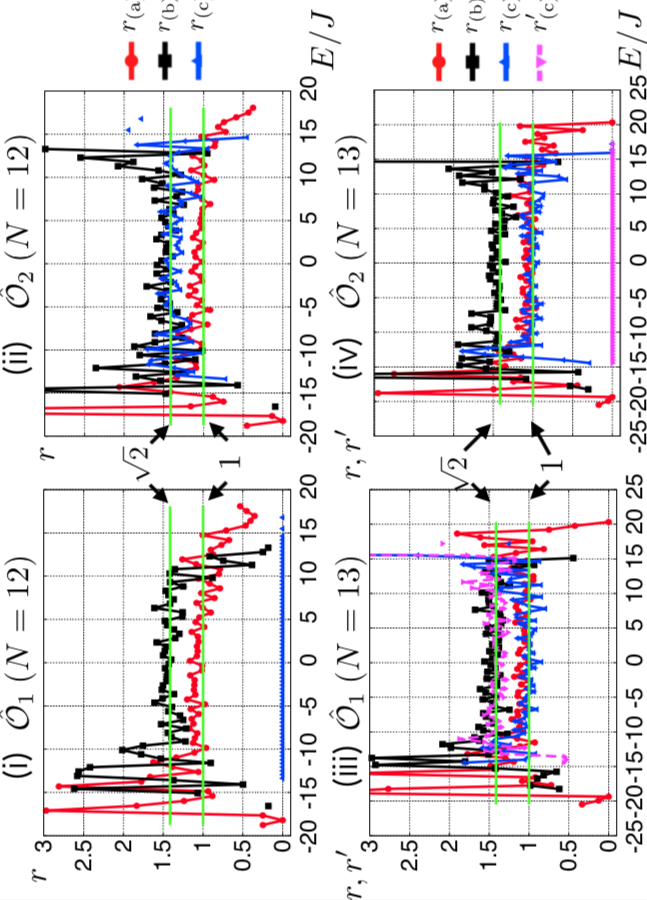}
\caption{Ratios of the standard deviations of the matrix elements as a function of energy (we take $F=6N$) in \EQ{eqratio}.
The results are shown for the case with (i) $\hat{\mc{O}}_1,N=12$, (ii) $\hat{\mc{O}}_2,N=12$, (iii) $\hat{\mc{O}}_1,N=13$, and (iv) $\hat{\mc{O}}_2,N=13$.
Each graph shows the results of $r$ for models (a) (circle), (b) (square), and (c) (upward triangle).
For $N=13$, we also show the results of $r'$ for model (c) (downward triangle).
By analyzing the data from the viewpoint of symmetry, we can see that the RMT conjecture is valid for these few-body operators.
(We note that some data points are missing for the edges of the spectrum because only few eigenstates exist there.)
}\label{ratio}
\end{center}
\end{figure}
We show the results of $r$ calculated for models (a), (b), and (c) with $N=12$ as a function of energy in Figs. \ref{ratio}(i) and (ii).
We have calculated standard deviations from the eigenstates within the range $[E-\omega_s,E+\omega_{s}]$, which is obtained by dividing the entire spectrum into $F$ regions.
Here, $\omega_{s}:=\frac{E_\mr{max}-E_\mr{min}}{2F}$ ($E_\text{max(min)}$ is the maximum (mimimum) energy eigenvalue) and we assume $\omega_{s}<\omega_\mr{sh}$.\footnote[2]{We have confirmed that the small change of $\omega_s$ does not affect the discussion below.}
Figure \ref{ratio}(i) shows the results for $\hmo_1$.
For a wide range of spectrum (except for the edges), $r_\mr{(a)}\simeq 1, r_\mr{(b)}\simeq \sqrt{2}$ and $r_\mr{(c)}=0$ are obtained, where the subscript indicates the type of the model.
Similarly, from the results for $\mc{\hat{O}}_2$ shown in Figure \ref{ratio}(ii), we obtain
$r_\mr{(a)}\simeq 1, r_\mr{(b)}\simeq \sqrt{2}$, and  $r_\mr{(c)}\simeq \sqrt{2}$.
These results are consistent with the RMT conjecture that predicts
$r_\mr{A,even}=1, r_\mr{AI,even}=\sqrt{2}, r_\mr{AI,odd}=0$.

We next show the results of $r$ calculated for models (a), (b), and (c) (and $r'$ for model (c)) with $N=13$ as a function of energy in Figs. \ref{ratio}(iii) and (iv).
Figure \ref{ratio}(iii) shows that the results for $\hmo_1$ with $N=13$ are almost the same as those with $N=12$ if we consider models (a) or (b).
On the other hand, we have $r_\mr{(c)}\simeq 1$ and $r'_\mr{(c)}\simeq \sqrt{2}$ for model (c).
Figure \ref{ratio}(iv) shows that the results for $\hmo_2$ with $N=13$ are again almost the same as those with $N=12$ if we consider models (a) and (b).
For model (c), we have $r_\mr{(c)}\simeq 1$ and $r'_\mr{(c)}=0$.
These results are consistent with the RMT conjecture that predicts
$r_\mr{AII,even}=1, r_\mr{AII,odd}=1,r'_\mr{AII,even}=0$, and $r'_\mr{AII,odd}=\sqrt{2}$.

\begin{figure}
\begin{center}
\includegraphics[width=5cm,angle=-90]{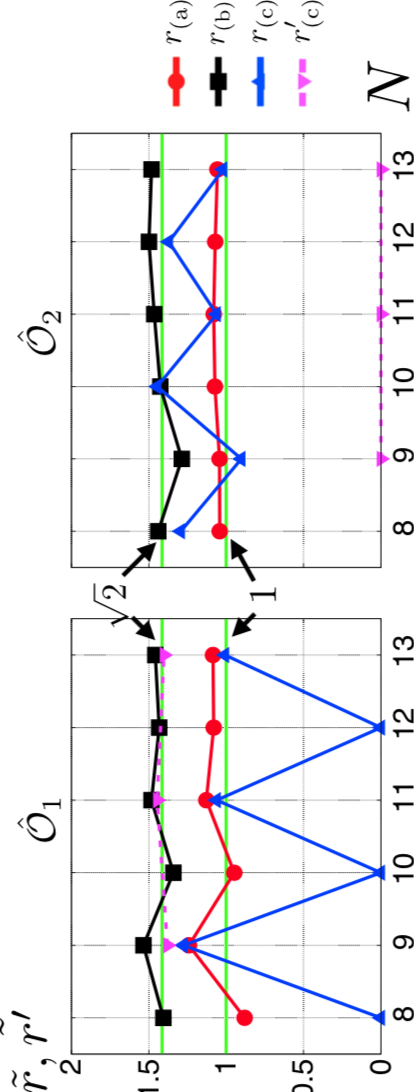}
\caption{$N$-dependences of $\tilde{r}$ and $\tilde{r'}$ for models (a),(b), and (c), where the averages are obtained from the middle one-third  of the spectrum, and $F=3N$ is used for $N\leq 10$ and $F=6N$ is used for $N\geq 11$.
The ratio $\tilde{r}$ is $1$ and $\sqrt{2}$ for models (a) and (b) independent of $N$, respectively, since model (a)/(b) belongs to Class A/AI irrespective of $N$.
On the other hand, for model (c), $\tilde{r}$ and $\tilde{r'}$ depend on the parity of $N$ because the model belongs to Class AI/AII when $N$ is even/odd.
}\label{fssss}
\end{center}
\end{figure}

To show that these results indeed depend only on the parity of $N$, we show the $N$-dependences of $\tilde{r}$ and $\tilde{r'}$ in Fig. \ref{fssss}, where $\tilde{r}$ and $\tilde{r'}$ are the average values of $r(E)$ and $r'(E)$ in the middle of the spectrum, respectively.
As shown in the graphs, $\tilde{r}$ is $1$ and $\sqrt{2}$ for models (a) and (b) independent of $N$, respectively, since model (a)/(b) belongs to Class A/AI irrespective of $N$. (Note that both $\hmo_1$ and $\hmo_2$ are even operators.)
On the other hand, for model (c), $\tilde{r}$ and $\tilde{r'}$ depend on the parity of $N$ because the model belongs to Class AI/AII when $N$ is even/odd.

These results indicate that the ratio of standard deviations in nonintegrable systems for few-body observables is consistent with the conjecture of the random matrix model, even when the anti-unitary symmetry of the systems and the observables are varied.
In contrast to the previous study~\cite{Steinigeweg13} where the authors claim that the diagonal and off-diagonal matrix elements seem unrelated, our results indicate that they are actually related.
This fact strengthens the validity of the RMT predictions that may be related to the underlying mechanism of the ETH (for the small energy shell).

\subsubsection{Probability densities of the off-diagonal elements}
\begin{figure}
\begin{center}
\includegraphics[width=12cm]{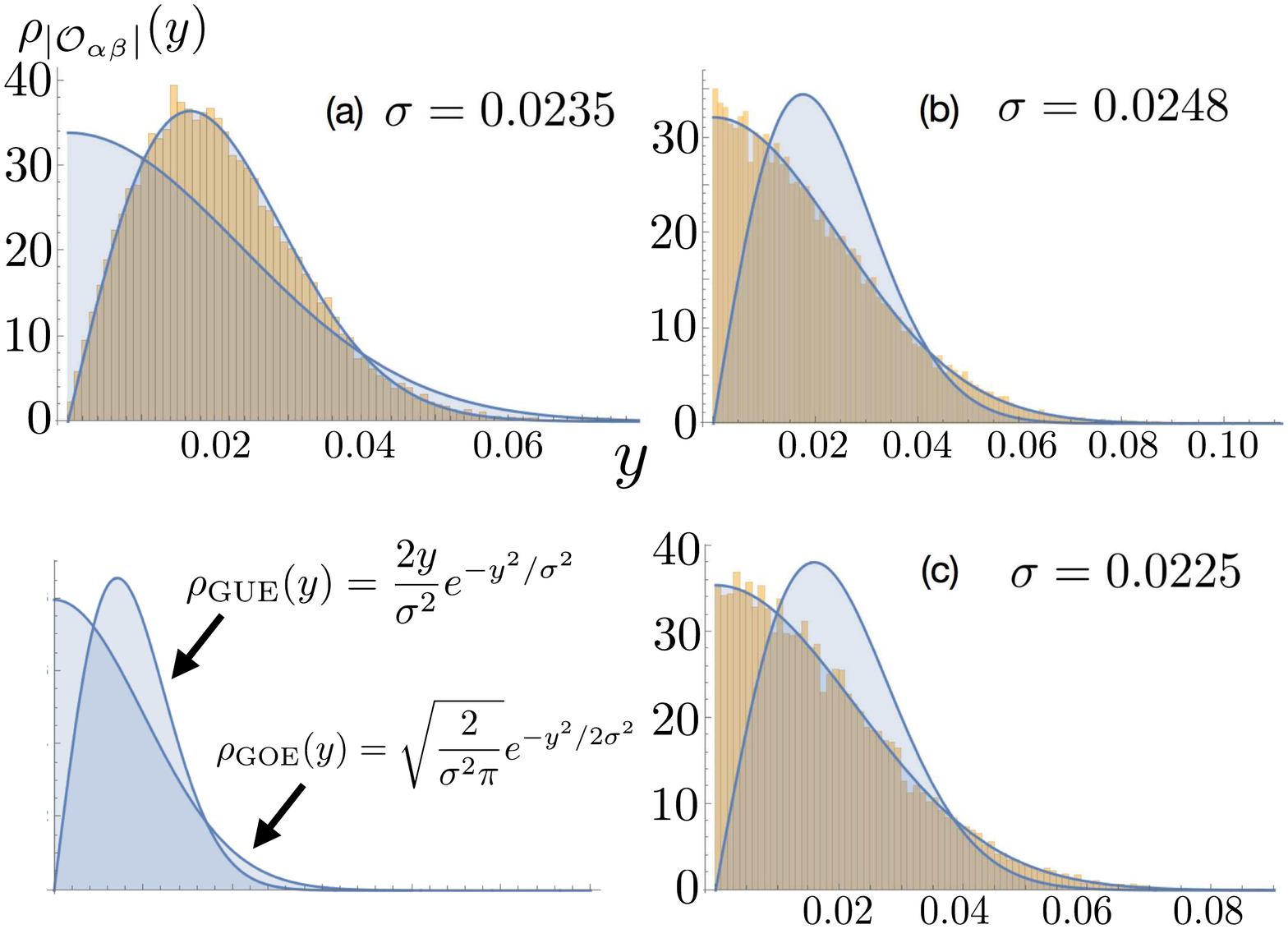}
\caption{Probability densities of $|\mc{O}_{\alpha\beta}|\:(\alpha,\beta\in\mc{T})$ for models (a) (upper left), (b) (upper right), and (c) (bottom right).
We take $N=12$ and $d_{s}=200$.
As a reference, we also show the predictions of random matrices in \EQ{gaussy}.
}\label{o1d_abc}
\end{center}
\end{figure}

Next, we consider the probability densities of the off-diagonal matrix elements for few-body observables and show that they obey the conjecture of the random matrix models.
For simplicity, we consider $\hmo=\hmo_1$ and $N=12$.
In Figure \ref{o1d_abc}, we show the probability density of $|\mc{O}_{\alpha\beta}|\:(\alpha,\beta\in\mc{T})$ for models (a), (b), and (c).
Here, we take an energy shell such that
\aln{
\mc{T}=\lrm{\alpha: \frac{D-d_{s}}{2}\leq \alpha<\frac{D+d_{s}}{2}},
}
where $D$ is the dimension of the entire Hilbert space.
As a reference, we also show the predictions of random matrices in \EQ{gaussy}.
We note that $\sigma=\Delta \mc{O}_\mr{od}$ can be calculated numerically.
We can see that the probability density for model (a) obeys the statistics corresponding to the GUE 
and that the probability density for models (b) and (c) obeys the statistics corresponding to the GOE.
These results are consistent with the conjecture of the random matrix model.

\subsection{Many-body correlations}
In this subsection, we investigate the matrix elements of (nonsingular) many-body operators.
We especially consider $l$-body spin correlations that are defined by
\aln{\label{oiyo}
\hmo_l:=\prod_{n=1}^l\hat{\sigma}_n^z \:\:\:(l\geq 3).
}
For models (a) and (b), $\hmo_l$ is always an even operator.
For model (c), $\hmo_l$ is even if $l$ is even, and $\hmo_l$ is odd if $l$ is odd.

\begin{figure}
\begin{center}
\includegraphics[width=11cm,angle=90]{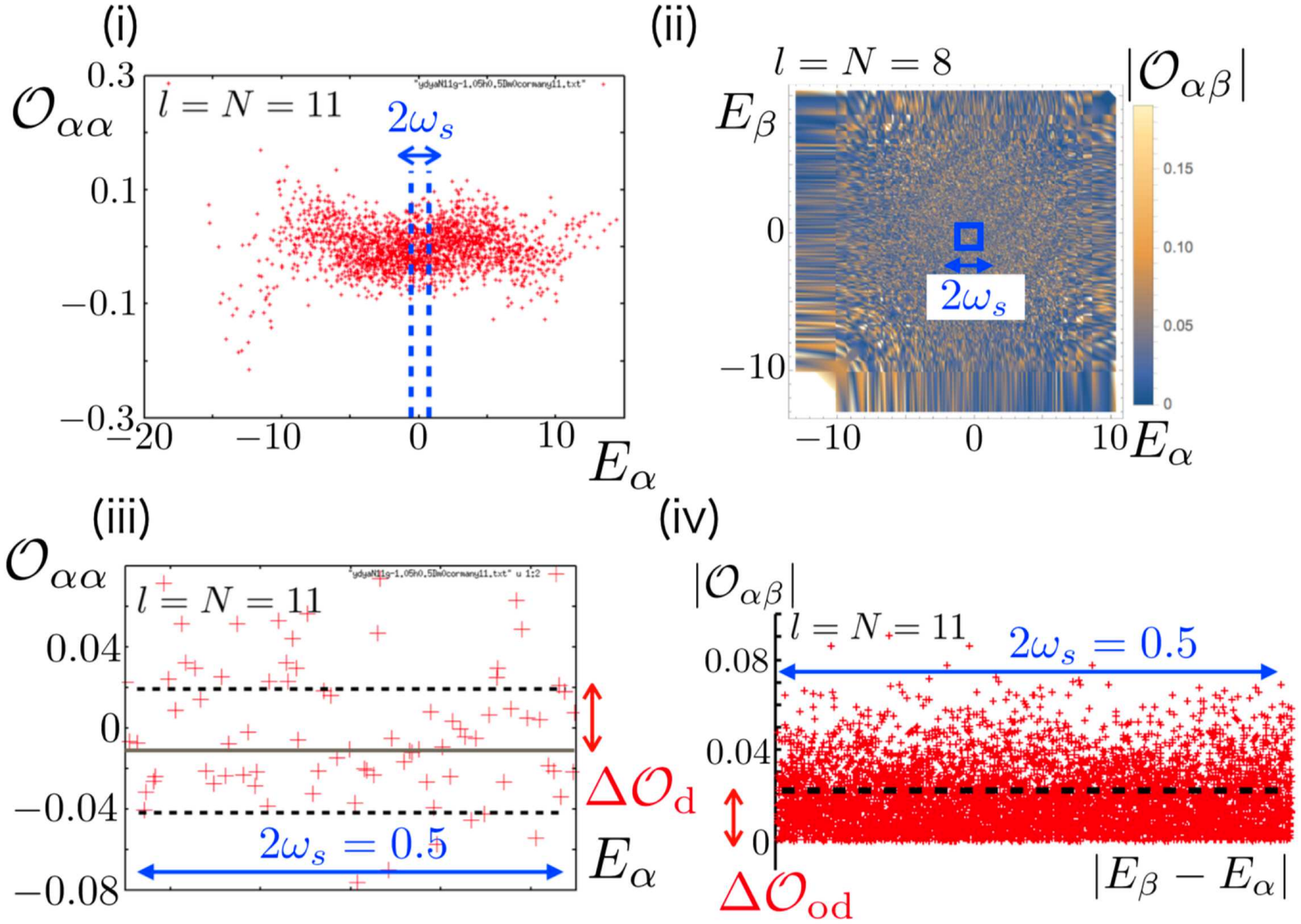}
\caption{Diagonal and off-diagonal matrix elements of $\hmo=\hmo_l$ for model (b). We use $N=l=11$ for figures (i), (iii), (iv), and $N=l=8$ for figure (ii).
(i) Diagonal matrix elements $\mc{O}_{\alpha\alpha}$ for all of the eigenstates as a function of $E_\alpha$.
(ii) Density plot of the absolute values of the off-diagonal matrix elements as a function of $E_\alpha$ and $E_\beta$.
(iii) Plot of diagonal matrix elements in an energy shell with width $2\omega_{s}=0.5$.
(iv) Plot of off-diagonal matrix elements in an energy shell with width $2\omega_{s}=0.5$.
}\label{ol_b}
\end{center}
\end{figure}

As for few-body observables, we first show the example of diagonal and off-diagonal matrix elements of $\hmo=\hmo_l$ for model (b) in Figs. \ref{ol_b}(i)-(iv).
Figure (i) shows the diagonal matrix elements $\mc{O}_{\alpha\alpha}$ for all of the eigenstates as a function of $E_\alpha$.
Similarly, Figure (ii) shows the density plot of the absolute value of the off-diagonal matrix elements as a function of $E_\alpha$ and $E_\beta$.
Both of these figures show that the behavior of the matrix elements depends on the global energy.
However, compared with the case of few-body observable (see Fig. \ref{o1_b}), the dependence of the global energy is evident neither for diagonal nor off-diagonal matrix elements.
Anyway, we take some small energy shell with width $2\omega_{s}=0.5$ and plot matrix elements.
Then, we obtain Figure (iii) for the diagonal matrix elements and Figure (iv) for the off-diagonal matrix elements as a function of $|E_\beta-E_\alpha|$.
Using the eigenstates within the energy shell, we can again consider $r=\frac{\Delta\mc{O}_\mr{d}}{\Delta\mc{O}_\mr{od}}$ and the probability densities of $\mc{O}_{\alpha\beta}$.

\subsubsection{The universal ratios}

\begin{figure}
\begin{center}
\includegraphics[width=14cm]{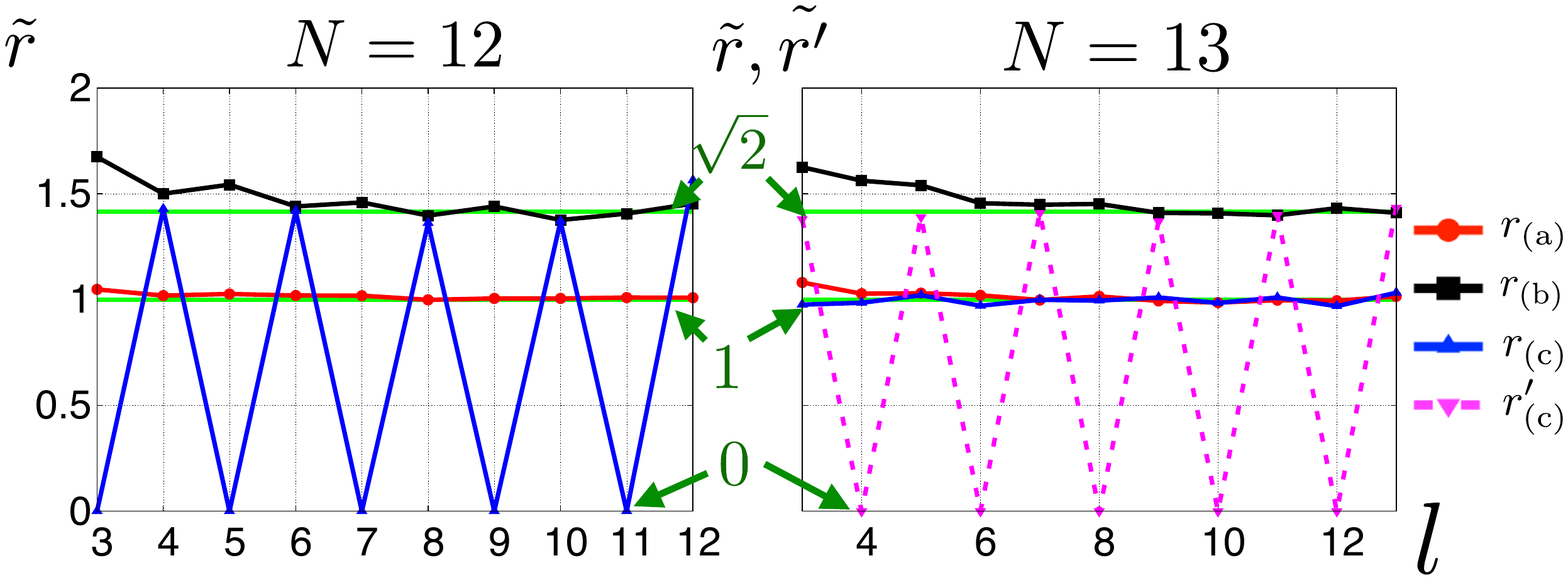}
\caption{$l$-dependences of $\tilde{r}$ and $\tilde{r'}$ for models (a), (b), and (c), where the averages are obtained from the middle one-third of the spectrum ($F=6N$ is used).
The ratio $\tilde{r}$ is $1$ and $\sqrt{2}$ for models (a) and (b) independent of $N$, respectively, since model (a)/(b) belongs to Class A/AI irrespective of $N$ and $\hmo_l$ is always even.
On the other hand, for model (c), $\tilde{r}$ and $\tilde{r'}$ depend on the parity of $N$ and $l$ because of the changes of the symmetry.
}\label{old_abc}
\end{center}
\end{figure}

We show $l$-dependence of $\tilde{r}$ and $\tilde{r'}$ for $N=12$ and $N=13$ in Fig. \ref{old_abc}.
As shown in the graphs, $\tilde{r}$ and $\tilde{r'}$ become universal even for the large $l$'s that are comparable to $N$, namely for many-body correlations.
Indeed, $\tilde{r}$ is $1$ and $\sqrt{2}$ for models (a) and (b) independent of $N$, respectively, since model (a)/(b) belongs to Class A/AI irrespective of $N$ and $\hmo_l$ is always even.
On the other hand, for model (c), $\tilde{r}$ and $\tilde{r'}$ depend on the parity of $N$ and $l$ because of the changes of the symmetry.

\subsubsection{Probability densities of the off-diagonal elements}
\begin{figure}
\begin{center}
\includegraphics[width=11cm,angle=-90]{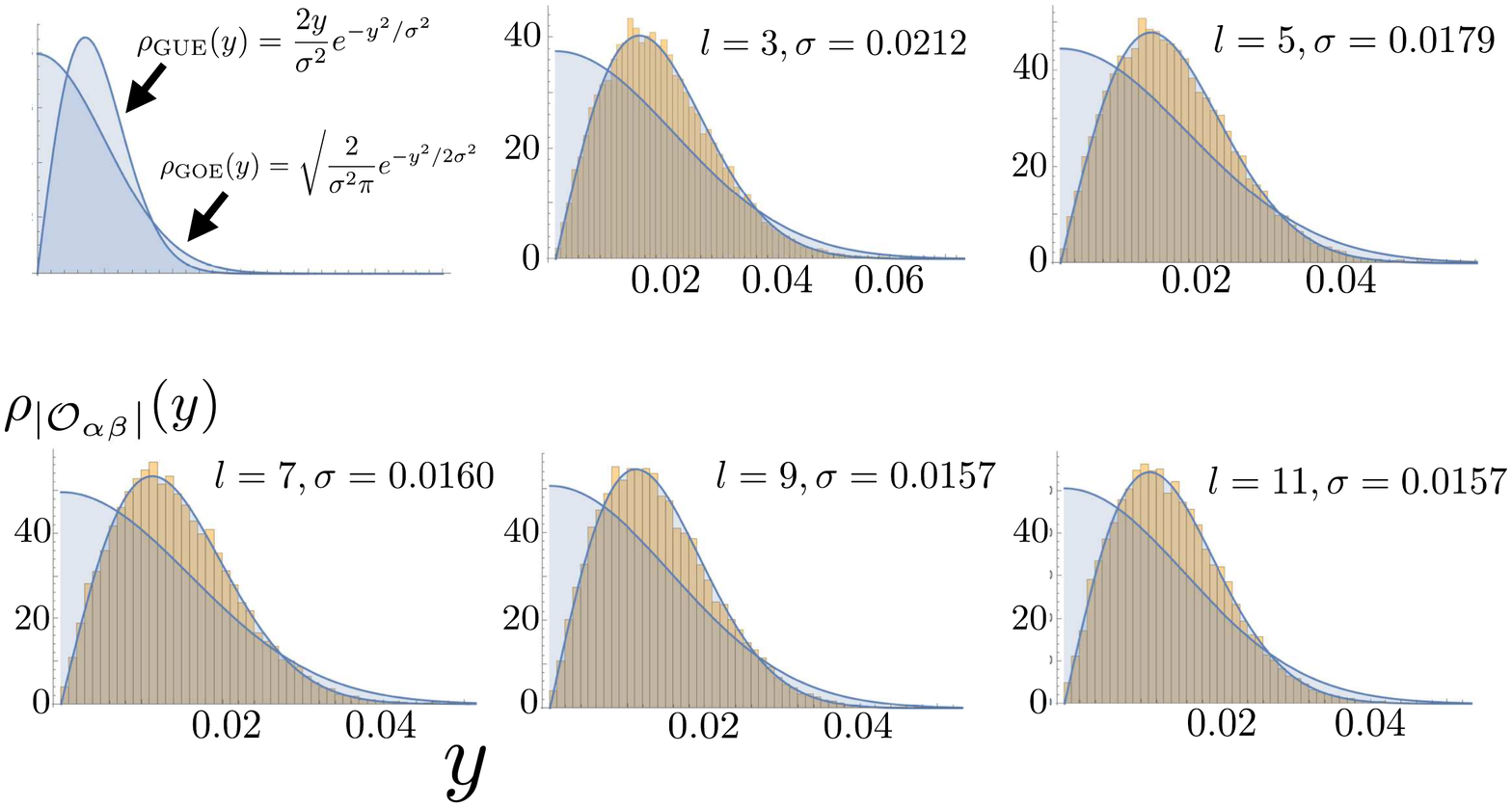}
\caption{Probability densities of $|\mc{O}_{\alpha\beta}|\:(\alpha,\beta\in\mc{T})$ for model (a) with $N=12$.
We take $d_{s}=200$.
As a reference, we also show the predictions of random matrices in \EQ{gaussy}.
We can see that the probability density obeys the statistics corresponding to the GUE for all $l$,
which indicates that the conjecture from RMT is valid even for the many-body observables.
}\label{oldd_b}
\end{center}
\end{figure}

Next, we consider the probability densities of the off-diagonal matrix elements of many-body observables $\hmo=\hmo_l$ and show that they also obey the conjecture of the random matrix models.
For simplicity, we consider model (a) and $N=12$.
In Figure \ref{oldd_b}, we show the probability density of $|\mc{O}_{\alpha\beta}|\:(\alpha,\beta\in\mc{T})$ for $l=3,5,7,9$, and 11.
As a reference, we also show the predictions from RMT in \EQ{gaussy}.
We can see that the probability density obeys the statistics corresponding to the GUE for all $l$.
This result indicates that the conjecture of the random matrix model is valid even for the many-body observables.

The results of the ratios and the probability densities indicate that the RMT conjecture about the ETH and its finite-size corrections within the small energy shell may be valid even for many-body observables.
In fact, numerical simulations suggest that the ETH seems true for many-body operators, too  (see Appendix \ref{sec:ethmany}).
Since the ETH seems true, even many-body observables are expected to relax to stationary values that are  describable by the microcanonical ensemble.
This fact is beyond the conventional notion of the thermal equilibrium introduced in Chapter \ref{ch:Eq}, namely MITE and MATE.\footnote[1]{In Ref.~\cite{Hosur16}, it is reported that many-body operators may satisfy the ETH for diagonal matrix elements. However, it is difficult to interpret the possible origin of the ETH from their measure of justifying the ETH.}
Our results imply that local, macroscopic or few-body nature of observables is not a necessary condition for proving the (strong) ETH in a nonintegrable system (in contrast to the suggestion by Ref.~\cite{Mueller15}).
We rather have to investigate why the random matrix description is valid in describing the finite-size corrections of the ETH for actual situations.

\subsection{Density matrices corresponding to pure states}
Here, we investigate the statistics of singular operators.
As a singular operator, we take a density matrix corresponding to a pure state with a form $\hmo=\ket{\psi}\bra{\psi}$.\footnote[2]{Note that $\hmo$ is also the most singular observable even after projecting onto $\mc{H}_\mr{sh}$.
Indeed, since
\aln{
\mc{\hat{P}}_\mr{sh}\ket{\psi}\bra{\psi}\mc{\hat{P}}_\mr{sh}=\braket{\psi|\mc{\hat{P}}_\mr{sh}|\psi}\frac{\mc{\hat{P}}_\mr{sh}\ket{\psi}}{\sqrt{\braket{\psi|\mc{\hat{P}}_\mr{sh}|\psi}}}\frac{\bra{\psi}\mc{\hat{P}}_\mr{sh}}{\sqrt{\braket{\psi|\mc{\hat{P}}_\mr{sh}|\psi}}},
}
it can be regarded as the form of \EQ{katati} with $a=\braket{\psi|\mc{\hat{P}}_\mr{sh}|\psi}$.
}
We note that the matrix elements $\mc{O}_{\alpha\beta}$ are relevant for the $\rho_{\alpha\beta}$ that appeared in the previous chapters if we regard $\ket{\psi}$ as an initial pure state.

If we take an eigenstate of $\hH$ as $\ket{\psi}$, the ETH does not hold true as we have seen in \SEC{sec:mot}.
In this case, the conjecture of the random matrix models is not valid, which means that there is always a counterexample of this conjecture even in nonintegrable systems.
So, is there any pure state that satisfies the conjecture of the random matrix models?

To see this is the case, we take a $\gamma$-th energy eigenstate of another Hamiltonian $\hH_0$ as $\ket{\psi}$, where $\hH_0$ is a Hamiltonian with $h=0.05, D=0$ in \EQ{spinchain}.\footnote[3]{We assume that the bond-dependent interaction is the same for $\hH$ and $\hH_0$. Then $\mc{O}_{\alpha\beta}$ is relevant for the quench setup where the Hamiltonian is suddenly changed from $\hH_0$ to $\hH$.}
For simplicity, we take $\gamma=D/2+1$ (a highly excited state) and $\gamma=1$ (a ground state).
We note that $\hmo$ is an even operator for both values of $\gamma$ in model (b).\footnote[4]{Since $\hat{K}\hH_0\hat{K}^{-1}=\hH_0$ and $\hH_0$ has no degeneracies (which is confirmed with numerics), we can assume that an eigenstate $\ket{E_\gamma}_0$ satifies $\hat{K}\ket{E_\gamma}_0=\ket{E_\gamma}_0$.}

\begin{figure}
\begin{center}
\includegraphics[width=12cm]{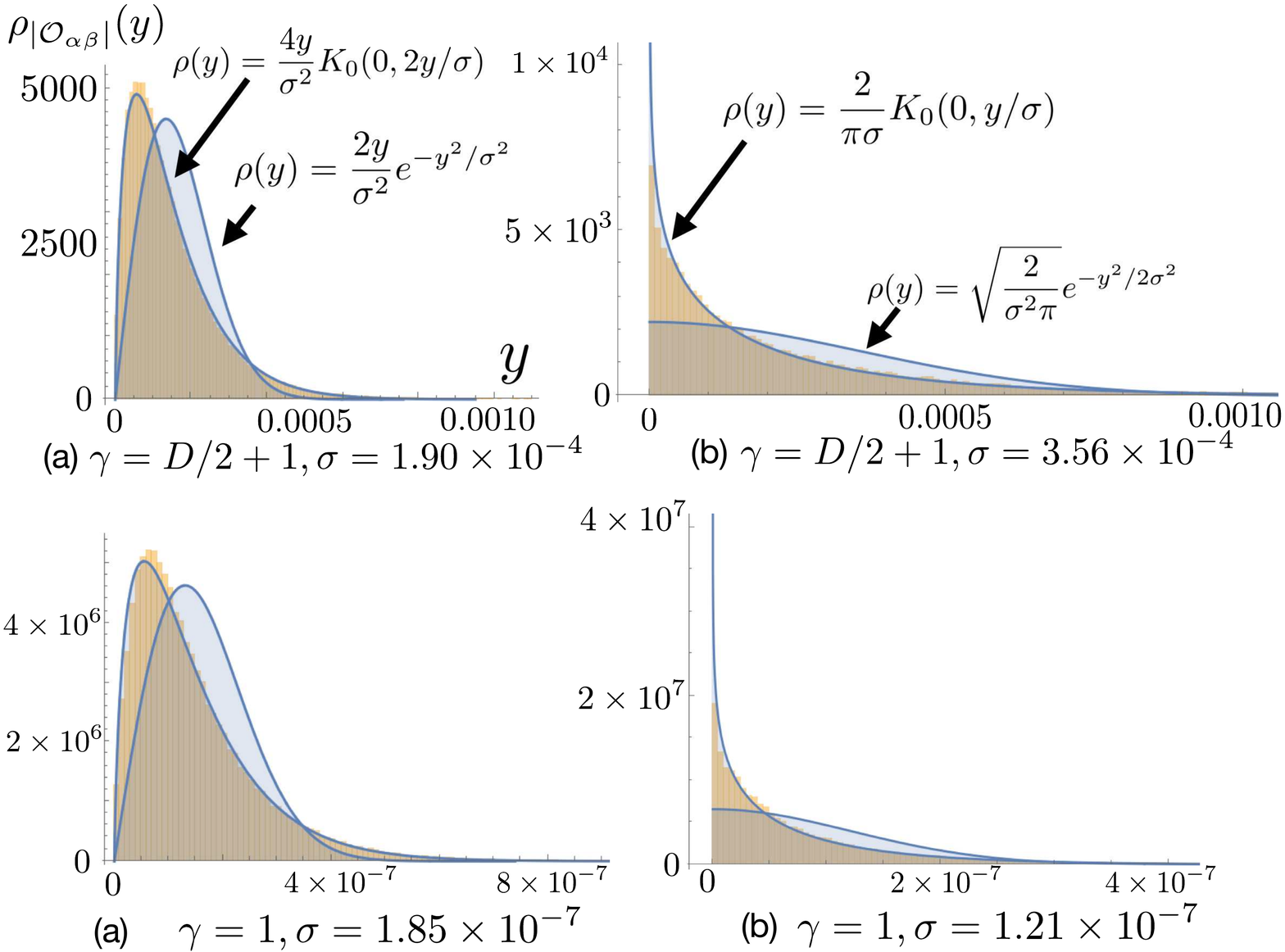}
\caption{Probability densities of $|\mc{O}_{\alpha\beta}|\:(\alpha,\beta\in\mc{T})$ for model (a) with $\gamma=D/2+1$ (upper left), (b) with $\gamma=D/2+1$ (upper right), (a) with $\gamma=1$ (bottom left), and (b) with $\gamma=1$ (bottom right).
We take $N=13$ and $d_{s}=400$.
As a reference, we also show the predictions of random matrices in Eqs. (\ref{gaussy}) and (\ref{k0y}).
}\label{que_ab}
\end{center}
\end{figure}

In Fig. \ref{que_ab}, we show the probability densities of the off-diagonal matrix elements in models (a) and (b) with $N=13$ for different values of $\gamma$.
As a reference, we also show the predictions of random matrices for nonsingular and the most singular cases (see Eqs. (\ref{gaussy}) and (\ref{k0y})).
We can see that for both values of $\gamma$, the probability density for model (a)/(b) obeys the GUE/GOE statistics for the most singular observables in \EQ{k0y} rather than the statistics for nonsingular observables in \EQ{gaussy}.
These results are consistent with the conjecture of the random matrix model.

We note that the universal ratios are also found for these singular operators.
Indeed, for model (a) with $N=13$, we have $\tilde{r}=0.981$ and $\tilde{r}=0.984$ for $\gamma=D/2+1=4097$ and $\gamma=1$, respectively.
They are consistent with the RMT prediction of $r_\mr{A}=1$.
For model (b) with $N=13$, we have $\tilde{r}=1.41$ and $\tilde{r}=1.43$ for $\gamma=D/2+1=4097$ and $\gamma=1$, respectively.
They are consistent with the RMT prediction of $r_\mr{AI,even}=\sqrt{2}$.

\subsection{A simple counterexample}
\begin{figure}
\begin{center}
\includegraphics[width=12cm,angle=90]{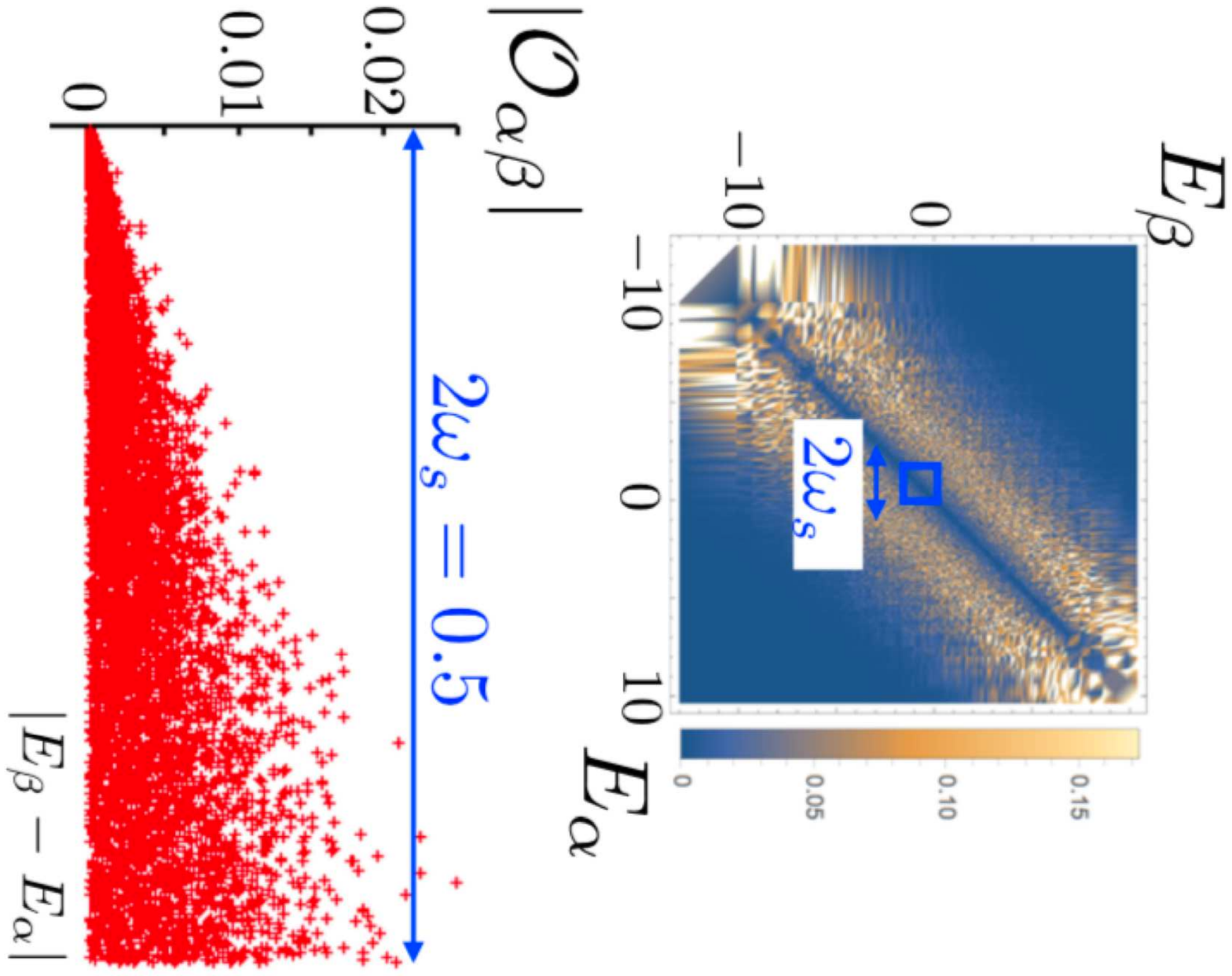}
\caption{(upper figure) Off-diagonal matrix elements of $\hmo=\hmo_y$ for model (b) with $N=8$ as a function of $E_\alpha$ and $E_\beta$.
(bottom figure) The off-diagonal matrix elements are plotted over an energy shell with width $2\omega_{s}=0.5$ for $N=11$.
}\label{county}
\end{center}
\end{figure}

Previous results show that the conjecture of the random matrix models applies to a wide class of observables with various symmetries including many-body or singular ones. Nevertheless, we can easily find counterexamples of the conjecture in Fig. \ref{conjecture} (and Srednicki's conjecture in \EQ{corre}) among simple observables.
To illustrate this fact, we show the off-diagonal matrix elements of $\hmo_y=\hat{\sigma}_{[N/2]+1}^y$ for model (b) in Figs. \ref{county}.
The upper figure shows the density plot of the off-diagonal matrix elements for all of the eigenstates as a function of $E_\alpha$ and $E_\beta$.
As for the cases with $\hmo_1$ and $\hmo_2$, the typical magnitude of $\mc{O}_{\alpha\beta}$ rapidly decays when $|E_\alpha-E_\beta|$ becomes large.
On the other hand, contrary to the previous examples, the typical magnitude is small for $|E_\alpha-E_\beta|\simeq 0$ as well.
To investigate this in detail, we take some small energy shell with width $2\omega_{s}=0.5$ and plot matrix elements as a function of $|E_\alpha-E_\beta|$ in the bottom figure of Fig. \ref{county}.
The figure shows that the typical magnitude of $|\mc{O}_{\alpha\beta}|$ vanishes with $|E_\alpha-E_\beta|\ra 0$.
This is in contradiction to the conjecture in Fig. \ref{conjecture} and Srednicki's conjecture in \EQ{corre}, both of which predict the plateau-like structure of $|\mc{O}_{\alpha\beta}|$ for $\omega<\omega_\mr{sh}$.\footnote{Indeed, even if we make $\omega_{s}$ smaller, no plateau-like structure is obtained.}

\begin{figure}
\begin{center}
\includegraphics[width=25cm]{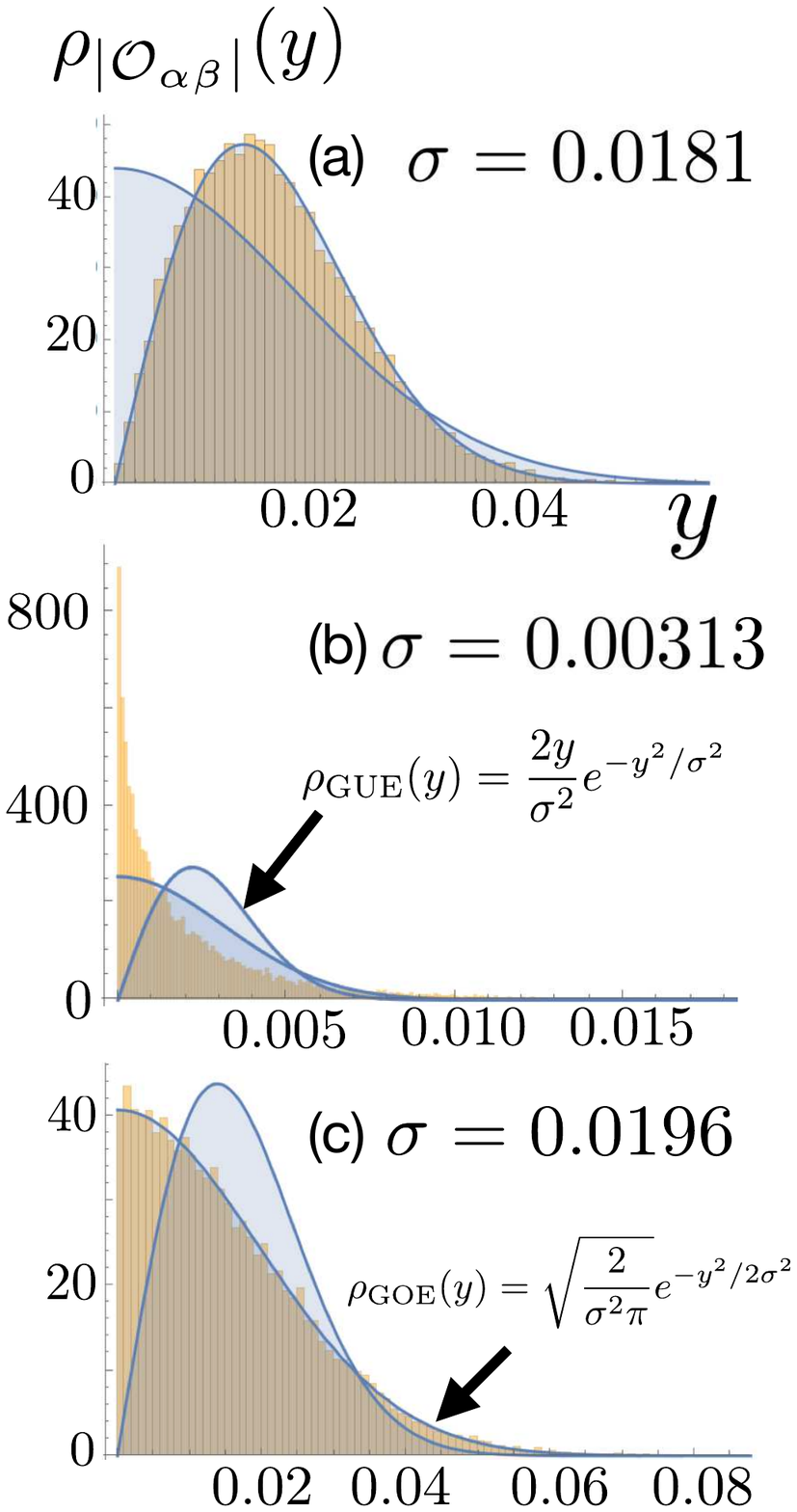}
\caption{Probability densities of $|\mc{O}_{\alpha\beta}|\:(\alpha,\beta\in\mc{T})$ for models (a) (up), (b) (middle), and (c) (bottom).
We take $N=12$ and $d_{s}=200$.
As a reference, we also show the predictions of random matrices in \EQ{gaussy}.
The prediction of the random matrix models breaks down in model (b), whereas it holds true in models (a) and (c).
}\label{count_abc}
\end{center}
\end{figure}

We plot the probability densities of the off-diagonal matrix elements $|\braket{E_\alpha|\hmo_y|E_\beta}|$ for models (a), (b), and (c) in Fig. \ref{count_abc}.
As shown in the figure, the prediction of the random matrix models breaks down in model (b), whereas it holds true in models (a) and (c).

These results can be understood by realizing that $\hmo=\hat{\sigma}_{[N/2]+1}^y=-\frac{i}{2h'} [\hH,\hat{\sigma}_{[N/2]+1}^z]$ for model (b), where $h'$ is defined in \EQ{spinchain}.
We obtain
\aln{\label{kokan}
|\mc{O}_{\alpha\beta}|= \lrv{\frac{E_\alpha-E_\beta}{2h'}}|\braket{E_\alpha|\hat{\sigma}_{[N/2]+1}^z|E_\beta}|.
}
Since $|\braket{E_\alpha|\hat{\sigma}_{[N/2]+1}^z|E_\beta}|$ behaves as in Fig. \ref{o1_b}(iv) for $\omega<\omega_{s}$ (i.e., form a plateau-like structure), $|\mc{O}_{\alpha\beta}|$ behaves as shown in Fig. \ref{county}.
This is the case only for model (b) because we do not have the relation in \EQ{kokan} for models (a) and (c).

The important point is that even simple (i.e., few-body and local) operators can break the RMT conjecture for matrix elements, and that such operators can always be easily constructed.
In fact, by taking a commutator between the Hamiltonian and a local observable, we can obtain another local observable $\hmo$ (if the Hamiltonian is composed of local interactions).
Such an observable $\hmo$ is expected to break the RMT conjecture, since the relation similar to \EQ{kokan} is obtained.

\section{Conclusions and Discussions}\label{sec:sumrmt}
We have shown that RMT can predict the finite-size corrections of the ETH (within some energy shell) in nonintegrable systems and for a wide class of observables, including many-body operators.
We have first refined and generalized the RMT predictions for the finite-size corrections of the ETH (see Fig. \ref{conjecture}).
We have seen that the ratios between standard deviations of diagonal and off-diagonal matrix elements become universal ones that depend only on anti-unitary symmetries of the Hamiltonian and those of the observable.
We have also shown that the probability densities of the off-diagonal matrix elements $\rho_{|\mc{O}_{\alpha\beta}|}(y)$ are different depending on the singularity of the observable as well as the symmetries of the Hamiltonian.
Next, we have numerically investigated matrix-element statistics of various observables in nonintegrable systems that only conserve energy.
We have demonstrated that the finite-size corrections of the ETH are  in excellent agreement with the predictions of RMT for a wide class of observables with various symmetries, including many-body correlations and singular operators.
We have also remarked that counterexamples can always be constructed even among simple observables.

Our results suggest that for a wide class of observables, the ETH holds  with the mechanism related to RMT, which also tells us its finite-size corrections.
Unlike the ETH for MITE or MATE, we show that even many-body operators can satisfy the ETH due to that mechanism.\footnote[1]{We note that not all many-body operators satisfy the ETH as we have discussed in \SEC{sec:mot}.}
This is expected because the crucial assumption is the behavior of $U_{\alpha i}$, which is not directly related to few-body or macroscopic properties of observables.
We thus expect that for the achievement of a rigorous proof of the ETH, it is important to investigate why the behavior of $U_{\alpha i}$ is mimicked by RMT.
We note that counterexamples are easily constructed by taking $\hmo=\ket{E_\gamma}\bra{E_\gamma}$ or by taking commutators of the Hamiltonian and another observable.
These counterexamples are somehow ``related" to the Hamiltonian.
From this observation, it is important to understand such ``relations" quantitatively for clarifying the criteria for the validity of the conjecture of the RMT model.\footnote{We note that the ETH may be valid for counterexamples of the RMT conjecture. Thus, the conjecture of the RMT model is not a necessary condition.}

Let us comment on future perspectives.
First, as we have mentioned above, quantifying the criteria for the breakdown of the RMT conjecture seems unavoidable for understanding the mechanisms of the ETH and its finite-size corrections.
Operators that are written as $\hmo=[\hH,[\hH,\cdots,[\hH,\hmo']\cdots]]$ and operators that are conserved $[\hmo,\hH]=0$ will break the RMT conjecture, but how about operators that approximately satify such relations?
Commutators that involve the Hamiltonian may be utilized for quantifying them, but this is a future problem.
Second, we need to extend our results in Subsection \ref{sec:distd} to the off-diagonal matrix elements $\braket{E_\alpha|\hmo|E_\beta}$ with $|E_\alpha-E_\beta|>\omega_\mr{sh}$.
We believe that within a small energy shell $\lrm{(E_\alpha,E_\beta):|E_\alpha-E_1|<\omega_\mr{sh,1},|E_\beta-E_2|<\omega_\mr{sh,2}}$, the probability densities of $O_{\alpha\beta}$ will obey the similar statistics as discussed in Subsection \ref{sec:distd}.\footnote[2]{The RMT prediction should be extended for such cases. We expect to do this by a similar technique done in Subsection \ref{sec:distd}.}
Finally, it is interesting if we can relate our findings about the deviations from the ETH to measurable fluctuations (e.g., temporal fluctuations of the expectation value in \EQ{vart}) in isolated small systems.
As we have mentioned in Chapter \ref{ch:Stat}, the dynamics in such small systems are realized in Ref.~\cite{Kaufman16}.
We expect that the RMT conjecture may help us understand finite-size effects that are not captured by the standard statistical mechanics.

\chapter{Generalized Gibbs ensemble (GGE) in integrable systems}\label{ch:GGE}
In this chapter, we review the generalized Gibbs ensemble (GGE) in integrable systems.
After introducing a general form of the GGE and raising several open questions,
 we consider simple integrable systems that can be mapped to free-quasiparticle systems.
Then, we stress the importance of the locality of conserved quantities in constructing the GGE.
In particular, we introduce the notion of the so-called ``truncated GGE."
Finally, we briefly review the results for interacting integrable systems that are solvable by the Bethe ansatz.

\section{Non-thermal stationary states due to conserved quantities}
As we saw in the previous chapters, isolated systems can equilibrate if we only consider a restricted set of observables.
One of the open questions is how to characterize the stationary state from the information of the initial state.
Under certain conditions, we can assume that the stationary state can effectively be described by the diagonal ensemble introduced in Chapter \ref{ch:Eq}: $\hrho_\mr{d}=\sum_{\alpha=1}^D |c_\alpha|^2\ket{E_\alpha}\bra{E_\alpha}$.
However, the diagonal ensemble is constructed from $D$  parameters, unlike the microcanonical ensemble that requires only macroscopic energy for its construction.
What we want is a statistical ensemble that is constructed from a few parameters, such as the microcanonical, canonical and grandcanonical ensemble.

Nonintegrablity of the system is expected to play a key role in determining whether the stationary state is described by the usual (micro)canonical ensemble.
In a nonintegrable system that conserves energy alone, we expect that the canonical ensemble describes the stationary state because of the ETH (see Chapters \ref{ch:Eq} and \ref{ch:ETH}).
On the other hand, the ETH does not hold true in integrable systems due to the existence of many conserved quantities.
In this case, the stationary states are not described by the canonical ensemble in general (note that the equilibration often occurs without the ETH because of the large effective dimension).
Are the stationary states in integrable systems describable by some other statistical ensembles?

The generalized Gibbs ensemble (GGE) is a candidate for a statistical ensemble that describes the stationary state in an integrable system.
Let us denote the set of conserved quantities by $\{\hat{I}_m\}\:(m=1,2,\cdots)$.
 For simplicity, we assume that $\hat{I}_m$'s commute with one another.
The GGE is defined as~\cite{Jaynes57,Jaynes57II,Rigol07}
\aln{
\hrho_\mr{GGE}:=\frac{e^{-\sum_m\lambda_m\hat{I}_m}}{Z_\mr{GGE}},
}
where $Z_\mr{GGE}:=\Tr[e^{-\sum_m\lambda_m\hat{I}_m}]$ and $\lambda_m$'s are determined from the initial information of the conserved quantities:
\aln{
\Tr[\hrho_0\hat{I}_m]=\Tr[\hrho_\mr{GGE}\hat{I}_m].
}  
Then, assuming the diagonal ensemble, we want the GGE to satisfy the following relation for (the sum of) few-body observables $\hmo$ in the thermodynamic limit:\footnote{In the followings, we will only consider  (sums of) few-body operators as observables. Furthermore, we will often consider spatially local observables.}
\aln{
\Tr[\hrho_\mr{d}\hmo]=\Tr[\hrho_\mr{GGE}\hmo].
}
In the form of the microscopic thermal equilibrium (MITE), we want
\aln{\label{diaggge}
\Tr_{S^c}[\hrho_\mr{d}]=\Tr_{S^c}[\hrho_\mr{GGE}]
}
for a small subsystem $S$.

We remark that there is another way to define the stationary state in the thermodynamic limit.
While \EQ{diaggge} conderns the diagonal ensemble for a finite system with the size $L$ and then take the thermodynamic limit $L\ra\infty$, we can first consider the thermodynamic limit and then the long-time limit:
\aln{
\hrho_{\infty,S}:=\lim_{t\ra\infty}\lim_{L\ra\infty}\Tr_{S^c}[\hrho(t)].
}
Then we can examine if 
\aln{\label{gyaku}
\hrho_{\infty,S}=\lim_{L\ra\infty}\Tr_{S^c}[\hrho_\mr{GGE}]
}
holds true.
The former definition in \EQ{diaggge} is adopted in, e.g., Refs.~\cite{Rigol07,Rigol09,Vidmar16},
and the latter definition in \EQ{gyaku} is adopted in, e.g.,  Refs.~\cite{Cazalilla06,Calabrese07,Essler16Q}.\footnote[1]{We note that the former definition can treat finite-size effects, and the latter definition is convenient for treating nonequilibrium steady states~\cite{Bhaseen15,Bertini16D,Bertini16T}.}
Two definitions are equivalent if 
\aln{
\lim_{L\ra\infty}\Tr_{S^c}[\hrho_\mr{d}]=\hrho_{\infty,S}.
}
Although this condition is nontrivial, it is proven for certain systems~\cite{Essler16Q}.
We note that we adopted the definition similar to the former (using the (micro)canonical ensemble instead of the GGE) in the previous chapters.
We will also use the former definition in our work in Chapter \ref{ch:Gn}.

The applicability of the GGE has not completely been understood yet.
We raise two open questions.
\begin{enumerate}
\item
To what kind of systems and observables is the GGE applicable?
Do we need specific initial states to justify the GGE?
\item
How should we choose the minimal set of $\{\hat{I}_m\}$?
Are there more important conserved quantities?
\end{enumerate}
Many previous studies investigated these questions using various integrable models.
We will review some of them in the following sections.

\section{The GGE in essentially free systems}
First we consider an integrable system whose Hamiltonian can be mapped to a quadratic form:
\aln{
\hH=\sum_k\epsilon_k\hat{b}_k^\dag\hat{b}_k,
}
where $\hat{b}_k$ is an annihilation operator of some quasiparticle and $\epsilon_k$ is the dispersion relation.
After the notable work using a one-dimensional lattice system with hard-core bosons~\cite{Rigol07,Rigol09}, the GGE has extensively been investigated in these essentially free-quasiparticle systems, including transverse-field Ising models~\cite{Calabrese11,Cazalilla12,Fagotti13}, XY models~\cite{Cazalilla12}, Luttinger liquids~\cite{Cazalilla06,Cazalilla12}, a system of hard-core anyons~\cite{Wright14}, and quantum field theories~\cite{Calabrese07,Sotiriadis14,Sotiriadis16}.

Let us illustrate a simple example in which the GGE is applicable, following Ref.~\cite{Essler16Q}.
We consider a one-dimensional fermionic paring model as follows:
\aln{
\hH_f(\Delta, \mu)=-J\sum_{i=1}^L(\hat{c}_i^\dag \hat{c}_{i+1}+\mr{h.c.})-\mu\sum_{i=1}^L\hat{c}_i^\dag \hat{c}_i
+\sum_{i=1}^L(\hat{c}_i^\dag \hat{c}_{i+1}^\dag+\mr{h.c.}),
}
where $\hat{c}_i$ is an annihilation operator\footnote{The anticommutation relations $\{\hat{c}_i,\hat{c}_j\}=\{\hat{c}_i^\dag,\hat{c}_j^\dag\}=0$ and $\{\hat{c}_i,\hat{c}_j^\dag\}=\delta_{ij}$ are satisfied.} of a fermion at the site $i$ and we impose a periodic boundary condition.
By the Fourier transformation $\hat{c}_i=\frac{1}{\sqrt{L}}\sum_ke^{-ikx_i}\hat{c}(k)$, where $x_i=i$ is the coordinate of site $i$ (we set the lattice constant to unity), we obtain
\aln{
\hH_f(\Delta,\mu)&=-\sum_k\lrl{(2J\cos(k)+\mu)\hat{c}^\dag(k)\hat{c}(k)+i\Delta\sin(k)(\hat{c}^\dag(k)\hat{c}^\dag(-k)-\hat{c}(-k)\hat{c}(k))}\NON
&=\frac{1}{2}\sum_{k}
\begin{pmatrix}
 \hat{c}^\dag(k) &
 \hat{c}(-k)
\end{pmatrix}
\begin{pmatrix}
 -(2J\cos(k)+\mu) & -2i\Delta\sin(k) \\
 2i\Delta\sin(k) & (2J\cos(k)+\mu) 
\end{pmatrix}
\begin{pmatrix}
 \hat{c}(k) \\
 \hat{c}^\dag(-k)
\end{pmatrix} +\mr{const}.
}
Using the Bogoliubov transformation 
\aln{
\begin{pmatrix}
 \hat{\alpha}(k) \\
 \hat{\alpha}^\dag(-k)
\end{pmatrix}
=
\begin{pmatrix}
 \cos(\theta_k/2) & -i\sin(\theta_k/2) \\
 -i\sin(\theta_k/2) & \cos(\theta_k/2) 
\end{pmatrix}
\begin{pmatrix}
 \hat{c}(k) \\
 \hat{c}^\dag(-k)
\end{pmatrix},
}
we have the diagonalized Hamiltonian
\aln{
\hH_f(\Delta,\mu)=\sum_{k\geq 0}\epsilon(k)\hat{\alpha}^\dag(k)\hat{\alpha}(k)+\mr{const}.
}
Here 
\aln{
\epsilon(k)=\sqrt{(2J\cos(k)+\mu)^2+4\Delta^2\sin^2(k)}
}
and
\aln{
e^{i\theta_k}=\frac{-2J\cos(k)-\mu+2i\Delta \sin(k)}{\epsilon(k)}.
}

We consider a quench from a pre-Hamiltonian $\hH_0=\hH_f(\Delta_0,\mu)\:\:(\Delta_0\neq 0)$ to a post-Hamiltonian $\hH=\hH_f(0,\mu)$ at time $t=0$.
Note that $\hH=-\sum_{k}(2J\cos(k)+\mu)\hat{c}^\dag(k)\hat{c}(k)$.
We take an initial state as the Bogoliubov fermion vacuum:
\aln{
\ket{\psi_0}=\ket{0},\:\:\:\hat{\alpha}(k)\ket{0}=0\:\text{ for all } k.
}
Using the Heisenberg representation $\hat{c}(k,t)=e^{i\hH t}\hat{c}(k)e^{-i\hH t}$, we obtain
\aln{
\hat{c}(k,t)=e^{-i(-2J\cos(k)-\mu)t}\hat{c}(k)=e^{-i(-2J\cos(k)-\mu)t}\lrl{\cos(\theta_k/2)\hat{\alpha}(k)+i\sin(\theta_k/2)\hat{\alpha}^\dag(-k)}.
}
Then, we can calculate the two-point functions at $t>0$ as
\aln{
\braket{\psi(t)|\hat{c}^\dag(k)\hat{c}(q)|\psi(t)}&=\braket{0|\hat{c}^\dag(k,t)\hat{c}(q,t)|0}=\delta_{kq}\sin^2(\theta_k/2),\\
\braket{\psi(t)|\hat{c}(k)\hat{c}(q)|\psi(t)}&=\delta_{k,-q}\frac{i}{2}(\sin\theta_k)e^{-2i(-2J\cos(k)-\mu)t}.
}
In the position space we have
\aln{
\braket{\psi(t)|\hat{c}^\dag_{j+l}\hat{c}_j|\psi(t)}&=\frac{1}{L}\sum_ke^{ikl}\sin^2(\theta_k/2)=:f_L(l),\\
\braket{\psi(t)|\hat{c}_{j+l}\hat{c}_j|\psi(t)}&=\frac{1}{L}\sum_ke^{-ikl}\frac{i}{2}(\sin\theta_k)e^{-2i(-2J\cos(k)-\mu)t}=:g_L(l,t).
}
The crucial aspect of our initial state is that we can use Wick's theorem to calculate the multi-point correlation functions. For example,
\aln{
\braket{\psi(t)|\hat{c}^\dag_{j}\hat{c}^\dag_{l}\hat{c}_m\hat{c}_n|\psi(t)}=g_L^*(l-j,t)g_L(n-m,t)-f_L(j-n)f_L(l-m)+f_L(l-n)f_L(j-m).
}
In the thermodynamic limit and the long-time limit, we have
\aln{
f_L(l)&\ra \int_0^{2\pi}\frac{dk}{2\pi}e^{ikl}\sin^2(\theta_k/2),\\
g_L(l,t)&\ra 0.
}
We thus conclude that the stationary state of a given subsystem is completely characterized by $\lim_{L\ra\infty}f_L(l)$.

The GGE is constructed from the set of mode occupation numbers $\hat{n}_k:=\hat{c}_k^\dag \hat{c}_k$ with the quasi-momentum $k=0,\frac{2\pi}{L},\frac{4\pi}{L},\cdots,\frac{2\pi(L-1)}{L}$ as
\aln{\label{ggemom}
\hrho_\mr{GGE}=\frac{e^{-\sum_k\lambda_k \hat{n}_k}}{Z_\mr{GGE}},
}
where $Z_\mr{GGE}=\Tr\lrl{e^{-\sum_k\lambda_k \hat{n}_k}}$.
Here $\lambda_k$ is determined by
\aln{\label{icchi}
\Tr\lrl{\hrho_\mr{GGE}\hat{n}_k}&=\frac{e^{-\lambda_k}}{1+e^{-\lambda_k}}\NON
&=\braket{\psi_0|\hat{n}_k|\psi_0}=\sin^2(\theta_k/2).
}
Since the GGE has a quadratic form, Wick's theorem holds true.
Then, due to the relation in \EQ{icchi}, the stationary state and the GGE give the same multi-point correlation functions.
Therefore, we prove that the stationary-state expectation values of observables on a small subsystem are correctly predicted by the GGE.

The method for using Wick's theorem was first introduced in Ref.~\cite{Cazalilla12}, which discussed transverse-field Ising models, Luttinger models, and XY models.
To justify Wick's theorem of the initial states, we can take the canonical ensemble or the ground state of Hamiltonians that are written as quadratic forms.
Further, in Ref.~\cite{Sotiriadis14}, the authors proved the validity of the GGE in \EQ{ggemom} for massive free quantum field theories by assuming the cluster decomposition properties of the initial state.
This extends the applicability of the GGE to non-Gaussian initial states.
However, it was also shown that the GGE in \EQ{ggemom} fails for massless free field theories if the initial state is not Gaussian~\cite{Sotiriadis16}.

\section{Importance of the locality of conserved quantities and the truncated GGE}\label{sec:tGGE}
In this section we discuss the importance of the locality of conserved quantities in describing the expectation values  of local observables.
While we have constructed the GGE from $L$ mode occupation numbers in \EQ{ggemom}, we can construct the GGE using an extensive conserved quantities that can be written as sums of local operators.
Moreover, numerical simulations suggest that if conserved quantities become more local, they become more important in describing local observables.

To illustrate the notion of the locality of conserved quantities, consider the fermionic Hamiltonian $\hH(\Delta=0,\mu)$ introduced in the previous section.
As we have done in constructing the GGE in \EQ{ggemom}, we can use the mode occupation number $\hat{n}_k=\hat{c}_k^\dag\hat{c}_k$ to construct the GGE.
On the other hand, we can also consider an equivalent set of extensive conserved quantities as follows:
\aln{
\hat{I}_{n,+}=2J\sum_k\cos(x_nk)\hat{c}^\dag(k)\hat{c}(k)&=J\sum_i(\hat{c}_i^\dag \hat{c}_{i+n}+\mr{h.c.}),\\
\hat{I}_{n,-}=2J\sum_k\sin(x_nk)\hat{c}^\dag(k)\hat{c}(k)&=iJ\sum_i(\hat{c}_i^\dag \hat{c}_{i+n}-\mr{h.c.}).
}
Each conserved quantity is an extensive quantity that is the sum of local operators.
We especially call these local operators as $(n+1)$-local operators, since $\hat{c}_i^\dag \hat{c}_{i+n}$ has a support on $(n+1)$-neighboring sites.\footnote{We note that an operator in \EQ{locloc} is $l_0$-local  if $\mu_{i_0}\neq 0$ and $\mu_{i_0+l_0-1}\neq 0$.}
Then the GGE can be reconstructed from these conserved quantities as
\aln{\label{ggeloc}
\hrho_\mr{GGE}=\frac{e^{-\sum_n(\mu_{n,+} \hat{I}_{n,+}+\mu_{n,-} \hat{I}_{n,-})}}{Z_\mr{GGE}}.
}
Here $\mu_{n,\pm}$'s are determined from the initial condition of the conserved quantities.
We note that there are cases where the equivalence between mode occupation numbers and extensive  conserved quantities does not exist~\cite{Essler15}.

The importance of the locality of conserved quantities was first realized by Fagotti and Essler~\cite{Fagotti13} using the 1D transverse-field Ising model in the thermodynamic limit:
\aln{
\hH=-J\sum_{i=-\infty}^\infty[\hat{\sigma}_i^x\hat{\sigma}_{i+1}^{x}+h\hat{\sigma}_i^z].
}
This model is diagonalized with the Jordan-Wigner transformation followed by the Bogoliubov transformation.
Even though this procedure is complicated because the Jordan-Wigner transformation is a nonlocal transformation, we can find a set of extensive conserved quantities instead of the mode occupation numbers as
\aln{
\hat{I}_{n,+}&=-J(U_{n+1}+U_{1-n})+hJ(U_n+U_{-n}),\NON
\hat{I}_{n,-}&=J(V_{n+1}+V_{-1-n}),
}
where
\aln{
U_{n>0}&=\frac{1}{2}\sum_{i=-\infty}^\infty \hat{\sigma}_i^x\lrs{\prod_{l=1}^{n-1}\hat{\sigma}_{i+l}^z}\hat{\sigma}_{i+n}^x,\NON
U_{0}&=-\frac{1}{2}\sum_{i=-\infty}^\infty \hat{\sigma}_i^z,\NON
U_{n<0}&=\frac{1}{2}\sum_{i=-\infty}^\infty \hat{\sigma}_i^y\lrs{\prod_{l=1}^{|n|-1}\hat{\sigma}_{i+l}^z}\hat{\sigma}_{i+|n|}^y,\NON
V_{n>0}&=\frac{1}{2}\sum_{i=-\infty}^\infty \hat{\sigma}_i^x\lrs{\prod_{l=1}^{n-1}\hat{\sigma}_{i+l}^z}\hat{\sigma}_{i+n}^y,\NON
V_{n<0}&=-\frac{1}{2}\sum_{i=-\infty}^\infty \hat{\sigma}_i^y\lrs{\prod_{l=1}^{|n|-1}\hat{\sigma}_{i+l}^z}\hat{\sigma}_{i+|n|}^x.\NON
}
The important point is that $\hat{I}^{(n,\pm)}$'s are $(n+2)$-local because they can be written as the sums of $(n+2)$-neighboring spin correlations.
The GGE is constructed as
\aln{
\hrho_\mr{GGE}=\frac{1}{Z_\mr{GGE}}\exp\lrl{-\sum_{n=0}^\infty\sum_{\sigma=\pm}\lambda_{n,\sigma}\hat{I}_{n,\sigma}}.
}
The authors in Ref.~\cite{Fagotti13} have confirmed that this ensemble well describes the stationary state after a certain quench.

Next, we want to reduce the number of the conserved quantities in the GGE without losing the validity of describing the stationary state for the subsystem with the size $l$.
In Ref.~\cite{Fagotti13}, the authors introduced what is called the ``truncated generalized Gibbs ensemble" (tGGE):
\aln{
\hrho_{\mr{tGGE}}^{(y)}=\frac{1}{Z_{\mr{tGGE}}^{(y)}}\exp\lrl{-\sum_{n=0}^{y-1}\sum_{\sigma=\pm}\lambda_{n,\sigma}^{(y)}\hat{I}_{n,\sigma}},
}
where $Z_{\mr{tGGE}}^{(y)}=\Tr\lrl{e^{-\sum_{n=0}^{y-1}\sum_{\sigma=\pm}\lambda_{n,\sigma}^{(y)}\hat{I}_{n,\sigma}}}$, and we note that $\lambda_{y,n,\sigma}\neq \lambda_{n,\sigma}$ in general.
In the limit $y\ra \infty$, $\hrho_{\mr{tGGE}}^{(y)}$ is equivalent to $\hrho_\mr{GGE}$.

\begin{figure}
\begin{center}
\includegraphics[width=10cm,angle=-90]{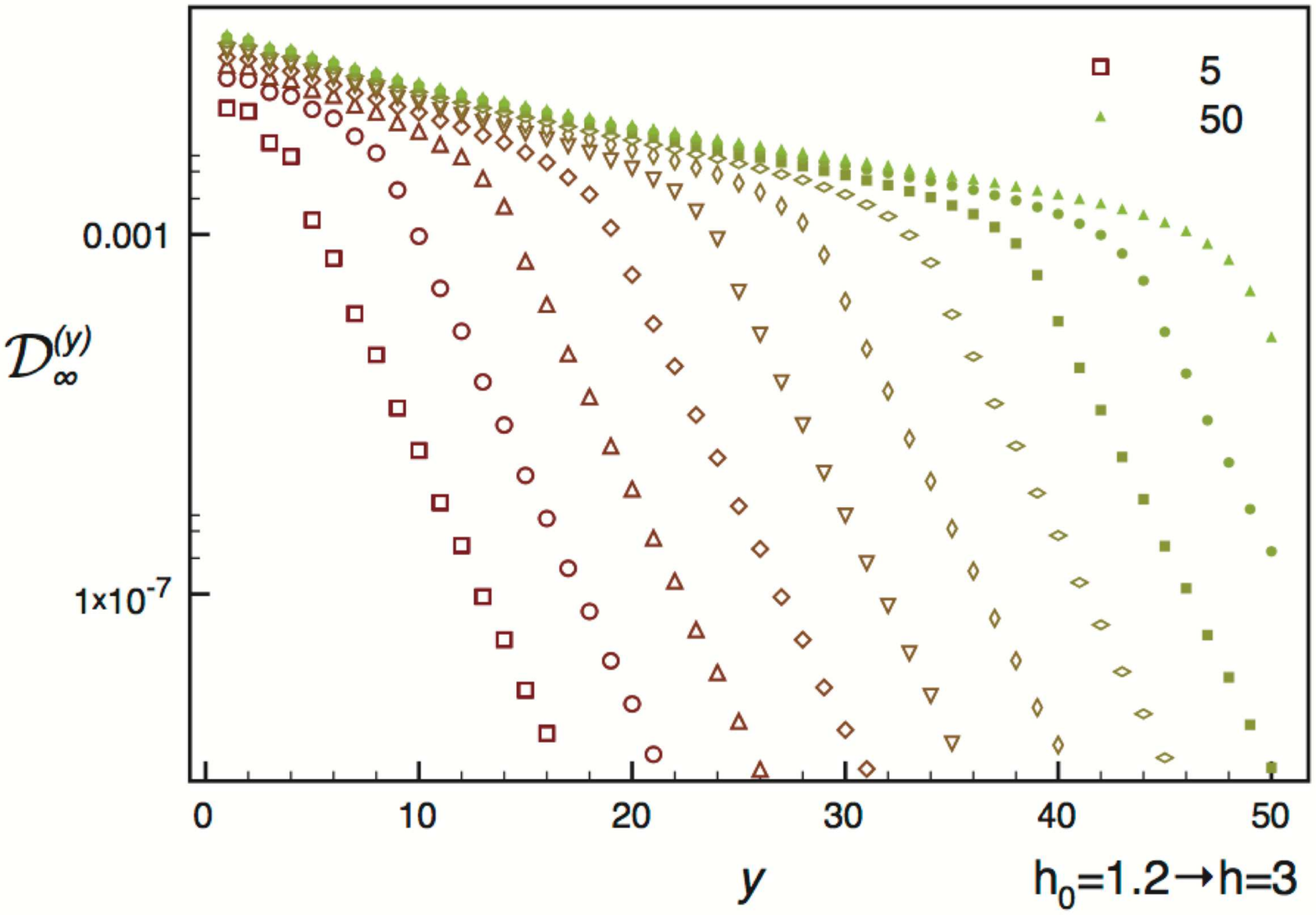}
\caption{
Distances between $\hrho_{\mr{tGGE},l}^{(y)}$ and $\hrho_{\mr{GGE},l}$ plotted as a function of $y$.
The different plots denote the different subsystem sizes with $l = 5 \text{ (the leftmost)}, 10, 15, \cdots,  50$ (the rightmost).
The GGE and tGGE are constructed in the quench from $h=h_0 = 1.2$ to $h = 3$.
For $l\lesssim y$, the distances start to decrease exponentially with respect to $y$.
Reproduced from Fig. 7 of~\cite{Fagotti13}. \copyright 2013 American Physical Society.}
\label{fagotti}
\end{center}
\end{figure}

The authors in Ref.~\cite{Fagotti13} investigated how close $\hrho_{\mr{tGGE},l}^{(y)}$ and $\hrho_{\mr{GGE},l}$ are for various $y$, where the subscript $l$ means that the density matrices are reduced to the subsystem with the size $l$.
Figure \ref{fagotti} shows the distance $\mc{D}_\infty^{(y)}=\mc{D}(\hrho_{\mr{tGGE},l}^{(y)},\hrho_{\mr{GGE},l})$ between $\hrho_{\mr{tGGE},l}^{(y)}$ and $\hrho_{\mr{GGE},l}$ as a function of $y$ for various $l$, where $\mc{D}(\hrho_1,\hrho_2)$ is defined as\footnote{They used the Frobenius norm that is defined by $||\hat{A}||_F:=\sqrt{\Tr[\hat{A}^\dag\hat{A}]}$.}
\aln{
\mc{D}(\hrho_1,\hrho_2):=\frac{||\hrho_1-\hrho_2||_F}{\sqrt{||\hrho_1||_F^2+||\hrho_2||_F^2}}.
}
The different plots denote the different subsystem sizes with $l = 5$ \text{ (the leftmost)}, 10, 15, $\cdots,  50$ (the rightmost).
The figure shows that the distances start to decrease rapidly only for $l\lesssim y$.
This result implies that the conserved quantities $\hat{I}_{n,\sigma}$ that satisfy $n\lesssim l$ are the most important in  describing the subsystem with the size $l$, and that the less local operators with $n\gg l$ play negligible roles.
Simply put, we can say that if conserved quantities become more local (in a sense that they can be written as sums of local operators), they become more important in constructing the GGE that can describe local observables.

\section{The GGE in interacting systems solved by the Bethe ansatz}
The GGE in interacting systems that can be solved by the Bethe ansatz is also investigated.
In contrast to essentially free-quasiparticle systems that we have reviewed in the previous two sections, 
we cannot use the mode occupation numbers to construct the GGE in interacting systems.
In fact, the applicability of the GGE for such systems is an open question.
However, efforts have been made for XXZ models~\cite{Pozsgay14CQ,Wouters14,Ilievski15}, Lieb-Liniger models~\cite{Mossel12,Caux12,Kormos13}, quantum field theories~\cite{Mussardo13,Essler15}, and so on.
In this section,  we briefly review the recent development on the stationary states of the XXZ models. 

In general, the eigenstates $\ket{E_\alpha}$ of the Bethe-ansatz-solvable systems are characterized by a set of complex quantum numbers $\{\lambda_k\}_{k}$, as $\ket{E_\alpha}=\ket{\{\lambda_k\}_{k}}$.
The quantum numbers $\{\lambda_k\}_{k}$ are called rapidities and obtained by the so-called Bethe equations.
For example, consider a 1D XXZ model as
\aln{
\hH=\frac{J}{4}\sum_{i=1}^N[\hat{\sigma}_i^x\hat{\sigma}_{i+1}^x+\hat{\sigma}_i^y\hat{\sigma}_{i+1}^y+\Delta(\hat{\sigma}_i^z\hat{\sigma}_{i+1}^z-1)],
}
where $J>0$ and $\Delta =\cosh\eta\geq 1$.
If we consider the fixed sector with a magnetization $S_\mr{tot}^z=\frac{N}{2}-M$, the Bethe equations for $\{\lambda_k\}_{k=1}^M$ can be written as~\cite{Korepin97,Ilievski15}
\aln{\label{beeq}
\lrs{\frac{\sin(\lambda_j+i\eta/2)}{\sin(\lambda_j-i\eta/2)}}^N=-\prod_{k=1}^M\frac{\sin(\lambda_j-\lambda_k+i\eta)}{\sin(\lambda_j-\lambda_k-i\eta)}
}
for all $j$.
Note that these equations are very complicated because they are nonlinear and $\lambda_j$'s are related to one another.

The Bethe equations in \EQ{beeq} are obtained either from the so-called coordinate Bethe ansatz or the algebraic Bethe ansatz.
In the coordinate Bethe ansatz, we assume a specific form of the wave functions (called the Bethe ansatz wavefunctions) as the eigenstates of the Hamiltonian~\cite{Korepin97}.
Then, by imposing a periodic boundary condition, we obtain the Bethe equations.
In the algebraic Bethe ansatz, we introduce some operators (called the R-matrix, the L-matrix, and the monodromy matrix) that satisfy certain algebras (i.e., the Yang-Baxter equations).
We then construct the so-called transfer matrix $T(\lambda)$ from the monodromy matrix such that $[T(\lambda),T(\mu)]=0$ is satisfied for different two rapidities $\lambda$ and $\mu$.
If we choose the appropriate R-matrix, the transfer matrix becomes a generating function of a certain Hamiltonian (e.g., the Heisenberg, the XXZ or the Lieb-liniger Hamiltonian) and other extensive conserved quantities that are written as the sums of local operators.
In this case, the transfer matrix and these conserved quantities (including the Hamiltonian) are simultaneously diagonalized.
Finally, we construct the eigenstates of the transfer matrix using a creation operator that is determined from the monodoromy matrix.
The Bethe equations in \EQ{beeq} are obtained as the consistency condition of this construction.

In Refs.~\cite{Pozsgay14CQ,Wouters14}, the failure of the ``naive" GGE in describing the stationary state is reported.
In Ref.~\cite{Wouters14}, the authors construct the GGE using the extensive conserved quantities:
\aln{
\hat{I}_n=\sum_{j}\hat{\mc{I}}_{j,j+1,\cdots,j+n},
}
where $\hat{\mc{I}}_{j,j+1,\cdots,j+n}$ is an $(n+1)$-local operator.
We note that $\hat{I}_n$ is obtained from the $n$th derivative of the transfer matrix.
By comparing the expectation values of observables for the GGE and those for the stationary state,\footnote{In general, it is not easy to treat the Bethe equations in \EQ{beeq}. However, the so-called generalized thermodynamic Bethe ansatz (gTBA) is developed to approximately examine properties of the system in the thermodynamic limit~\cite{Mossel12}. In this approach, we replace the set of the rapidities with a distribution function of the rapidities. Then, we find the equation (i.e., the gTBA equation) for the distribution function from the knowledge of the ensembles under consideration. For such ensembles, we can take either the GGE or the stationary-state ensemble. The method for using the latter ensemble is called a quench action method~\cite{Caux13,Caux16}.}
they found that there is a discrepancy between these two results.
This result indicates that the GGE constructed only from the extensive sums of local operators cannot be used for the XXZ models.

A more improved GGE is proposed and investigated in Ref.~\cite{Ilievski15} by taking the so-called quasi-local operators into account.
While local operators have supports exactly on the finite number of neighboring sites, for quasi-local operators we allow quantities whose overlaps with less local sites are present but decay sufficiently fast (for an exact definition, see Ref.~\cite{Ilievski15QC}).
In XXZ models, it is shown that extensive conserved quantities that can be written as sums of such quasi-local operators are constructed by applying the algebraic Bethe ansatz in a more technical way than usual~\cite{Ilievski15QC,Ilievski15}.\footnote{More concretely, the possible class of the $L$ matrices and the $R$ matrices are extended.}
The authors in Ref.~\cite{Ilievski15} find that the GGE constructed from the extensive conserved quantities that are written as the sums of quasi-local as well as local operators well describes the expectation values of local observables in the stationary state.
This result implies that the GGE is valid if we can include all conserved quantities that have significant overlaps with local observables.\footnote{We note that the importance of quasi-local operators has already been recognized in the context of the nonequilibrium transport in XXZ chains~\cite{Prosen13} and the GGE in quantum field theories~\cite{Essler15}.}
We note, however, that a recent paper in Ref.~\cite{Ilievski16} claims that even quasi-local operators may not be enough to characterize the stationary state in XXZ models.
Thus, the applicability of the GGE in these interacting integrable models is still an open question.

\section{Conclustion}
We have reviewed the applicability of the GGE for two types of integrable systems: systems that can be mapped to free-quasiparticle systems and systems that are solved by the Bethe ansatz.
The Hamiltonians of the systems of both types have eigenstates that are characterized by sets of certain quantum numbers.
In contrast, it seems more important to pay attention to conserved quantities that can be written as sums of (quasi-)local operators.
In fact, the locality of conserved quantities seems crucial in constructing the GGE that describes the expectation values of local observables in the stationary state, as the success of the truncated GGE indicates.
Even though controversial discussions exist, many researchers believe that the GGE constructed from (quasi-)local operators will be valid in certain initial conditions.

\chapter{Generalized Gibbs ensemble in a nonintegrable system with an extensive number of local symmetries}\label{ch:Gn}

\section{Motivation}
In the previous chapters, we have seen that conserved quantities play an important role for thermalization in isolated quantum systems.
In nonintegrable systems that conserve energy alone, the stationary state is expected to be described by the (micro)canonical ensemble because of the ETH.
On the other hand, the stationary state cannot be described by the canonical ensemble in systems that are integrable or many-body localized, since there exist many nontrivial conserved quantities in these systems.

As we have seen in Chapter \ref{ch:GGE}, the GGE is the promising candidate for describing stationary states in integrable systems.
These integrable systems have sets of conserved quantities from which each energy eigenstate can be identified.
We note that this feature is also expected to exist in systems that show the ``fully" many-body localization (see Subsection \ref{sec:dameda}).
In this case, each energy eigenstate is expected to be characterized by the localized bits (see \EQ{FMBL})~\cite{Vosk13,Serbyn13,Huse14}.

To clarify the importance of conserved quantities for the appearance of non-thermal stationary states, it is interesting to study models with less numbers of conserved quantities than the usual integrable systems.
Previous studies showed two extreme cases: the stationary state seems to be described by the canonical ensemble if the system conserves only energy, and the GGE is necessary when sufficiently many conserved quantities exist so that every eigenstate is identified.
Then, it is of interest how many conserved quantities the system should possess for the appearance of the stationary states that are described by the GGE, not by the canonical ensemble.
We note that such systems are nonintegrable in a sense that the sets of conserved quantities cannot characterize each energy eigenstate.

In this chapter, we discuss our work based on Ref.~\cite{Hamazaki16G}.
We show that the stationary state is described by the GGE if the system has an extensive number of local symmetries, even when it is a nonintegrable system.
We have investigated a nonintegrable model of hard-core bosons with an extensive number of local $\mathbb{Z}_2$ symmetries by the exact-diagonalization analysis.
We show that the expectation values of observables in the stationary state are described by the GGE rather than the canonical ensemble.
In this case, the usual ETH does not hold true.
Instead, the ETH for each symmetry sector, which we call the restricted ETH (rETH), holds true and we argue that the rETH plays an important role for our system to approach the GGE.
We have also examined a model that has only one global $\mathbb{Z}_2$ symmetry, and a model with a size-independent number of local $\mathbb{Z}_2$ symmetries.
We show that the usual canonical ensemble well describes the stationary states and that we do not have to use the GGE for these two models.

\section{A model with an extensive number of local symmetries}\label{defmodel}
\begin{figure}
\begin{center}
\includegraphics[width=11cm]{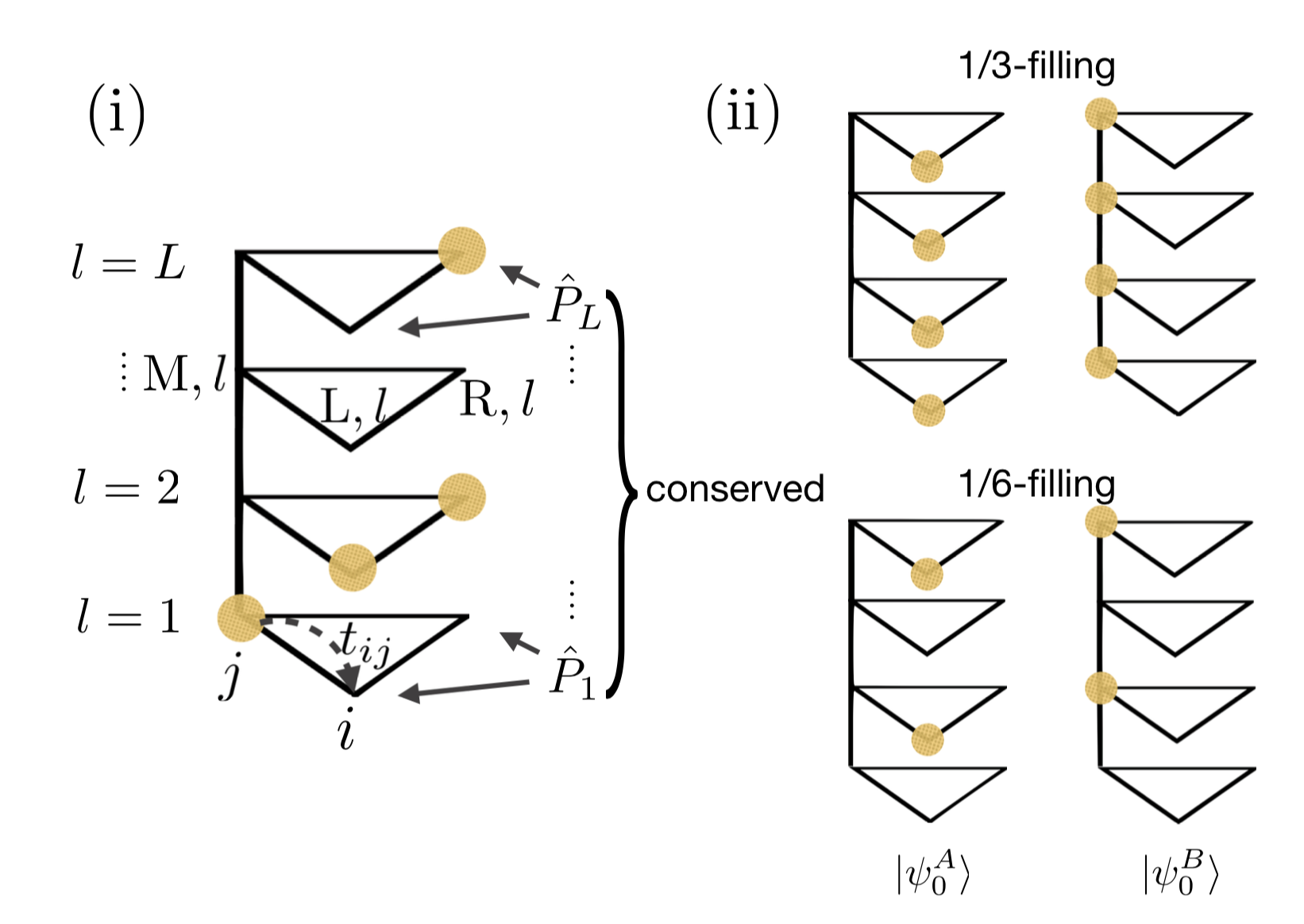}
\caption{(i) A model of hard-core bosons with $L=N_b=N_s/3=4$ (1/3-filling).
Bosons can hop between neighboring sites connected with the solid bonds with the hopping energy $t_{ij}$.
Every layer labeled by $l\:(0\leq l \leq L)$ has a local $\mathbb{Z}_2$ symmetry that corresponds to a swap of two sites L and R.
We denote the swapping operator for the $l$th layer by $\hat{P}_l$.
(ii) Two initial states $\ket{\psi^A_0}$ (left) and $\ket{\psi^B_0}$ (right), where hard-core bosons are placed at $(\mathrm{L},l)$ and $(\mathrm{M},l)$, respectively. For the 1/3-filling (up), we choose $l=1,2,\dots,L$, and for the 1/6-filling (down), we choose $l=2,4,\dots,L$.
Reproduced from Fig. 1 of~\cite{Hamazaki16G}. \copyright 2016 American Physical Society.
}
\label{bose}
\end{center}
\end{figure}

As shown in Fig. \ref{bose}(i), we consider a nonintegrable model of $N_b$ hard-core bosons on $N_s$-number of lattice sites that are arranged in $L$ layers in the shape of triangles $(N_s=3L)$.
Each site $i\:(1\leq i\leq N)$ is labeled by two indices $(s,l)$, where
$l\:(=1,2,\dots, L)$ labels the layer and $s\:(=\mathrm{L,M,R})$ labels the position in each layer. 

The Hamiltonian can be written as
\begin{equation}\label{Ham}
\hat{H}=-\sum_{\langle i,j \rangle}t_{ij}(\hat{b}^\dag_i \hat{b}_j +\text{h.c.}),
\end{equation} 
where $\hat{b}_i$ is the annihilation operator of a hard-core boson at the site $i$, $t_{ij}\in\mathbb{R}$ is the  hopping energy between two sites $i$ and $j$,
and $\langle i,j\rangle\:(i<j)$ represents a pair of neighboring sites.

We assume that 
\begin{equation}\label{symcond}
t_{\text{M}l,\text{L}l}=t_{\text{M}l,\text{R}l}
\end{equation}
is satisfied for the hopping energy $t_{ij}$,
which allows the system to have a local $\mathbb{Z}_2$ symmetry with the corresponding operator $\hat{P}_l\:(1\leq l \leq L)$ for each layer.
This operator swaps two sites (L,$l$) and (R,$l$) in Fig. \ref{bose} (i), and satisfies 
\aln{
\hat{P}_l\hat{b}_{\text{L}l}\hat{P}^\dag_l&=\hat{b}_{\text{R}l},\\ 
\hat{P}_l\hat{b}_{\text{R}l}\hat{P}^\dag_l&=\hat{b}_{\text{L}l}.
}
We note that $\hat{P}_l$ can be written as 
\aln{
\hat{P_l}= \hat{I} +\hat{b}_{\text{L}l}^\dag \hat{b}_{\text{R}l}
+\hat{b}_{\text{R}l}^\dag \hat{b}_{\text{L}l} - (\hat{b}_{\text{L}l}^\dag \hat{b}_{\text{L}l}-\hat{b}_{\text{R}l}^\dag \hat{b}_{\text{R}l})^2,
}
where $[\hat{H},\hat{P}_l]=0$ and $\hat{P}_l^2=1$ are satisfied.
We call the eigenvalues of $\hat{P}_l$, namely $q_l=\pm 1$, as positive and negative $\mathbb{Z}_2$ parities.
If we map the hard-core bosons to the spin 1/2 operators, we can interpret 
 $\hat{P}_l$ as the projection operator onto the spin singlet $(q_l=-1)$ and triplet $(q_l=+1)$ states that involve the spins on (L,$l$) and (R,$l$).

By this construction, the system has a symmetry group that can be written as 
\aln{
G=\bigotimes_{l=1}^L \mathbb{Z}_2.
}
Since $G$ is abelian, we can divide the set of energy eigenstates into the $|G|=2^L$ symmetry sectors \cite{Georgi} that are determined by a set of $\mathbb{Z}_2$ parities $\mathbf{q}:=(q_l)_{l=1}^L$.
If we denote the symmetry sectors by $\mathbf{q}$, the entire Hilbert space $\mathcal{H}$ of the system is decomposed as 
\aln{
\mathcal{H}=\bigoplus_{\mathbf{q}}\mathcal{H}_\mathbf{q}.
}

To remove unwanted symmetries and degeneracies, we add randomness to $t_{ij}$.
We assume that $t_{\mathrm{M}l,\mathrm{L}l}$(=$t_{\mathrm{M}l,\mathrm{R}l}$), $t_{\mathrm{L}l,\mathrm{R}l}$, and $t_{\mathrm{L}l,\mathrm{R}(l+1)}$ can  be written as 
\begin{align}\label{rand}
t_{ij}={t_\mathrm{hop}}(1+\epsilon_{ij}),
\end{align}
where the randomness $\epsilon_{ij}$ is uniformly chosen from
$[-0.5,0.5]$.
We have confirmed that this randomness romoves all degeneracies and most of the symmetries except for the $\mathbb{Z}_2$ symmetry.

\begin{figure}
\begin{center}
\includegraphics[width=11cm,angle=-90]{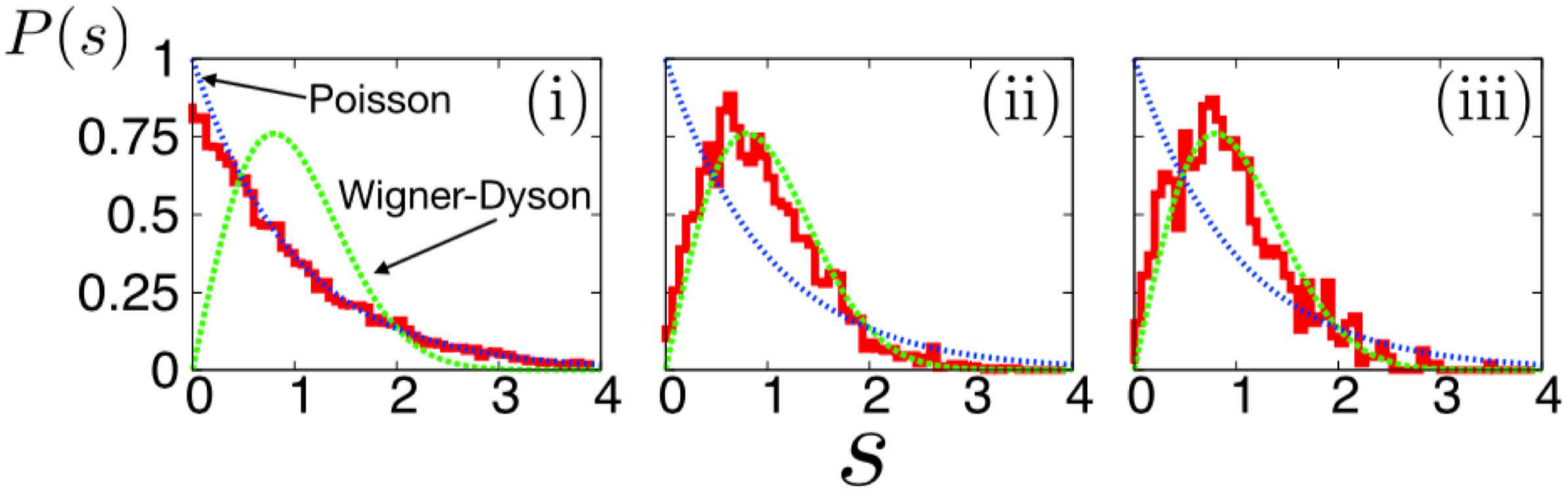}
\caption{(solid) Level-spacing statistics calculated for (i) the entire spectrum, (ii) $\mathcal{H}_{\mathbf{q}_1}$, and (iii) $\mathcal{H}_{\mathbf{q}_2}$. The Poisson ($e^{-s}$) and Wigner-Dyson ($\frac{\pi}{2}se^{-\frac{\pi}{4}s^2}$) distributions are superposed for the sake of comparison.
Reproduced from Fig. 9 of~\cite{Hamazaki16G}. \copyright 2016 American Physical Society.
}
\label{strlevelall}
\end{center}
\end{figure}

We note that the level-spacing statistics obeys the Wigner-Dyson statistics within each symmetry sector $\mathcal{H}_\mathbf{q}$ that contains sufficiently many eigenstates.
As we have seen in Chapter \ref{ch:ETH}, in nonintegrable systems that conserve energy alone, the level-spacing statistics is expected to resemble the Wigner-Dyson statistics, $P_\mathrm{WD}(s)=\frac{\pi}{2}se^{-\frac{\pi}{4}s^2}$.\footnote{Note that we use the statistics for the GOE because the Hamiltonian in Eq. (\ref{Ham}) is invariant under the time-reversal operation.}
On the other hand, the statistics in systems with additional conserved quantities obeys the one without level repulsions such as the Poisson statistics $P_\mathrm{P}(s)=e^{-s}$.
Figure \ref{strlevelall} (i) shows the level-spacing statistics for the entire spectrum in our model.
It is close to the Poisson statistics, not to the Wigner-Dyson statistics.
This result reflects the fact that our model has $\mathbb{Z}_2$ symmetries~\cite{Santos10b}.
Figures \ref{strlevelall} (ii) and (iii) show the level-spacing statistics of the eigenstates that belong to the sectors with $\mathbf{q}_1:=(+1,+1,...,+1)$ and $\mathbf{q}_2:=(-1,+1,...,+1)$, respectively.
They obey the Wigner-Dyson statistics rather than the Poisson statistics.

\section{Time evolutions from two initial states}\label{quench}
As initial states, we consider two cases as $\ket{\psi_0}=\ket{\psi^A_0}$ and $\ket{\psi^B_0}$, where bosons are placed at $(\mathrm{L},l)$ and $(\mathrm{M},l)$, respectively (see Fig. \ref{bose} (ii)).
We will call time evolutions from these initial states as Case A and Case B.
We consider the cases of 1/3-filling, where $N_b=L$ and $l=1,2,\dots,L$,  and 1/6-filling, where $N_b=L/2$ and $l=2,4,\dots,L$.
While $\ket{\psi^A_0}$ extends over the different $\mathcal{H}_\mathbf{q}$'s, 
$\ket{\psi^B_0}$ belongs to only one sector $\mathcal{H}_{\mathbf{q}_1}$, where $\mathbf{q}_1=(+1,+1,\dots,+1)$ (see Appendix \ref{sec:proj}).

The state at time $t$ is obtained as $\ket{\psi(t)}=e^{-\frac{i\hat{H}t}{\hbar}}\ket{\psi_0} =\sum_\alpha c_\alpha e^{-\frac{i{E}_\alpha t}{\hbar}}\ket{E_\alpha}$, where $c_\alpha = \braket{E_\alpha|\psi_0}$.
The long-time average of a local observable ${\hat{\mathcal{O}}}$ is described by the diagonal ensemble under certain conditions (see Chapter \ref{ch:Eq}):
\begin{align}\label{diag}
\lim_{T\rightarrow \infty}\frac{1}{T}\int_0^T{\braket{\psi(t)|{\hat{\mathcal{O}}}|\psi(t)}} = \mathrm{Tr}[ \hat{\rho}_d {\hat{\mathcal{O}}} ],
\end{align}
where $\hat{\rho}_\mathrm{d}:=\sum_\alpha |c_\alpha|^2 \ket{E_\alpha}\bra{E_\alpha}$.

We define the canonical ensemble and the GGE that may describe the stationary state with a few parameters.
We define the canonical ensemble as
\begin{align}\label{canonicale}
\hat{\rho}_\mathrm{can} = \frac{1}{Z_\mathrm{can}} e^{-\beta \hat{H}},
\end{align}
where $Z_\mathrm{can} =\mathrm{Tr}[e^{-\beta \hat{H}}]$ and the inverse temperature $\beta$ is determined from the total energy
$
E_0 := \braket{\psi_0|\hat{H}|\psi_0}=\mathrm{Tr}[\hat{H}\hat{\rho}_\mathrm{can}].
$
On the other hand, we define the GGE as
\begin{align}\label{GGE}
\hat{\rho}_\mathrm{GGE} &= \frac{1}{Z_\mathrm{GGE}} e^{-\tilde{\beta }\hat{H}-\sum_{l=1}^L \lambda_l \hat{P}_l},
\end{align}
where $Z_\mathrm{GGE} =\mathrm{Tr}[e^{-\tilde{\beta }\hat{H}-\sum_{l=1}^L \lambda_l \hat{P}_l}]$.
Here $\tilde{\beta}$ and $\lambda_l\:(1\leq l\leq L)$ are uniquely determined from the initial conditions as follows:
\aln{
\braket{\psi_0|\hat{H}|\psi_0}&=\mathrm{Tr}[\hat{H}\hat{\rho}_\mathrm{GGE}],\\
\braket{\psi_0|\hat{P}_l|\psi_0}&=\mathrm{Tr}[\hat{P_l}\hat{\rho}_\mathrm{GGE}]\:\:\:(1\leq l \leq L).
}
Note that our definition of the GGE uses an extensive number of truly local conserved operators, whereas the usual GGE in integrable systems takes the sum of (quasi)-local operators as conserved quantities.

We note that both of the initial states have the total conserved energy 
\aln{
E_0=\braket{\psi^A_0|\hat{H}|\psi^A_0}=\braket{\psi^B_0|\hat{H}|\psi^B_0}=0,
}
which leads to the infinite temperature ($\beta=0$) in the canonical ensemble.
To show this, we should solve the equation for $\beta$,
\aln{
0=E_0=\frac{1}{Z_\mathrm{can}}\sum_\alpha E_\alpha e^{-\beta E_\alpha},
}
where $Z_\mr{can}=\sum_\alpha e^{-\beta E_\alpha}$.
Since the right-hand side of this equation monotonically decreases with respect to $\beta$, 
we have a unique solution.
Moreover, for the Hamiltonian in Eq. (\ref{Ham}), we can show
\aln{
\mathrm{Tr}[\hat{H}]=-\sum_{\braket{ij}}t_{ij}\mathrm{Tr}[\hat{b}_i^\dag\hat{b}_j+\mathrm{h.c.}]=0.
}
Here we have used $\mathrm{Tr}[\hat{b}^\dag_i \hat{b}_j]= 0$ for $i\neq j$, which can be understood by treating the trace with the Fock basis on the sites.
Consequently, we obtain $\mathrm{Tr}[\hat{H}]=\sum_\alpha E_\alpha=0$, which leads to $\beta=0$.
Note that the canonical ensemble at the infinite temperature is proportional to the identity operator as
$
\hat{\rho}_\mathrm{can}=\frac{1}{D}
$, where $D:= \mathrm{dim}[\mathcal{H}]$ is the dimension of the entire Hilbert space.

\begin{figure}
\begin{center}
\includegraphics[width=11cm,angle=-90]{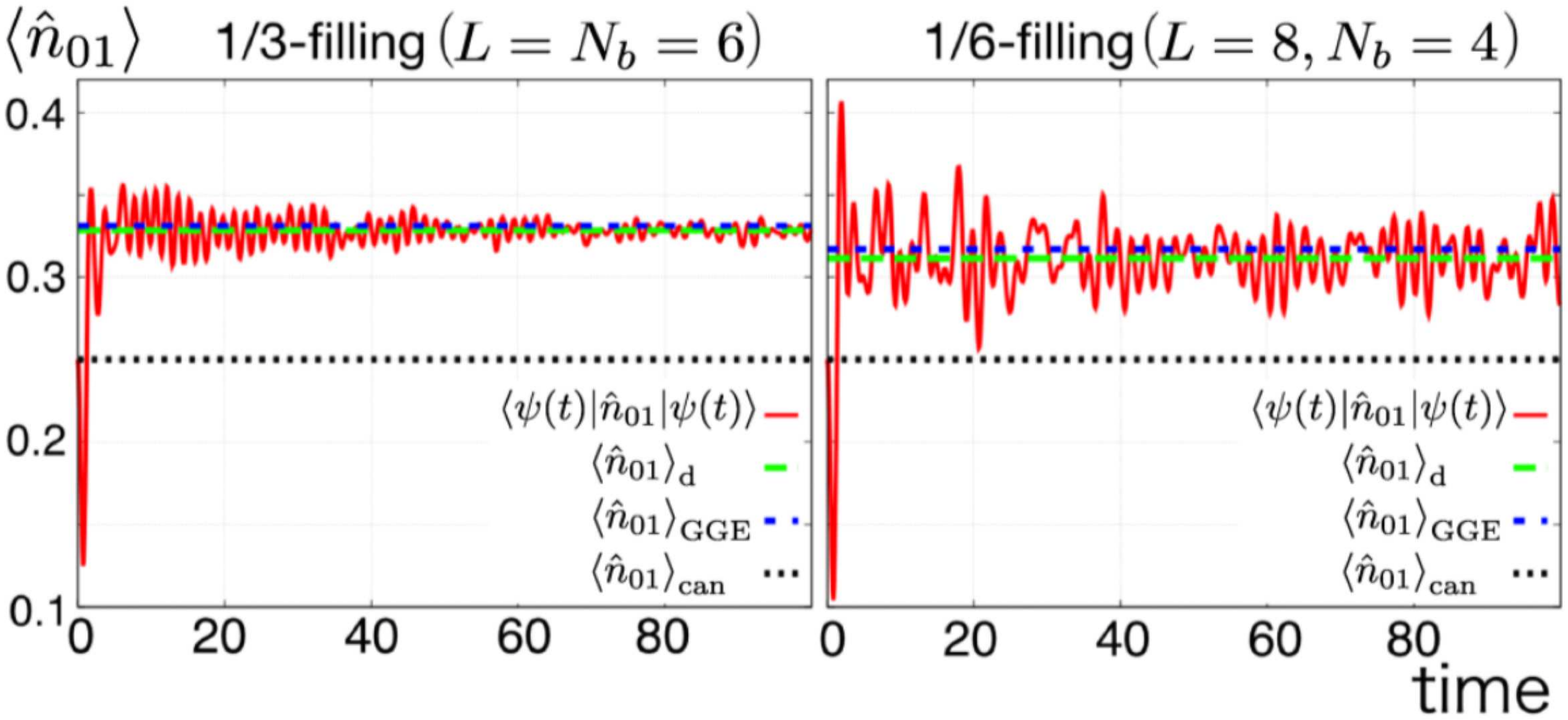}
\caption{Time evolutions of the expectation values of $\hat{n}_{01}$ for Case A (solid curve). We measure the time in units of $\hbar/t_\mathrm{hop}$.
The left and right figures respectively show the result for the 1/3-filling ($L=N_b=6$) and that for the 1/6-filling ($L=8, N_b=4$).
The stationary state is well described by the prediction of the diagonal ensemble (yellow long dashed line).
The GGE (blue short dashed line) also describes the stationary state well, but the canonical ensemble (black dotted line) does not.
Reproduced from Fig. 2 of~\cite{Hamazaki16G}. \copyright 2016 American Physical Society.
}
\label{timeevolve}
\end{center}
\end{figure}

As observables, we consider the Fourier transform of the hard-core boson operators and a (renormalized) mode occupation number with $\mathbf{k}=(k_x,k_y,k_z)$.
Then we take a marginal distribution of the occupation number by integrating out $k_z$, as
$\hat{n}(k_x,k_y)=\frac{1}{2^2N_b}\sum_{{i,j}}\delta_{z_i,z_j}e^{-i\mathbf{k\cdot (r_i-r_j)}}\hat{b}^\dag_i \hat{b}_j$.
Here we denote $\mathbf{r_i}=(x_i,y_i,z_i)$ by the coordinate of the site $i$ (the lattice constant is set to unity).
We will especially consider $\hat{n}_{00}:= \hat{n}(0,0),{n}_{01}:= \hat{n}(0,\pi)$, and $\hat{n}_{11}:= \hat{n}(\pi,\pi)$ in the following discussions.
We note that these observables are macroscopic in the sense that they can be written as the averages of local operators on each layer.

We show typical time evolutions of the expectation value of $\hat{n}_{01}$ for Case A in Figure \ref{timeevolve}.
The left and right figures respectively show the result of the 1/3-filling ($L=N_b=6$) and that of the 1/6-filling ($L=8, N_b=4$).
We also show the predictions of the diagonal ensemble, the canonical ensemble, and the GGE, which are respectively given by $\braket{\hat{n}_{01}}_\mathrm{d}:= \mathrm{Tr}[\hat{n}_{01}\hat{\rho}_\mathrm{d}]$, $\braket{\hat{n}_{01}}_\mathrm{can}:= \mathrm{Tr}[\hat{n}_{01}\hat{\rho}_\mathrm{can}]$ and $\braket{\hat{n}_{01}}_\mathrm{GGE}:= \mathrm{Tr}[\hat{n}_{01}\hat{\rho}_\mathrm{GGE}]$.
We can see that the expectation values for large $t$ are well described by the prediction of the diagonal ensemble with small temporal fluctuations.
As shown in the figure, we find that the GGE coincides with the prediction of the diagonal ensemble (Fig. \ref{timeevolve}) very well, whereas the canonical ensemble does not.
In fact, the canonical ensemble at $\beta=0$ always gives\footnote{
For example, for ${\hat{n}_{01}}$, we obtain
\begin{align}
\braket{\hat{n}_{01}}_\mathrm{can} =
\frac{1}{2^2N_bD}\sum_{i,j}e^{-i\mathbf{k\cdot (r_i-r_j)}}\delta_{z_i,z_j}\mathrm{Tr}[\hat{b}^\dag_i \hat{b}_j].
\end{align}
Since the trace for $i\neq j$ vanishes, it becomes $\frac{1}{2^2N_bD}\sum_{i}\mathrm{Tr}[\hat{b}^\dag_i \hat{b}_i]$.
Then, by treating the trace using the energy eigenstates, we have
\begin{align}
\braket{\hat{n}_{01}}_\mathrm{can} =\frac{1}{2^2N_bD}\sum_{i}\sum_\alpha \braket{E_\alpha|\hat{b}^\dag_i \hat{b}_i|E_\alpha}=\frac{1}{4},
\end{align}
where we have used $\sum_i \braket{E_\alpha|\hat{b}^\dag_i \hat{b}_i|E_\alpha}=N_b$.
Similarly, we obtain $\braket{\hat{n}_{00}}_\mathrm{can}=\braket{\hat{n}_{11}}_\mathrm{can}=\frac{1}{4}$.
}
\begin{align}
\braket{\hat{n}_{00}}_\mathrm{can}=\braket{\hat{n}_{01}}_\mathrm{can}=\braket{\hat{n}_{11}}_\mathrm{can}=\frac{1}{4},
\end{align}
which is not equal to $\braket{\hat{n}_{00}}_\mathrm{d},\braket{\hat{n}_{01}}_\mathrm{d}$, and $\braket{\hat{n}_{11}}_\mathrm{d}$ in general.
This result highlights our key finding independent of the value of fillings: we need the GGE to describe the stationary state in a nonintegrable system with an extensive number of local symmetries.
We will verify this observation in more detail by focusing on the case of the 1/3-filling ($L=N_b$) in the following sections.

\section{Verification of the GGE by the finite-size scaling analysis}\label{atagge}
In this section, we quantitatively analyze how well the GGE describes the stationary state compared with the canonical ensemble by the finite-size scaling analysis.
We define the relative difference between the canonical ensemble/GGE and the diagonal ensemble as follows:
\begin{align}\label{rela}
\overline{\delta n_\mathrm{can}}&:=\overline{\left|\frac{\braket{\hat{n}}_\mathrm{d}-\braket{\hat{n}}_\mathrm{can}}{{\braket{\hat{n}}_\mathrm{d}}}\right|}, \nonumber \\
\overline{\delta n_\mathrm{GGE}}&:=\overline{\left|\frac{\braket{\hat{n}}_\mathrm{d}-\braket{\hat{n}}_\mathrm{GGE}}{\braket{\hat{n}}_\mathrm{d}}\right|}.
\end{align}
Here $\hat{n}$ represents $\hat{n}_{00},\hat{n}_{01}$ or $\hat{n}_{11}$, and $\overline{\cdot\cdot\cdot}$ denotes the average over 20 sample Hamiltonians with different values of randomness in $t_{ij}$ (see Eq. (\ref{rand})).

\begin{figure}
\begin{center}
\includegraphics[width=14cm]{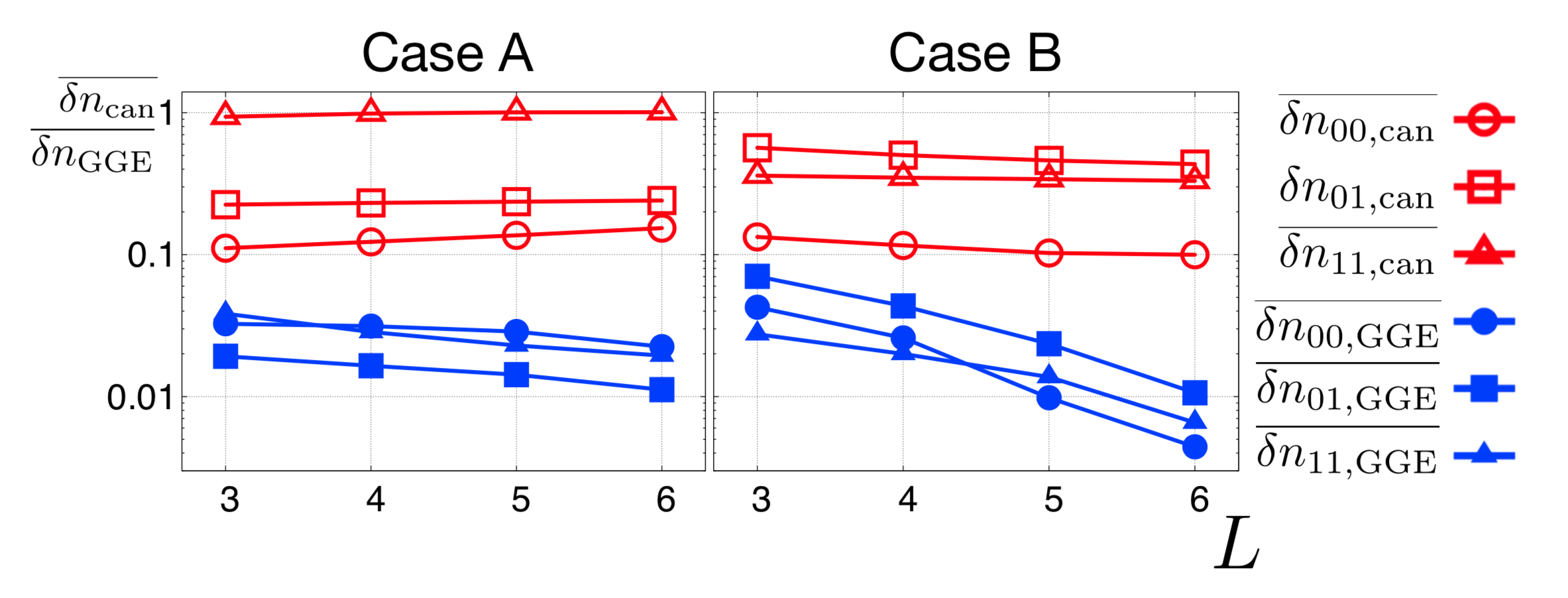}
\caption{Relative difference of the canonical ensemble (open) and the GGE (filled) compared with the diagonal ensemble (see Eq. (\ref{rela}))  for $\hat{n}_{00}\text{ (circle)},\hat{n}_{01}\text{ (square)}$, and $\hat{n}_{11}\text{ (triangle)}$. The left and right figures respectively show the results for Case A and Case B.
For both of the initial states, the relative difference for the canonical ensemble does not decrease with the system size $L$, whereas it rapidly decreases for the GGE.
Reproduced from Fig. 3 of~\cite{Hamazaki16G}. \copyright 2016 American Physical Society.
}
\label{diff}
\end{center}
\end{figure}

As shown in Fig. \ref{diff}, the relative difference of the GGE is about ten times smaller than that of the canonical ensemble (note that the graph is displayed using the semi-log scale).
We note that the relative difference stays more than 10\% for the canonical ensemble even if we increase $L$, whereas it rapidly decreases with $L$ for the GGE.

Figure \ref{diff} also shows that there is some difference between Case A and Case B if we consider the $L$-dependence of the relative difference of the GGE:
the relative difference decreases less rapidly in Case A than in Case B with increasing the system size. 
This results from the mixing of the symmetry sectors with negative parities in Case A.
We will examine this difference in detail in the next section.

\section{Verification of the ETH for each symmetry sector}\label{RETH}
In this section, we analyze the ETH for diagonal matrix elements of observables $\hmo$ and seek for the reason why the GGE works well and the canonical ensemble does not work in our model.
We note that we will only treat the ETH for diagonal matrix elements and call it just as ``the ETH" in the following discussions.
As we have seen in Chapters \ref{ch:Eq} and \ref{ch:ETH}, the ETH states that $\braket{E_\alpha|\hat{\mathcal{O}}|E_\alpha}$ is equal to the spectral average within a small energy shell in the thermodynamic limit.
We will call $\mc{O}_{\alpha\alpha}=\braket{E_\alpha|\hat{\mathcal{O}}|E_\alpha}$  as the eigenstate expectation value (EEV).
When the $|c_\alpha|$'s have a sharp peak around the average energy,
the ETH justifies the microcanonical ensemble and the canonical ensemble (see Chapter \ref{ch:Eq}).\footnote{We assume the equivalence of the microcanonical ensemble and the canonical ensemble.}

\begin{figure}
\begin{center}
\includegraphics[width=11cm,angle=-90]{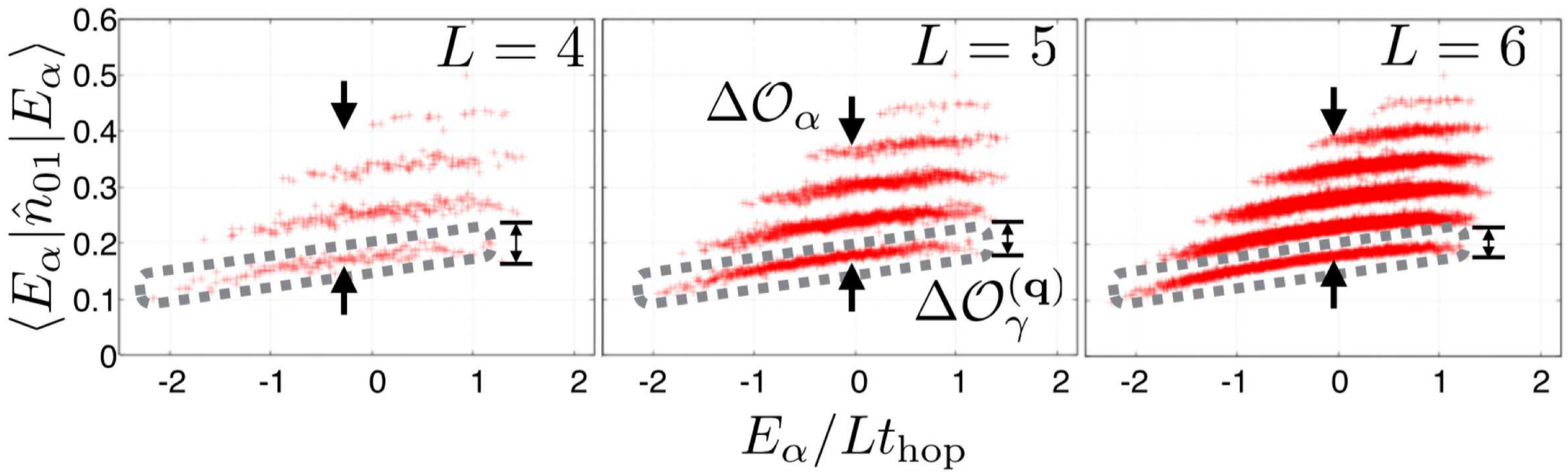}
\caption{$L$-dependence of the EEVs for $\hat{n}_{01}$ plotted for $L=4$ (left), 5 (middle) and 6 (right). The EEV fluctuations, $\Delta \mathcal{O}_\alpha$, which are shown by a pair of arrows, do not decrease with increasing $L$. 
We also show the subset of the EEVs belonging to $\mathcal{H}_{\mathbf{q}_1}$ in the region encircled by the dotted curves.
The EEV fluctuations in the restricted subset, $\Delta \mathcal{O}_{\gamma}^{(\mathbf{q}_1)}$, which are indicated by updown arrows, decrease with increasing the system size.
Reproduced from Fig. 4 of~\cite{Hamazaki16G}. \copyright 2016 American Physical Society.
}
\label{eth}
\end{center}
\end{figure}

Figure \ref{eth} shows the EEVs for $\hat{n}_{01}$, indicating that the ETH does not hold true for the entire spectrum.
The fluctuations of EEVs (EEV fluctuations) $\Delta \mathcal{O}_\alpha$ indicated by a pair of arrows in Fig. \ref{eth} do not decrease even when $L$ becomes larger.\footnote{We note that $\Delta \mathcal{O}_\alpha$ is regarded as the second term on the right-hand side of \EQ{corre} for $\alpha=\beta$.}
We note that  similar results are found for $\hat{n}_{00}$ and $\hat{n}_{11}$.

Nevertheless, we have found that the EEV fluctuations decrease if we consider only eigenstates that are restricted to each symmetry sector.
For example,  in Fig. \ref{eth}, each region encircled by dotted curves shows the restricted eigenstates that belong to $\mathcal{H}_{\mathbf{q}_1}$.
In this sector, the EEV fluctuations seem to decrease with increasing the system size.
To be more precise, we define the EEV fluctuation $\Delta \mathcal{O}_\gamma^{(\mathbf{q})}$ in sector $\mathcal{H}_\mathbf{q}$ by
\begin{align}\label{ethfs}
\braket{E_{\gamma}^{(\mathbf{q})}|\hat{\mathcal{O}}|E_{\gamma}^{(\mathbf{q})}}=\braket{\hat{\mathcal{O}}}_\mathcal{T}^{(\mathbf{q})}(E_{\gamma}^{(\mathbf{q})})+\Delta \mathcal{O}_{\gamma}^{(\mathbf{q})},
\end{align}
where $\ket{E_{\gamma}^{(\mathbf{q})}}$ is an energy eigenstate in $\mathcal{H}_{\mathbf{q}}$ with an energy $E_{\gamma}^{(\mathbf{q})}$, and $\gamma\:(1\leq \gamma\leq \mathrm{dim}[\mathcal{H}_{\mathbf{q}}])$ is a label of the eigenstate.
We have also defined the spectral average in the sector $\mathcal{H}_\mathbf{q}$ within a small energy shell (cf. \EQ{save}):
\begin{align}\label{gmic}
\braket{\hat{\mathcal{O}}}_\mathcal{T}^{(\mathbf{q})}(E):=\frac{1}{\mathcal{N}^{(\mathbf{q})}_{E,\omega_s}}\sum_{|E-E_{\gamma}^{(\mathbf{q})}|<\omega_s}\braket{E_{\gamma}^{(\mathbf{q})}|\hat{\mathcal{O}}|E_{\gamma}^{(\mathbf{q})}},
\end{align}
where $\mathcal{N}^{(\mathbf{q})}_{E,\omega_s}$ is the number of the energy eigenstates in $\mathcal{H}_\mathbf{q}$ within the energy shell $[E-\omega_s,E+\omega_s]$.
We also define the average of (the generalized version of) the microcanonical ensemble $\braket{\hat{\mathcal{O}}}_\mathrm{mic}^{(\mathbf{q})}(E)$ in the sector $\mathcal{H}_\mathbf{q}$ by replacing $\omega_s$ in \EQ{gmic} with the microcanonical energy width $\Delta E$.\footnote{We note that, in a marcroscopic system, $\Delta E$ may be subextensive with the system size, whereas $\omega_s (<\omega_\mr{sh})$ is expected to remain small in many cases~\cite{DAlessio16} (also see Chapter \ref{ch:ETH}).}

\begin{figure}
\begin{center}
\includegraphics[width=10cm,angle=-90]{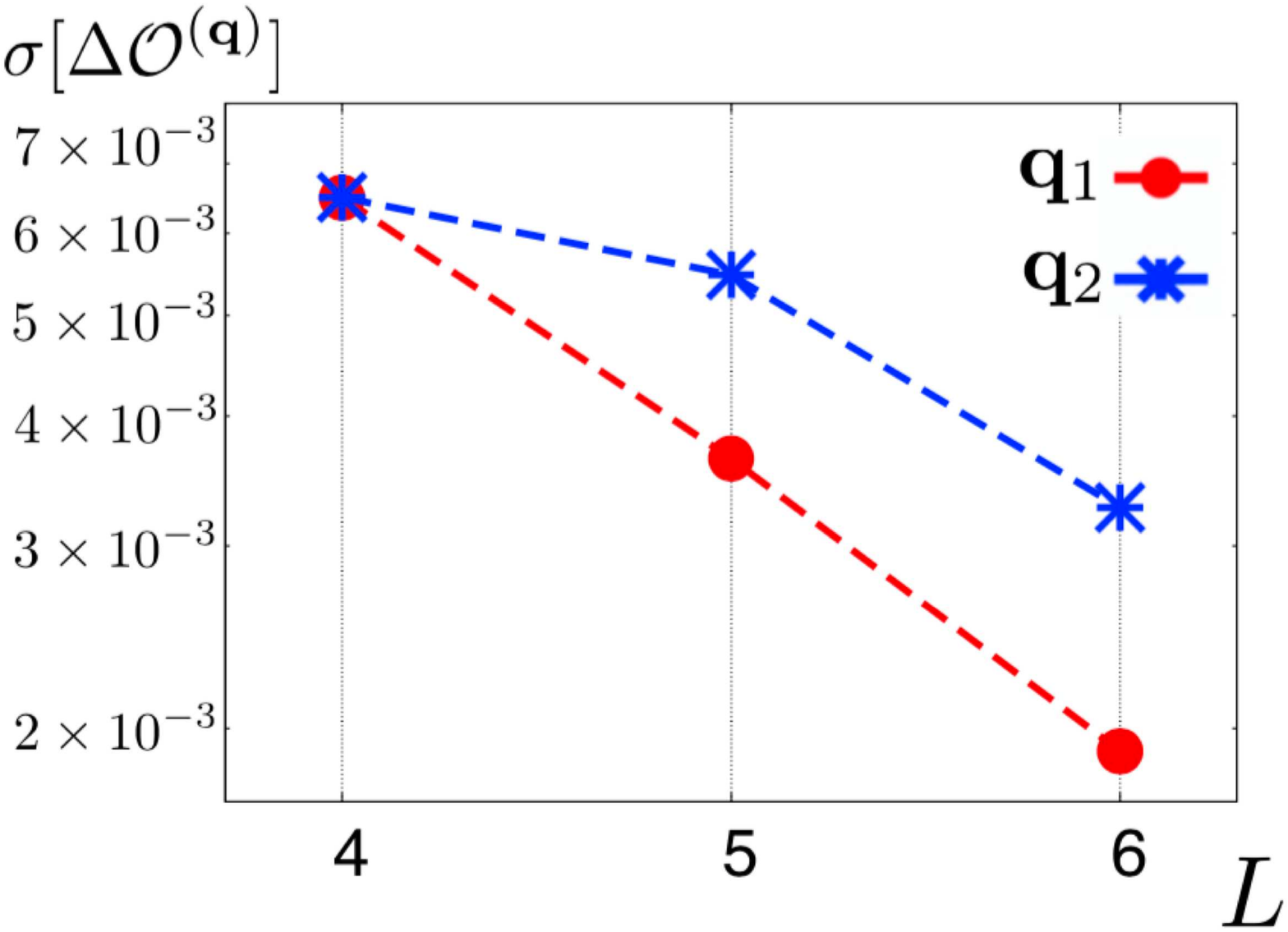}
\caption{$L$-dependence of the standard deviation $\sigma[{\Delta {\mathcal{O}}^{(\mathbf{q})}}]$ of $\braket{E_{\gamma}^{(\mathbf{q})}|\hat{\mathcal{O}}|E_{\gamma}^{(\mathbf{q})}}$ within the energy shell $[E-\omega_s,E+\omega_s]$ with $\omega_s=0.18L$ and $E=0$. 
We show $\sigma[{\Delta {\mathcal{O}}^{(\mathbf{q})}}]$ for $\mathcal{H}_{\mathbf{q}_1}$ (circle) and $\mathcal{H}_{\mathbf{q}_2}$ (asterisk).
Both of them rapidly decrease with increasing the system size $L$, which indicates that the ETH is valid for each symmetry sector.
Reproduced from Fig. 5 of~\cite{Hamazaki16G}. \copyright 2016 American Physical Society.
}
\label{ethyuragi}
\end{center}
\end{figure}

Figure \ref{ethyuragi} illustrates the validity of the ETH for each symmetry sector.
We evaluate the typical magnitude of $\Delta \mathcal{O}_{\gamma}^{(\mathbf{q})}$ with $\sigma[{\Delta {\mathcal{O}}^{(\mathbf{q})}}]$,  where $\sigma[{\Delta {\mathcal{O}}^{(\mathbf{q})}}]$ is the standard deviation of $\braket{E_{\gamma}^{(\mathbf{q})}|\hat{\mathcal{O}}|E_{\gamma}^{(\mathbf{q})}}$ within the energy shell $[E-\omega_s,E+\omega_s]$ for $\mathbf{q}_1$ and $\mathbf{q}_2:=(-1, +1, ..., +1)$.\footnote{We note that $\sigma[{\Delta {\mathcal{O}}^{(\mathbf{q})}}]$ is a generalized version of the first equation in \EQ{vnokei}.}
As shown in the figure, both $\sigma[{\Delta {\mathcal{O}}^{(\mathbf{q_1})}}]$ and $\sigma[{\Delta {\mathcal{O}}^{(\mathbf{q_2})}}]$ rapidly decrease with increasing the system size $L$.\footnote{Strictly speaking, the decay of the standard deviations is directly related to the weak ETH. However, (exponentially) fast decrease of them is observed in systems where the strong ETH is also expected to hold~\cite{Beugeling14}. We also note that the EEV fluctuations are evaluated by the deviations from linear fittings within the small energy shell (see the first footnote of Subsection \ref{tatatatata}).}

Assuming the ETH for each sector, the diagonal ensemble is approximately written as a statistical mixture of the microcanonical ensembles in all sectors:
\begin{align}
\mathrm{Tr}[\hat{\mathcal{O}}\hat{\rho}_\mathrm{d}] 
&\simeq\sum_\mathbf{q} p_\mathbf{q} \braket{\hat{\mathcal{O}}}_\mathrm{mic}^{(\mathbf{q})}(E_\mathbf{q}), \label{gethtogge}
\end{align}
which is obtained by applying the derivation in \EQ{henkeida} for each sector.
Here,
\begin{align}\label{opn}
p_\mathbf{q}=\sum_{{\gamma}} |c_{\gamma}^{(\mathbf{q})}|^2
=\braket{\psi_0|\mathcal{\hat{P}}_\mathbf{q}|\psi_0} 
\end{align}
represents the occupation ratio of the sector $\mathcal{H}_\mathbf{q}$, $\mathcal{\hat{P}}_\mathbf{q}$ is the  projection operator onto the sector $\mathcal{H}_\mathbf{q}$, and $c_{\gamma}^{(\mathbf{q})}:= \braket{E_{\gamma}^{(\mathbf{q})}|\psi_0}$. 
Moreover, we define
\begin{align}
E_\mathbf{q}=\frac{1}{p_\mathbf{q}}\sum_{\gamma} |c_{\gamma}^{(\mathbf{q})}|^2E_{\gamma}^{(\mathbf{q})}
=\frac{1}{p_\mathbf{q}}\braket{\psi_0|\mathcal{\hat{P}}_\mathbf{q}\hat{H}\mathcal{\hat{P}}_\mathbf{q}|\psi_0} 
\end{align}
as the average energy in sector $\mathcal{H}_\mathbf{q}$.
To derive Eq. (\ref{gethtogge}), we have assumed that $|c_{\gamma}^{(\mathbf{q})}|$'s have a sharp peak around $E_\mathbf{q}$ (cf. \EQ{henkeida}). 
Note that Eq. (\ref{gethtogge}) depends on $2|G|=2^{L+1}$ parameters $p_\mathbf{q}$ and $E_\mathbf{q}$,  whereas the diagonal ensemble depends on $\mathrm{dim} [\mathcal{H}]=\frac{(3L)!}{L!(2L)!}(\gg 2|G|)$ parameters.

Now, we define what we call the ``restricted GGE (rGGE)" with $2^{L+1}$ conserved quantities from which  $p_\mathbf{q}$ and $E_\mathbf{q}$ are determined.
By taking $\hat{Q}_0:= \hat{H}, \hat{Q_l}:= \hat{P}_l\:(1\leq l \leq L)$ and their higher-order correlations as such conserved quantities, we construct the rGGE as
\begin{align}\label{GGGE}
\hat{\rho}_\mathrm{rGGE}
=\frac{1}{Z_\mathrm{rGGE}}e^{-\sum_{l=0}^L \kappa_l \hat{Q}_l -\sum_{l < m} \kappa_{lm} \hat{Q}_l\hat{Q}_m - \cdot\cdot\cdot},
\end{align}
where $Z_{\mathrm{rGGE}}:= \mathrm{Tr}[\exp(\cdots)]$ (see Refs. \cite{Kollar08,Goldstein13,Sels14} for similar concepts).
Parameters $\{\kappa_{lm\cdot\cdot\cdot}\}$ are uniquely determined from the initial condition as
\aln{
\braket{\psi_0|\hat{Q}_l\hat{Q}_m\dots|\psi_0}=\mathrm{Tr}[\hat{\rho}_\mathrm{rGGE}\hat{Q}_l\hat{Q}_m \dots ].
}
From Eq. (\ref{GGGE}) we obtain
$\mathrm{Tr}[\hat{\rho}_\mathrm{rGGE}\mathcal{\hat{P}}_\mathbf{q}]=p_\mathbf{q}$ and $\frac{1}{p_\mathbf{q}}\mathrm{Tr}[\hat{\rho}_\mathrm{rGGE}\mathcal{\hat{P}}_\mathbf{q}\hat{H}\mathcal{\hat{P}}_\mathbf{q}]=E_\mathbf{q}$, which justifies the rGGE as the ensemble that describes a stationary state.

We conjecture that the GGE defined in Eq. (\ref{GGE}) can approximate the rGGE if we consider observables that can be written as the sums (or averages) of operators whose supports lie in each layer.
As we have seen in \SEC{sec:tGGE}, a related conjecture (i.e., the conjecture of the truncated GGE) was made in Ref. \cite{Fagotti13}, which states that we can remove those conserved quantities that are less local than the observables in constructing the GGE.
In our model, the multiple products of $\hat{Q}_l$'s in Eq. (\ref{GGGE}) have supports over the multiple layers.
They are thus expected to be excluded from the rGGE for $\hat{n}_{00},\hat{n}_{01}$, and $\hat{n}_{11}$, which can be written as the averages of the local operators that have supports in each layer.

Before closing this section, we briefly explain why $\overline{\delta n_\mathrm{GGE}}$ is less sensitive to the change of $L$ for Case A than for Case B.
In usual nonintegrable systems, the EEV fluctuations $\Delta \mathcal{O}_\alpha$ rapidly decrease with increasing $\mathrm{dim}[\mathcal{H}]$ as we have seen in Chapter \ref{ch:ETH}.
We expect that the restricted EEV fluctuations $\Delta \mathcal{O}_{\gamma}^{(\mathbf{q})}$ also  decrease with increasing $\mathrm{dim}[\mathcal{H}_\mathbf{q}]$.
When more negative $\mathbb{Z}_2$ parities ($q_l=-1$) exist in the sectors, they have smaller Hilbert  dimensions, which results in a larger $\Delta \mathcal{O}_{\gamma}^{(\mathbf{q})}$.
Then the EEV fluctuations remain larger for Case A because of the sectors with negative $\mathbb{Z}_2$ parities when $L$ increases.
Thus, $\overline{\delta n_\mathrm{GGE}}$ is less sensitive to $L$ in Case A than in Case B.

\section{Models with fewer local symmetries}\label{sec:otm}

\begin{figure}
\begin{center}
\includegraphics[width=11cm,angle=-90]{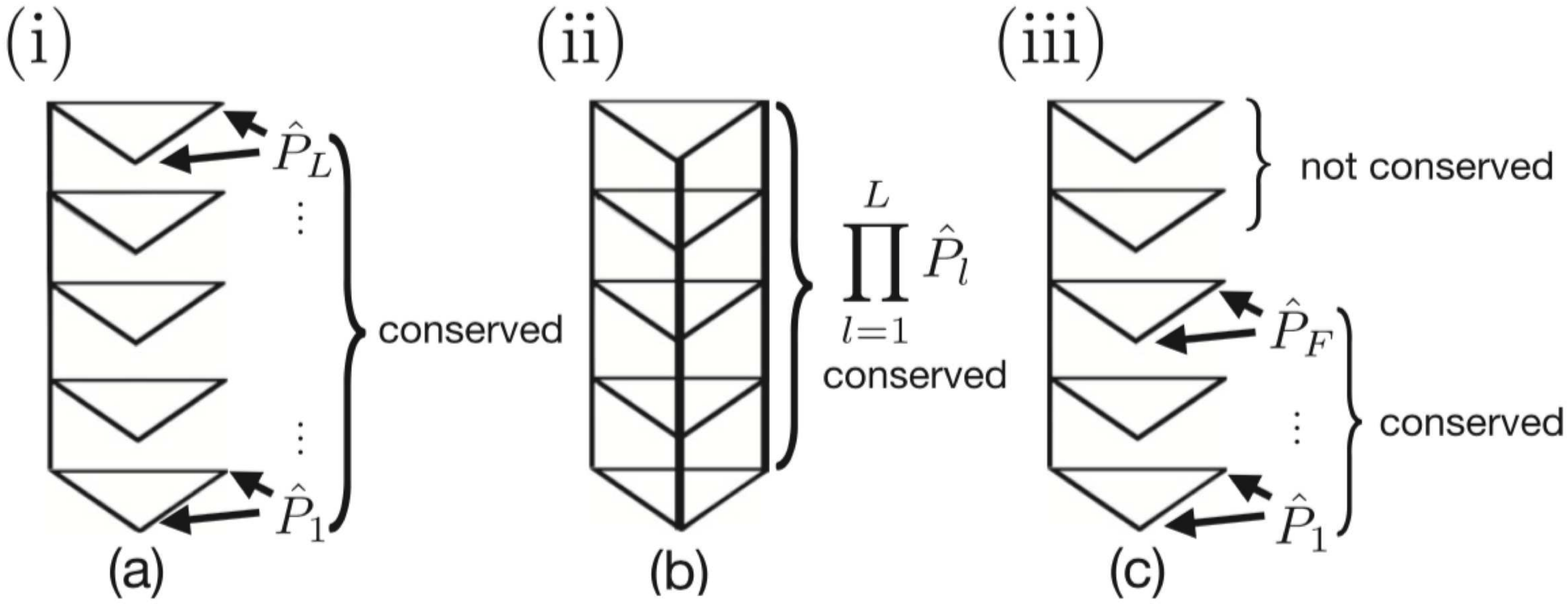}
\caption{Three models with different types of conserved quantities. (i) Model (a), which is the same as the model in Fig. \ref{bose}.
(ii) Model (b). Compared with (a), bosons can also hop between the L (or R) sites of the neighboring layers.
This model has only one global $\mathbb{Z}_2$ symmetry $\prod_{l=1}^L\hat{P}_l$, instead of the local $\mathbb{Z}_2$ symmetries.
(iii) Model (c),
which has local  $\mathbb{Z}_2$ symmetries only at the layers with $1\leq l\leq F$ (the case of $F=3$ is shown) due to the additional randomness introduced in the other layers.
Reproduced from Fig. 6 of~\cite{Hamazaki16G}. \copyright 2016 American Physical Society.
}
\label{othermodels}
\end{center}
\end{figure}

In this section, we show that the canonical ensemble well describes our macroscopic observables and that the GGE is not necessary when the number of the local symmetries does not increase with increasing $L$.
To demonstrate this, we first introduce two models with fewer local $\mathbb{Z}_2$ symmetries. 

In Figure \ref{othermodels} (ii), we show model (b), which has only one global $\mathbb{Z}_2$ symmetry.
The difference from model (a) is that bosons can hop vertically between the L (or R) sites of the neighboring layers.
We assume that $t_{\mathrm{L}l,\mathrm{L}(l+1)}=t_{\mathrm{R}l,\mathrm{R}(l+1)}\neq 0$,
which leads to a global conserved quantity $\prod_{l=1}^L\hat{P}_l$.
This operator swaps the sites  R and L on each layer simultaneously.

In Fig. \ref{othermodels} (iii), we show model (c), which has an $L$-independent number $F\:(F=0,1,2,3)$ of local $\mathbb{Z}_2$ symmetries.
In this model $t_{\mathrm{M}l,\mathrm{L}l}=t_{\mathrm{M}l,\mathrm{R}l}$ is satisfied only for $l\leq F$ due to the additional randomness introduced in the other layers.
Then, it has local $\mathbb{Z}_2$ symmetries only at the layers with $1\leq l\leq F$ for $F>0$.
In addition, model (c) with $F=0$ is a usual nonintegrable system that conserves only energy.

\begin{figure}
\begin{center}
\includegraphics[width=11cm,angle=-90]{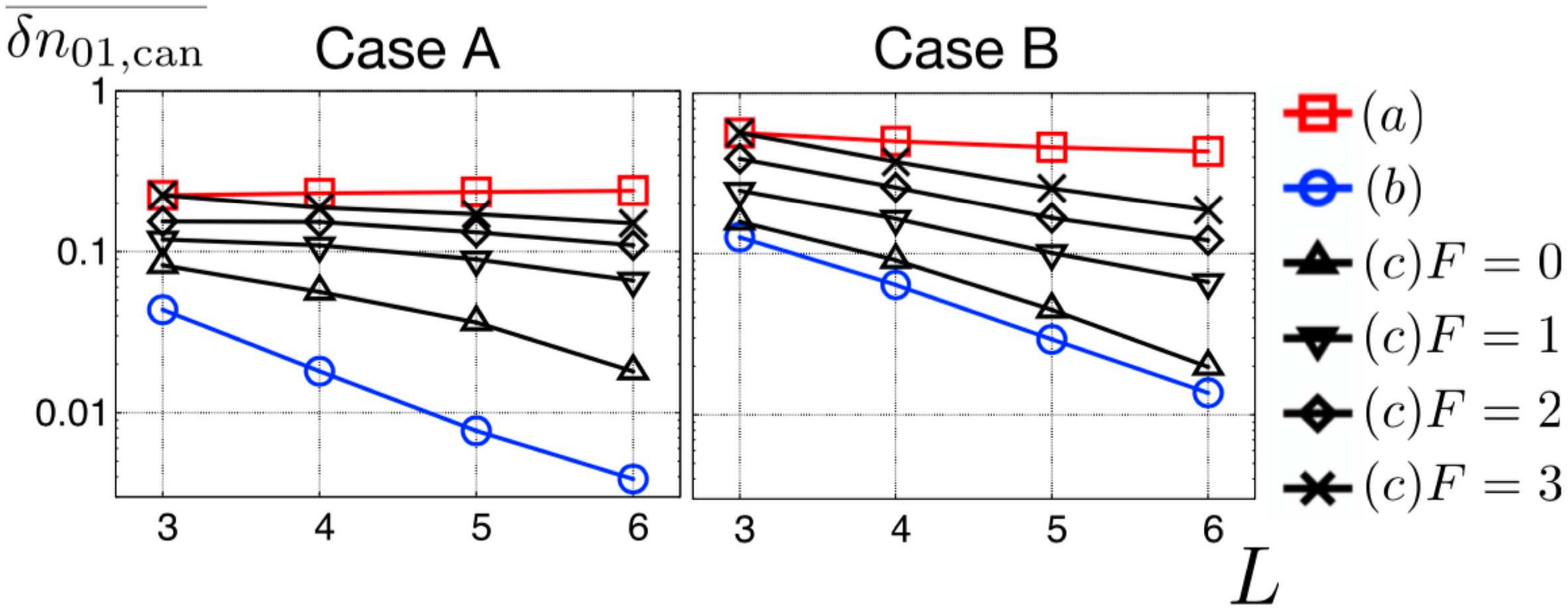}
\caption{Relative difference of the canonical ensemble for $\hat{n}_{01}$ compared with the diagonal ensemble in model (a) (square), model (b) (circle), and model (c) with $F=$0 (triangle), 1 (downward triangle), 2 (diamond) and 3 (cross). The left and right figures respectively show the result for  Case A and Case B.
The relative difference decreases with increasing the system size for both models (b) and (c).
We note that the decrease is rather slow for (c) with $F\geq 1$ (see Appendix \ref{sec:gethom}).
Reproduced from Fig. 7 of~\cite{Hamazaki16G}. \copyright 2016 American Physical Society.
}
\label{bc0can}
\end{center}
\end{figure}

Figure \ref{bc0can} shows the validity of the canonical ensemble in the models (b) and (c) by showing the $L$-dependence of $\overline{\delta n_{01,\mathrm{can}}}$.
In the models (b) and (c) with $F=0$, $\overline{\delta n_{01,\mathrm{can}}}$ rapidly decreases with increasing the system size down to about one tenth  compared with (a) at $L=6$.
These results justify use of the canonical ensemble in these models.
In the models (c), the $L$-dependence is much less evident for $F\geq 1$ than $F=0$.
Nevertheless, $\overline{\delta n_{01,\mathrm{can}}}$ decreases even for $F=3$, which again implies the validity of the canonical ensemble.
Similar results are obtained for other macroscopic observables such as $\overline{\delta n_{00,\mathrm{can}}}$ and $\overline{\delta n_{11,\mathrm{can}}}$.
We attribute these results to the usual ETH, which holds weakly for $F\geq 1$ (see Appendix \ref{sec:gethom}).

Figure \ref{waruyo} illustrates the $F$-dependence of $\overline{\delta n_{\mathrm{can}}}$ with $L=6$.
The figure shows that the canonical ensemble works better when the value of $F$ (or equivalently, $F/L$) is smaller.
This result implies that the expectation values of macroscopic observables in the stationary state can be  predicted by the canonical ensemble if the number of symmetries are much less than the system size.

\begin{figure}
\begin{center}
\includegraphics[width=11cm,angle=-90]{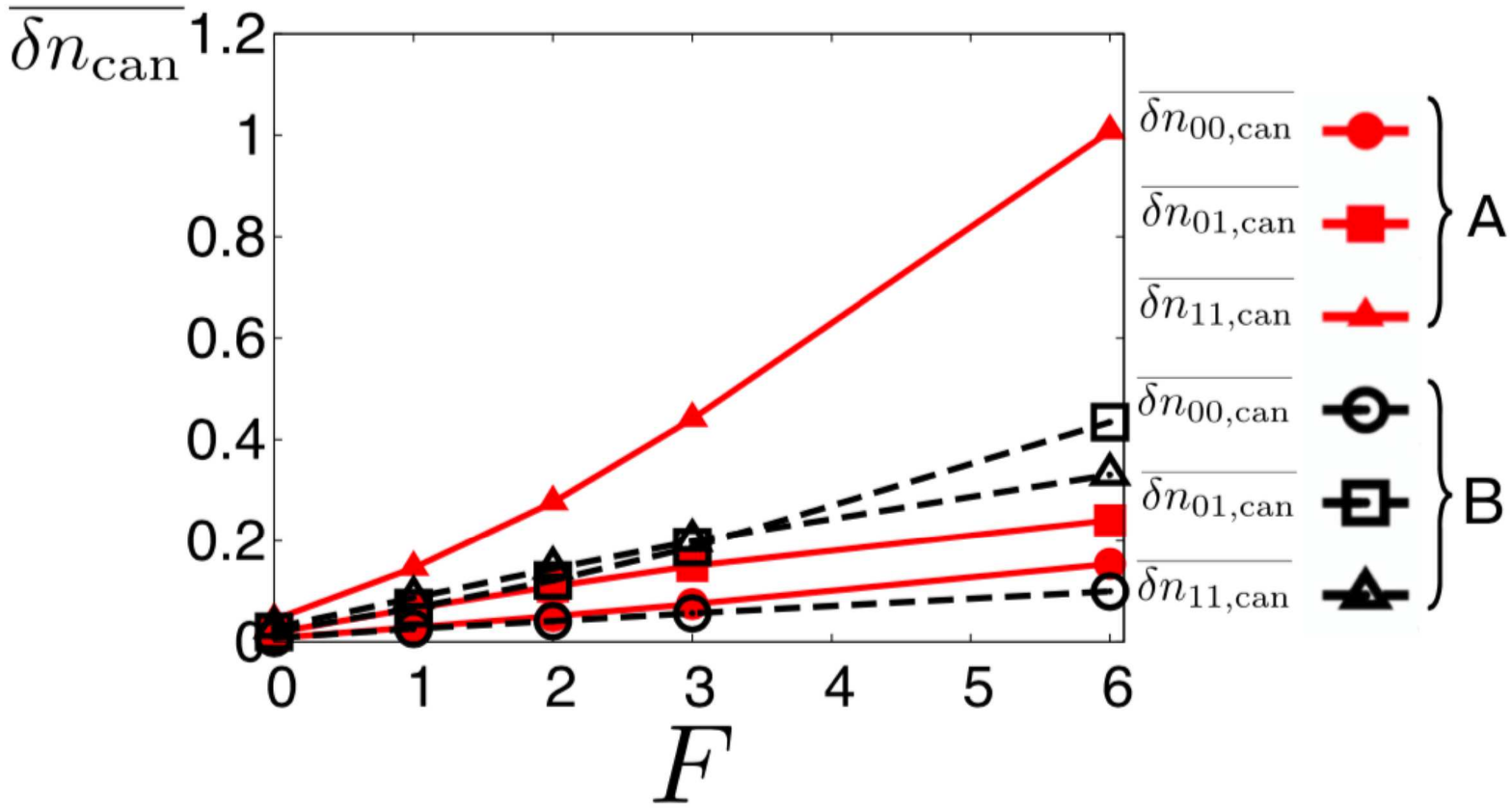}
\caption{$F$-dependence of $\overline{\delta n_{\mathrm{can}}}$ for $\hat{n}_{00}\text{ (circle)},\hat{n}_{01}\text{ (square)}$ and $\hat{n}_{11}\text{ (triangle)}$ with $L=6$. 
Here $F=0,1,2,3$ are the results for the models (c) and $F=6$ is the result for model (a).
We show both  Case A (filled) and Case B (open).
The relative difference of the canonical ensemble decreases with decreasing $F$.
Reproduced from Fig. 8 of~\cite{Hamazaki16G}. \copyright 2016 American Physical Society.
}
\label{waruyo}
\end{center}
\end{figure}

\section{Conclusions and discussions}
Let us summarize this chapter and make some discussions.
We have shown that stationary states for the nonintegrable model with an extensive number of local $\mathbb{Z}_2$ symmetries (Fig. \ref{bose}) can be described by the GGE and not by the canonical ensemble.
We find that the ETH breaks down for the entire spectrum, but it holds true for each symmetry sector.
We have discussed that this restricted ETH leads to the GGE if we neglect multiple correlations among local conserved quantities.
Next, by studying the models with only one global $\mathbb{Z}_2$ symmetry or the $L$-independent number of local $\mathbb{Z}_2$ symmetries,
we find that the canonical ensemble works well for predicting the expectation values of the macroscopic observables in these models.
Our results have clarified that the GGE is necessary to describe stationary states if the system has an extensive number of local symmetries, even if they do not label every energy eigenstate. 

We have several open problems about the relation between our GGE and stationary states.
First, the initial states that we have used are almost homogeneous over different layers, which makes $\lambda_l$'s in \EQ{GGE} close to one another.
It is of interest to investigate whether \EQ{GGE} is valid even for inhomogeneous initial states.
Second, model (a) has an extensive number of the {\it most} local conserved quantities $\hat{P}_l$, from which we can construct the GGE that describes the observables defined in each layer.
On the other hand, {\it in total}, this model has more than extensive number of the local conserved quantities such as  $\hat{P}_1\hat{P}_2$, which may affect the expectation values of less local observables.
Therefore, it is an open problem how far we can truncate the rGGE to predict the expectation values of given observables in stationary states.
The third problem is to clarify how many symmetries are enough to prevent macroscopic observables from relaxing to the stationary states that can be described by the canonical ensemble.
In other words, what will stationary state be when the system has local symmetries which increase with increasing $L$ in a non-extensive manner?
Since $L$ increases much faster than the number of local symmetries in this case, this question cannot be answered by our exact diagonalization analysis.
These questions are left for the future investigation.

\chapter{Conclusions and Future prospects}\label{ch:Con}

\section{Conclusions}
In this thesis, we have investigated how and when isolated quantum systems approach thermal equilibrium with an emphasis on the nonintegrability of systems.
Previous studies have strongly indicated that nonintegrable systems that conserve only energy approach stationary states that are described by the (micro)canonical ensemble.
The eigenstate thermalization hypothesis (ETH) is one possible candidate that justifies thermalization in isolated quantum systems. 
However, the rigorous proofs and definite criteria for the validity of the ETH are far from trivial.
Thus, for understanding the mechanisms of the ETH, it is important to provide clues by possible analytical explanations and numerical simulations.

In the first part of our work, we have shown that random matrix theory can predict the ETH and its finite-size corrections (within some energy shell) in nonintegrable systems and for a wide class of observables.
We have first refined and generalized the RMT predictions to investigate  finite-size corrections of the ETH.
We have especially focused on two types of quantities of matrix elements:
one is the ratios between standard deviations of diagonal and off-diagonal matrix elements, 
and the other is the probability densities of the off-diagonal matrix elements.
The RMT predicts that these quantities have universal features that depend on the anti-unitary symmetries of Hamiltonians, the anti-unitary symmetries of observables, and the ``singularities" of observables.\footnote{Here singular observables are those that do not satisfy \EQ{norio}.}
Next, we have numerically investigated matrix-element statistics of various observables in nonintegrable systems that only conserve energy. 
We have demonstrated that the finite-size corrections of the ETH are in excellent agreement with the predictions of RMT for a wide class of observables with various symmetries, including many-body correlations and singular operators.
We have also remarked, however, that counterexamples can always be constructed even among simple observables.

Nonintegrable systems with additional conserved quantities have  been investigated much less.
It is expected that stationary states are effectively described by the GGE in integrable systems, which have conserved quantities that determine every energy eigenstate.
Then, it is of interest whether nonintegrable systems relax to non-thermal stationary states (possibly described by the GGE) if they have additional conserved quantities due to symmetries.

In the second part of our work, we have shown that stationary states for a nonintegrable model with an extensive number of local $\mbb{Z}_2$ symmetries can effectively be described by the GGE and not by the canonical ensemble.
For this model, the ETH holds true only for each symmetry sector instead of the entire spectrum.
We have discussed that this restricted ETH leads to the GGE if we neglect multiple correlations among local conserved quantities.
We have also studied the models with only one global $\mbb{Z}_2$ symmetry or the $L$-independent number of local $\mbb{Z}_2$ symmetries.
We find that the canonical ensemble works well for predicting the expectation values of the macroscopic observables in these models.
Our results have clarified that the GGE is necessary to describe stationary states in the presence of an extensive number of local symmetries, even if they do not label each energy eigenstate.

\section{Future prospects}
Before concluding this thesis, we would like to discuss several future prospects.

First, we believe that the success of the RMT model in nonintegrable systems for various operators will enable us to investigate finite-size corrections of quantum statistical mechanics.
Although we have considered only matrix elements of observables in this thesis, we expect that our method can  similarly be applied to more directly observable quantities in small nonintegrable quantum systems such as temporal fluctuations after a quench which have in fact experimentally been observed~\cite{Kaufman16}.
The foundations of statistical mechanics in such systems may be important for the basics of quantum thermodynamics in small systems~\cite{Goold16}.
We also expect that properly applying RMT might reveal nontrivial aspects of nonequilibrium dynamics, which has attracted a lot of attention from high-energy physics to condensed matter physics~\cite{Polkovnikov11,Eisert15,Calabrese06,Maldacena16,Kormos16}.

Second, we wonder how the notion of the GGE can be developed and applied to characterize nonequilibrium  dynamics in macroscopic systems.
It is known that if prethermalization occurs due to approximate conserved quantities, the prethermalized plateau is well described by the GGE constructed from these quantities~\cite{Kollar11}.
Recent studies have also shown that the GGE can be applied to describe nonequilibrium stationary states~\cite{Bhaseen15,Bertini16D,Bertini16T} and ``generalized hydrodynamics"~\cite{Castro16} in integrable systems.
Our work has suggested that these works may be generalized to systems with sufficiently many conserved quantities irrespective of integrability.
We expect that by properly choosing approximately conserved quantities, nonequilibrium dynamics can be captured even in nonintegrable systems.

\appendix
%
\chapter{Review of random matrix theory (RMT)}\label{ch:RMT}
In this appendix we give a brief review of random matrix theory (RMT).
We first briefly review the history of RMT.
Next, we explain the general definitions and classifications of the Gaussian random matrices.
Then we explain the statistics that universally emerges in RMT, as a supplement to the main text.
Finally, we review the notion of the ergodicity of random matrices.
More detailed calculations and miscellaneous topics are given in Refs.~\cite{Haake,Mehta,Brody81,Beenakker97,Guhr98,Beenakker15,Borgonovi16}.

\section{History}
Random matrix theory (RMT) was first applied to physics in 1951~\cite{Wigner51} by Eugene P. Wigner, who investigated the excitation spectrum of nuclei.
He conjectured that statistical properties of the eigenvalues of complex nuclei can be described by the eigenvalues of randomly generated matrices.
After Wigner's seminal work, Dyson published a series of  papers~\cite{Dyson62S1,Dyson62S2,Dyson62S3,Dyson62T,Dyson62B} on the mathematical formulations and generalizations of random matrices in 1962.
Of particular importance is what is called the ``\textit{threefold way}," the classification of random matrices in terms of an anti-unitary symmetry operator that commutes with the Hamiltonian.
He also formulated other important concepts, such as the circular ensembles or the Brownian motion of energy levels~\cite{Haake}.

After its basics were established by Dyson, RMT was actively applied in the field of nuclear physics in the 1960's and the  1970's.
On one hand, experimentalists verified that the level fluctuations in the spectrum of nuclei coincide with RMT prediction.
On the other hand, theorists made efforts to develop RMT for a better description of the experiments.
For example, they introduced the notion of the random S matrices and the embedded random ensembles.
Although the applications of RMT were mainly made in  nuclear physics in these decades, we remark that RMT was also applied to the field of ecological systems~\cite{May72}.

RMT greatly developed both in foundations and applications in the 1980's and the 1990's.
One important finding was the connection between RMT and mesoscopic physics in quantum transport phenomena (including the theory of the universal conductance fluctuations) or disordered systems (including the Anderson localization transition).
Another notable application was made in the field of  quantum chaos, where it was conjectured that the spectrum of a quantum Sinai billiard and the spectra of more general quantum chaotic systems are described by those of random matrices (i.e., the Bohigas-Giannoni-Schmit conjecture).
As a mathematical development, the technique of the supersymmetry was introduced, which enabled one to easily calculate the propagator of a random Hamiltonian.
We also remark that the classification of random matrices in terms of symmetries was enlarged by Altland and Zirnbauer in 1997~\cite{Altland97} (known as the ``\textit{tenfold way}").

RMT has further been utilized in describing various fields of physics recently.
For example, distributions of height fluctuations that appear in the Kardar-Parisi-Zhang equations~\cite{Kardar86} have turned out to be described by the Tracy-Widom distributions\footnote{The Tracy-Widom distribution is a distribution of the maximum eigenvalue of the random matrix.} of RMT~\cite{Prahofer00}.
Topological insulators and superconductors are classified and investigated using the Altland-Zirnbauer tenfold way~\cite{Schnyder08}.
Last but not least, RMT is closely related to the eigenstate thermalization hypothesis (ETH) in isolated quantum systems~\cite{DAlessio16} (see Chapter \ref{ch:ETH}).
It is expected that the complexity of the energy eigenstates of nonintegrable systems are modeled by RMT.
RMT is also expected to be related to the transition between the delocalized phase (where the ETH holds true) and the many-body localized phase (where the ETH does not hold)~\cite{Serbyn16S}.

\section{Definitions and classifications}
\subsection{Gaussian ensembles}
Let $\hH$ be a $D\times D$ random matrix which is chosen from the probability $P(\hH)[d\hH]$, where $[d\hH]$ is the volume element of $d\hH$.
In the Gaussian ensembles, each matrix element $H_{ij}$ is independent and identically follows a Gaussian distribution.
Moreover, $P(\hH)$ is invariant under certain symmetry transformations, which we will explain below.

\subsubsection{Threefold way by Dyson}
Dyson classified Gaussian random matrices in terms of an anti-unitary operator $\hT$ that commutes with $\hH$.
If there exists no such $\hT$, the ensemble is called the Gaussian unitary ensemble (GUE) because it is invariant under any $D\times D$ unitary matrix.
If there exists $\hT$ that satisfies $\hT^2=1$, the ensemble is called the Gaussian orthogonal ensemble (GOE), since it is invariant under any $D\times D$ orthogonal matrix.
Finally, if $\hT^2=-1$, the ensemble is called the Gaussian symplectic ensemble (GSE), which is invariant under any symplectic transformation.

Firstly, matrices in the GUE have $D^2$ independent degrees of freedom.
The probability measure for the GUE can be written as follows:
\aln{
P(\hH)[d\hH]=c_2\exp\lrl{-\sum_iH_{ii}^2-2\sum_{i>j}|H_{ij}|^2}\prod_idH_{ii}\prod_{i>j}d^2H_{ij},
}
where $d^2H_{ij}=d\mr{Re}[H_{ij}]d\mr{Im}[H_{ij}]$ and $c_2$ is a normalization constant.

Secondly, matrices in the GOE have $D(D+1)/2$ independent degrees of freedom because each element can be taken as real variables (i.e., $H_{ij}=H_{ij}^*=H_{ji}$).
The probability measure for the GOE can be written as follows:
\aln{
P(\hH)[d\hH]=c_1\exp\lrl{-\frac{1}{2}\sum_iH_{ii}^2-\sum_{i>j}H_{ij}^2}\prod_idH_{ii}\prod_{i>j}dH_{ij},
}
where $c_1$ is a normalization constant.

Finally, matrices in the GSE are written in terms of the quaternion notation as
\aln{
\hH=\hat{h}^{(0)}\otimes \hat{\mbb{I}}_{2\times 2}-i\sum_{\gamma=1}^3\hat{h}^{(\gamma)}\otimes \hat{\sigma}^{(\gamma)},
}
where $\hat{\sigma}^{(\gamma)}$'s are the Pauli matrices and $\hat{h}^{(\mu)}$'s $\:(\mu=0,1,2,3)$ are $(D/2)\times(D/2)$ matrices (we assume that $D$ is even in this case).
Let us assume that $\hT$ is the time-reversal operator (similar discussions can be applied to other anti-unitary symmetries).
Then the $\hT$-invariance and the Hermiticity lead to the conditions $h_{nm}^{(\mu)}=h_{nm}^{(\mu)*}\:(\mu=0,1,2,3)$, $h_{mn}^{(0)}=h_{nm}^{(0)}$ and $h_{mn}^{(\gamma)}=-h_{nm}^{(\gamma)}\:(\gamma=1,2,3)$; we thus have $D(D-1)/2$ independent variables.
The probability measure for the GSE can be written as follows:
\aln{
P(\hH)[d\hH]=c_4\exp\lrl{-2\sum_n(h_{nn}^{(0)})^2-4\sum_{n>m}\sum_{\mu=0}^3(h_{nm}^{(\mu)})^2}\prod_{n=1}^{D/2}dh^{(0)}_{nn}\prod_{n>m}\prod_{\mu=0}^3 dh_{nm}^{(\mu)},
}
where $c_4$ is a normalization constant.
Note that we have used the notation that enables us to write down 
\aln{
P(\hH)=c_\beta\exp\lrl{-\frac{\beta}{2} \Tr \hH^2}
}
for all symmetry classes ($\beta=1,2,$ and 4 for the GOE, the GUE, and the GSE, respectively).

Next, we consider the distributions of eigenvalues and eigenvectors of random matrices.
Let us first consider the case with the GUE.
In this case the Hamiltonian can be diagonalized by a unitary matrix $U$ as
\aln{
\hH=UEU^\dag
}
where $E=\mr{diag}(x_1,\cdots,x_D)$ represents a set of the eigenvalues.
By taking the derivative, we obtain
\aln{
d\hH&=dUEU^\dag+UEdU^\dag+UdEU^\dag,
}
and then
\aln{
U^\dag d\hH U&=U^\dag dUE+EdU^\dag U+dE\NON
&=U^\dag dUE-EU^\dag dU+dE,
}
where we have used $dU^\dag U=-U^\dag dU$.
Each matrix element on the right-hand side of this equation can be written down as
\aln{
(U^\dag dUE-EU^\dag dU+dE)_{ij}=
 \begin{cases}
    dx_{ii} & (i=j),\\
    (x_j-x_i)(U^\dag dU)_{ij} & (i>j).
  \end{cases}
}
Since $[d\hH]$ is the invariant measure of the unitary transformation, namely $[U^\dag d\hH U]=[d\hH]$ (for a proof, see Ref.~\cite{西垣}),
we obtain
\aln{
[d\hH]=\prod_{i>j}|x_i-x_j|^2\prod_idx_i\prod_{i>j}\mr{Re}(U^\dag dU)_{ij}\mr{Im}(U^\dag dU)_{ij}.
}
Here we note that the complex nature of the matrix elements leads to the factor $|x_i-x_j|^2$, which represents the quadratic level repulsion.
In fact, by integrating out the variables for eigenvectors and taking the Gaussian weight, we obtain the eigenvalue distributions for the GUE as follows:
\aln{
p(x_1,\cdots,x_D)\propto \prod_{i>j}|x_i-x_j|^2\prod_i e^{-x_i^2}.
}
The eigenvector part, $\prod_{i>j}\mr{Re}(U^\dag dU)_{ij}\mr{Im}(U^\dag dU)_{ij}$, is invariant under an arbitrary unitary transformation, so is the Haar measure on the unitary group.
Similar considerations hold true for the case with the GOE and the GSE.
For the GOE, we find
\aln{
[d\hH]=\prod_{i>j}|x_i-x_j|\prod_idx_i\prod_{i>j}(O^T dO)_{ij},
}
where $\prod_{i>j}(O^T dO)_{ij}$ is the Haar measure on the orthogonal group.
We can obtain the result for the GSE, too, and finally obtain the unified formula for the distributions of eigenvalues as
\aln{\label{betabeta}
p(x_1,\cdots,x_D)\propto \prod_{i>j}|x_i-x_j|^\beta\prod_i e^{-\frac{\beta}{2}x_i^2}.
}

\subsubsection{Tenfold way by Altland and Zirnbauer}
Do we have universality classes other than the GUE, the GOE, and the GSE?
In fact, in the context of the QCD, the so-called chiral random matrix ensembles (the chGUE, the chGOE, and the chGSE) were found.
Matrices in these ensembles have chiral symmetries: the corresponding operator is unitary and anti-commutes with the Hamiltonians.
Altland and Zirnbauer then added four more classes and completed the ten classes focusing on the role of symmetries.
These classes are distinguished by an anti-unitary symmetry operator $\hat{T}$ that commutes with the Hamiltonian (which we call a time-reversal symmetry in this subsection), 
an anti-unitary symmetry operator $\hat{\Pi}$ that anti-commutes with the Hamiltonian (a particle-hole symmetry), and
a unitary symmetry operator $\hat{C}$ that anti-commutes with the Hamiltonian (a chiral/sublattice symmetry).

\begin{figure}
\begin{center}
\includegraphics[width=11cm,angle=-90]{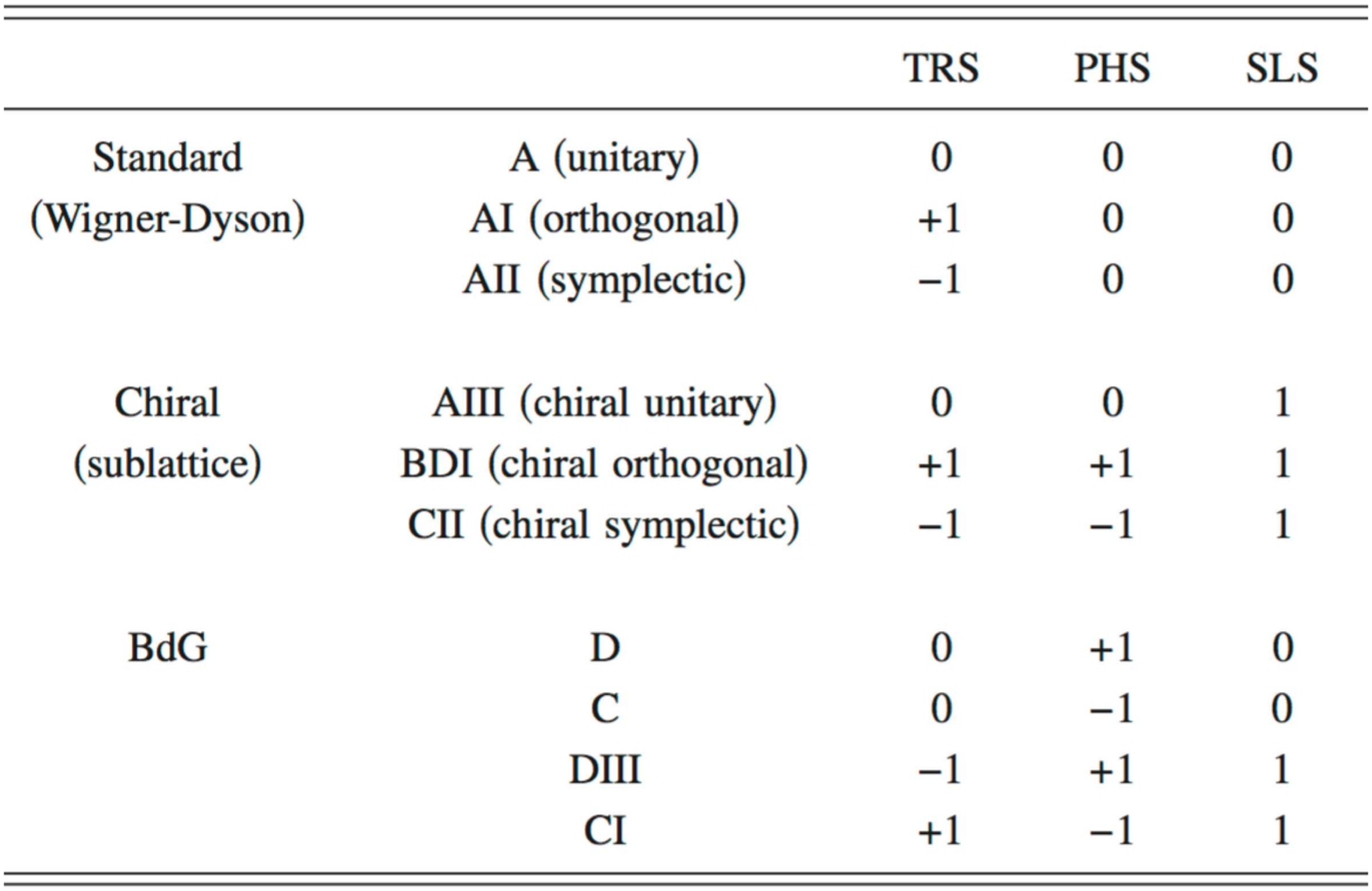}
\caption{
Classification of random matrices by the time-reversal symmetry (TRS), particle-hole symmetry (PHS), and sublattice (chiral) symmetry (SLS).
The numbers $\pm 1$ in the table represents the values of the square of the symmetry operators ($\hat{T}^2,\hat{\Pi}^2$ and $\hat{C}^2$); if no such symmetry exists, the number is 0.
Reproduced from Fig. 1 of~\cite{Schnyder08}. \copyright 2008 American Physical Society.
}
\label{univtop}
\end{center}
\end{figure}

In Fig. \ref{univtop}, we show the ten classifications with respect to these symmetries.
Here, we assume that the values of $\hat{T}^2$ and  $\hat{\Pi}^2$ will be either $+1$ or $-1$ (times the identity ooperator).
We note that we do not have to consider the case where two symmetries of the same type exist.
For example, if two anti-unitary symmetry operators $\hat{T}_1$ and $\hat{T}_2$ exist, $\hat{T}_1\hat{T}_2$ also becomes a symmetry operator that commutes with the Hamiltonian.
Since $\hat{T}_1\hat{T}_2$ is unitary, we can decompose the Hamiltonian into irreducible blocks by this symmetry and treat these individual blocks again.
This discussion can be used for $\hat{\Pi}$ and $\hat{C}$ as well.
Similarly, if $\hat{\Pi}$ and $\hat{T}$ exist, $\hat{T}\hat{\Pi}$ becomes a unitary symmetry operator that anti-commutes with the Hamiltonian, which plays the role of $\hat{C}$.
Thus, the presence of the time-reversal and particle-hole symmetry necessarily leads to the chiral symmetry.
We also note that these classes are called Class A, AI, ..., CI, which are adapted from the mathematical terminology due to \'{E}lie Cartan.

\section{Statistics in Gaussian random matrices}
In this section, we explain some details of the statistics of RMT to complement the main text.

\subsection{Level-spacing statistics}
As we have seen in Chapter \ref{ch:ETH}, the level-spacing distributions of the random matrix are  approximately described by \EQ{stat}.
In fact, this is the result obtained from \EQ{betabeta} with $D=2$.
We can write the level-spacing distribution as
\aln{
P(s)=C_\beta\int_{-\infty}^\infty\int_{-\infty}^\infty dx'_1dx'_2 |x'_1-x'_2|^\beta e^{-A_\beta({x_1'}^2+{x_2'}^2)}\delta(s-|x_1'-x_2'|),
}
where $x_1'$ and $x_2'$ are renormalized levels.
From the normalization conditions $\int ds P(s)=\int ds sP(s)=1$, we can determine $C_\beta$ and $A_\beta$, which results in \EQ{stat}.
This is the result for $N=2$, but it is known that it can approximate the level-spacing distribution for $N\gg 1$ as well~\cite{Haake,Mehta}.

\subsection{Distributions of eigenstates}
Let us consider a single eigenstate $\ket{E_\alpha}$ and its components with respect to a fixed basis set $\{\ket{a_i}\}\:(1\leq i\leq d)$.
In the case of the GUE, the joint probability of finding $z_1=\mr{Re}[\braket{a_1|E_\alpha}],z_2=\mr{Im}[\braket{a_1|E_\alpha}],\cdots, z_{2d-1}=\mr{Re}[\braket{a_d|E_\alpha}],z_{2d}=\mr{Im}[\braket{a_d|E_\alpha}]$ is
\aln{
P_\mr{GUE}^{(2d)}(z_1,\cdots,z_{2d})\propto\delta \lrs{1-\sum_{n=1}^{2d}z_n^2}.
}
By integrating out $z_{2l+1},\cdots,z_{2d}$, we obtain the marginal distribution of $z_1,\cdots,z_{2l}$ as
\aln{
P^{(2d,2l)}_\mr{GUE}(z_1,\cdots,z_{2l})=\pi^{-l}\frac{\Gamma(d)}{\Gamma(d-l)}\lrs{1-\sum_{n=1}^{2l} z_n^2}^{d-1-l}.
}
In particular, we can take $l=1$ and find the distribution of $y=z_1^2+z_2^2=|\braket{a_1|E_\alpha}|^2$ as
\aln{
\rho_{|\braket{a_1|E_\alpha}|^2}(y)=de^{-dy}
}
in the large-$d$ limit.
This distribution is called the Porter-Thomas distribution.

Similarly, for the case with GOE, we can consider the joint probability of finding 
$z_1=\braket{a_1|E_\alpha},\cdots,z_{d}=\braket{a_d|E_\alpha}$ as\footnote{We take the basis set such that each $z_l$ becomes real.}
\aln{
P_\mr{GOE}^{(d)}(z_1,\cdots,z_{d})\propto\delta \lrs{1-\sum_{n=1}^{d}z_n^2}.
}
Consequently, we have
\aln{
P^{(d,l)}_\mr{GOE}(z_1,\cdots,z_{l})=\pi^{-l/2}\frac{\Gamma(d/2)}{\Gamma(d/2-l/2)}\lrs{1-\sum_{n=1}^{l} z_n^2}^{d/2-1-l/2}
}
and the Porter-Thomas distribution (the probability of finding $y=z_1^2=|\braket{a_1|E_\alpha}|^2$) for a large $d$ as
\aln{
\rho_{|\braket{a_1|E_\alpha}|^2}(y)=\sqrt{\frac{d}{2\pi y}}e^{-dy/2}.
}
We briefly consider the eigenvector statistics for the GSE in Appendix \ref{sec:gseE}.

\section{Ergodicity of Gaussian random matrices}\label{sec:ergo}
In this section, we review the so-called ``ergodicity" of random matrices, which connects the spectral average and the ensemble average.
We first consider the general framework and then prove the ergodicity for the second moments of off-diagonal matrix elements.

The ergodicity of random matrices means that, for most of the fixed Hamiltonians that are sampled from certain random ensemble, the spectral average is approximated by the ensemble average.\footnote{As we have mentioned in Chapter \ref{ch:ETH}, the ergodicity of random matrices is different from the usual terminology of  the ergodicity of dynamical systems, which states that the phase-space average is equal to the long-time average.}
Let us begin with a function $g_\alpha$, which depends on a single energy eigenstate $\ket{E_\alpha}$ (or its eigenvalue $E_\alpha$). 
We can consider the spectral average of $g_\alpha$
\aln{
\braket{g_\alpha}_\mc{T}:=\frac{1}{d_s}\sum_{\alpha\in\mc{T}}g_\alpha,
}
where $\mc{T}$ denotes a set of the labels of the samplings and $d_s$ is the number of the samplings.
We can also consider the ensemble average as
\aln{
\av{g_\alpha}:=\int g_\alpha P(\hH)d\hH.
}
We note that the ensemble average is often easy to calculate analytically by RMT.

To prove the ergodicity, we first have to show 
\aln{\label{averen}
\av{\braket{g_\alpha}_\mc{T}}=\av{g_\alpha},
}
which is valid if $\av{g_\alpha}$ is constant in $\alpha\in\mc{T}$.
Next, we need to show
\aln{\label{varren}
\av{\braket{g_\alpha}_\mc{T}^2}-\av{\braket{g_\alpha}_\mc{T}}^2\ra 0
}
for $d,d_s\ra\infty$.
If Eqs. (\ref{averen}) and (\ref{varren}) are satisfied in this limit, we have
\aln{
\braket{g_\alpha}_\mc{T}\simeq\av{g_\alpha}
}
for most of the Hamiltonians in the random ensemble.

We prove the ergodicity for the second moments of the off-diagonal matrix elements.
For simplicity, we consider a nonsingular observable $\hat{O}$ and $d\times d$ random matrices in the GUE.
Since we have seen that $|O_{\alpha\beta}|^2 \sim d^{-1}$ in the main text, we define
$g_{\alpha\beta}=d|O_{\alpha\beta}|^2\:(\alpha\neq\beta)$ to get a nontrivial result.
In this case, we define the spectral average
\aln{
\braket{g_{\alpha\beta}}_{\mc{T}\mc{T}}=\frac{1}{d_s(d_s-1)}\sum_{\alpha,\beta\in\mc{T};\alpha\neq\beta}g_{\alpha\beta},
}
where we assume that there is no degeneracy in the spectrum.

First, \EQ{averen} is valid because
\aln{
\av{\braket{g_{\alpha\beta}}_{\mc{T}\mc{T}}}=\frac{1}{d_s(d_s-1)}\sum_{\alpha,\beta\in\mc{T};\alpha\neq\beta}\av{g_{\alpha\beta}}=\av{g_{\alpha\beta}},
}
where 
\aln{
\av{g_{\alpha\beta}}=\frac{d^2}{d^2-1}\lrl{\frac{1}{d}\sum_ia_i^2-\lrs{\frac{1}{d}\sum_ia_i}^2}\ra\frac{1}{d}\sum_ia_i^2-\lrs{\frac{1}{d}\sum_ia_i}^2
}
for a large $d$.

Next, for \EQ{varren}, we have
\aln{\label{uho}
\av{\braket{g_{\alpha\beta}}_{\mc{T}\mc{T}}^2}-\av{\braket{g_{\alpha\beta}}_{\mc{T}\mc{T}}}^2=&
\frac{1}{d_s^2(d_s-1)^2}\sum_{\alpha,\beta\in\mc{T};\alpha\neq\beta}\sum_{\gamma,\delta\in\mc{T};\gamma\neq\delta}\lrs{\av{g_{\alpha\beta}g_{\gamma\delta}}-\av{g_{\alpha\beta}}\:\av{g_{\gamma\delta}}}\NON
=&
\frac{2}{d_s(d_s-1)}\lrs{\av{g_{\alpha\beta}^2}-\av{g_{\alpha\beta}}^2}+\frac{4(d_s-2)}{d_s(d_s-1)}\lrs{\av{g_{\alpha\beta}g_{\alpha\beta'}}-\av{g_{\alpha\beta}}\:\av{g_{\alpha\beta'}}}\NON
&+\frac{(d_s-2)(d_s-3)}{d_s(d_s-1)}\lrs{\av{g_{\alpha\beta}g_{\alpha'\beta'}}-\av{g_{\alpha\beta}}\:\av{g_{\alpha'\beta'}}},
}
where $\alpha\neq\alpha'$ and $\beta\neq\beta'$.
Since $\lrs{\av{g_{\alpha\beta}g_{\gamma\delta}}-\av{g_{\alpha\beta}}\:\av{g_{\gamma\delta}}}$ is at most of order one, the first and second terms in the final expression of \EQ{uho} vanish when $d_s$ is large.
Moreover, when $d$ is large, the correlation between $g_{\alpha\beta}$ and $g_{\alpha'\beta'}$ is expected to vanish: $\av{g_{\alpha\beta}g_{\alpha'\beta'}}=\av{g_{\alpha\beta}}\:\av{g_{\alpha'\beta'}}+\mr{o}(1)$~\cite{Brody81}.\footnote{To show this, we apply the method in Subsection \ref{sec:distd}: we first move $\ket{E_\alpha}$ in the $(d-3)$-dimensional Hilbert space that is orthogonal to $\ket{E_\beta},\ket{E_{\alpha'}}$, and $\ket{E_{\beta'}}$.
Then, by moving $\ket{E_\beta}$ in the $(d-2)$-dimensional Hilbert space that is orthogonal to $\ket{E_{\alpha'}}$ and $\ket{E_{\beta'}}$, we obtain the result.
}
Then \EQ{varren} holds true in the limit $d,d_s\ra\infty$.
We expect that the ergodicity holds true similarly for higher moments  and distributions of diagonal and off-diagonal matrix elements in other classes of random matrices.
Some other aspects of the ergodicity (e.g., level densities and level-spacing distributions) are reviewed in Ref.~\cite{Brody81}.

\chapter{Detailed derivations in the main text}\label{ch:Det}

\section{Derivation of Eq. (\ref{typMITE})}\label{sec:typMITE}
We follow the proof of Refs.~\cite{杉田06,IkedaD} for finite-dimensional lattice systems.
Let us denote the basis set of operators in the subsystem $S$ by $\{\hat{A}_l\}_{l=1}^{d_S^2}$, where $d_S:=\dim[\mc{H}_S]$. 
We can assume the orthonormality condition
\aln{
\Tr_S[\hat{A}_l\hat{A}_k]=d_S\delta_{lk}\hat{\mbb{I}}_S.
}
A given operator $\hrho$ on $S$ can be expanded as
\aln{
\hrho=\frac{1}{d_S}\sum_l\Tr_S[\hrho\hat{A}_l]\hat{A}_l.
}
Then we have
\aln{
||\hrho_S-\hrho_{\mr{mic},S}||_\mr{op}^2&\leq ||\hrho_S-\hrho_{\mr{mic},S}||_{F}^2\NON
&=\Tr_S[(\hrho_S-\hrho_{\mr{mic},S})^2]\NON
&=\frac{1}{d_S}\sum_l\lrm{\Tr_S[(\hrho_S-\hrho_{\mr{mic},S})\hat{A}_l]}^2\NON
&=\frac{1}{d_S}\sum_l\lrm{\braket{\psi|\hat{A}_l\otimes \hat{\mbb{I}}_{S^c}|\psi}-\mbb{E}[\braket{\psi|\hat{A}_l\otimes \hat{\mbb{I}}_{S^c}|\psi}]}^2,
}
where $||\hrho||_F:=\sqrt{\Tr[\hrho^2]}$ is the Frobenius norm and we have used the relation $||\hrho||_\mr{op}\leq ||\hrho||_F$.
Therefore,
\aln{
\mbb{E}\lrl{||\hrho_S-\hrho_{\mr{mic},S}||_\mr{op}^2}&\leq
\frac{1}{d_S}\sum_l\mbb{V}\lrl{\braket{\psi|\hat{A}_l\otimes \hat{\mbb{I}}_{S^c}|\psi}}\NON
&\leq\frac{1}{d_S}\sum_l\frac{||\hat{A}_l\otimes \hat{\mbb{I}}_{S^c}||^2_\mr{op}}{d+1}\NON
&\leq\frac{d_S^2}{d+1}.
}
Using Markov's inequality, we obtain \EQ{typMITE}.

\section{Derivation of Eq. (\ref{ukky})}\label{sec:dtood}
We follow the supplement of Ref.~\cite{Reimann15}. First notice an inequality 
\aln{\label{hutou}
\mbb{P}\lrl{\sum_\phi f_\phi \geq\sum_\phi \epsilon_\phi}&=
\mbb{E}\lrl{\theta\lrs{\sum_\phi f_\phi -\sum_\phi \epsilon_\phi}}\NON
&\leq \sum_\phi\mbb{E}\lrl{\theta\lrs{f_\phi -\epsilon_\phi}}=\sum_\phi\mbb{P}\lrl{f_\phi\geq\epsilon_\phi}.
}
Next, for a fixed pair $\rho\neq\sigma$, we define the following four vectors:
\aln{
\ket{\phi_1}&:=\lrs{\ket{\rho}+\ket{\sigma}}/\sqrt{2},\NON
\ket{\phi_2}&:=\lrs{\ket{\rho}-\ket{\sigma}}/\sqrt{2},\NON
\ket{\phi_3}&:=\lrs{\ket{\rho}+i\ket{\sigma}}/\sqrt{2},\NON
\ket{\phi_4}&:=\lrs{\ket{\rho}-i\ket{\sigma}}/\sqrt{2} .
}
Then we find
\aln{
\braket{\phi_1|\hat{O}|\phi_1}-\braket{\phi_2|\hat{O}|\phi_2}-i\braket{\phi_3|\hat{O}|\phi_3}+i\braket{\phi_4|\hat{O}|\phi_4}=2\braket{\sigma|\hat{O}|\rho}
}
and hence
\aln{
\braket{\sigma|\hat{O}|\rho}&=\lrs{\delta_1-\delta_2-i\delta_3+i\delta_4}/2,\\
\delta_i &:=\braket{\phi_i|\hat{O}|\phi_i}-\mbb{E}\lrl{\braket{\phi_i|\hat{O}|\phi_i}}.
}
This leads to $\lrv{\braket{\sigma|\hat{O}|\rho}}\leq \sum_{\phi=1}^4|\delta_\phi|/2$ and thus
\aln{
\mbb{P}\lrl{|\braket{\sigma|\hat{O}|\rho}|\geq \epsilon}\leq \mbb{P}\lrl{\sum_{\phi=1}^4|\delta_\phi|\geq 2\epsilon}\leq
\sum_{\phi=1}^4 \mbb{P}\lrl{|\delta_\phi|\geq \epsilon/2},
}
where we have used Eq. (\ref{hutou}).

\section{Derivation of Eqs. (\ref{gseE}-\ref{gseE3})}\label{sec:gseE}
We first consider the statistics of the eigenstates for the Gaussian symplectic ensemble (GSE).
We expand an eigenstate $\ket{E_\alpha}$ with respect to the symplectic basis set $\ket{a_1},\ket{\tilde{a_1}},\ket{a_2},\ket{\tilde{a_2}},\cdots \ket{a_{d/2}}, \ket{\tilde{a_{d/2}}}$, where $\ket{\tilde{a_{i'}}}=\hT\ket{a_{i'}}\:(i'=1,\cdots,d/2)$.
In this case, the joint (marginal) probability distribution for finding
$z_1=\mr{Re}[\braket{a_1|E_\alpha}],z_2=\mr{Im}[\braket{a_1|E_\alpha}], z_3=\mr{Re}[\braket{\tilde{a_1}|E_\alpha}],z_4=\mr{Im}[\braket{\tilde{a}_1|E_\alpha}],\cdots, z_{2l-1}=\mr{Re}[\braket{\tilde{a_{l/2}}|E_\alpha}],z_{2l}=\mr{Im}[\braket{\tilde{a_{l/2}}|E_\alpha}]$ is\footnote{Precisely speaking, in the case of the GSE, we have room to choose the Kramers pair after we sample a Hamiltonian. In other words, we have the freedom to rotate two degenerate eigenstates as $\ket{a_{i'}}\rightarrow s\ket{a_{i'}}+t\ket{\tilde{a_{i'}}},\ket{\tilde{a_{i'}}}\rightarrow -t^*\ket{a_{i'}}+s^*\ket{\tilde{a_{i'}}}\:\:(|s|^2+|t|^2=1)$. We assume that the random average is invariant under this rotation in the degenerate space.}
\aln{
P^{(2d,2l)}_\mr{GSE}(z_1,\cdots,z_{2l})=\pi^{-l}\frac{\Gamma(d)}{\Gamma(d-l)}\lrs{1-\sum_{n=1}^{2l} z_n^2}^{d-1-l}.
}
This distribution is equivalent to the case with the GUE, from which we obtain moments such as
\aln{
\av{|\braket{E_\alpha|a_{i'}}|^2}&=\av{|\braket{E_\alpha|\tilde{a_{i'}}}|^2}=\frac{1}{d},\NON
\av{|\braket{E_\alpha|a_{i'}}|^4}&=\av{|\braket{E_\alpha|\tilde{a_{i'}}}|^4}=\frac{2}{d(d+1)},\NON
}
and 
\aln{
\av{|\braket{E_\alpha|a_{i'}}|^2|\braket{E_\alpha|a_{j'}}|^2}=\av{|\braket{E_\alpha|a_{i'}}|^2|\braket{E_\alpha|\tilde{a_{i'}}}|^2}=\frac{1}{d(d+1)}.
}

Next, consider the case where two eigenstates ($\ket{E_\alpha},\ket{E_\beta}$ or $\ket{E_\alpha},\ket{\tilde{E_\alpha}}$) are involved.
By switching the roles of $\ket{a_{i'}}$ and $\ket{E_\alpha}$ in the previous results, we get
\aln{
\av{|\braket{E_\alpha|a_{i'}}|^2|\braket{E_\beta|a_{i'}}|^2}=\av{|\braket{E_\alpha|a_{i'}}|^2|\braket{\tilde{E_\alpha}|a_{i'}}|^2}=\frac{1}{d(d+1)}.
}
Moreover, noting that $\braket{E_\alpha|\tilde{a_{i'}}}=-\braket{a_{i'}|\tilde{E_\alpha}}$,
we get
\aln{
\av{|\braket{E_\alpha|a_{i'}}|^2|\braket{E_\beta|\tilde{a_{i'}}}|^2}=\av{|\braket{E_\alpha|a_{i'}}|^2|\braket{\tilde{E_\beta}|{a_{i'}}}|^2}
=\frac{1}{d(d+1)}.
}

We also consider the equality
\aln{
0=\sum_i\braket{E_\alpha|a_i}\braket{a_i|E_\beta}
}
and its random average
\aln{
0=\sum_{i'} \av{\braket{E_\alpha|a_{i'}}\braket{a_{i'}|E_\beta}}+ \av{\braket{E_\alpha|\tilde{a_{i'}}}\braket{\tilde{a_{i'}}|E_\beta}}.
}
From these equations, we obtain 
\aln{
\av{\braket{E_\alpha|a_{i'}}\braket{a_{i'}|E_\beta}}=\av{\braket{E_\alpha|\tilde{a_{i'}}}\braket{\tilde{a_{i'}}|E_\beta}}=0
}
because we assume that the random average is invariant under the rotation in the degenerate space (see the footnote).
Similarly, we get 
\aln{
\av{\braket{E_\alpha|a_{i'}}\braket{a_{i'}|\tilde{E_\alpha}}}=\av{\braket{E_\alpha|\tilde{a_{i'}}}\braket{\tilde{a_{i'}}|\tilde{E_\alpha}}}=0.
}

Finally, we consider the random averages that are related to four different inner products.
Since $\ket{E_\beta}$ and $\ket{\tilde{E_\beta}}$ are not statistically distinct with respect to $\ket{E_\alpha}$, we have
\aln{
\av{\braket{E_\alpha|a_{i'}} \braket{a_{i'}|E_\beta} \braket{\tilde{a_{i'}}|E_\alpha} \braket{E_\beta|\tilde{a_{i'}}}}
=\av{\braket{E_\alpha|a_{i'}} \braket{a_{i'}|\tilde{E_\beta}} \braket{\tilde{a_{i'}}|E_\alpha} \braket{\tilde{E_\beta}|\tilde{a_{i'}}}}.
}
We note that the right-hand side of this equation is
\aln{
-\av{\braket{E_\alpha|a_{i'}} \braket{a_{i'}|E_\beta} \braket{\tilde{a_{i'}}|E_\alpha} \braket{E_\beta|\tilde{a_{i'}}}}.
}
From this, we obtain
\aln{
\av{\braket{E_\alpha|a_{i'}} \braket{a_{i'}|E_\beta} \braket{\tilde{a_{i'}}|E_\alpha} \braket{E_\beta|\tilde{a_{i'}}}}=0.
}
Similarly, we obtain
\aln{
\av{\braket{E_\alpha|a_{i'}} \braket{a_{i'}|\tilde{E_\alpha}} \braket{\tilde{E_\alpha}|a_{j'}} \braket{{a_{j'}}|E_\alpha}}=0
}
for $i'\neq j'$.
Next, we consider
\aln{\label{sikir}
0=&\sum_{ij}\braket{E_\alpha| a_i}\braket{a_i|E_\beta}\braket{a_j|E_\alpha}\braket{E_\beta|a_j}\NON
=&\sum_i |\braket{E_\alpha|a_i}|^2|\braket{E_\beta|a_i}|^2+\sum_{i'} (\braket{E_\alpha|a_{i'}} \braket{a_{i'}|E_\beta} \braket{\tilde{a_{i'}}|E_\alpha} \braket{E_\beta|\tilde{a_{i'}}}+\mr{h.c.})\NON
&+\sum_{i,j;a_i\neq a_j}\braket{E_\alpha|a_i} \braket{a_i|E_\beta} \braket{a_j|E_\alpha} \braket{E_\beta|a_j},
}
where $i=1,2,\cdots,d$ and $\ket{\tilde{a_{i'}}}=\ket{a_{d-i'+1}}$.
The three sums consist of $d, d/2\times 2$, and $d(d-2)$ terms, respectively.
Taking the average of this equation, we get
\aln{
0=d\times \frac{1}{d(d+1)}+d(d-2)\times \av{\braket{E_\alpha|a_i} \braket{a_i|E_\beta} \braket{a_j|E_\alpha} \braket{E_\beta|a_j}}
}
and then
\aln{
\av{\braket{E_\alpha|a_i} \braket{a_i|E_\beta} \braket{a_j|E_\alpha} \braket{E_\beta|a_j}}=-\frac{1}{d(d+1)(d-2)}\:\:(a_i\neq a_j).
}

Using these formula, we calculate the ensemble average and the variance of the matrix elements.
For even observables $\hat{O}\:([\hat{O},\hat{T}]=0)$, the diagonal term can be similarly calculated as in the case for the GUE, and the matrix elements with respect to the Kramers pairs vanish because of the symmetry (see the main text).
For the variance of the off-diagonal matrix elements, we have
\aln{
 \av{|O_{\alpha\beta}|^2}
=&\frac{1}{d(d+1)}\sum_i a_i^2 -\frac{1}{d(d+1)(d-2)}\sum_{a_i\neq a_j}a_ia_j \NON
=&\frac{1}{(d+1)(d-2)}\sum_i a_i^2-\frac{1}{d(d+1)(d-2)}\lrs{\sum_i a_i}^2\NON
=&\frac{d}{(d+1)(d-2)}\lrl{\frac{1}{d}\sum_i a_i^2-\lrs{\frac{1}{d}\sum_i a_i}^2},
}

Next, consider odd observables $\hat{O}\:(\{\hat{O},\hat{T}\}=0)$.
The average for diagonal matrix elements is
\aln{
\av{\braket{E_\alpha|\hat{{O}}|E_\alpha}}=\sum_{a_{i'}>0}a_{i'}(\av{|\braket{E_\alpha|a_{i'}}|^2}-\av{|\braket{E_\alpha|\tilde{a_{i'}}}|^2})
=0.
}
For the variance, we have
\aln{
 \av{|O_{\alpha\alpha}|^2}=&
\sum_{a_{i'},a_{j'}>0}a_{i'}a_{j'} (
\av{|\braket{ E_\alpha|a_{i'}}|^2|\braket{ E_\alpha|a_{j'} }|^2}+\av{|\braket{E_\alpha |\tilde{a_{i'}} }|^2|\braket{E_\alpha |\tilde{a_{j'}} }|^2}\NON
&\:\:\:\:\:\:\:\:\:\:\:\:\:\:\:\:-\av{|\braket{ E_\alpha|\tilde{a_{i'}} }|^2|\braket{E_\alpha | a_{j'}}|^2}-\av{|\braket{E_\alpha |a_{i'} }|^2|\braket{ E_\alpha| \tilde{a_{j'}}}|^2}
)\NON
=&\frac{2}{d(d+1)}\sum_{a_{i'}>0} a_{i'}^2=\frac{1}{d(d+1)}\sum_{i} a_i^2.
}
For the matrix elements $\braket{E_\alpha|\hat{{O}}|\tilde{E_\alpha}}$, we have
the average $\av{\braket{E_\alpha|\hat{O}|\tilde{E_\alpha}}}=0$.
The variance can be obtained as
\aln{
 \av{|O_{\alpha\tilde{\alpha}}|^2}=&
\av{\lrv{2\sum_{a_{i'}>0}a_i\braket{E_\alpha|a_{i'}}\braket{a_{i'}|\tilde{E_\alpha}}}^2}\NON
&=4\sum_{a_{i'},a_{j'}>0}a_{i'}a_{j'}\av{\braket{E_\alpha|a_{i'}}\braket{a_{i'}|\tilde{E_\alpha}}\braket{\tilde{E_\alpha}|a_{j'}}\braket{a_{j'}|E_\alpha}}\NON
=&\frac{4}{d(d+1)}\sum_{a_{i'}>0}a_{i'}^2\NON
=&\frac{2}{d(d+1)}\sum_{i}a_i^2.
}
Finally, for the off-diagonal matrix elements with respect to the eigenstates with different energy, we can find that the average is zero and that the variance is
\aln{
 \av{|O_{\alpha\beta}|^2}=& \sum_{a_{i'},a_{j'}>0}a_{i'}a_{j'}(
\av{\braket{E_\alpha|a_{i'}}\braket{a_{i'}|E_\beta}\braket{E_\beta|a_{j'}}\braket{a_{j'}|E_\alpha}}
-\av{\braket{E_\alpha|a_{i'}}\braket{a_{i'}|E_\beta}\braket{E_\beta|\tilde{a_{j'}}}\braket{\tilde{a_{j'}}|E_\alpha}}\NON
&-\av{\braket{E_\alpha|\tilde{a_{i'}}}\braket{\tilde{a_{i'}}|E_\beta}\braket{E_\beta|a_{j'}}\braket{a_{j'}|E_\alpha}}
+\av{\braket{E_\alpha|\tilde{a_{i'}}}\braket{\tilde{a_{i'}}|E_\beta}\braket{E_\beta|\tilde{a_{j'}}}\braket{\tilde{a_{j'}}|E_\alpha}}
)\NON
=&\sum_{a_{i'}>0}a_{i'}^2\frac{2}{d(d+1)}\NON
=&\sum_{i}a_i^2\frac{1}{d(d+1)}.
}

\section{Justification of $\sigma^2\simeq \mc{V}/d$ in the RMT model (Subsection \ref{sec:distd})}\label{sec:bunsan}
From \EQ{ohayo}, we have
\aln{
\frac{\mc{V}}{d}&=\frac{1}{d}\lrl{\frac{1}{d}\sum_ia_i^2-\lrs{\frac{1}{d}\sum_ia_i}^2}(1+\mr{O}(d^{-1}))\NON
&=\av{|O_{\alpha\beta}|^2}\times (1+\mr{O}(d^{-1})).
}
Then, if we assume the ergodicity of random matrices (see Appendix \ref{sec:ergo}), we can replace $\av{|O_{\alpha\beta}|^2}$ with the spectral average.
Thus, we have $\frac{\mc{V}}{d}\simeq \sigma^2$.

\section{Occupation ratios of each symmetry sector in \SEC{quench}}\label{sec:proj}
In this section, we calculate the occupation ratios $p_{\mathbf{q}}$ in Eq. (\ref{opn}) for the 1/3-filling.
We first note that the relation
\begin{align}
\ket{\psi}=\sum_{q_1,\dots,q_l=\pm 1}\left\{\left[\frac{1}{2^L}\prod_{l=1}^L(1+q_{l}\hat{P}_l)\right]\ket{\psi}\right\}
\end{align}
holds true.
We then note that the state in the curly brackets on the right-hand side is an eigenstate of the symmetry operators $(\hat{P}_1, ...,\hat{P}_L)$ with the eigenvalues $(q_1,...,q_L)$.
Thus the normalized projection operator onto $\mbf{q}$ is written as $\hat{\mathcal{P}}_{\mathbf{q}}=\frac{1}{2^L}\prod_{l=1}^L(1+q_{l}\hat{P}_l)$.
Since $\hat{P_l}$ is an operator that swaps two sites on the $l$-th layer, we obtain $\braket{\psi_0^A|\hat{{P}}_{l_1}\hat{{P}}_{l_2}\dots|\psi_0^A}=0$ in Case A and $\braket{\psi_0^B|\hat{{P}}_{l_1}\hat{{P}}_{l_2}\dots|\psi_0^B}=1$ in Case B $(l_1< l_2<\dots)$.
Expanding $\hat{\mathcal{P}}_\mathbf{q}$ and using the above results, we obtain $p_\mathbf{q}=\frac{1}{2^L}\prod_{l=1}^L(1+0)$ for Case A and $p_\mathbf{q}=\frac{1}{2^L}\prod_{l=1}^L(1+q_{l})$ for Case B. Therefore, we obtain
\begin{align}
&\text{(Case A)}\:\:\:p_{\mathbf{q}}=\frac{1}{2^L}\:\:\:(\mathbf{q}=(q_1,...,q_L)), \\
&\text{(Case B)}\:\:\:p_{\mathbf{q}}=
\begin{cases}\label{forB}
1 & (\mathbf{q}_1=(+1,...,+1)), \\ 
0 & (\text{otherwise)}.
\end{cases}
\end{align}
We note that the result for Case B is the same as in Eq. (\ref{forB}) even for the case of the 1/6-filling.

\chapter{Miscellaneous topics}\label{ch:Mis}


\section{Tasaki's MATE}\label{sec:tmate}
It is often difficult in practice to construct the set of mutually commuting observables $\lrm{\hat{M}_1,\cdots,\hat{M}_K}$ from $\lrm{\hat{M}'_1,\cdots,\hat{M}'_K}$.
Tasaki~\cite{Tasaki16} avoids the step of making the observables commute and defines the equilibrium subspace in a bit different manner (his definition is called TMATE~\cite{Goldstein15}).
First, the microcanonical equilibrium values of the original observables $\lrm{\hat{M}'_1,\cdots,\hat{M}'_K}$ are defined as 
\aln{
V_j=\mr{Tr} [\hrho_\mr{mic}\hat{M}'_j]\:\:(1\leq j\leq K).
}
We define the following projection operator for each observable
\aln{
\hat{\mc{P}}^T_j:=\sum_{V_j-\Delta \mu'_j\leq \mu'_j \leq V_j+\Delta \mu'_j}\ket{\mu'_j}\bra{\mu'_j},
}
which is the projection associated with the eigenvalues $\mu'_j$ (the corresponding eigenvector is denoted as $\ket{\mu'_j}$) of $\hat{M}'_j$ that lie within some resolutions $\Delta \mu'_j$.
Then the equilibrium subspace can be defined as
\aln{
\mr{TMATE}=\bigcap_{j=1}^K\lrm{\ket{\psi}\in \mc{H}_\mr{mic} \:|\braket{\psi|\hat{\mc{P}}^T_j|\psi}\simeq 1}.
}
MATE and TMATE are similar to each other.

\section{The numerical verification of the ETH for many-body correlations in Chapter \ref{ch:Obs}}\label{sec:ethmany}
In this section, we numerically show that the ETH seems to hold true even for many-body correlations in \EQ{oiyo}.
For simplicity, we show the result of the ETH for diagonal matrix elements (we have also checked the ETH for off-diagonal matrix elements).

\begin{figure}
\begin{center}
\includegraphics[width=12cm]{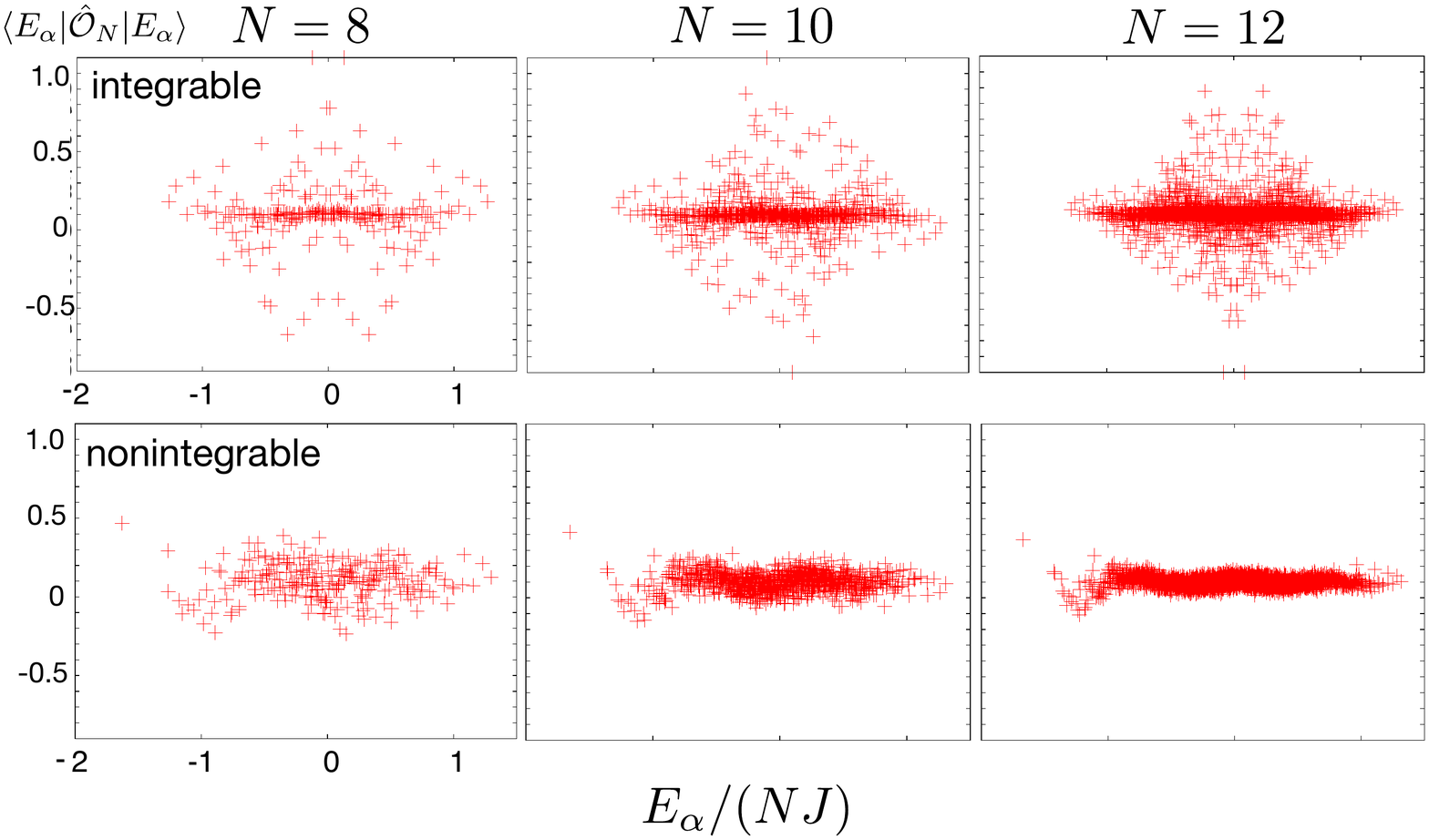}
\caption{
$N$-dependences of the EEVs for $\hmo_N$ in integrable (upper row) and nonintegrable (lower row) systems plotted for $N=8$ (left), $N=10$ (middle), and $N=12$ (right).
The fluctuations of the EEVs decrease with increasing $N$ for nonintegrable systems, whereas
they remain large in integrable systems.
}
\label{maneth}
\end{center}
\end{figure}

In Fig. \ref{maneth}, we show the eigenstate expectation values (EEVs) $\braket{E_\alpha|\hat{\mc{O}}_N|E_\alpha}$ for the many-body correlations $\hat{\mc{O}}_N$ (i.e., we take $l=N$ in \EQ{oiyo}) in integrable and nonintegrable systems.
For an integrable system, we take a disorder-free transverse-field Ising model with the open boundary condition whose Hamiltonian can be written as
\aln{
\hH=-\sum_{i=1}^{N-1}J\hat{\sigma}_i^z\hat{\sigma}_{i+1}^z-\sum_{i=1}^Nh'\hat{\sigma}_i^x,
}
where we take $J=1$ and $h'=-1.05$.
For a nonintegrable system, we take model (b) defined in Chapter \ref{ch:Obs}.
Figure \ref{maneth} shows that the fluctuations of the EEVs rapidly decrease with $N$ for nonintegrable systems, whereas
they remain large in integrable systems.\footnote{We note that for the integrable model, the EEVs have  certain symmetric structures. Namely, we obtain $\braket{E_\alpha|\hat{\mc{O}}_N|E_\alpha}=\braket{-E_\alpha|\hat{\mc{O}}_N|-E_\alpha}$ for $N=8,12$ and $\braket{E_\alpha|\hat{\mc{O}}_N|E_\alpha}=-\braket{-E_\alpha|\hat{\mc{O}}_N|-E_\alpha}$ for $N=10$. This symmetry is due to the chiral symmetry operator $\hat{C}$ that transforms the Pauli operators as $\hat{\sigma}_i^x\ra -\hat{\sigma}_i^x, \hat{\sigma}_i^y\ra \hat{\sigma}_i^y$ and $\hat{\sigma}_i^z\ra (-1)^i\hat{\sigma}_i^z$. Since $\{\hH, \hat{C}\}=0$, we have a pair of eigenstates $\ket{E_\alpha}$ and $\ket{-E_\alpha}=\hat{C}\ket{E_\alpha}$, where $\hH\ket{-E_\alpha}=-E_\alpha\ket{-E_\alpha}$ is satisfied. This symmetry leads to the condition 
\aln{
\braket{E_\alpha|\hat{\mc{O}}_N|E_\alpha}=(-1)^{N/2}\braket{-E_\alpha|\hat{\mc{O}}_N|-E_\alpha},
}
which explains the numerical results.
}
This result implies that the ETH does and does not hold true in nonintegrable and integrable systems, respectively, even for many-body correlations.
We note that we obtain similar results for other values of $l$.\footnote{We have shown the scaling for $\hmo_N$ with respect to $N$. We can make another scaling, where only the system size changes with a fixed observable $\hmo_l$. In any case, we find that the EEV fluctuations decrease with increasing the system size $N$ for nonintegrable systems.}

\section{The ETH for the models (b) and (c) in \SEC{sec:otm}}\label{sec:gethom}
\begin{figure}
\begin{center}
\includegraphics[width=16cm,angle=-90]{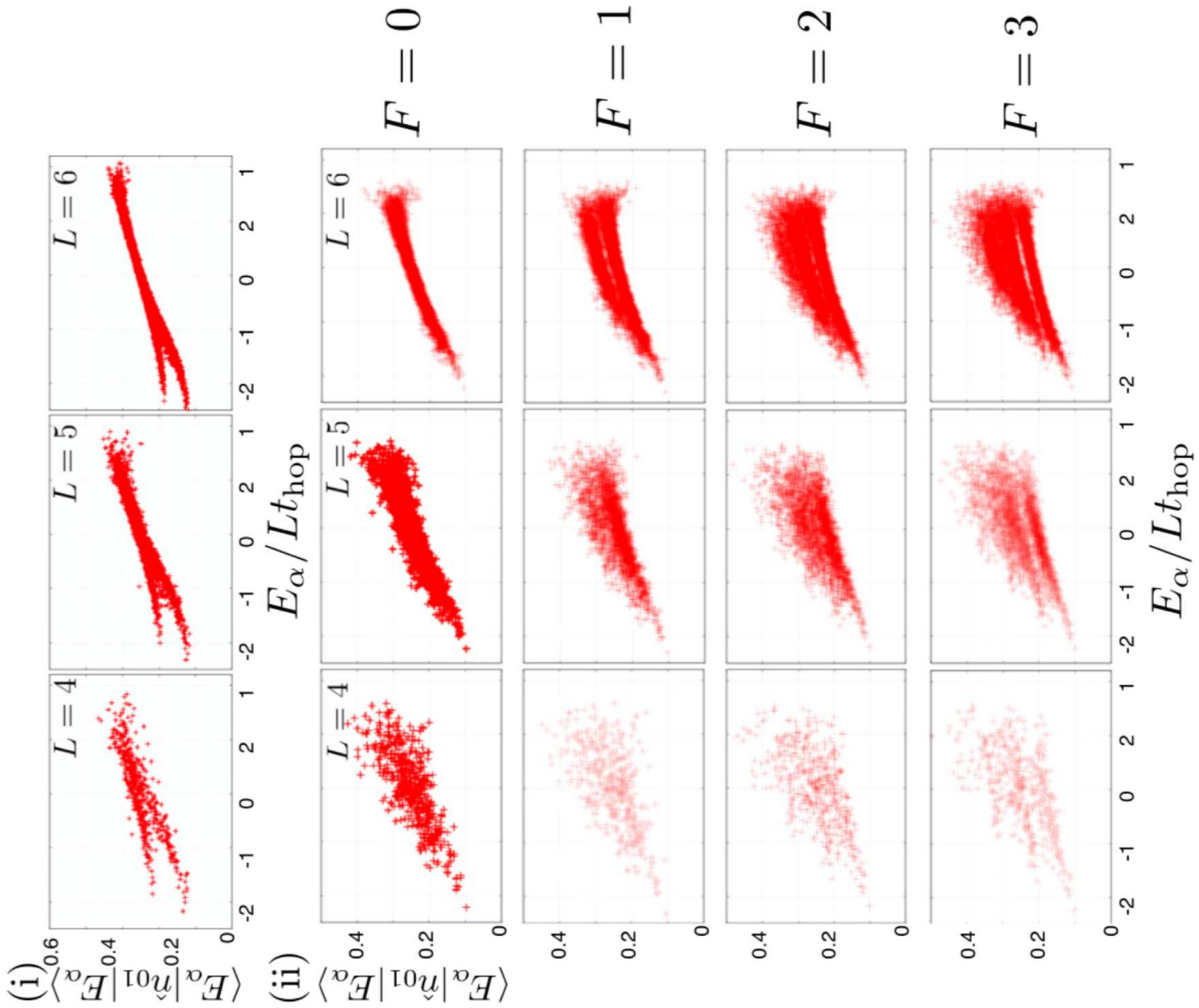}
\caption{(i) $L$-dependence of the EEVs for $\hat{n}_{01}$ in model (b).
Although the EEVs are split into two branches, the splitting moves to the edge of the spectrum with increasing the system size.
Therefore, the EEV fluctuations decrease with increasing $L$ in the middle of the spectrum.
(ii) $L$-dependence of the EEVs for $\hat{n}_{01}$ in model (c).
The number of the local symmetries with $F=0,1,2,3$ increases from the top to the bottom.
The EEV fluctuations decrease with increasing the system size $L$, but rather slowly for the large values of $F$. 
Reproduced from Fig. 10 of~\cite{Hamazaki16G}. \copyright 2016 American Physical Society.
}
\label{ethb}
\end{center}
\end{figure}

\begin{figure}
\begin{center}
\includegraphics[width=11cm,angle=-90]{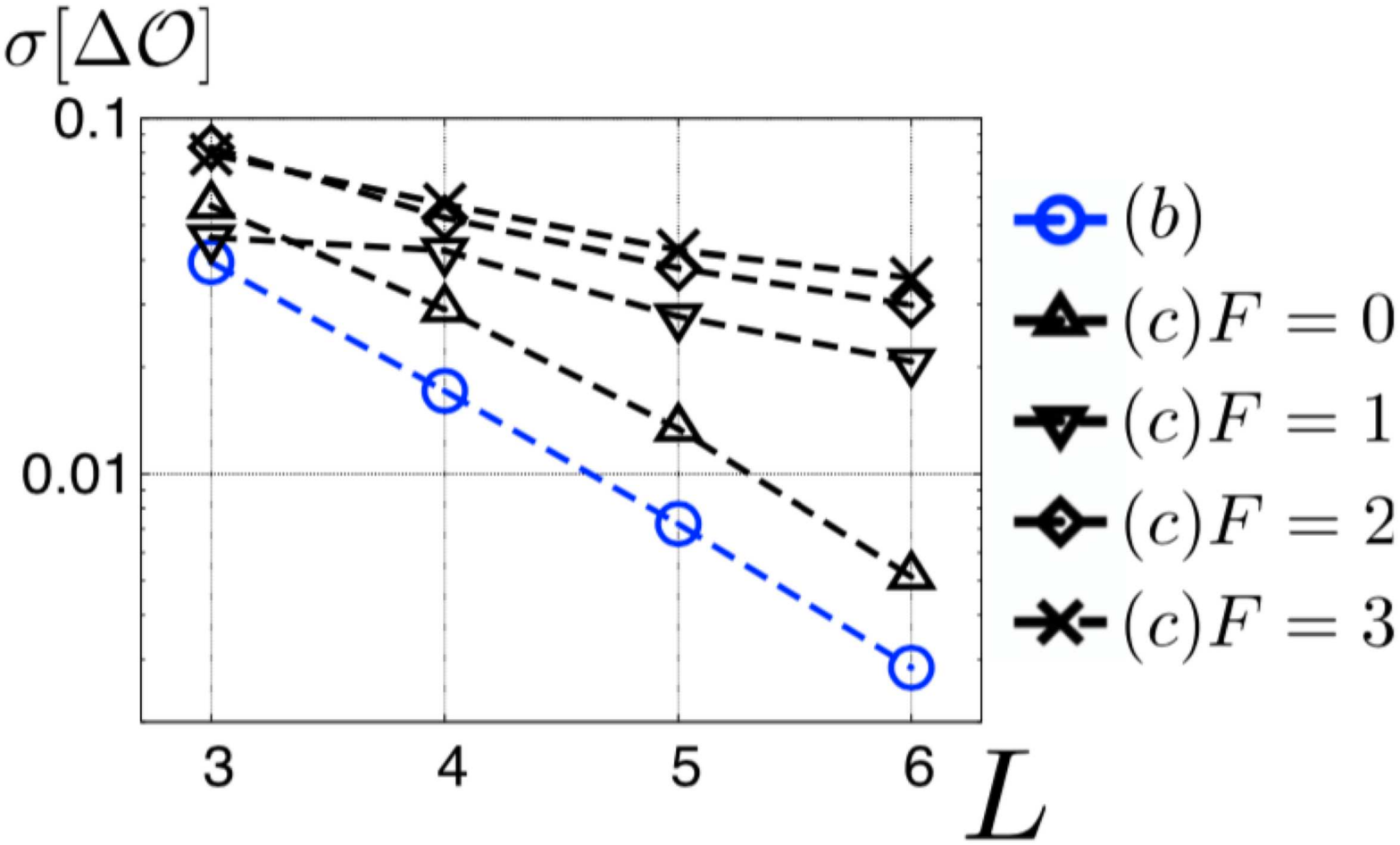}
\caption{$L$-dependence of the standard deviation $\sigma[\Delta \mathcal{O}]$ of $\braket{E_{\alpha}|\hat{\mathcal{O}}|E_{\alpha}}$ within the small energy shell $[E-\omega_s,E+\omega_s]$, where we take $\omega_s=0.18L$ and $E=0$. 
We show $\sigma[\Delta \mathcal{O}]$ in the models (b) (circle), (c) $F=0$ (upward triangle), 1 (downward triangle), 2 (diamond), and 3 (cross).
While $\sigma[\Delta \mathcal{O}]$ decreases with the system size $L$ for all values of $F$, its $L$-dependence is much less sensitive for $F\geq 1$ than for $F=0$.
Reproduced from Fig. 11 of~\cite{Hamazaki16G}. \copyright 2016 American Physical Society.
}
\label{otheryuragi}
\end{center}
\end{figure}

In Figs. \ref{ethb} (i) and (ii), we show the EEVs for $\hat{n}_{01}$ in the models (b) and (c), respectively.
In Fig. \ref{otheryuragi}, we also show $\sigma[\Delta \mathcal{O}]$, a typical magnitude of the EEV fluctuations $\Delta \mathcal{O}_\alpha$ in the middle of the spectrum.
We define  $\sigma[\Delta \mathcal{O}]$ as the standard deviation of $\braket{E_{\alpha}|\hat{\mathcal{O}}|E_{\alpha}}$ within a small energy shell $[E-\omega_s,E+\omega_s]$.\footnote{We note that this is equivalent to $\Delta\mc{O}_\mr{d}$ in Chapter \ref{ch:Obs}.}
In Fig. \ref{ethb} (i), while the splittings of the EEVs are seen due to the global $\mathbb{Z}_2$ symmetry, this splitting seems to move to the edge of the spectrum with increasing the system size.
Therefore, the ETH is expected to hold true in the thermodynamic limit especially in the middle of the spectrum (see Fig. \ref{otheryuragi}).
Next, Figs. \ref{ethb} (ii) and \ref{otheryuragi} show that even though $\Delta \mathcal{O}_\alpha$ and $\sigma[\Delta \mathcal{O}]$ decrease with increasing the system size, their $L$-dependences are weaker for $F\geq 1$ than for $F=0$.
This result is consistent with the behavior of the relative difference in model (c): the $L$-dependence is much less sensitive for $F\geq 1$ than for $F=0$.

%
\addcontentsline{toc}{chapter}{Acknowledgements}
\chapter*{Acknowledgements}
First of all, I would like to express my deepest gratitude to my supervisor, Professor Masahito Ueda.
He is both an extraordinary physicist and an exceptional teacher.
Every discussion with him was full of thoughtful criticism and encouragement: 
for leading me to seek for the theme that has an impact on the development of physics, he has patiently tried to clarify my clumsy ideas without ever denying them and given me insightful advices.
He has also spared a lot of time for reading the manuscript of this thesis and giving me enormous comments.
I also thank him for recommending me the main subject of this thesis: thermalization in isolated quantum systems.

I am also deeply grateful to my collaborator, Assistant Prof. Tatsuhiko N. Ikeda.
Even though he had been writing his own dissertation, he was willing to  discuss the GGE in nonintegrable systems reviewed in Chapter \ref{ch:Gn} with me.
He taught me a lot of things without which this thesis would be far from completed: the basics and advancement of thermalization in isolated quantum systems, technical methods for calculations, the computer programming, and how to write a paper and give a presentation. 
Most of all, his attitude toward research has impressed me a lot.

I have been fortunate to ask questions and have discussions with many others.
Assistant Prof. Shunsuke Furukawa in our group has taught me many things from the computer programming to condensed matter physics.
Whenever I ask him (often stupid) questions, he kindly taught me in a clear manner.
We would also like to thank great researchers that tackle the problem of dynamics in quantum many-body systems, especially Assistant Prof. Takashi Mori, Dr. Sho Sugiura, Dr. Kazuya Fujimoto,  Mr. Yuto Ashida, Ms. Mamiko Tatsuta, and Mr. Zongping Gong.
Discussions with them have always been helpful and stimulating.
Moreover, their intriguing works have motivated me to become a better physicist.
We also thank all the members in Masahito Ueda Group for providing me an exciting and enjoyable environment for doing my research.

Finally, I acknowledge the financial support and lectures through the Program for Leading Graduate Schools (ALPS).

%

\bibliography{refer_them}

\end{document}